\documentclass[reprint,rmp,aps,longbibliography]{revtex4-1}

\usepackage[usenames,dvipsnames]{color}
\usepackage[colorlinks=true, citecolor=blue, linkcolor=Plum, urlcolor=RedOrange]{hyperref}
\usepackage{graphicx, subfigure}
\usepackage{amsmath, amsfonts, amssymb}
\usepackage[format=plain, justification=centerlast, small, bf]{caption}

\begin{document}

\newcommand{\R}[1]{\textcolor{red}{#1}}

\title{Gravitational Radiation Detection with Laser Interferometry}
\author{Rana X Adhikari}
\email{rana@caltech.edu}
\affiliation{Division of Physics, Math, and Astronomy, \\
                California Institute of Technology, \\
                Pasadena, CA 91125, USA}
\preprint{LIGO-P1200121-v1}
%\date{\today}

\begin{abstract}
Gravitational-wave detection has been pursued relentlessly for over 40 years. With the imminent
operation of a new generation of laser interferometers, it is expected that detections
will become a common occurrence. The research into more ambitious detectors promises
to allow the field to move beyond detection and into the realm of precision science
using gravitational radiation. In this article, I review the state of the art for the detectors and
describe an outlook for the coming decades. 
\end{abstract}
\pacs{04.30.-w, 04.80.Nn, 95.55.Ym, 07.60.Ly, 42.62.Eh, 07.05.Dz, 07.05.Mh, 91.30.-f}

\maketitle
\tableofcontents

%   #####    what is left to fix   ##############

%  5) conclusion

% comments from
%  Giles  /
%  Saulson /
%  Haixing /
%  Julien /
%  Harald /
%  Gabriele /
%  DHS /
%  Alastair /
%  DMC /

% --- make arm cavity cartoons have red light

% estimate of h - dimensional anal

% FP - give formula of how to optimize Finesse & length

%  1)  check SUS k_el v. the Saulson 1990 formula

% fix up shit in the Pulsar and Sat timing section

%=====================================================
% --------------------  during LSC review

% --  GW Spectrum plot

%  --- NN: bugs, tumbleweed, new baluns
%              is there an older reference than QPR? Ask Rai.

%  ---  Metrology: add micro-roughness images, Cheryl's new picture

%  ---  substrate / coating thermal noise section: 
%                add a couple sentences about phi_para

% ---  Electric fields: expand or delete

%  ---  cryo SUS
%        - add a plot showing cooling power as a function of temperature for radiation
%        - and also the cooling power of fibers v. T for a few different diameters assuming
%          that the length is 1m. Also include sapphire and diamond?        

%  --- match fonts for all plots
% ---  make plots show up in the right places

%  --- URLS in refs

% ===============================================================
\section{Introduction}
\label{sec:intro}
Nearly a century ago, Einstein predicted the existence of gravitational radiation as a
consequence of his General Theory of 
Relativity (GR)~\cite{Einstein:1916b, einstein1997collected, Einstein:1918a}. 
For the next
several decades, the existence and properties of gravitational radiation were hotly contested
within the theoretical community but remained out of observational reach.
In 1974, Hulse and Taylor~\cite{TFM1979} discovered
a pulsar in a binary neutron star system. They soon realized that this system serves 
as an excellent laboratory to test GR. The decrease in the orbital energy of the binary system 
was found to match the theoretical predictions.
During the following decades several other binary pulsars with orbital periods of less than a day
have been discovered and the combined data show that the measured
energy loss matches exquisitely well with the calculated loss due to the emission of 
gravitational radiation. Gravitational waves are real.

Modern efforts to detect gravitational radiation on the Earth focus on the use of laser 
interferometry~\cite{Rai:QPR}. Lab scale
research throughout the last several decades of the 20$^{\rm th}$ century 
led to the construction of a worldwide network of kilometer scale 
interferometers~\cite{LIGO:Science, PF:RPP2009, LuEA2006, Tat2008, VIR2008}.

Several excellent 
monographs~\cite{Saulson:book, Rai:RMP, BaWe:PhysToday, lrr-2011-5, 
Aufmuth:2005vv, Cella:2011, Braginsky:2008ea, Giazotto:1989dr, lrr-2010-1}
have been written on the techniques of gravitational wave detection 
by laser interferometry.
In this review, we will discuss the current state of gravitational wave detectors, 
describing in detail the fundamental limits to their astrophysical reach, and then 
present prospects for the future.

% ===============================================================
\section{Gravitational Waves}
\label{sec:GW}
In the weak-field approximation of General Relativity, the space-time metric, $g_{\mu \nu}$, 
can be described as~\cite{MTW1973}
\begin{equation}
g_{\mu \nu} \simeq \eta_{\mu \nu} + h_{\mu \nu}
\label{eq:gmunu}
\end{equation}
where $\eta_{\mu\nu}$ is the Minkowski metric describing flat space and $h_{\mu\nu}$ is the
perturbation to the metric due to the gravitational wave. In the transverse-traceless gauge,
this can be understood as a \textit{strain} in space-time:
\begin{equation}
h_{\mu \nu}(z,t) = 
  \begin{pmatrix}
     0    & 0            &  0             & 0 \\
     0    & -h_{+}     &  h_{\times}  & 0 \\
     0    & h_{\times} &  h_{+}        & 0 \\
     0    & 0            &  0             & 0 
  \end{pmatrix} 
 \label{eq:hmatrix}
\end{equation}
where the two independent polarizations of the wave have amplitudes
 $h_{+}$ and $h_{\times}$, respectively.

\subsection{Response of Interferometer to Space-Time Strain}
In order to relate this perturbed metric with laboratory observables, we can examine
how some precision measurement apparatus will respond to such a strain.
To illustrate this we can set up two free masses, one located at
the origin and one located a distance, $x = L$, from the origin. We can measure the
separation between these two masses by sending a laser beam from the origin
to bounce off of the far mass and measure the phase of the return beam relative to the source.
The accumulated round trip phase is:
\begin{equation}
\Phi_{rt}(t_{rt}) = \int\limits_{0}^{t_{rt}} 2 \pi \nu \, dt
     \label{eq:roundtripflat}
\end{equation}
where $t_{rt}$ is the time it takes for the light to make one round trip and
$\nu$ is the frequency of the light. In the absence of gravitational radiation, we can do the
integral by changing it into an integral over length. To do this we use the 
flat space metric, $\eta_{\mu \nu}$, to relate space and time for light
($t_{rt} = 2 L/c$ and $dt = dx / c$). 

In the presence of a gravitational wave, we instead use Eq.~\ref{eq:gmunu} to calculate
the space-time interval; the perturbed round trip phase is
\begin{equation}
\Phi_{rt}(t_{rt}) = 2 \frac{2 \pi \nu}{c} \int\limits_{0}^{L} \sqrt{|g_{xx}|} \, dx
                    \simeq 2 (1 - h_{+}/2) \frac{2 \pi L}{\lambda}
         \label{eq:roundtrip}
\end{equation}
in the case of a ''plus'' oriented wave with a period much longer than the round trip
light travel time. Repeating this integral, but doing the
integration now along the y-axis, we get that 
$\Phi_{rt} \simeq 2 (1 + h_{+}/2) (2 \pi L / \lambda)$. The difference
in the phase shift between the two arms is then
$\Delta \Phi \simeq 2 h_{+} (2 \pi L / \lambda)$. 

\begin{figure}[h]
   \centering
   \includegraphics[width=\columnwidth]{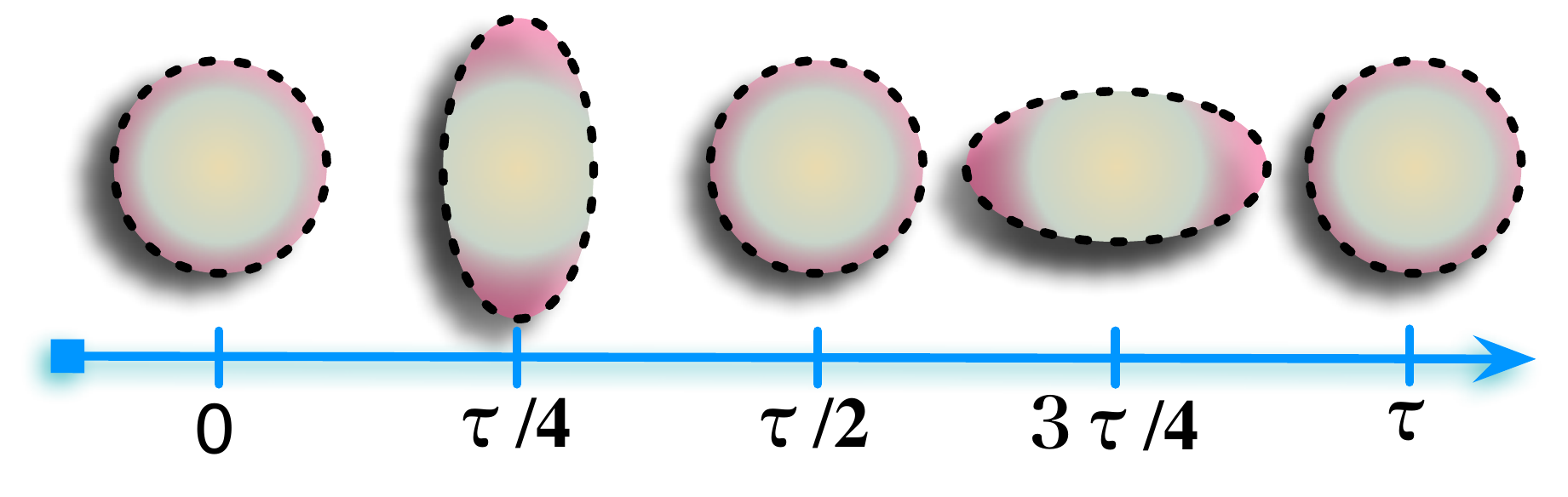} 
   \caption{(Color online) Exaggerated example of the effect of a GW on a ring of test particles. The GW is
                  coming from above, is 'plus' polarized, and has a period $\tau$. As the wave
                  passes, the ring is alternately stretched and compressed. This quadrupolar strain
                  pattern matches well to the geometry of a Michelson interferometer.}
   \label{fig:CropCircles}
\end{figure}

Interpreting the phase shifts as length variations means that the apparent 
length of each arm is stretched and compressed as the gravitational wave passes. 
A diagram of this is shown in Fig.~\ref{fig:CropCircles}. The
length change is proportional to the original distance between the masses,
\begin{equation}
\frac{\Delta L}{L} = \frac{1}{2} h_+
    \label{eq:h}
\end{equation}
which is why a gravitational wave is said to cause a strain in space. In contrast,
the term 'gravity wave' is usually used to refer to waves in fluids
or solids where the restoring force is due to gravity.

The strain along the interferometer arms for a gravitational wave from an arbitrary direction
(in spherical coordinates centered on the detector) is~\cite{Nelson:PRD92}:
\begin{eqnarray}
h_{xx} &=& - \cos{\theta} \sin{2 \phi} \; h_{\times} + (\cos^2{\theta} \cos^2{\phi} - \sin^2{\phi}) h_{+} \\
h_{yy} &=&  \cos{\theta} \sin{2 \phi} \; h_{\times} + (\cos^2{\theta} \sin^2{\phi} - \cos^2{\phi}) h_{+}
\end{eqnarray}

\begin{figure*}[ht]
        \subfigure{
            \label{fig:hplus}
            \includegraphics[width=0.7\columnwidth]{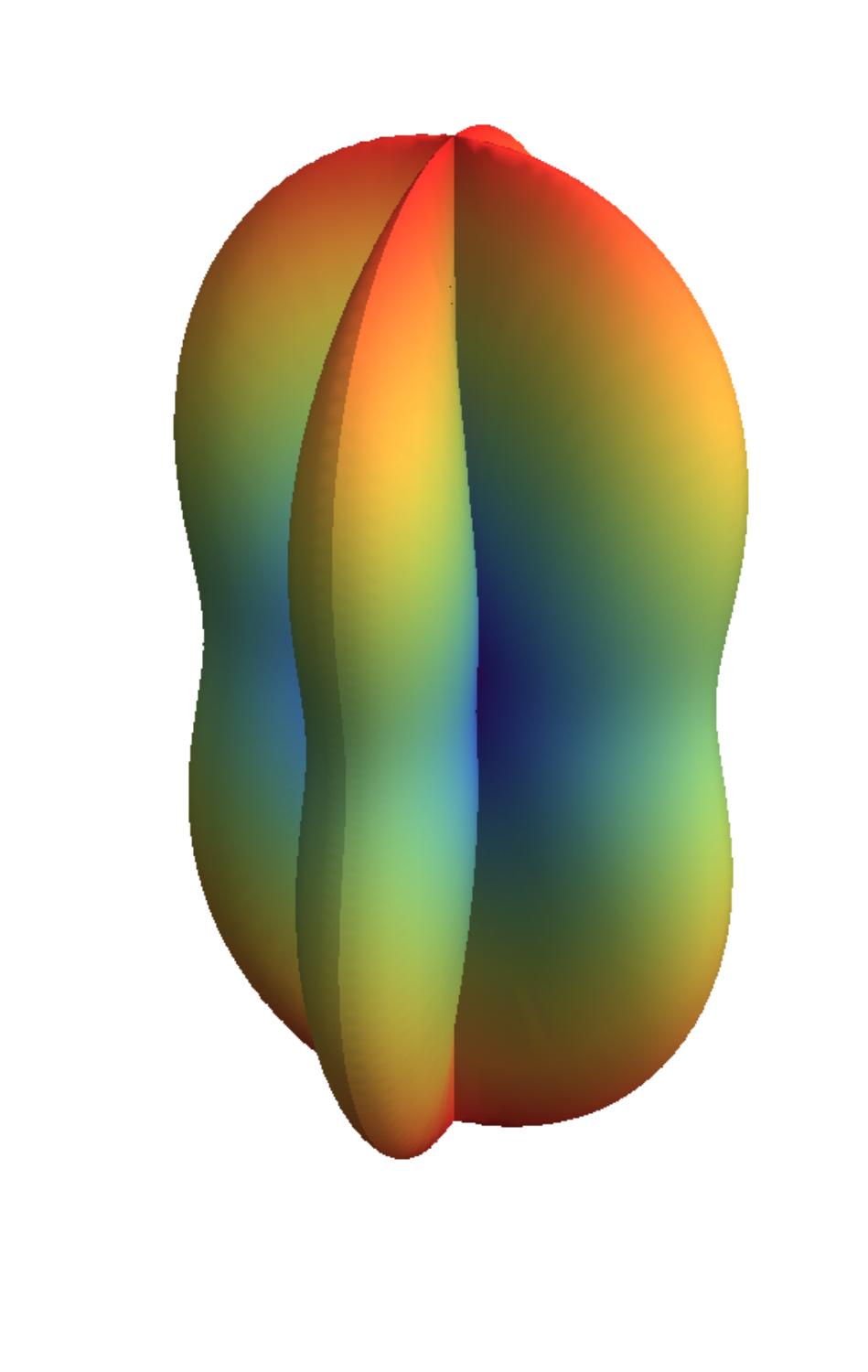}
           } 
       \subfigure{
            \label{fig:hcross}
            \includegraphics[width=0.52\columnwidth]{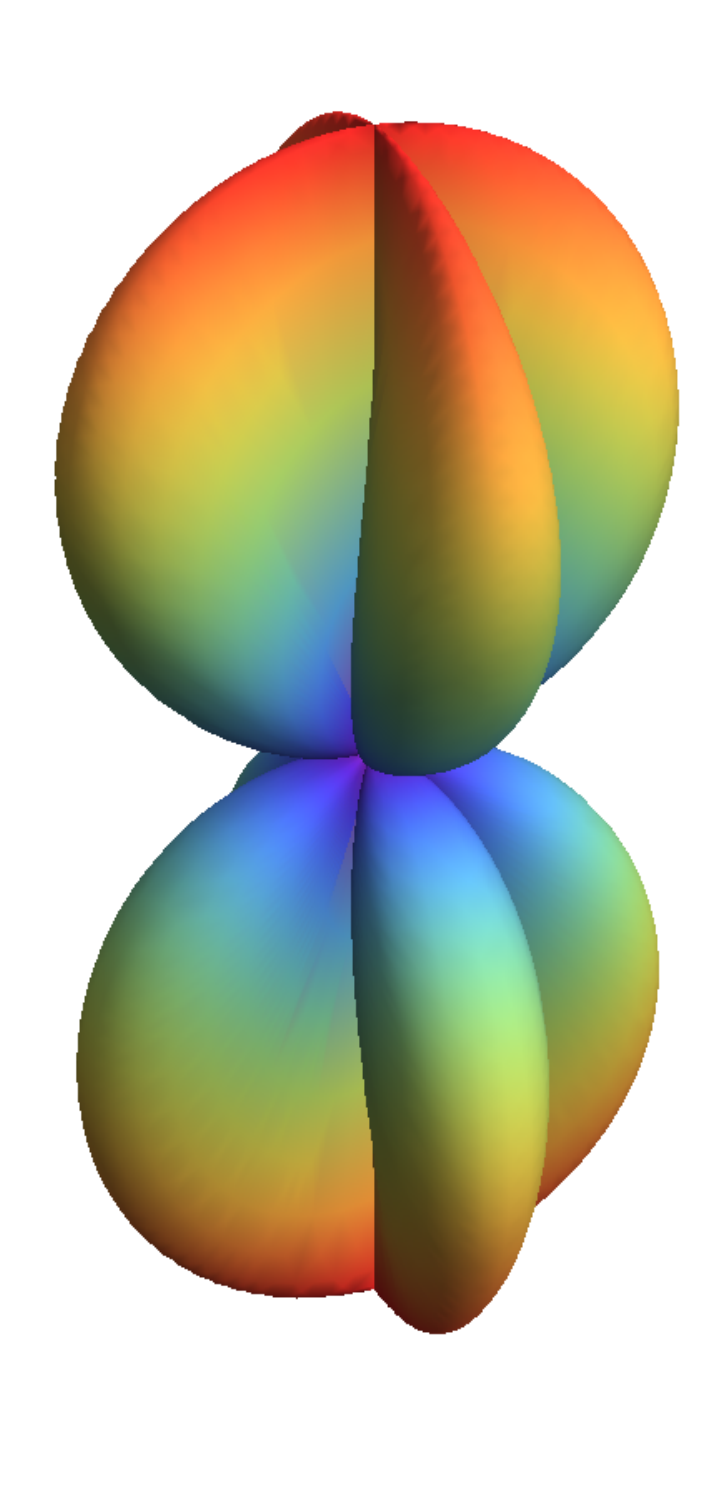}
           } 
       \subfigure{
            \label{fig:hpeanut}
            \includegraphics[width=0.7\columnwidth]{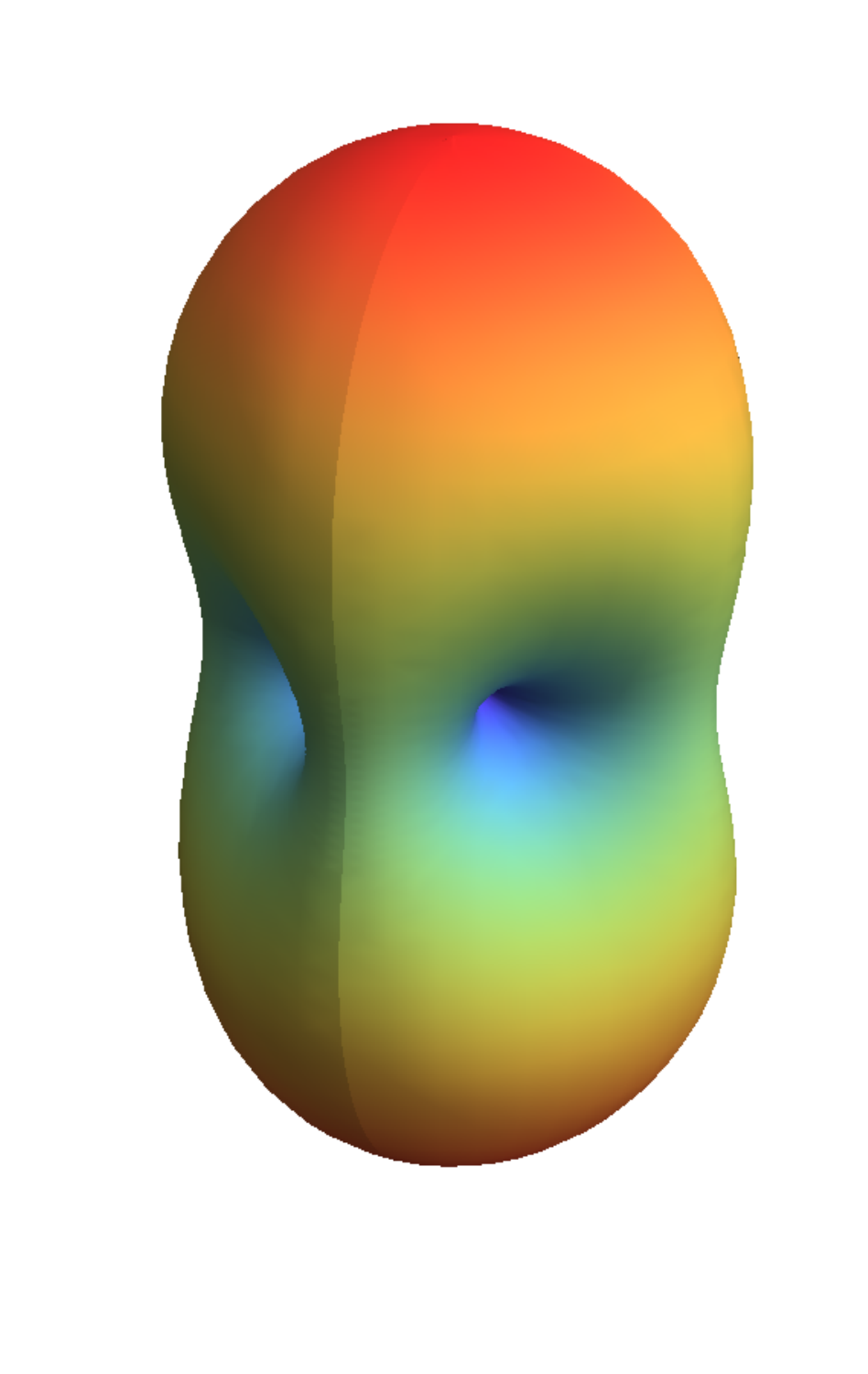}
           }
   \caption{(Color online) Interferometer Antenna Response for (+) polarization [left], 
                 ($\times$) polarization [middle], and unpolarized waves [right]. Color indicates
                 increasing sensitivity from indigo to red.}
   \label{fig:peanuts}
\end{figure*}

The interferometer response in the low frequency approximation (time scales much longer
than the one way light travel time) is proportional to 
$|h_{yy}-h_{xx}|$. Fig.~\ref{fig:peanuts} shows this DC 
response for $+$ waves, for $\times$ waves, and for unpolarized
waves (a quadrature sum of the two cases). In the coordinate system used in these
plots, the interferometer is located at the origin with the arms parallel to the
x and y axes.

\subsection{Brief Overview of Sources}
\label{sec:sources}
All terrestrial detectors of gravitational waves are focused roughly on the audio
frequency band due to technological limits of the detectors and probable source
characteristics. In order to verify all of the properties of the waves, one would like
to follow in the footsteps of Heinrich Hertz by generating and then detecting the
gravitational waves. However, due to the relatively high rigidity of spacetime, 
it is not feasible to generate measurable amounts of gravitational radiation in the 
lab~\cite{Romero:1981} by conventional means or even through the use of nuclear explosives
arranged to produce quadrupolar mass-energy accelerations~\cite{Nukes:1974}. 
Therefore, we look to astrophysical and
cosmological sources to provide the radiation. In this way, the hunt for gravitational
radiation leads to the development of a new branch of astronomy. Previous 
overviews~\cite{CuTh2002, 300:years}  have covered the list of known sources as well as describing
the astrophysical and cosmological science that can be extracted from 
them~\cite{Sathya:LRR}.

\subsubsection{Pulsars}
\label{sec:pulsars}
One of the earliest predicted sources of gravitational radiation were the recently discovered 
pulsars~\cite{Bell:1968}. The extremely stable period of pulsation of these rotating neutron
stars tells us that the energy lost to gravitational radiation must be small~\cite{Ipser:1971} at best. 
The compensating factor that makes detection a possibility is the periodic nature of the signal; 
after correcting for the Doppler modulations from the detector motions relative to the 
source~\cite{S5:Pulsars, BCCS1998},
one can improve the signal-to-noise ratio by the square root of the integration time. 

Observations~\cite{Deepto:2003} of a 'speed limit' for pulsars seem to support the 
theory~\cite{Bildsten:1998} that gravitational radiation works to brake the spin of the fastest pulsars 
before they are ripped apart by their relativistic spins. Expectations
from neutron star models indicate that the ellipticity may range 
from $10^{-9} - 10^{-6}$~\cite{Bildsten:2000,Owen:2006}
for conventional neutron stars and somewhat larger for more exotic stars~\cite{Owen:2005}.

In order to greatly improve the sensitivity of the pulsar searches, the Einstein@Home~\cite{EatHweb}
project distributes some of the LIGO data to the home computers of an international team of
volunteers. Although no gravitational waves have been detected so far, this project has detected
pulsars using electromagnetic astronomical data~\cite{EatH:2011}.

\subsubsection{Transients}
\label{sec:CBC}
The signal which all ground based detectors are aimed towards is the inspiral and merger
of compact binary objects: neutron stars (NS) and black holes (BH). Perhaps 
1/3 to 1/2 of the stars in the universe have companions~\cite{Lada:2006}. Through various
mechanisms, some small fraction of these can evolve into a NS/NS, NS/BH, or BH/BH binary
(white dwarfs are not quite so compact; mass transfer between the stars 
begins~\cite{Lobo:WD2005, FaPh2003} well before
the inspiral signal enters the accessible band of 
the ground based detectors). These compact binaries will eventually merge after they have
released their orbital energy through gravitational radiation. The Hulse-Taylor binary
is one such binary; it is expected to merge in $\sim3 \times 10^8$ years.
Estimates of the binary merger rates~\cite{Sterl:Minimal, BKB2002}
using bounds from astrophysical observations as well as predictions from population
synthesis models vary by a few orders of magnitude. For the upcoming second generation
interferometric detectors, the compact binary detection rate may be as low as 1/year or as
high as 3/day~\cite{LSC:rates}. A combination of extensive analytic methods~\cite{Bala:7half}
and high accuracy numerical simulations~\cite{ninja2, Scheel:2009, Bela:2009Sim} have allowed for 
the calculation of accurate
waveforms by which one can search for these binary inspirals using matched template
methods~\cite{FINDCHIRP}.

%\subsubsection{Unmodeled Transients}
%\label{sec:bursts}
It is most likely that the largest fraction of gravitational wave sources have not yet been
modeled well enough to use a template based search. These will include sources such
as stellar collapse
leading to supernovae~\cite{Ott:SN2009}, the boiling of the cooling neutron star at
the end of the collapse~\cite{Lee:NS2001}, and soft gamma-ray repeaters~\cite{LIGO:SGR}.
The most exciting prospect in making a broadband search for gravitational
waves is to make a discovery of an entirely unexpected astrophysical 
phenomenon~\cite{CuTh2002, GWHENreview}.

\subsubsection{Cosmic Background Radiation}
\label{sec:SGWB}
\cite{Staro:1979} and others~\cite{Wise:GW, Rubakov:1982} pointed out that a period of
cosmic expansion in the early universe could produce a spectrum of
gravitational radiation. \cite{Bruce:1988} later derived the full spectrum
of gravitational waves expected from a standard inflationary universe
scenario. This model predicts a nearly white spectrum (in units of energy)
in the frequency band from $10^{-15} - 10^{10}$\,Hz~\cite{Turner:1997cs}.
This radiation from the early universe would travel to our detectors with
very little scattering along the way giving us a direct measurement of
the state of the universe at a time which is less than $10^{-30}$\,s after the 
Big Bang~\cite{Weinberg:GW2003}.
A review of prospects for detecting this inflationary background
as well as possible astrophysical foregrounds is given in~\cite{All1996}.

There are two observational constraints on the cosmological background
of gravitational waves. The relative abundances of the light elements
in the universe today constrains tightly any deviations from the standard
model in Big Bang Nucleosynthesis (BBN)~\cite{Peebles:Cosmo}. An excess
of gravitational radiation at the time of BBN would change the expansion
rate of the universe. The BBN model places an upper limit
of $\sim\,10^{-5}$ (in units of the closure density of the universe) on the
energy in this primordial gravitational radiation.
Certain exotic theories of the early universe predict higher frequency
gravitational radiation~\cite{Vuk:PreBigBang, Woodard:HFGW}; for some of 
those models, a recent search using the LIGO detectors makes a slightly 
tighter bound~\cite{LIGO:SGWB2009} than from the BBN model.

% ===============================================================
\section{Alternatives to Interferometric Detection}
\subsection{Acoustic Detectors}
Attempts to make a direct detection of gravitational radiation started 50~years
ago with Joseph Weber~\cite{Weber:GW,Weber:Bars}. Weber's claims of
detection were never confirmed~\cite{Tyson:Nothing, Kafka:1978, Tyson:1982}; a review of these
confirmation efforts is given in~\cite{Giffard:1978}. 

Nevertheless, the excitement generated in the early 1970's led,
in the following years, to the development of an active worldwide network
of acoustic 'bar' detectors with an ever increasing astrophysical reach. By the end of the
20$^{\rm th}$ century, the bars had reached strain sensitivities of
$3 - 7 \times 10^{-19} $ for $\sim\,1$\,ms bursts~\cite{Blair:RPP2000}. A summary of the
sensitivity of these detectors is shown in Table~\ref{t:bars}.

\begin{table}
\begin{tabular}[t]{@{\extracolsep{\fill}}l  l  p{10mm}  p{15mm}  p{15mm}}
\hline
\cline{1-2}
Detector    & Location        & Freq. (Hz) & Peak Strain (h$_c$)& Strain Noise (h(f))\\
\hline
ALLEGRO   & LSU                & 900                  & $7 \times 10^{-19}$& $7 \times 10^{-19}$ \\
EXPLORER  & CERN             & 900                  & $7 \times 10^{-19}$& $7 \times 10^{-19}$ \\
NIOBE   & UWA                   & 700                  &  $5 \times 10^{-19}$& $7 \times 10^{-19}$ \\
NAUTILUS   & Frascati        & 900                  & $6 \times 10^{-19}$& $7 \times 10^{-19}$ \\
AURIGA   & Legnaro          & 900                  & $3 \times 10^{-19}$& $7 \times 10^{-19}$ \\
\hline
\end{tabular}

\caption{Best sensitivity of acoustic bar detectors~\cite{Blair:RPP2000}. Sensitivity is
               characterized by minimal detectable strain in the bar bandwidth (peak strain) and
               also the strain noise spectral density at the frequency of best sensitivity.}
\label{t:bars}
\end{table}

\subsection{Pulsar Timing}
\label{sec:PulsarTiming}
In the late 1970's, Sazhin~\cite{Sazhin} and Detweiler~\cite{Detweiler:PTA} pointed out that 
the regular pulse periods of radio pulsars
could be used to search for gravitational radiation in the 10\,--\,100\,nHz band. 
For the past three decades, astronomers have used the ever improving timing available for
radio antennas and the ever increasing number of known 
pulsars~\cite{Hellings:PTA, Lorimer:LRR, Larry:PulsarTiming2009, Yuri:PulsarTiming2009}
to search for a stochastic GW background of cosmological origin as well as the mergers of
massive black holes.

\subsection{Artificial Satellite Timing}
\label{sec:Satellites}
Doppler tracking of man made spacecraft was proposed as a means of detecting low frequency
gravitational waves in 1975~\cite{EsWa1975}. A carrier signal is sent to the spacecraft from
the earth, a transponder on the spacecraft sends the signal back, and the frequency of the incoming
and outgoing signals are compared. The relative fractional frequency fluctuations, $y_2$, due to GWs
can be written as~\cite{Wah1987,Armstrong:SGW}:
\begin{equation}
y_2[t] = -\frac{1-\mu}{2} \bar{\Psi}[t] - \mu \bar{\Psi}[t - \frac{1+\mu}{2} T_2]
            + \frac{1 + \mu}{2} \bar{\Psi}[t - T_2]
\end{equation}
where $\mu$ is the projection of the gravitational wave unit wavevector onto the earth-satellite
unit vector, and $\bar{\Psi}$ is a function encoding the response of the satellite signal's
response to the two polarizations of gravitational waves. 
The best sensitivity using this method was achieved~\cite{Armstrong:SGW} using the 2001--2002
data tracking the Cassini satellite. The strain noise in the 0.01--10\,mHz band ranged from
$10^{-13} - 10^{-12} /\sqrt{\rm Hz}$. Prospects for improving this sensitivity have been 
explored~\cite{Armstrong:LRR}; improved frequency standards, subtraction of plasma dispersion,
and reduction of mechanical vibration in the terrestrial antenna may lead to as much as an order of
magnitude improvement. Until a dedicated laser interferometer mission can be launched, satellite
tracking will remain the most sensitive probe of gravitational waves in this frequency 
band~\cite{Armstrong:2005}.

\subsection{Polarization of the Microwave Background}
At the largest spatial scales (timescales of order the age of the universe), gravitational waves can be
observed by measuring the polarization of the cosmic background radiation itself~\cite{Caldwell:1998dq}.
The largest polarization signals are produced by the cosmological density fluctuations and would seem
to swamp the small signals expected by gravitational waves. Hope is not lost, however. The gravitational
waves produce a polarization vector field with a curl, whereas the scalar density perturbations do not~\cite{Hu:1997vp}.
Finding this signal would provide an unambiguous signal of cosmic inflation. There are many sources
of foreground~\cite{CMBpol:foreground} contamination which must be removed in order to 
extract the gravitational wave signal. These removal techniques are being actively developed by the teams
pursuing polarization signals in WMAP~\cite{WMAP:9year} and Planck~\cite{Planck:2013} data as well as the 
numerous ground based experiments which have been specifically designed to hunt for the
polarization signal. Space-based mission concepts such as CMBpol~\cite{CMBpol:2009} and 
PRISM~\cite{PRISM:concept} would be the ultimate word in the detection of these gravitational waves.

% ===============================================================
\section{Fundamentals of Interferometric Detectors}
   \label{sec:fund}
All of the large GW laser interferometers in the past, as well as those planned for the next
decade, are essentially Michelson interferometers (as opposed to e.g., Sagnac 
interferometers~\cite{Stanford:Sagnac1996}). As Eq.~\ref{eq:roundtrip} shows, the
measured optical phase shift is proportional to the Michelson arm length; with typical parameters
($L\,\sim$1\,km, $\lambda\,\sim\,1\,\rm \mu m$, $h\,\sim\,10^{-21}$) the phase shift is just 
10$^{-11}$~radians. In order to amplify the signal to detectable levels, one would like to increase
$L$ by a few orders of magnitude. Unfortunately, the interferometer arm lengths are limited to
a few~km due to practical constraints (chiefly available land and prohibitively high construction costs).
In order to improve the signal-to-noise ratio, the Michelson is enhanced using several compound
optical resonators.

\subsection{Delay Lines vs. Fabry-Perot}
\label{sec:DelayLines}
In order to artificially increase the Michelson arm length, one can bounce the light back and forth in the
arms to increase the interaction time with the gravitational wave, thereby increasing the optical phase shift.
With sufficiently large mirrors, one could construct a Herriott delay 
line~\cite{Herriott:65, DHS:Garching, Byer:Sagnac} with hundreds of bounces. 
Drever~\cite{Drever:FP1991, Ron:GR1983} proposed to instead use Fabry-Perot 
optical resonators in place of the delay lines.
These cavities have the advantage of combining all of the many 'bounces' of the delay 
line onto a single spot. This greatly reduces the size, and thereby, the cost, of the mirrors. 
An added complexity is that the Fabry-Perot cavity must be servo controlled to be within 
a small fraction of its resonance linewidth in order to operate linearly.

Nearly all of the modern interferometers now use Fabry-Perot cavities instead of delay lines due to issues
with scattered light in the latter~\cite{Schnupp:1985}. The technical servo control issues have been
largely solved over the past few decades using multi-degree of freedom 
extensions~\cite{Mavalvala:2001vn, TAMA:LSC, Acernese:2006gm, Hartmut:PhD}
of the Pound-Drever-Hall RF heterodyne cavity locking technique~\cite{PDH:1983}

Fluctuations in the 
alignment~\cite{Mavalvala:1998ur, Morrison:94theory, Morrison:94exp, 
Virgo:ASC, Grote:2002wl} 
and transverse beam size~\cite{Mueller:2000ul} are sensed in a similar fashion.

\begin{figure}[h]
   \centering
   \includegraphics[width=\columnwidth]{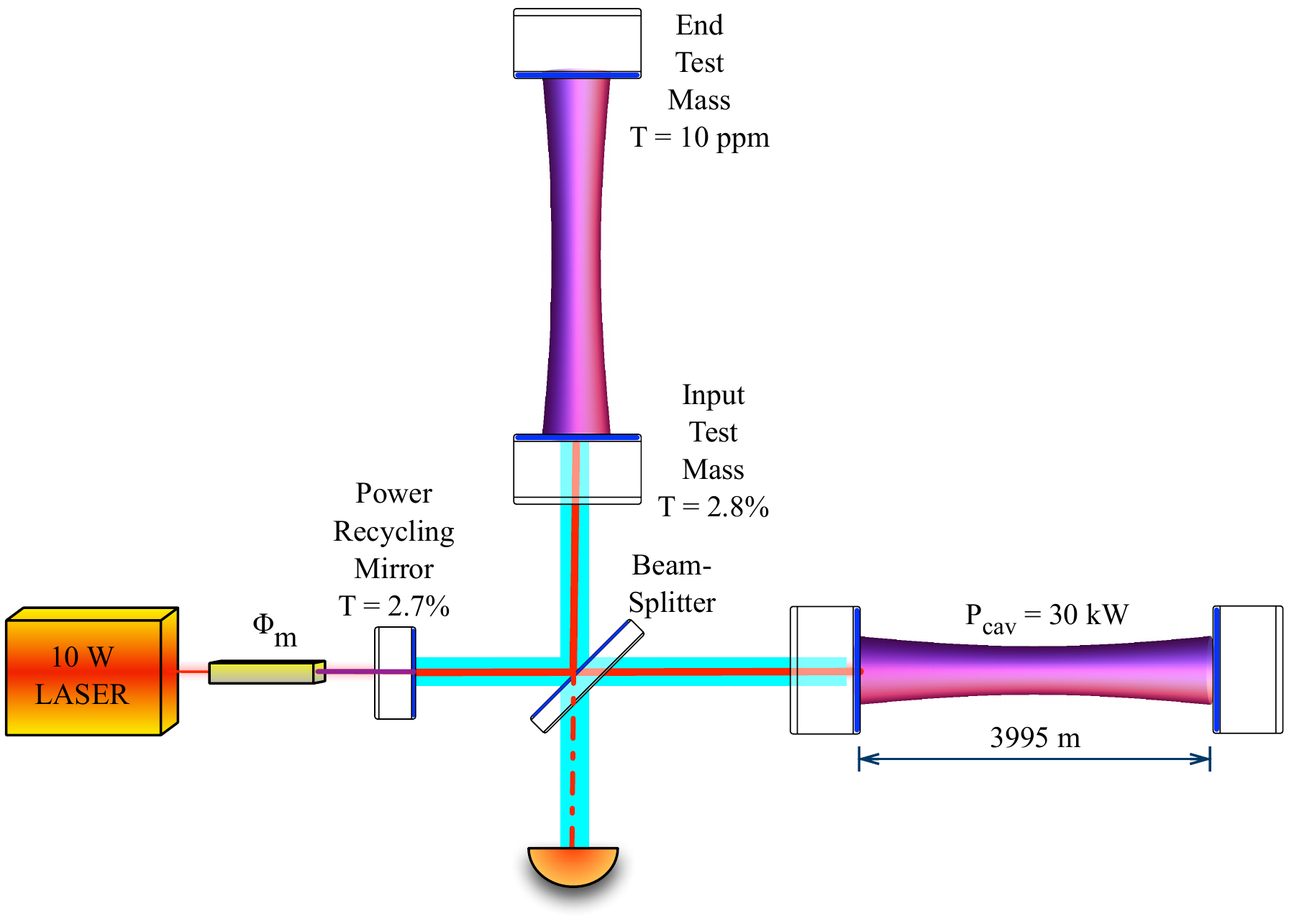} 
   \caption{(Color online) Schematic of the initial LIGO interferometers. The input beam is 
     phase modulated
                 and then built up resonantly in the power recycling cavity. The (cyan) phase modulation
                 sidebands resonate only in the power recycling cavity.
                  The light incident on
                 the photodetector at the bottom of the diagram carries the GW signal.}
   \label{fig:PRFPMI}
\end{figure}

\subsection{Power Recycling}
\label{sec:PR}
The interferometer arm cavities are adjusted in length microscopically such that the fields from 
each arm
interfere destructively at the Michelson anti-symmetric port. This causes almost all of the laser
light to return towards the laser. By placing a partially transmitting mirror between the laser
and the Michelson beamsplitter, this return light can be made to return towards the beamsplitter
interfering constructively with the incoming laser light. The finite transmissivity of this so-called
``power recycling~\cite{Drever:PR} mirror'' is chosen to nearly equal the total scattering losses
from the Michelson's optics and thereby provide optimum power coupling from the laser
source into the interferometer arms (GW transducer). In this sense, one can think of the power
recycling mirror providing an impedance match to the rest of the 
interferometer~\cite{Fritschel:1992jk}. The modern
GW interferometers with Fabry-Perot arm cavities have been able to increase the laser power impinging on the
beamsplitter by a factor of $\sim$\,65 by using this method (cf. Fig.~\ref{fig:PRFPMI}). 
The GEO600 detector has achieved a power
gain of 1000 using power recycling~\cite{blair2012advanced}.

\subsection{Signal Recycling and Extraction}
\label{sec:SR}
Just as a mirror on the symmetric side of the beamsplitter can coherently amplify the power stored
in the interferometer, a carefully placed mirror at the anti-symmetric side of the beamsplitter can
amplify differential signals (e.g., Fig.~\ref{fig:DRFPMI}). This technique is 
called \textit{Signal Recycling}~\cite{Mee1988, Miz1995} and
can be used to resonantly build up the GW signal. The GEO600 interferometer~\cite{Hartmut:2010} 
has been using signal recycling for the past several years successfully. 

%This signal recycling 
%comes at a price: beyond a certain point, increasing the finesse of the 
%signal recycling cavity improves the signal only at low frequencies. Higher frequency GW signals
%are filtered by the narrow coupled cavity resonance.

The alternative strategy (which is often used in practice) is to use a kind of 'anti-recycling'. To reduce
the thermal loading due to bulk absorption in the input test masses, the Fabry-Perot
arm cavities are made to have a very high finesse. For the same arm cavity power, this lowers
the power level in the optics of the power recycling cavity and allows for a high power 
to be stored
in the cavity with minimal thermal distortions. This narrow linewidth cavity would then normally
only amplify the low frequency
GW signals (and not the signals above the arm cavity pole frequency) and thereby seems 
like a nonsensical design choice. 
However, by adjusting the microscopic position of
the 'signal recycling mirror' to form a resonant cavity with the input test mass mirrors,
the effective linewidth of the combined system (the differential arm cavity mode + the signal
recycling cavity) is broadened. The signal recycling mirror's function has been transformed from
signal recycling to resonantly extracting the GW sidebands. This technique is referred to as
\textit{Resonant Sideband Extraction} (RSE)~\cite{Mizuno:RSE1993, SensingStrain:2003}.

These two configurations, signal recycling and resonant sideband extraction, are the extrema
of a continuous space of detuning for the signal recycling cavity. The microscopic tuning of
this cavity allows for great flexibility in shaping the detector's frequency response 
(cf. Fig.~\ref{fig:aLIGO_cases}).

% ===============================================================
\section{Sensitivity Limits of Laser Interferometers}
\label{sec:noise}
Laser interferometers are limited by two broad classes of noise: displacement noise and phase noise.

Displacement (or \textit{force}) noises work by directly moving the interferometer mirrors. Most of these forces
are filtered by the mechanical response of the mirror and its suspension and so are strongly attenuated
above several~Hz. Many of these force noises can be mitigated by increasing the mass of the mirror.

Phase noises produce fluctuations in the phase of the optical field used to read out the GW strain. 
These noise sources are modified only by the opto-mechanical response of the interferometer
(in nearly the same way as the gravitational-wave strain), and therefore have no strong frequency
dependence.

\subsection{Phase Noise}

\subsubsection{Quantum Vacuum Fluctuations}
\label{sec:Quantum}
A ``fundamental'' limit to the sensing of optical phase shifts comes from the stochastic fluctuations in the
arrival times of photons at the photodetector. Before 1980, the
picture was that a laser interferometer
could, at best, be limited by the Poisson statistics. In this picture the signal-to-noise ratio (SNR) for
optical sensing would vary as $1/\sqrt{\rm P}$ (where P is the input
laser power) and the fluctuating radiation pressure on the mirror would vary
as $\sqrt{\rm P}$. This description is similar to that of the 'Heisenberg microscope' used to describe uncertainty
in introductory physics courses~\cite{Feynman:III}.

A more precise characterization of the quantum 
measurement limits was derived by Caves~\cite{Cav1980,CaSc1985} 
and others~\cite{Lou1981} in the early 1980's. In this picture, the noise arises from the beat between
the fluctuations of the vacuum ground state of the electromagnetic field and the stable laser light: vacuum fields 
entering from the anti-symmetric port
split at the beamsplitter, producing differential forces on the arm cavity mirrors (for in-phase fluctuations)
and phase fluctuations (for fields that are in the quadrature phase).
Vacuum fields at frequencies far from the laser frequency are rejected by the arm cavities, 
return to the photodetector, and beat with
the static field present at the anti-symmetric port (due to both, intentional and unintentional, asymmetries in the arms).

Increasing the laser power leads to a reduction in the measurement uncertainty for the mirror position but
increases the amount of momentum perturbations. These momentum perturbations produce position
fluctuations after a finite amount of time. Similarly, reducing the laser power reduces the momentum noise
but also decreases the positional precision. For a given set of parameters, the laser power may be
optimized to give the optimum strain sensitivity at a particular frequency.
A detailed analysis of this quantum limit for a free mass
leads to the so-called 'Standard Quantum Limit' (SQL)~\cite{BrKh1999a}:
\begin{equation}
S_x(f) = \frac{2 \hbar}{m (2 \pi f)^2}
\end{equation}
The SQL represents the envelope of minima in the strain noise as the laser power is tuned assuming
that the amplitude and phase fluctuations from the vacuum fields are uncorrelated.

At the turn of the century, our understanding of quantum noise in interferometers was revolutionized
by the work of Buonanno and Chen~\cite{BuCh2001, BuCh2002} and \cite{KLMTV2001}. They showed
that the combination of high power and a signal recycling cavity can build up significant
quantum correlations within the interferometer. The correlation of the vacuum fluctuations can then
allow for significant back action evasion in limited frequency ranges: microscopic detuning of the
signal recycling cavity leads to a radiation pressure driven restoring force. This 'optical spring'
can be tuned~\cite{Osamu:2006} via the cavity detuning to optimize the response to different astrophysical
sources. Further development of these
Quantum Non-Demolition (QND) techniques with application to 3$^{\rm rd}$ generation detectors
is presented in Sec.~\ref{sec:QND}.

\subsubsection{Scattering from Residual Gas}
Fluctuations in the column density of gas in the interferometer arms produce noise in the measured
optical phase~\cite{Zucker:Gas,TAMA:Gas}. For a single species of molecule, the power spectral 
density of apparent strain fluctuations is:
\begin{equation}
S_h(f) = \frac{(4 \pi \alpha)^2 \rho}{v_0 L^2} \int \limits_0^L \frac{exp[-2 \pi f \omega(z)/v_0]}{\omega(z)}dz
\end{equation}
where $\alpha$ is the polarizability, $\omega$ is the beam radius, $L$ is the interferometer arm length,
$\rho$ is the number density, and $v_0$ is the most probable speed for the particle. 
Taking $\rm H_2$ as an example, it is only necessary to reach a residual pressure of $\sim\,10^{-9}$~Torr
to reduce the induced strain noise from this molecule to $\sim\,10^{-25}/\sqrt{\rm Hz}$. The partial pressure
required for highly polarizable substances, such as hydrocarbons and water, is much more stringent.

\subsubsection{Backscatter}
\label{sec:scatter}
Imperfections in the mirror shape at spatial scales larger than $\sim$1~mm 
(called ``figure error''; cf. Fig.~\ref{fig:aLIGO_mirror}) can scatter
the light incident on the mirrors into small angles that deposit the light into the long beam tubes. Imperfections
at smaller spatial scales (called ``micro-roughness'') will produce a diffuse scatter of the light directly
into the nearby vacuum chambers. A small fraction of these scattered light fields is scattered back to the mirror 
and can then recombine with the circulating field via the mirror 
imperfections~\cite{Kip:scatter1989, Kip:Scatter95, Winkler:94, Stefano:scatter}.

Seismically driven motions of the vacuum system can in this way produce phase and 
amplitude fluctuations of the light field within the interferometer~\cite{Schilling:1981, Sam:Scatter2012}. 
Works prior to 2012, 
included only the terms leading to phase modulation of the interferometer's stored field. 
Since the relative phase between the scatterers and the interferometer field is random, there 
should be an equal contribution to both the phase and amplitude quadratures. With the increasingly 
high power levels in modern interferometers, the amplitude component turns out to be \textit{dominant
at low frequencies via the influence of radiation pressure} on the mirror motion.
This mechanism is analogous to that of the quantum noise in that the amplitude noise
becomes dominant at low frequencies.

Careful engineering of dark, polished, scattered-light beam traps throughout the long 
vacuum tubes and in the vicinity of the mirrors are expected to suppress the influence of the 
scattered light to below the current quantum backaction limits. Backscatter from the photodetectors
used for signal detection can be mitigated by moving the detectors onto a quiet, in-vacuum platform
and/or using external phase modulators~\cite{GEO:scatter2008}.

\subsection{Displacement Noise}
All of the following effects produce motion of the test mass through stochastic 
fluctuation of forces. As such, the power from these types of noise are concentrated at 
lower frequencies and are not important for higher frequency astrophysical sources (e.g.,
supernovae, millisecond pulsars, binary neutron star mergers and ringdowns).

\begin{figure}[h]
   \centering
   \includegraphics[width=\columnwidth, trim=0 0 -27mm 0, clip]{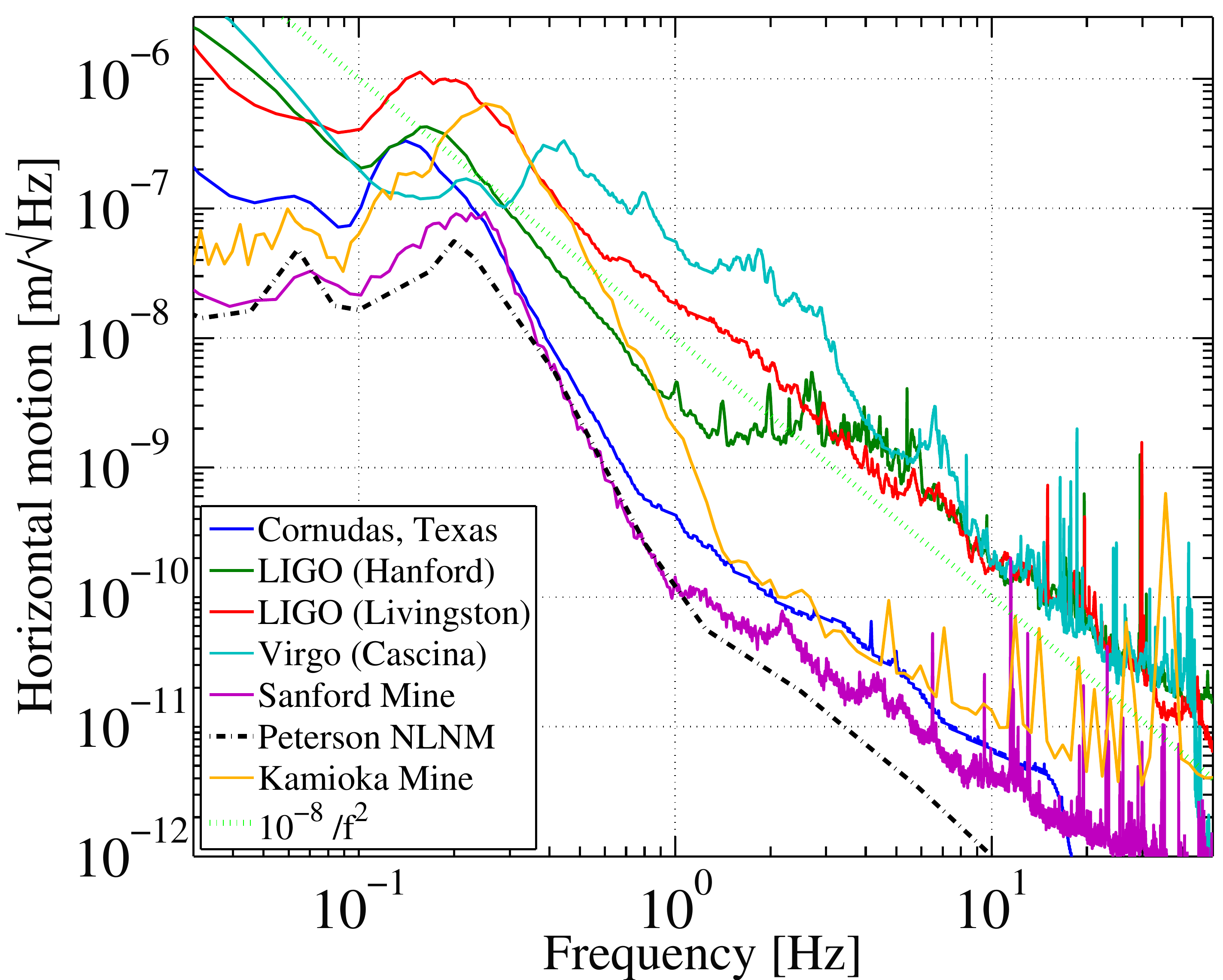} 
   \caption{(Color online). Shown are the seismic vibration spectral density for some of the 
                   relatively quiet sites of the current GW detector network. Also shown are 
                   two promising locations for future low frequency detectors in the U.S.: the 
                   4100\,ft. level of the Sanford Underground
                   Lab and a surface site near El Paso, TX. The USGS New Low Noise Model~\cite{Pet1993}
                   is included as a reference. All the spectra here 
                   (with the exception of Kamioka~\cite{Aso:Seismic}) are estimated using Welch's method
                   but with median instead of mean averaging so as to better reject 
                   non-Gaussian transients.}
   \label{fig:SeismicNoise}
\end{figure}

\subsubsection{Mirror Thermal Noise}
\label{sec:mirror_thermal}
When considering fundamental sources of displacement noise for macroscopic objects (such as 
the interferometer mirrors) we are reminded that the equipartition theorem demands that there 
be $k_B T$ of energy per mode in any solid that is in thermal equilibrium. In order to determine 
what the apparent displacement noise fluctuations are for the mirrors, we can compute the
 spectral density of 
fluctuations from each mode and then sum them up while including weighting factors for
the effective mass in each mode as well as the spatial overlap between the mechanical eigenmode 
and the laser field distribution~\cite{Gillespie:1995}.
This method is quite complicated and converges slowly with increasing mode number. 

An alternative approach~\cite{gonzalez:207, Levin:Direct}, is to directly apply 
Callen's Fluctuation-Dissipation Theorem~\cite{CaWe1951, Kubo:FDT, Callen:1959} 
to the mirror for the laser beam shape in question.  Here the power spectrum of apparent 
displacement fluctuations is:
\begin{equation}
S_x(f) = \frac{k_B T}{\pi^2 f^2} \left| Re \big[ Y(f) \big]\right| 
\label{eq:FDT}
\end{equation}
where $T$ is the temperature of the mirror and $Y(f) \equiv \dot{x}(f)/F(f)$ 
is the complex mechanical admittance (inverse of impedance) associated 
with the optical readout beam profile. The
meaning of this is the following: in order to determine the level of apparent RMS mirror fluctuation
due to thermal forces, we need only to apply a sinusoidal driving force, $F(f)$, and then 'measure' the 
response. In the case that there is no dissipation, the mechanical response of the system will be 
entirely in-phase (\textit{modulo} 180 degrees) with the applied force. 
As with a classical electronic circuit, this orthogonal phase response is proportional to the
dissipation: the phase shift between excitation and response is the loss angle ($\phi$) 
or equivalently, $1/Q$ (the Quality factor) of the material.

Following Levin's approach for the mirror thermal noise we can express the displacement 
noise power spectrum as:
\begin{equation}
S_x(f) = \frac{2 k_B T}{\pi^{3/2} f} \frac{(1 - \sigma)}{\omega E} \phi_{\rm sub}
\label{eq:mirrorTN}
\end{equation}
where $\omega$ is the spot size ($1/e^2$ radius) of the beam, $E$ is
the Young's modulus of the mirror substrate, $\sigma$ is the
scalar Poisson's ratio for the substrate, and $\phi_{\rm sub}$ is the loss angle. The best samples 
of fused silica, sapphire, and silicon can have loss angles as low as $10^{-8}$ or better 
and, as such, do not limit the sensitivity of modern detectors 
(cf. Fig.~\ref{fig:aLIGO_gwinc}).

\paragraph{Mirror Coating Thermal Noise}
\label{sec:coatings}
In fact, the dominant source of thermal noise of the mirror surface is the mechanical dissipation 
in the dielectric, thin film coating on the mirror surface and \textit{not the
bulk mirror material}. These coatings, which have very good optical qualities, are quite poor from 
the internal friction standpoint.
The dependence of this loss on type of material, number of layers, and layer
structure has been studied 
extensively \cite{Harry:CQG2002, Penn:CQG2003, BrVy2003, Harry:AO2006, Sheila:2005, 
Harry:CQG2007, Goro:2011, Ting:Brownian2012, Bassiri2012, evans2012reduced, Flaminio:2010uk}.

In addition to the Brownian noise, the thermodynamic temperature fluctuations in the 
coating also produce
noise~\cite{BrVy2003}. In this case, however, there is an additional complication:
the temperature fluctuations also give rise to fluctuations in the index of refraction of the 
dielectric thin-films.
Since this thermo-refractive noise has the same source as the thermo-elastic noise, 
they must add (or subtract) coherently~\cite{Matt:TOnoise}. A judicious choice 
of the coating layer structure can be used to mostly cancel the effects from these 
fundamental temperature fluctuations~\cite{harry2012optical}.

\subsubsection{Suspension Thermal Noise}
\label{sec:sus_thermal}
A simple example of the power of the Fluctuation-Dissipation theorem for calculating 
thermal noise is the damped harmonic oscillator~\cite{Saulson:ThermalNoise}. 
In this case, the admittance is simply
\begin{equation}
Y(f) = \frac{i}{2 \pi m} \frac{f}{f_0^2 + i f_0 f/Q -f^2}
  \label{eq:dsho}
\end{equation}
where $f_0$ is the resonance frequency of the oscillator. Eq.~\ref{eq:dsho} gives the admittance, 
and thereby the thermal noise, for an oscillator
damped in a \textit{viscous} manner. In the absence of technical limits such as damping from gas
in the vicinity of the oscillator (which can be removed through standard vacuum techniques)
or friction at the top clamp~\cite{Cagnoli:1999ud, Kovalik:1993} of a pendulum, the mechanical 
losses in low loss springs and flexures
can often be characterized by a constant complex term in the spring 
constant, $k = k_0(1 + i \phi)$,
where $\phi$, the loss angle is also equal to $1/Q$. This case is often referred to 
as \textit{structural} damping.

The suspension for the mirrors of the interferometer must serve several purposes: isolate the
mirror from ground vibrations, decouple the mirror from the ground to allow it to move freely in
response to the gravitational waves, and hold the mirror without introducing extra thermal noise.
These needs are simultaneously met by suspending the mirror as a pendulum from a thin fiber.

In contrast to a standard mechanical spring, nearly all of the potential energy for the pendulum is stored in
the gravitational field~\cite{gonzalez:207, Logan:1993};
the pendulum's gravitational spring constant is given by the simple relation: $k_g = m g / l$,
where $m$ is the mirror mass, $g$ is the acceleration due to gravity and $l$ is the pendulum 
length. With this simple model,
there would be no damping and the pendulum would have an infinite $Q$. In reality, there 
is some energy
stored in the bending of the pendulum wire at the two ends. The spring constant for a pendulum
supported by $N$ wires is 
$k_{\rm wire} = N \sqrt{T E I} /  2 l^2$ \cite{Saulson:ThermalNoise, Gonzalez:2000vw}, where
$T$ is the tension, $E$ is the Young's modulus of the material, and $I$ is the moment of
inertia wire's cross section. As this spring
is much weaker than the gravitational spring, the overall loss angle of the pendulum can be
approximated as $\phi_{\rm pend} = \phi_{\rm wire} (k_{\rm wire} / k_g)$. This reduction factor (the so-called
'dissipation dilution' factor~\cite{Geppo:2000}) is what allows for having such a low level of
thermal noise in a pendulum.

\subsubsection{Seismic Vibrations}
\label{sec:seismic}
Seismic vibrations of the laboratory prove to be a low frequency limit for all terrestrial
laser interferometers. The largest strains of the Earth's surface over km scales are
due to the tidal gravity from the Moon and the Sun~\cite{Melchior:Tides}. These Earth
Tides produce length changes of 100--200 microns over a 4\,km baseline and for all
of the large interferometers are compensated by long-range actuators external to the
vacuum system.

In the absence of earthquakes, the next largest component of the ground motion is known
as the 'secondary microseism' and occurs at periods of 3--10 seconds~\cite{uSeism:1992}. 
This low frequency vibration can sometimes grow to an amplitude of several microns and
must be cancelled by an appropriate feedback system.
Above $\sim\,1$\,Hz, the typical vibration spectra for reasonably quiet sites can be 
approximated as~\cite{AkRi2009}
\begin{equation}
x_G = 10^{-8} \left(\frac{1~\rm Hz}{f} \right)^2  \frac{\rm m}{\sqrt{\rm Hz}}
\end{equation}
This power law is compared with sample noise spectra from the GW detector sites in
Fig.~\ref{fig:SeismicNoise}.
To reach astrophysically interesting strain 
sensitivities ($\sim 10^{-21} /\sqrt{\rm Hz}$) with a km~scale detector 
therefore requires suppressing the vibrations by a factor of 
\textit{at least}  $10^8$ at 10\,Hz and $10^6$ at 100\,Hz.  

The best seismic vibration sensors reach a level of 
$\sim\,10^{-13}~\rm m/\sqrt{\rm Hz}$~\cite{RiHu2010}. Incorporating such sensors
into active vibration isolation platforms~\cite{Chu:1999, Richman:1997} 
is useful in reducing the large,
low frequency motions and bringing the interferometer close to the desired operating point. The final
several orders of magnitude in suppression can only be achieved by using passive isolation. In all
of the laser interferometers to date, this passive isolation is roughly the same: a chain of 
masses and springs isolates the final test mass from the actively controlled platform.
\begin{figure}[h]
   \centering
   \includegraphics[width=\columnwidth, trim=0 0 -27mm 0, clip]{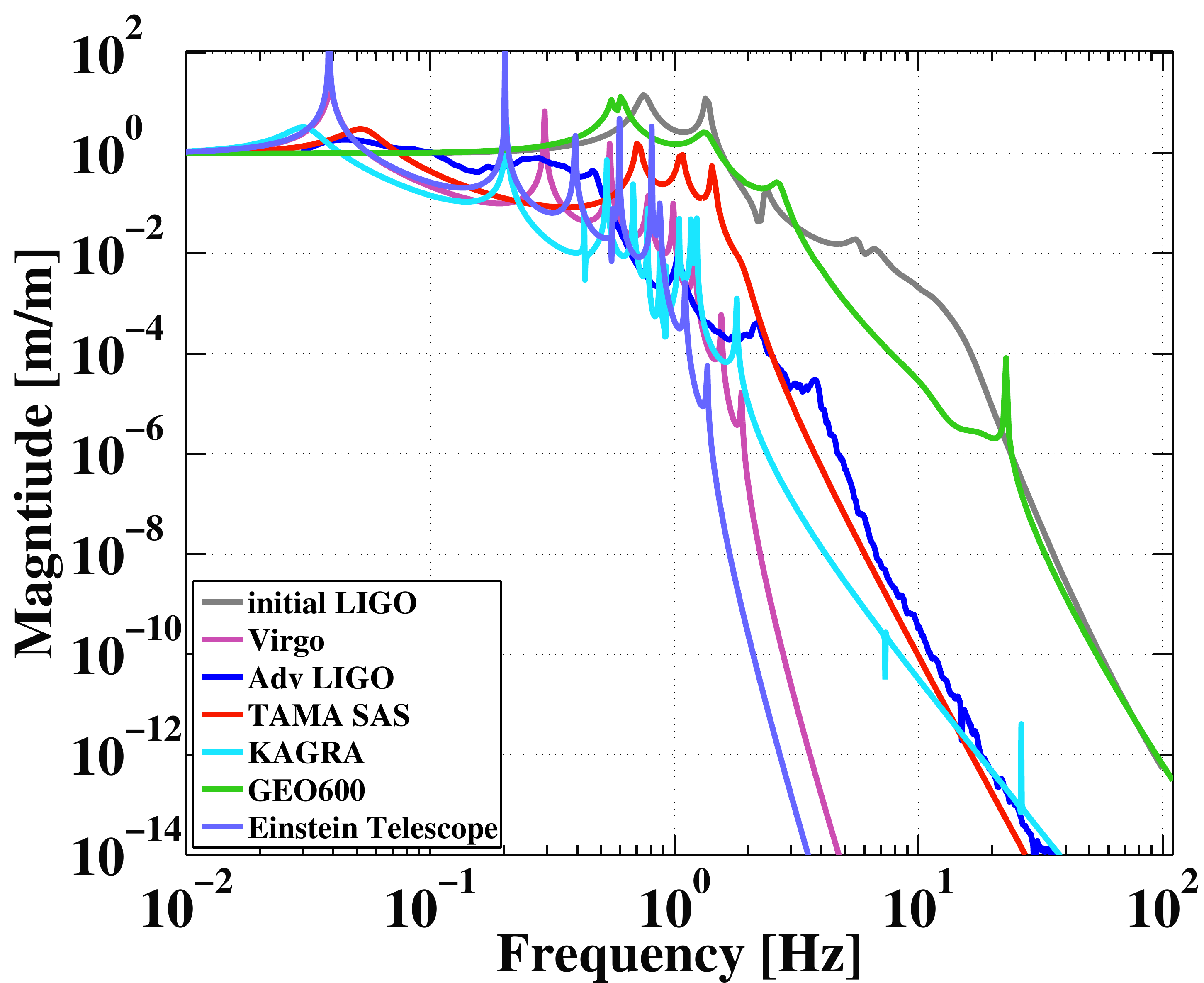} 
   \caption{(Color online). Vibration isolation for the initial LIGO~\cite{ponslet:432, Giaime:1996}, 
                   Virgo~\cite{Stefano:2001, Virgo:SA2010, Accadia:2011jh}, 
                   TAMA (with SAS)~\cite{Szabi:TAMASAS},
                   GEO600~\cite{Hartmut:PhD, Ken:GEOseismic, plissi:3055}, 
                   Adv. LIGO~\cite{aLIGO:Seismic2002},
                    KAGRA~\cite{Somiya:2011tb}, and the Einstein Telescope~\cite{ET2011}. In the KAGRA case, 
                    the mechanical links for cooling (included) are expected to limit the 
                    isolation performance above $\sim$\,1\,Hz~\cite{Takahashi:email}}
   \label{fig:VibrationIsolation}
\end{figure}

Fig.~\ref{fig:VibrationIsolation} shows the transfer function from horizontal motion of the
ground to motion of the test mass in the laser beam direction. In most cases, this allows one
to predict the motion of the test mass given the measurement of the ground noise. This
assumption must be corrected for the presence of active vibration isolation systems
incorporating seismometers with non-zero internal noise. The Advanced LIGO isolation
system is (intentionally) limited by the noise of these sensors in the 1--10 Hz region and so, instead of
a usual transfer function, the ratio of the modeled test mass motion to the ground motion is
shown.

A further complication comes from the non-trivial cross-couplings within the isolation systems.
Vertical motion and tilts~\cite{Giazotto:2011ux, Lantz:2009dw} of the ground couple to the test mass due 
to mechanical cross coupling
in the isolation platforms and mirror suspensions, as well as within the seismic sensors themselves.
As such, the transfer functions can only be considered to be approximations of the true vibration
isolation levels.

\subsubsection{Newtonian Gravity Noise}
\label{sec:NN}
Even with a much improved seismic vibration filtration system, there exists a fundamental
limit to terrestrial gravitational experiments~\cite{Rai:QPR}. Density fluctuations in the
atmosphere and surface waves on the ground can lead to fluctuations in the Newtonian
gravitational force (also called gravity gradient noise) on the test masses. Following
\cite{Saulson:NN}, the equivalent strain noise can be approximated as:
\begin{equation}
\delta h_{\rm NN}(f) = \frac{G}{\sqrt{3} \pi} \frac{\rho_E}{L} \frac{x_{\rm GND}(f)}{f^2}
         \label{eq:NNsimple}
\end{equation}
where $G$ is the gravitational constant, $\rho_E$ is the density of the nearby ground, 
$L$ is the interferometer arm length, and $x_{\rm GND}$ is the ambient ground noise.
More sophisticated treatments of the correlations among the seismic waves~\cite{HuTh1998, BeEA1998},
atmospheric perturbations~\cite{Cre2008} and anthropogenic influences~\cite{Kip:Dancing}
concluded that these Newtonian gravity fluctuations would
nearly limit the performance of the 2$^{\rm nd}$ generation detectors in the 5--15~Hz band.
%\R{some examples: tumbleweed outside of building and wasp around chamber}.

\begin{figure}[h]
   \centering
   \includegraphics[width=\columnwidth, trim=0 0 -27mm 0, clip]{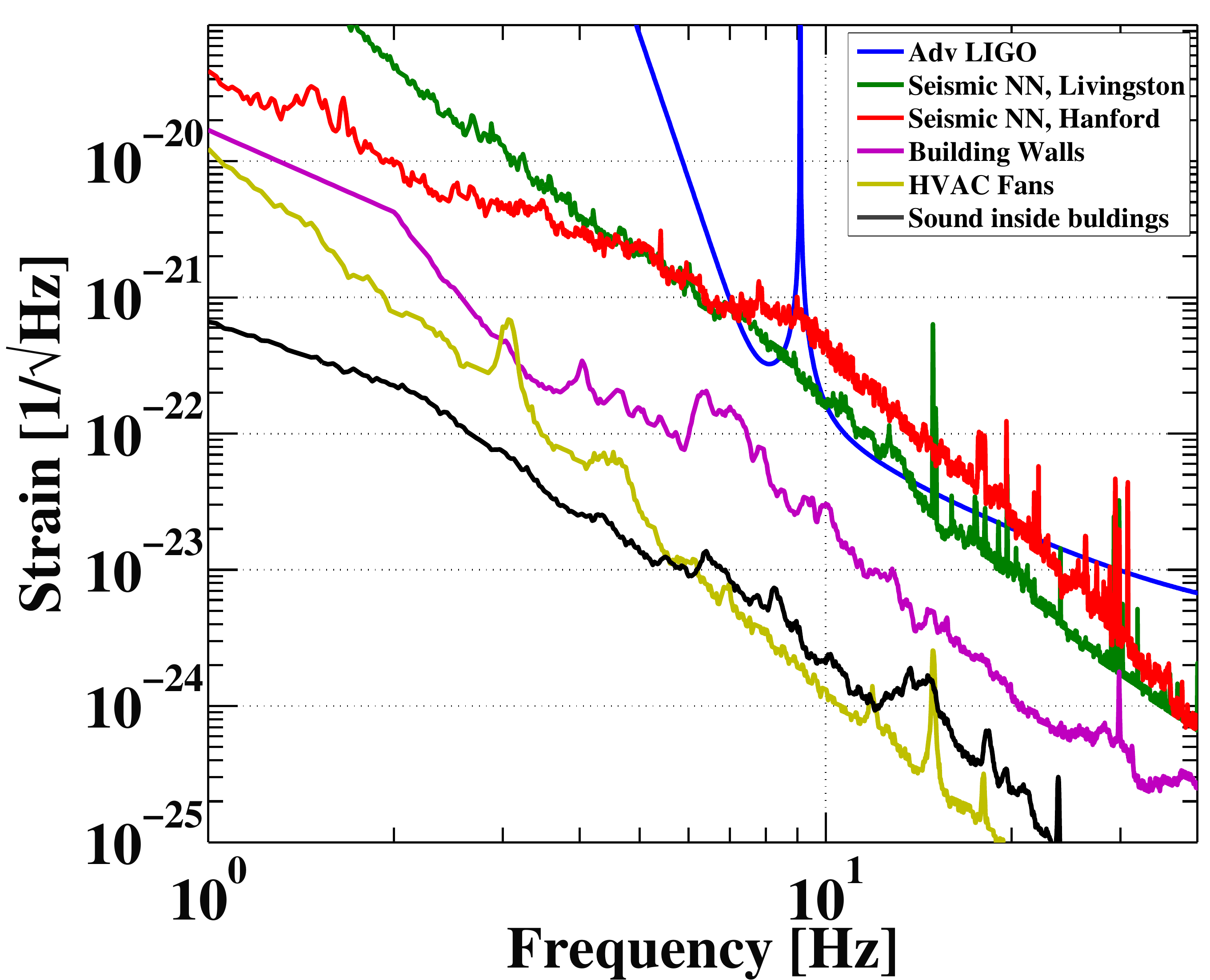} 
   \caption{(Color online). Estimate of the Newtonian gravity noise at the LIGO sites compared
                   to the fundamental thermodynamic and quantum limits for Advanced
                   LIGO. The dominant contribution is from surface waves on the nearby ground.
                   Vibrations of the building walls and acoustics within
                   the building are not very significant. The turbulence from wind outside of the building
                   primarily couples through the vibration of the building walls and is therefore 
                   already included in this
                   estimate.}
   \label{fig:aLIGO_NN}
\end{figure}

A detailed survey of the sources of vibration at the LIGO sites~\cite{DrHa2011} has taken into
account vibrating machinery, ambient acoustics, and resonances of the surrounding structures
in the laboratory; the resulting estimate is shown in Fig.~\ref{fig:aLIGO_NN}. Although the 
seismic and acoustic sources that produce these forces can
themselves be filtered out, there is no way to shield the test masses from their gravitational
forces. It is likely that the Newtonian noise will exceed the quantum backaction limits
at low frequencies (see Fig.~\ref{fig:aLIGO_gwinc}). Mitigation strategies are 
discussed below in Section~\ref{sec:NN_sub}.

\begin{table*}[ht]
    \begin{centering}
      \begin{tabular*}{0.85\textwidth}[t]{@{\extracolsep{\fill}}c | p{20mm} p{20mm} p{20mm} p{20mm} p{20mm}}
\hline
Detector              & TAMA   & GEO & Virgo & LIGO 2~km & LIGO 4~km\\
\hline
Arm Length [m]       &     300          & 600        & 3000  &  2009        & 3995        \\
Mirror Mass [kg]      &    1              &   5.5          & 21         &   10.5        &    10.5     \\
Beam Spot Size [cm]   &    0.85           &   2.4           & 2.1          &   3.5          & 3.5 \\
\# of seismic stages  &   0 + 1 + 4           & 1 + 3 + 3    & 0 + 1 + 6  &        1 + 4 + 1       & 1 + 4 + 1 \\   
Stored Power [kW]      &    1                &    5          &   20       &       30            &        50        \\
Strain Noise [$10^{-23}/\sqrt{\rm Hz}$]  &  150    & 20     & 6     &  5                & 2 \\
Sensitive Band [kHz]  & 0.3 -- 10           & 0.3 -- 5                  & 0.02 -- 3 & 0.06 -- 2 & 0.06 -- 2 \\
Location                & Japan & Germany & Italy & USA & USA \\
\hline
    \end{tabular*}
    \caption{Comparison of first generation interferometers.
                  The numbers in the '\# of seismic stages' row refer to the number of external active, 
                  internal passive, and pendulum suspension stages, respectively. 
                  For TAMA and LIGO, substantial hardware upgrades to the seismic isolation
                  took place during the commissioning phase - these numbers refer to the 
                  post-upgrade configurations.}
    \label{t:1G}
  \end{centering}
\end{table*}

\subsubsection{Electromagnetic Coupling}
\label{sec:EM}
In addition to the forces mentioned above, the mirrors of the interferometer may be
disturbed by spurious electromagnetic forces: ambient fluctuations of the local
electric and magnetic fields, as well as impacts from the background of cosmic rays.

\paragraph{Cosmic Rays}
At sea level, the stationary background of high-energy cosmic rays is dominated by
muons. Within the typical mirror volume of $\sim$0.02 m$^2$, there are $\sim10 -- 50$
muons passing through per second~\cite{Rai:QPR, Braginsky:2006ip}. The muons deposit
energy in the mirrors by exciting (or ionizing) the electrons bound by the molecules in the 
material. The Bethe formula~\cite{Gruyter:1997} tells us that most of the particles pass
through the mirror depositing $\sim$100~MeV in kinetic energy. The false alarm rate (rate of
apparent GW signals) due to these high energy muons is extremely low; they can practically 
be rejected by demanding a 
coincidence between remote interferometers for making gravitational wave detections. The 
background of low energy muons, however, leads to a stationary noise spectrum given 
by~\cite{Kazuhiro:Cosmic}:
\begin{equation}
h_{\rm cosmic}(f) \simeq \big[10^{-27} - 10^{-26} \big] \left(\frac{100~\rm Hz}{f}\right) \, \frac{1}{\rm \sqrt{Hz}} 
\label{eq:cosmic}
\end{equation}
with some variation between fused silica, sapphire, and silicon (the
most common mirror materials for GW detectors). This
noise source is 2--3 orders of magnitude lower than the standard quantum limit for all of
the present, and envisaged future detectors.

\paragraph{Ambient Magnetic Fields}
The ambient magnetic field fluctuation spectra are fairly broad. As measured at several sites
in the U.S. they show a characteristic $1/f$ behavior~\cite{Campbell:1965}; at the LIGO sites this $1/f$ character has
been observed with an amplitude of $B(f) \sim (10^{-11}/f)~\rm T/\sqrt{\rm Hz}$. In nearly all laboratories
on the earth, the dominant features in the spectrum are the harmonics of the AC power line
(60~Hz in the U.S.; 50~Hz at the GW detector sites in Italy, Japan, and Germany). At lower frequencies
(5--50~Hz), the dominant magnetic field fluctuations in the horizontal direction
are due to the Schumann resonances (Extremely Low Frequency
traveling waves within the Earth's surface-ionosphere cavity) and appear as a broad set of peaks
at multiples of $\sim$7~Hz~\cite{Balser:1960, AE:handbook}. 
The amplitude of these peaks changes diurnally and also with the
intensity of distant lightning activity.

These magnetic fluctuations couple into the interferometer chiefly through the magnets which are
used to actuate the interferometer's mirrors. In the 2$^{\rm nd}$
generation GEO and LIGO detectors, the magnets
have been removed from the test mass mirrors. The magnets on the next closest mirror 
in the suspension chain may also provide too strong of a coupling path depending upon
the magnitude of local ferromagnetic components (which cause gradients).

%\paragraph{Ambient Electric Fields}
%Less is known about the fluctuations of the electric fields within the vacuum chambers.
% How does an ambient field produce a force? Drives residual charges? What about the
% polarized test mass?
%\cite{Fairbank:1967}
%\R{expand or delete}

\paragraph{Surface Charge}
Surface charges on the arm cavity mirrors can produce spurious forces on the mirrors through
interaction with nearby conducting surfaces (e.g., the mirror suspension frames)
\cite{UW:Charging2010, Ugolini:2008, Charge:2004}. These charges may build up through the friction induced
by the movements of dust during the evacuation of the chambers. Random fluctuations of the
charges could produce force fluctuations comparable to the thermal and quantum limits for the
mirror, however, current estimates and measurements are not yet accurate enough to make the
case. To be safe, several mitigation strategies are being pursued, including irradiation of the
mirror surface with UV light~\cite{Sun:2006} and occasionally introducing small amounts 
of an ultra-pure ionized gas into the vacuum chambers.

%\cite{GPB:Patches2011}

% magnetic measurements
% http://adsabs.harvard.edu//abs/1965eqae.conf..495C

% electric field measurements
% http://prl.aps.org/abstract/PRL/v19/i18/p1049_1

%Schumann resonances push mirror via the magnetic actuators. LIGO using ESD. Virgo/KAGRA using magnets.
%E-field noise described in \cite{Rai:QPR}. Not a problem.

% ===============================================================
\section{First Generation Detectors}
\label{sec:1G}
The first generation of long-baseline interferometers formed the first broadband
worldwide network for gravitational wave detection. The network was consisted of
TAMA (300\,m) near Tokyo, Japan~\cite{TAMA:2008, TAMA:2009}; 
GEO (600\,m) near Hannover, Germany~\cite{Hartmut:2010, Hartmut:PhD}; 
Virgo (3\,km) near Pisa, Italy~\cite{Virgo:Review2012, Lisa:PhD}; 
and the LIGO 
interferometers~\cite{DHS:Detector2004, LIGO:Science, Tobin:DC, Rana:PhD, PF:RPP2009} --- a 
4\,km one in
Livingston, LA and in Hanford, WA both a 2\,km and a 4\,km interferometer in the
same vacuum system. Table~\ref{t:1G} lists some of the key parameters of these detectors.

The installation and initial commissioning of these detectors started in the late 1990's. Although
more advanced techniques were known at the time, these first generation instruments were
built with some conservatism and therefore had several similarities. TAMA300, Virgo, and
the three LIGO detectors were configured as power-recycled, Fabry-Perot Michelson interferometers.
The GEO600 interferometer was the least conservative of all and included three 'Advanced'
techniques: dual-recycling, triple suspensions, and fused silica fibers to hold the mirrors.

\begin{figure}[h]
   \centering
   \includegraphics[width=\columnwidth, trim=0 0 -25mm 0, clip]{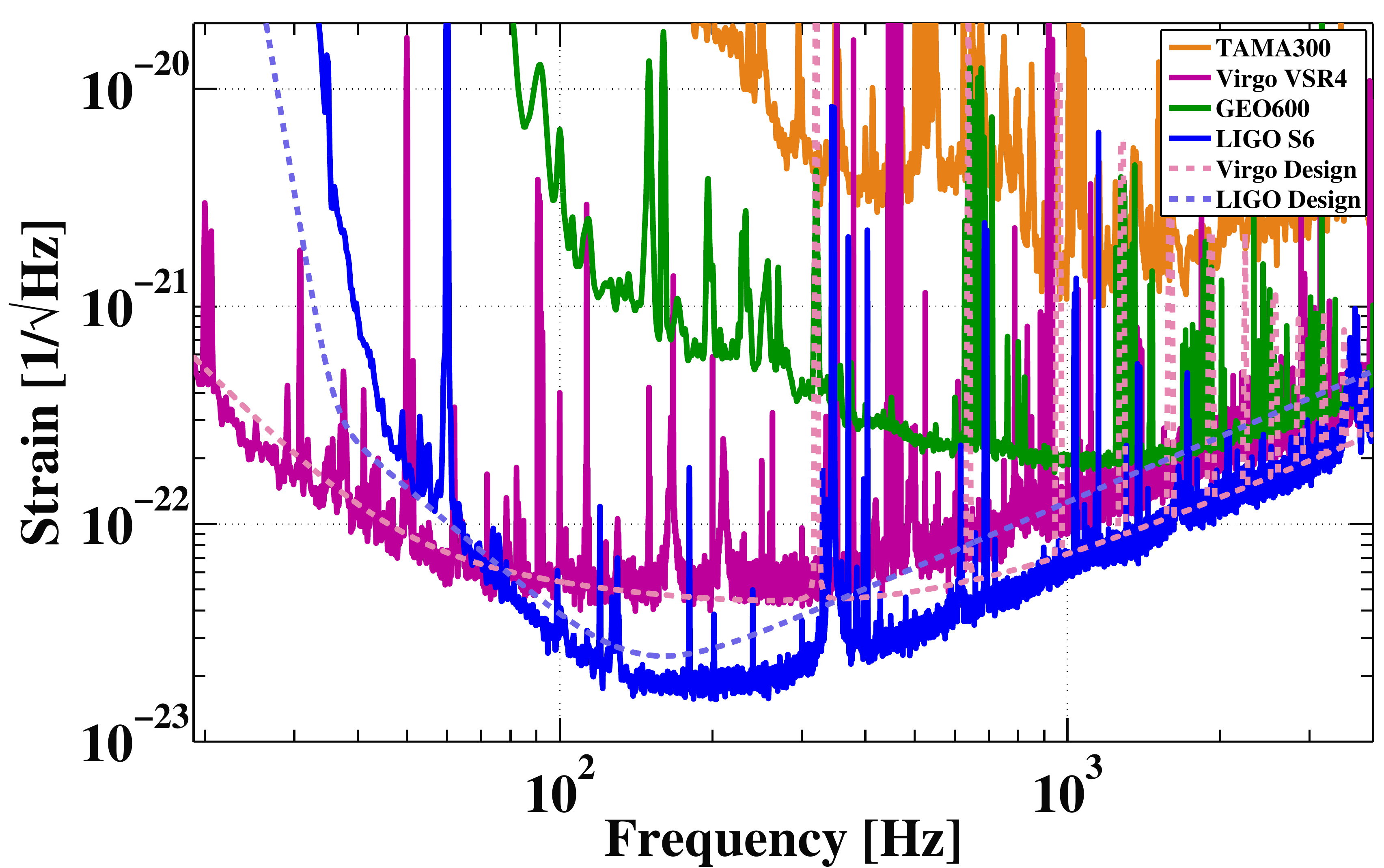}
   \caption{(Color online). Strain noise for the first
                  generation detectors. TAMA300, GEO600, Virgo+, and Enhanced LIGO.
                  Also shown (dashed) are the strain noise goals for the initial
                  Virgo and LIGO detectors.}
   \label{fig:Gen1_noise}
\end{figure}

\begin{figure*}[ht]
  \begin{center}
  \subfigure{
        \includegraphics[width=0.45\textwidth,height=6cm]{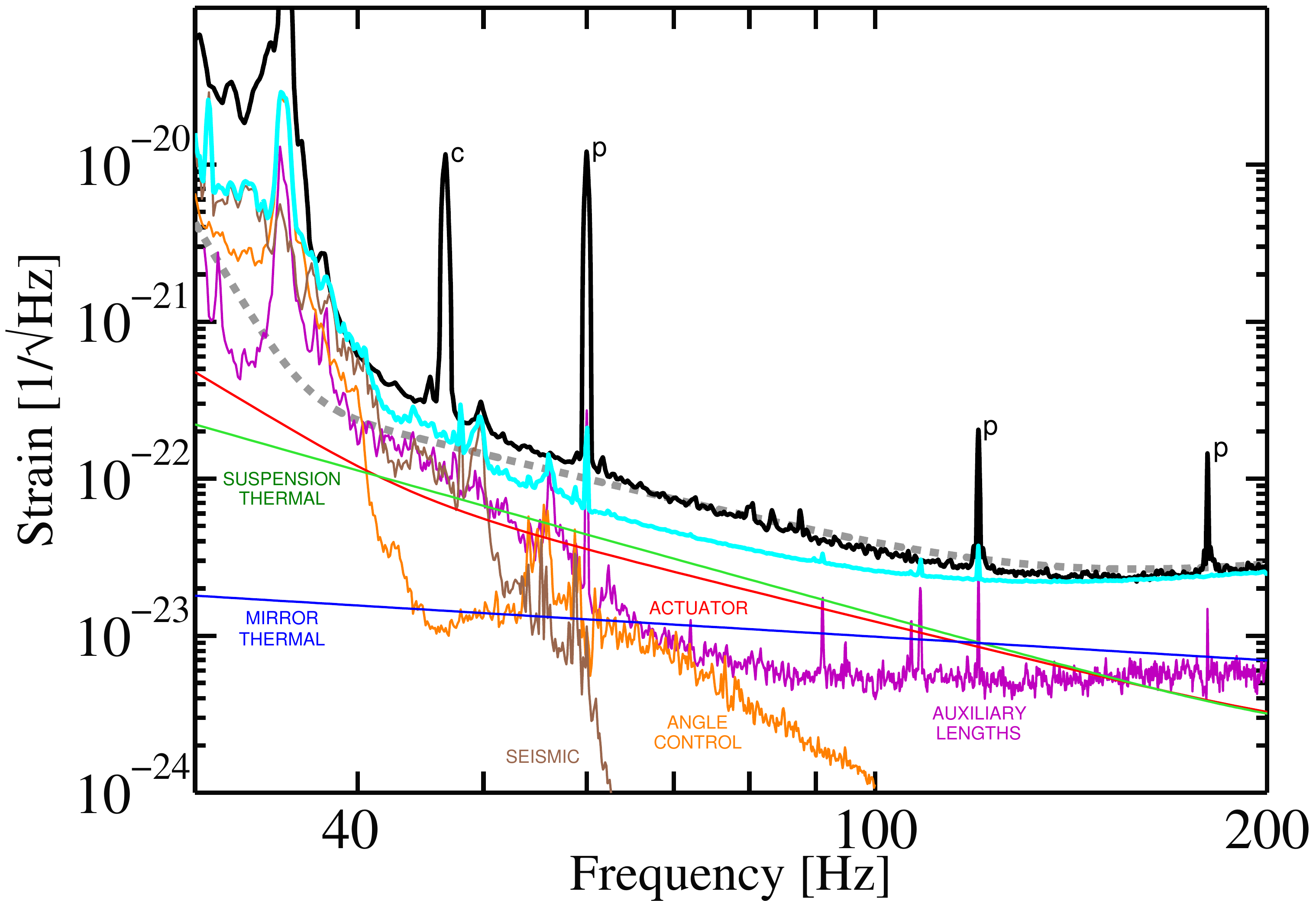}
                      \label{fig:iNBa}}
  \subfigure{
        \includegraphics[width=0.45\textwidth,height=6cm]{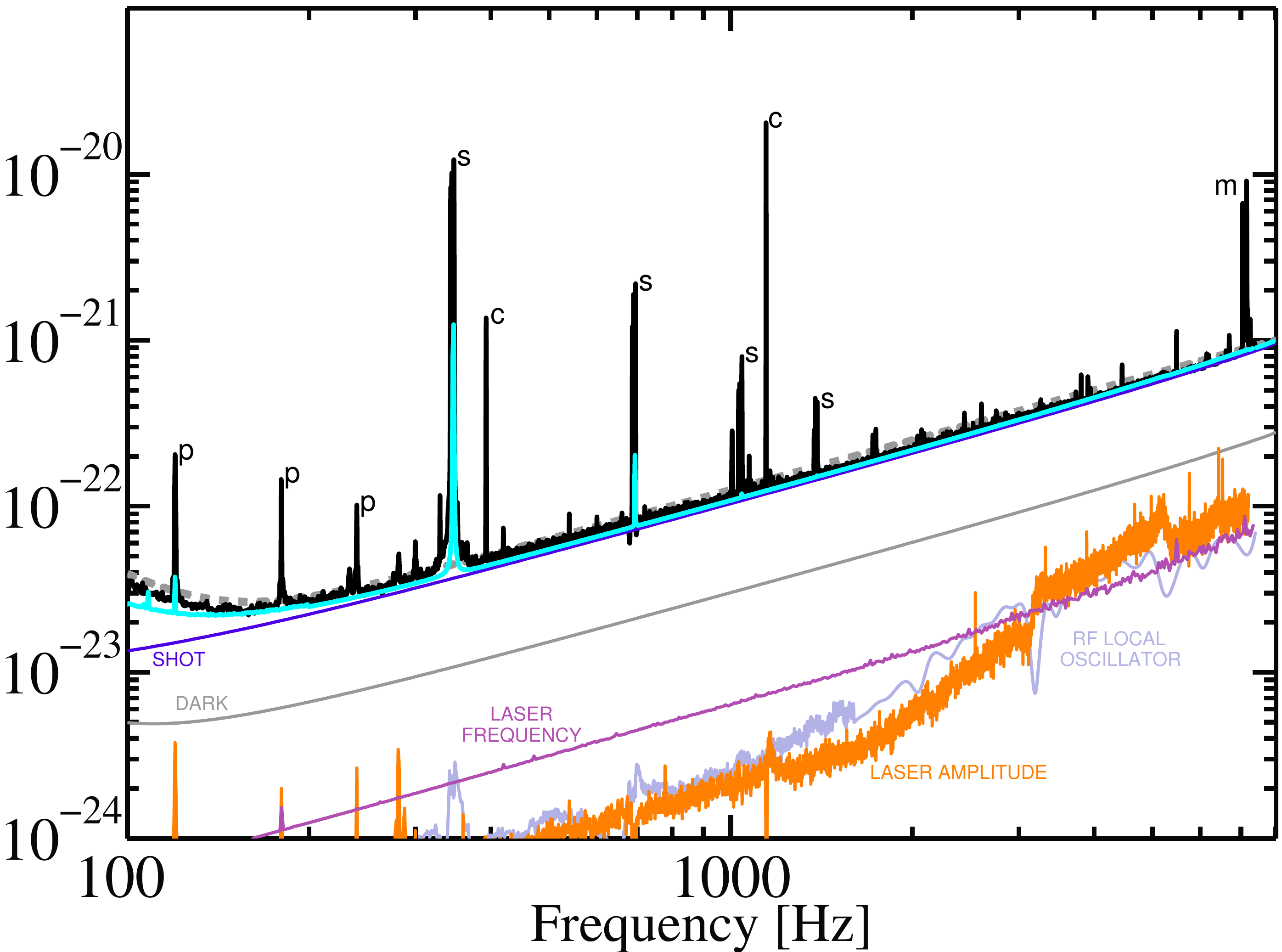}
                      \label{fig:iNBb}}
  \caption{(Color online) Noise Budget of the LIGO Hanford 4~km detector during the fifth 
    LIGO science run (S5)~\cite{PF:RPP2009}. The plot on the left shows mainly the noise
    sources that act as a force on the mirrors. The plot on the right shows noise sources 
    that appear as a phase noise on the light. The known peaks in the measured strain 
    data are indicated as (p) for power lines, (c) for calibration lines, (s) for the violin modes
    of the mirror suspension, and (m) for the mirror's internal eigenmodes. The CYAN trace 
    is a quadrature sum of all known noise sources and the BLACK trace is the measured 
    strain output of the interferometer. The discrepancy between these two 
    traces remains unexplained but is suspected to be due to excess friction in the 
    suspension wire attachments to the mirror. The dashed trace is the initial LIGO Science goal.}
  \end{center}
\end{figure*}

The reasonably good agreement between the initial design sensitivity goals and the final
performance of the LIGO and Virgo detectors (cf. Fig.~\ref{fig:Gen1_noise}) 
may lead to a false confidence in the accuracy
of those early estimates.
In reality, the commissioning period for all of the initial interferometers extended over
several years and greatly enhanced the understanding of the large interferometers. In
all cases, major hardware changes were made in order to bridge the gap between the
early performance and the science goals.

In the TAMA interferometer, the initial seismic isolation  was replaced with a
more elaborate (Virgo-like) system to greatly reduce the seismic noise in the 
1--100 Hz band~\cite{TAMA:2008}. 
For the GEO600 detector, an active seismic feedforward system,
a thermal compensation system, and scattered light mitigation techniques have been
installed over the years.
The Virgo interferometer was upgraded with an active thermal
lens correction system, isolation optics between the laser and the 
interferometer~\cite{Virgo:FI2008, Virgo:FI2010}, as
well as numerous control system upgrades~\cite{Virgo:LSC2011, Virgo:ASC2010}. 
LIGO also added a thermal compensation
system, as well as an active seismic isolation system for the Louisiana 
interferometer~\cite{abbott2004seismic, hardham2004multi}, acoustic isolation
chambers for the external optics, and an extensive upgrade to the digital control system.

Between the operation of the initial detectors and installation of the second generation
detectors, there was an additional scientific data taking run which followed major hardware
upgrades of the Virgo and LIGO detectors (Virgo+ and Enhanced LIGO) which
incorporated the noise analyses of the initial detectors (e.g. Figs.~\ref{fig:iNBa} and \ref{fig:iNBb}) and
several of the technologies in development for the second generation
machines.

In the rest of this Section, the most significant unexpected or non-ideal features are described
as well as the associated mitigation strategies.

\subsection{Excess Optical Loss}
With the use of power recycling, nearly all of the laser light is coupled into the
interferometer. Good matching between the interferometer arms ensures that only a 
small fraction ($\sim$few percent) escapes out of the anti-symmetric port. Most
of the laser power entering the interferometer is scattered into the surrounding vacuum system.
For all of the interferometers, the measured optical losses were significantly higher
than expected from the initial, table-top measurements~\cite{TAMA:Loss}. A small fraction 
of the losses came from absorption in the mirror substrate and on the high reflectivity dielectric
mirror coatings within the Fabry-Perot arms (in the case of LIGO, Virgo, and 
TAMA)~\cite{Ottaway:2006vb, GEO:Absorption, Aidan:2009}. Depending upon the level
of contamination, the absorption of the mirror surfaces ranged from 1--10 ppm, leading
to a wide range of problematic thermal gradients in the mirrors.

\paragraph{Scatter Losses}
As described below (Sec.~\ref{sec:mirrors}), perturbations in the mirror surface can
scatter light out of the interferometer. This scatter loss is the chief limit to the
power buildup within the resonant cavities. Although $\sim$ppm level losses
have been observed in small optical cavities~\cite{Ramin:Loss, uehara1995ultralow}, 
the round trip losses
in the Fabry-Perot arms of these large interferometers ranged from 100\,ppm (LIGO) to
300\,ppm (Virgo)~\cite{Virgo:Optical}. A small fraction of this was due to point defects 
(cf. Fig.~\ref{fig:aLIGO_mirror}) in the mirror coating. The largest fraction of the loss was
due to mirror surface perturbations at the scales of several centimeters.

\subsection{Optical Cross-Coupling}
All of these interferometers were designed with a high level of symmetry to passively
reject many
noise sources. Differential phase shifts in the interferometer arms (e.g., strain from a gravitational 
wave) directly produce a signal at the anti-symmetric port. Fluctuations of the incoming 
laser light or motions of the other mirrors also coupled through to the GW channel 
in (sometimes) unexpected ways and new techniques were developed to combat these issues.

\paragraph{Fluctuations of the Light}
The Michelson topology, in particular, is largely insensitive to amplitude
and frequency fluctuations of the illuminating laser light. By adjusting the length of the
interferometer arms microscopically, the anti-symmetric port is made to be nearly dark. In
this 'dark fringe' condition the common-mode rejection ratio for laser frequency noise was found
to be $\sim$200--1000 for the various interferometers, limited by the imbalance in scatter losses
between the arms. Laser power fluctuations can directly drive the mirrors through radiation pressure
and an imbalance of the power in the arms. Power fluctuations can also produce apparent
mirror fluctuations due to gain modulation of quasi-static offsets in the length control feedback 
loops of the
interferometer. Some of these operating point fluctuations are driven by seismic motion and so this
noise source comes from the \textit{product} of seismic motion and laser power fluctuations.

\paragraph{Local Oscillator Phase Noise}
In the TAMA, LIGO, and Virgo interferometers, the scheme that was used to read out differential
arm cavity strain is similar, mathematically, to the standard Pound-Drever-Hall technique
which is widely used with simple, rigid cavities. An important difference, however, is that both the
carrier field and the RF sidebands travel through a few optical cavities before being optically
recombined on the anti-symmetric port photodetector (cf. Fig.~\ref{fig:PRFPMI}).

In principle, phase noise of the oscillator used to generate the RF sidebands would
cancel during the demodulation of the heterodyne signal. The electronic local oscillator signal used
in the demodulation does not, however, experience the same temporal filtering that the optical sidebands do. 
Furthermore, the filtering
experienced by the optical sideband fields is not as simple as might be envisaged by modeling
the process by propagation of plane waves~\cite{Jordan:Light}. 
Each of the higher-order transverse modes of the 
sideband field experiences a different phase shift in the 
cavities~\cite{grote2008status, Stefan:Thesis}. In this way, the final recombined
signal depends in a detailed way on the mirror surface perturbations and, as explained 
below in Sec.~\ref{sec:TCS},
on the thermal state of the recycling cavity optics.

In order to reduce this noise to below the shot noise limits, multiple strategies were
employed: the mirror curvatures were adjusted with auxiliary heating lasers, the cavity
lengths were microscopically adjusted to match the optical and electronic paths, and finally,
an ultra-low-noise crystal oscillator~\cite{Wenzel:web} was used to reduce the source term 
by an order of magnitude (to a phase noise level of $< -160$\,dBc/Hz).

\paragraph{Motion of Auxiliary Mirrors}
Longitudinal motions of the other mirrors in the interferometer (e.g. the power recycling mirror and
the beamsplitter) couple to the GW readout weakly~\cite{Regehr:PhD}, 
but not so weakly that they can be completely neglected.

Motion of the beamsplitter (or more precisely, differential motion of the short Michelson interferometer
formed by the beamsplitter and the input test masses) couples in the usual way; the mirror motion
modulates the phase of the carrier field and produces a signal as if it was a gravitational wave.
This produces a weaker signal since it does not experience the resonant build-up of the 
arm cavities. However, this mirror has noise imposed on it by its feedback control
loop which is orders of magnitude above the shot noise limits of the GW channel. In order
to recover the quantum limited performance of the interferometer, this feedback noise was
filtered and injected into the end mirrors so as to cancel the initial noise 
injection~\cite{Mavalvala:2001vn}. This
\textit{feedforward} path was able to cancel the noise by a factor of $\sim30--100$ (for LIGO)
in the most sensitive frequency band. The Virgo feedforward system achieved several times
more cancellation by using an adaptive gain in this path~\cite{Virgo:LSC2010}.

The coupling of the power recycling mirror motion is less straightforward. This motion
produces a signal only through the existence of asymmetries. The imbalance
in the amplitude reflectivity of the arm cavities produces a carrier field at the anti-symmetric
port which is in the orthogonal phase from the gravitational wave signal sidebands. The
power recycling mirror motion modulates the phase of the RF sidebands and couples this
orthogonal phase field into the GW channel. This coupling was $\sim10\times$ smaller
than the Michelson coupling, but was dealt with in essentially the same way, although
the achieved cancellation factor was several times smaller.

Removing these noise sources allowed the interferometers to operate much closer to
their fundamental limits. An unpleasant side-effect is that the residual noise from these
processes is highly non-stationary, almost by definition. The static coupling path is
cancelled by these electronic cancellation paths, but time variation in the opto-mechanical
properties of the interferometers (due to temperature, seismic noise, beam pointing,
optical losses, etc.) produces large fluctuations in the residual coupling. The next
generation interferometers have the added complexity of also needing to cancel the
motion of the signal recycling cavity, but the added benefit of having much less
low frequency mirror motion resulting in less variation in the coupling constants.

% do the RF sidebands have to be unbalanced to get a PRC coupling???

% MICH is normal. We cancel it. Residual is non-stationary.

% PRC is unexpected. We cancel it a little bit. Residual is non-stationary.

% aLIGO will have PRC, MICH, and SRC.

% Virgo adaptive alpha beta

\subsection{Low Frequency Mirror Motion}
\label{sec:lowfmotion}
Simple estimates of the coupling of seismic vibration (e.g., Sec~\ref{sec:seismic}) to the
interferometer's strain channel assume that the coupling is essentially linear. During the
decade spent commissioning these interferometers, it became clear that this assumption
fails in a myriad of ways: large, low frequency motion produces noise in the GW
detection band.

\paragraph{Seismic Amplification}
As can be seen from Fig.~\ref{fig:VibrationIsolation}, below 1\,Hz, many of the isolation
systems amplify the ground noise. In the case of the passive systems this comes from the
lowest natural frequencies of the stacks and suspensions. In the case of the active systems,
this can come from the coupling of tilts into the active sensors or insufficient phase margin
in the control systems. As is well known from electronic filter design, it is necessary to
have some high resonances in the passband in order to have steep attenuation in the
stopband for reactive, low-pass filters. Such is also the case
for these mechanical vibration filters; damping the low frequency resonances would lead to
performance degradation in the GW band.
Unfortunately, this design tradeoff leads to an amplification of 
motion in the anthropogenic band which can be highly 
non-stationary~\cite{Daw:2004, VirgoSeis:2004, Saccorotti:2011hm, Accadia:2011ge}.

\paragraph{Noise from Damping}
In order to mitigate this problem, the suspension systems were designed to have some
capability of using 'cold damping': sensors local to each test mass can be used to sense
and suppress these high amplitude, low frequency motions.  This is only partially
successful. Although it is possible to reduce the motion somewhat, it proved impossible
to completely compensate the amplification without introducing excess noise into the 
GW signal band. The feedback filters must obey the Kramers-Kronig relations.

For the interferometer mirrors in the recycling cavity / Michelson area, the situation is more
complicated. 
Optics that are separated by much less than a seismic wavelength move coherently.
In the absence of active feedback systems, the differential motion among these optics is 
highly suppressed.
The noise of the local damping sensors has no such correlation, however. Attempting
to apply damping in such a situation actually amplifies the relative interferometric length
fluctuations at low frequencies. In practice, these issues require the delicate tailoring of 
the local damping
feedback filters and limits how strong the damping of the high Q mechanical resonances 
can be.

\subsection{Nonlinear Noise Generation}
These large, low-frequency motions all conspire to produce noise in the GW band through
several different nonlinear mechanisms.

%\paragraph{Nonlinearity of the Optical Readout}
%The fringe is cubic. DARM can be linearized, but its trouble for MICH/PRC.

\paragraph{Bilinear Angle to Length Conversion}
The large ground motions in the 0.1--1 Hz band produce angular fluctuations in the interferometer
mirrors through cross-couplings in the vibration isolation and suspension systems; the source
of the angular motion is chiefly horizontal motion of the ground and not tilt. These fluctuations
are partially cancelled by a complicated control 
system~\cite{Mavalvala:1998ur, Grote:2002wl, Virgo:ASC2006}
based on RF heterodyne detection of the optical wavefronts on quadrant photodetectors
and feedback through a MIMO (multiple input - multiple output) digital signal processing
system. The control system feeds some of the sensor noise in the GW detection band back into
the mirrors.
The mirror actuators are balanced so as to place the axes of rotation of the mirror at the center
of the laser beam spot position and this cancels the coupling of angular noise to interferometer
strain readout, to first order~\cite{Koji:glitch2006}.
Due to the residual low-frequency mirror motions, the resonating
laser beam moves around with respect to this null point by hundreds of microns. The angle to
strain coupling is therefore, non-stationary~\cite{Dooley:thesis}. 
During intense storms or times of high 
anthropogenic seismicity, the low-frequency noise of the detectors would become compromised
by this non-stationary noise source.

\paragraph{Actuator Nonlinearities}
In order to maintain the resonance condition of the interferometer, the control system must
compensate for the $\sim$micron scale motions below 1\,Hz while simultaneously 
introducing less than $10^{-19}$\,m of motion in the GW band around 100\,Hz. This
requires the mirror actuator to be highly linear: the upconversion of force noise must be
less than 1 part in $10^9$. While such a high dynamic range is just possible with modern
low noise electronics, it is not feasible to do so using magnetic actuators, due to the Barkhausen
effect~\cite{Durin:2004, bertotti1998hysteresis, Bittel:1969, barkhausen1919rauschen}. 
The low-frequency control forces are applied
to the mirror using magnet-coil pairs. The time-varying control forces, which are used to
compensate for the seismic motions, induce domain flips in the more loosely bound
domains of the magnets attached to the mirrors. In the NdFeB magnets used in LIGO
and TAMA, there were many weakly bound domains and the Barkhausen effect exhibited a
force noise upconversion of 1 part in $10^7$. The Virgo interferometer was instrumented
with SmCo magnets which have a much smaller Barkhausen effect. However, any
nearby ferromagnetic materials can lead to this fluctuating magnetic noise~\cite{Robert:Upconv}.
Future
interferometers are being designed to use multiple chain pendulums (as in Virgo and GEO)
so as to minimize the dynamic range requirements. To minimize the magnetic coupling, either
the magnets will be down-sized drastically or eliminated altogether in favor of electrostatic
actuators.

\subsection{Thermal Distortions}
\label{sec:TCS}
The small, but non-zero, optical absorption in the mirrors of the interferometers
produced significant thermal gradients within the optics. These gradients produced
distortions of the mirror surface (thermal expansion of the glass) as well as a significant
thermal lensing within the substrate (temperature dependence of the refractive 
index)~\cite{Winkler:TCS, Strain:TCS, Hello:1993, hello1990analytical}. The presence of low 
levels of contaminants on the optics' surfaces led to higher than anticipated levels of
absorption.

There are several mechanisms by which thermal distortions can lead to instability and
degraded noise performance in the interferometers. The simplest mechanism is through
reduction of signal; differential thermal lensing in the input test mass mirrors reduces the
spatial overlap of the GW signal sidebands with each other. This contrast defect also increases
the shot noise level at the anti-symmetric port. Thermal lensing in the recycling cavity
optics can also destabilize the angular control system by reducing the sensitivity to certain
degrees of freedom and destabilizing the feedback control matrix.

A particular optical design choice exacerbated some of these problems. The recycling
cavities were made much shorter (for practical reasons) than the long arms. With the
large beams required for low thermal noise, such short cavities are geometrically 
unstable (cavity g-factor~\cite{Siegman:Lasers} near unity)~\cite{Andri:PRCoffset}. 
Small thermal distortions were 
found to drive the system into instability due to the degeneracy among the higher order 
spatial modes.

In order to compensate for this effect, active thermal correction systems were installed
to smooth out the thermal 
gradients~\cite{Stefan:Thesis, Ryan:TCS, Hild:TCS, aVirgo:TCS, Bas:Virgo2010} as
well as to improve the fringe contrast at the anti-symmetric port.

Due to the troubles with degenerate cavities, the KAGRA and Advanced LIGO detectors 
are adding extra optics in their design to break the
modal degeneracy in the recycling cavities~\cite{Arain:2008fu, Granata:2010}.

% What to say about these lemons?
% \begin{itemize}
% \item Not enough low frequency seismic isolation...LIGO HEPI, Virgo ACC, TAMA SAS, GEO FF
% \item Excess loss in the optics - sort of a mystery. Check Hiro; any Virgo know-how?
%           Backscatter from aux places - partly due to dots on optics
% \item Osc Phase Noise unexpected in all? Why?
% \item Barkhausen
% \item LIGO: HEPI, TCS, Baffles, Acoustic Huts, Crystal Osc, OMC, DC Readout, digital, A2L, MICH-CORR
% \item Virgo: Damping, TCS, In-vac detectors, A2L, alpha, beta
% \item GEO: OMC, DC readout, TCS, in-vac AS port
% \item TAMA: SAS, A2L, coil adjustment, scattered light dumping, transmon dumping, staged commissioning, what else?
% \end{itemize}

% ===============================================================
\section{Second Generation Detectors}
\label{sec:2G}
The purpose of the second generation interferometers is to achieve such a strain sensitivity
that the detections of gravitational waves should become fairly regular, enabling the use
of these detectors as astronomical tools. They are roughly an order of magnitude more sensitive
than the first generation detectors (cf. FIg.~\ref{fig:Gen2_noise}). The world-wide 
network comprising Advanced LIGO~\cite{Gregg:aLIGO2010, aLIGO:web}, 
Advanced Virgo~\cite{aVirgo:web, aVirgo:TDR}, GEO-HF~\cite{Harald:2010un}, and 
KAGRA~\cite{Somiya:2011tb, LCGT:web} will all use power and signal recycling and 
a variant of tunable resonant sideband extraction.

\subsection{Monolithic Silica Suspensions}
\label{sec:monolithic}
\begin{figure}[h]
   \centering
   \includegraphics[width=\columnwidth, trim=0 0 0cm 0, clip]{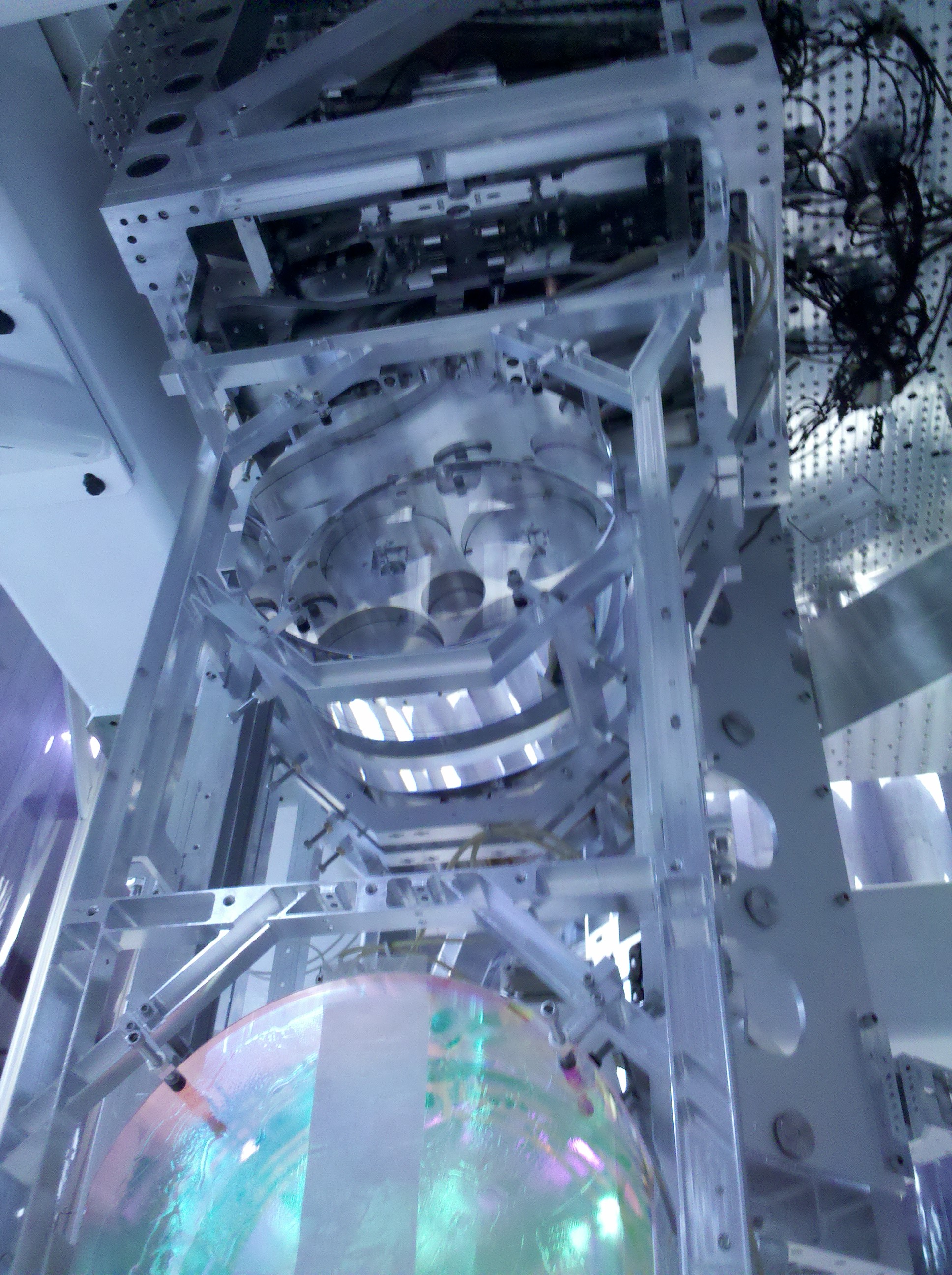} 
   \caption{(Color online) Advanced LIGO quadruple suspension. The final stage is a 40~kg mirror 
                  suspended by four laser-welded silica fibers.}
   \label{fig:QuadSUS}
\end{figure}

The first generation LIGO and Virgo interferometers were somewhat limited by thermal noise
in the mirror suspensions. This was partially due to the intrinsic dissipation of the steel
wires and partially due to excess friction in the wire attachments~\cite{PF:RPP2009}. In order
to avoid both of these problems, the new suspensions are nearly monolithic: instead of a steel
pendulum wire, a high quality fused silica 
(sapphire for KAGRA~\cite{LCGT:heatextract, uchiyama1998cryogenic}) fiber is drawn, and then 
bonded directly to silica attachments on the
mirror barrel~\cite{SUS:2002, Aston:2012, SUS:2012}. As most of the elastic energy (and
therefore the dissipation) is concentrated near the bending points at the 
ends (cf.\,Sec.~\ref{sec:sus_thermal}),
the cross sectional shape, near the ends, is optimized with respect to the 
noise~\cite{SUS:FEA2009}. 
This type of silica suspension has been used in the GEO600 and Virgo+ interferometers. An
example Advanced LIGO suspension is shown in Fig.~\ref{fig:QuadSUS}.

As with the mirror coatings (cf.\,Sec.~\ref{sec:mirror_thermal}), both the dissipation due to internal
friction as well as the thermo-elastic (Zener) damping need to be considered. For highly stressed
fibers, one must also consider the temperature dependence of the Young's modulus ($dY/dT$): the fundamental
thermodynamic temperature fluctuations which produce the usual thermoelastic noise via the
thermal expansion coefficient  also drive the stressed suspension fiber by changing the
Young's modulus~\cite{Phil:Nonlinear}. Fortuitously for LIGO and Virgo, fused silica has a positive
$dY/dT$; 
the result is that an appropriate fiber diameter can be chosen to cancel these 
thermoelastic effects.

Finally, studies of the fiber's mechanical loss as a function of fiber dimension have revealed that
the Q is limited by defects in the fiber's 
surface~\cite{gretarsson1999dissipation, Andri:Surface, Penn20063, heptonstall2010investigation, GIles:SUS2012} 
and not by the intrinsic mechanical dissipation of fused silica. Several decades of experience 
with surface treatment of quartz oscillators and quartz fibers have informed the current 
design for GW detectors. The fibers which are now used for the Advanced LIGO mirrors are pristine
with a small concentration of residual defects in the surface. These defects in addition to the losses
in the welded attachment point dominate the Brownian thermal noise in the suspension.
Techniques for evading this limit for the next generation are described in 
Sec.~\ref{sec:cryo_sus}.

\subsection{Mirror Metrology}
\label{sec:mirrors}
To support Gaussian beam shapes in the Fabry-Perot cavities, the mirrors are polished to have
spherical profiles. Deviations from the ideal shape reduce the overall interferometer performance
in a number of ways. Roughly speaking, perturbations at small spatial scales promptly scatter light 
out of the cavity. Larger scale defects distort the ideal TEM$_{00}$ eigenmode of the
arm cavities. To compute the power lost into wide angles, one needs to know only
the bidirectional reflectance distribution function (BRDF)~\cite{bass2009handbook} of the mirror, which is 
readily obtained from measurements of the mirror surface map~\cite{Hiro, Walsh:99}. To first order, this
distinction between small and large scales can be made in the following way: light scattered
from a mirror which falls off the opposing mirror of the cavity is lost and does not contribute to
the cavity mode distortion. In the LIGO case this corresponds to an angle of 
$\theta_{\rm lost} \sim r_{\rm mirror}/(L_{\rm FP})$ and a spatial scale 
of $x_{\rm rough} \sim \lambda/\theta_{\rm lost} \sim 2\, \rm cm$.

For the larger
spatial scales the situation is complex; the scattered field is captured on the far mirror and so it is not
precisely 'lost'. Rather, the resulting distortion in the cavity field results in an imperfect interference
at the Michelson anti-symmetric port.
At the smaller scales, however, a good approximation for the power lost due to surface roughness is
\begin{equation}
\frac{P_{\rm scatter}}{P_{\rm incident}} = \left(\frac{4 \pi \sigma}{\lambda} \right )^2
\label{eq:loss}
\end{equation}
where $\sigma$ is the RMS surface roughness and $\lambda$ is the laser wavelength.
Finally, at the smallest scales the dominant source of the loss is a random distribution of
point scatterers. The ultimate nature of these points has not been discovered as of this writing; the common
wisdom is that they are density or index defects in the dielectric coatings. The scatter from these
points is therefore treated as either Rayleigh or Mie scattering (depending upon the defect size).
\begin{figure}[h]
   \centering
   \includegraphics[angle=270, width=\columnwidth, trim=0cm 0 0cm 0, clip]{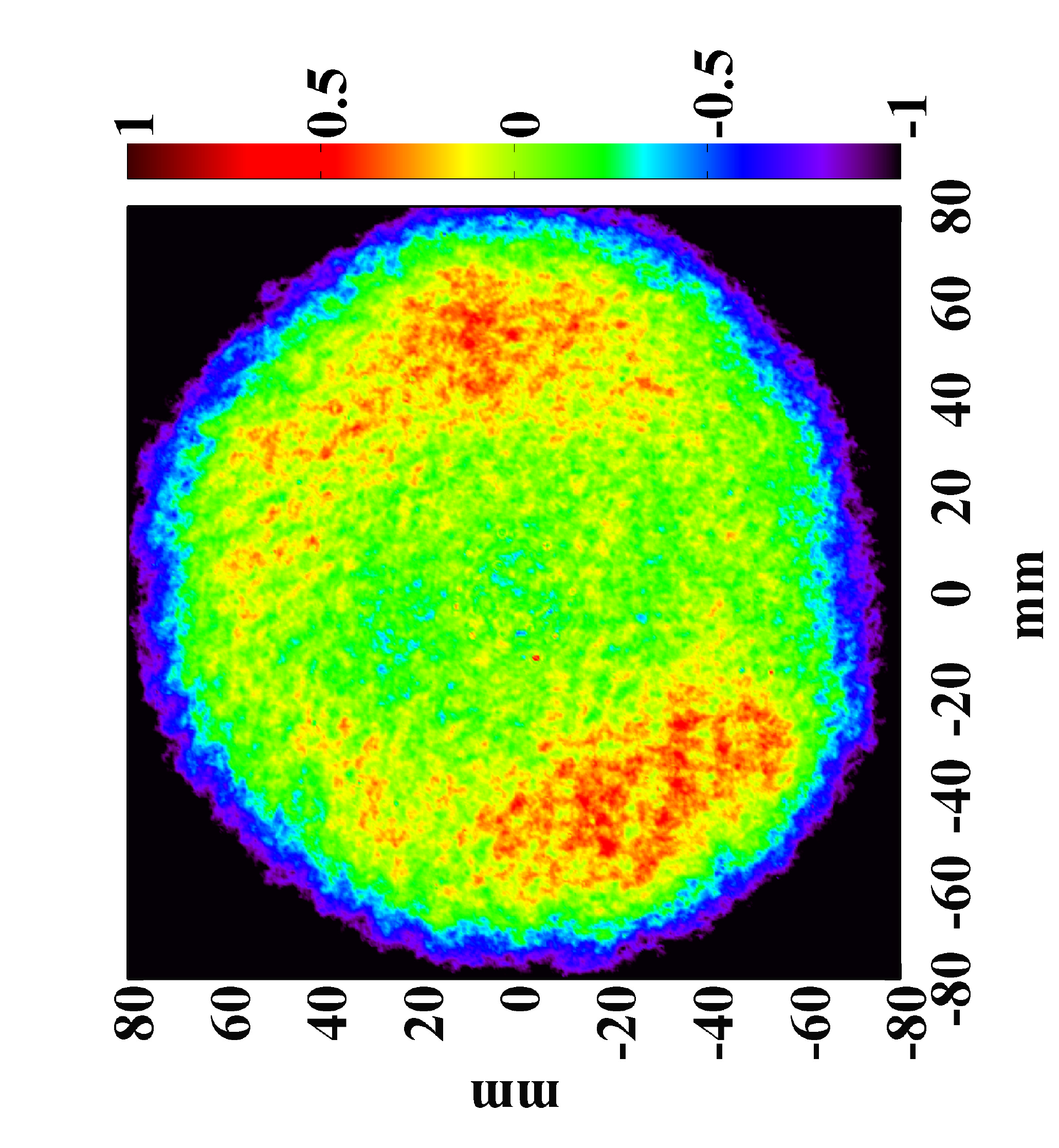} 
   \includegraphics[angle=0, width=\columnwidth, trim=-1cm -1cm -1cm -1cm, clip]{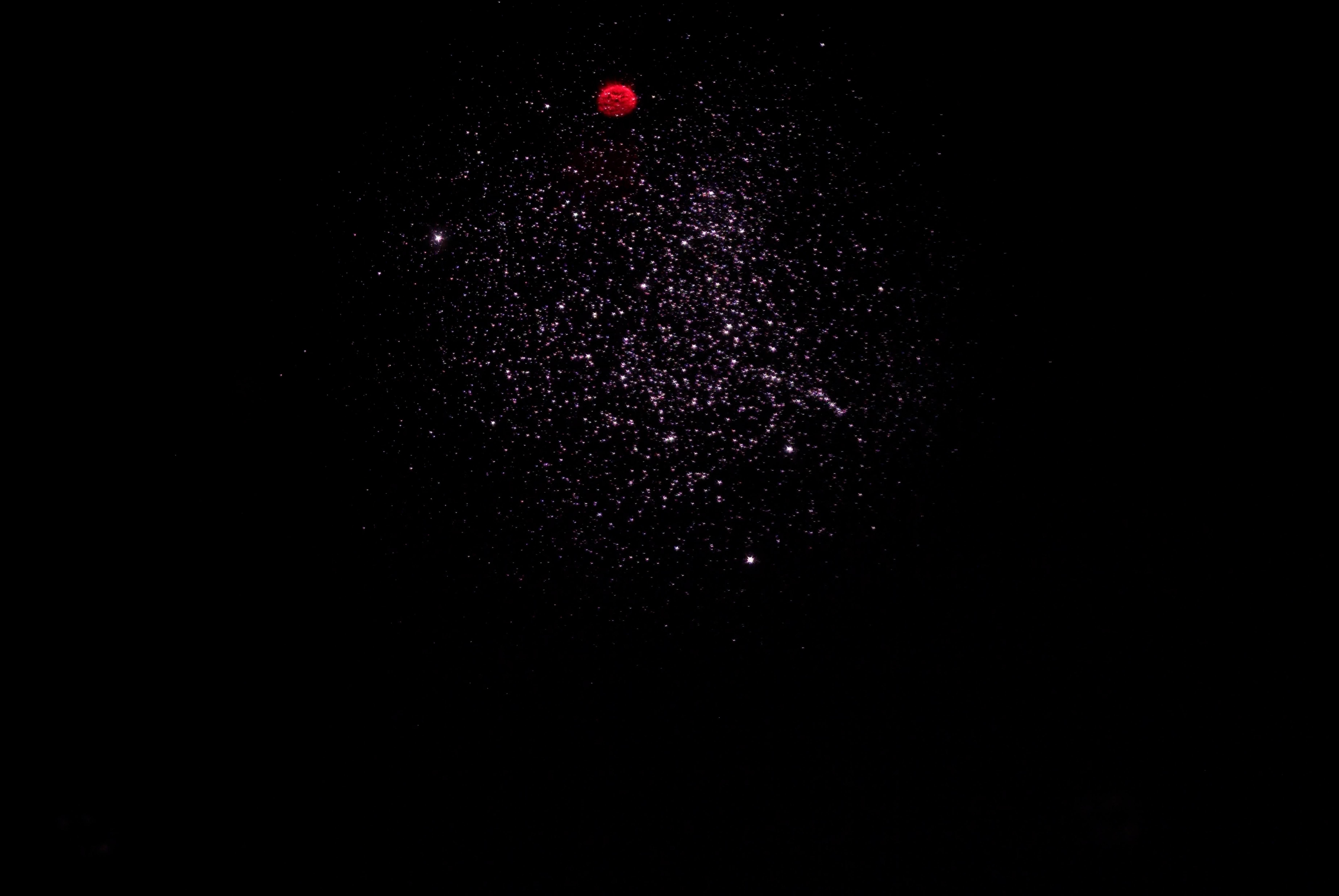} 
   \caption{(Color online) (top) Surface phase map (in units of nm) of one of the Advanced LIGO arm
                  cavity mirrors after applying the high reflectivity mirror coatings~\cite{Gari:Zygo}.
                  (bottom) infrared image of an initial LIGO arm cavity mirror taken with the cavity
                  locked, highlighting the abundance of point defects. The red oval is the diffuse scatter
                  from an auxiliary beam used for tracking the mirror angle~\cite{Cheryl:Picture}}
   \label{fig:aLIGO_mirror}
\end{figure}

Power loss limits the maximum achievable power recycling buildup, reduces the
maximum benefits achievable from QND techniques by degrading the quantum 
entanglement of the light (cf. Sec.~\ref{sec:QNDlosses}),
and introduces technical noise from backscatter (cf. Sec.~\ref{sec:scatter}). 
Over the past decade, 
an intense development effort has led to improvements in the mirror polish 
on both long and short scales. The combination of
extremely accurate metrology~\cite{Zygo:2011} of the mirror profile and the use of ion 
beam figuring has resulted in an order of magnitude smoother mirror (see 
Fig.~\ref{fig:aLIGO_mirror}) than the first
generation GW interferometers. It remains to be seen if a similarly good surface can
be achieved for sapphire (as is planned for 
KAGRA~\cite{uchiyama1999mechanical, somiya2012detector}) or silicon (which is being considered
for future cryogenic detectors).

\begin{figure}[h]
   \centering
   \includegraphics[width=\columnwidth]{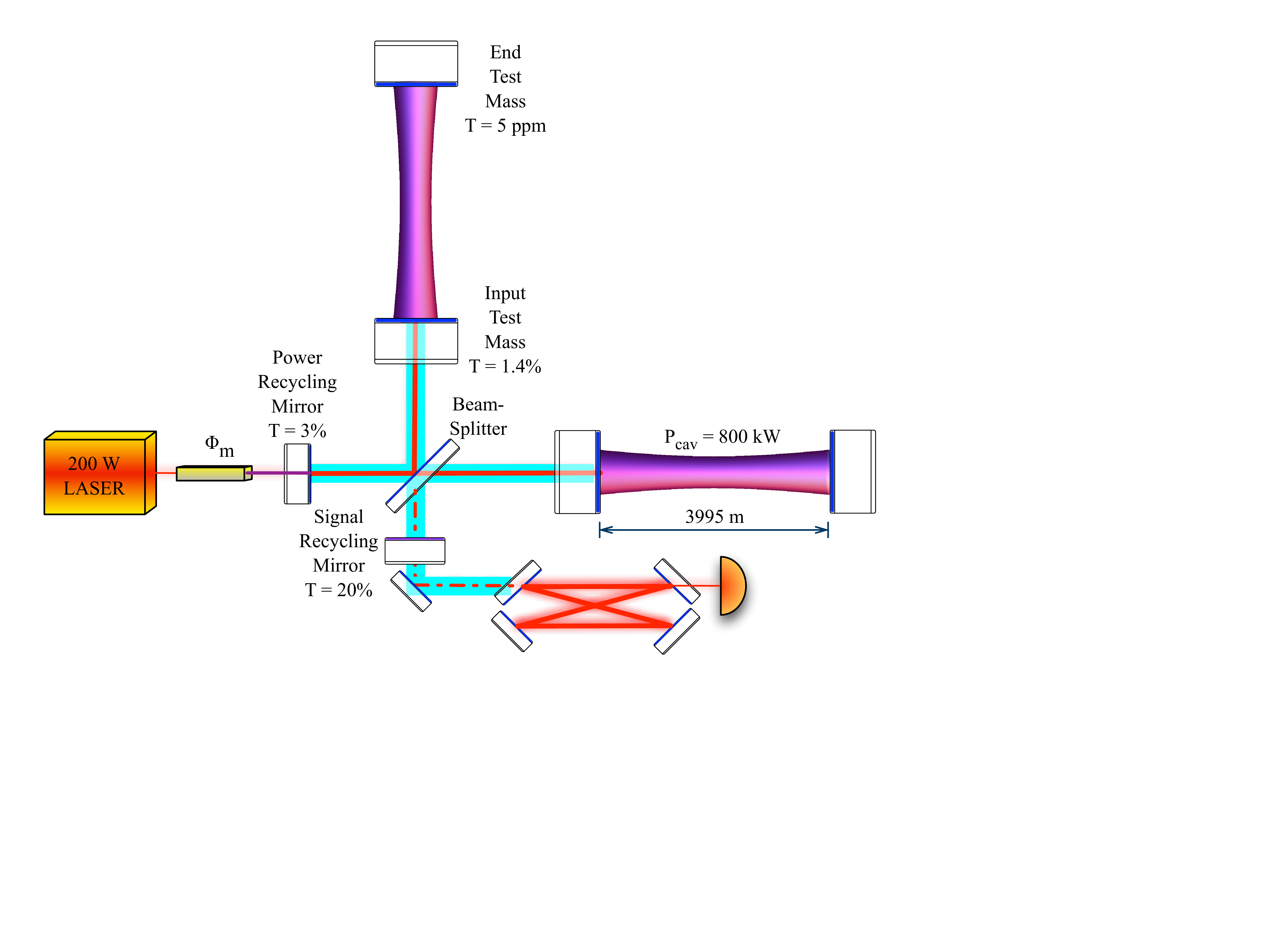} 
   \caption{(Color online) Schematic of the Advanced LIGO interferometers. The output
                  beam at the anti-symmetric port is filtered by a rigid bow-tie cavity to 
                   remove the RF sidebands
                  and the higher-order spatial modes that come from distortions in the optics.}
   \label{fig:DRFPMI}
\end{figure}

\subsection{Dual Recycling}
\label{sec:DR}
\begin{figure}[h]
   \centering
   \includegraphics[width=\columnwidth, trim=0 0 -3cm 0, clip]{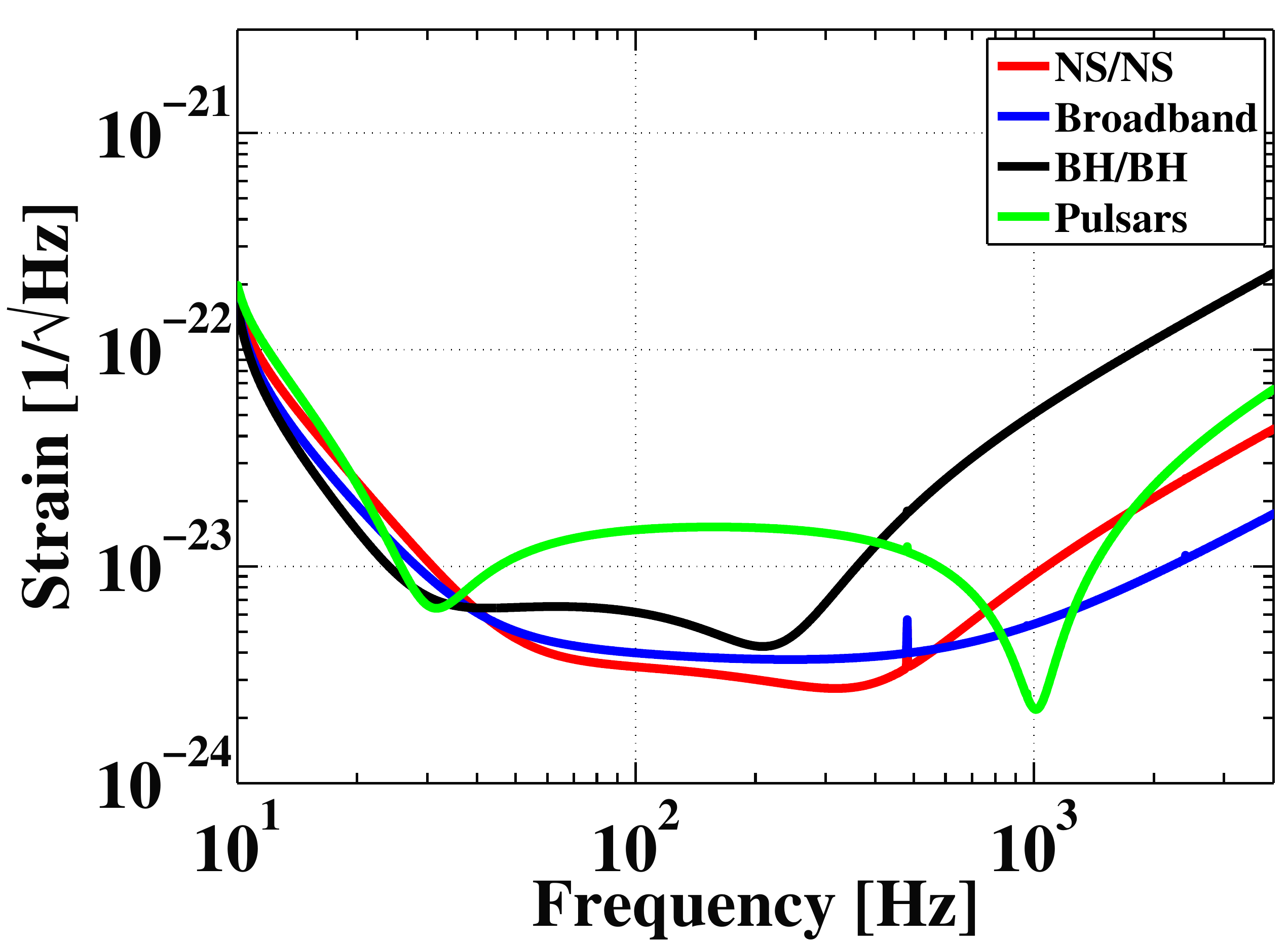}
   \caption{(Color online) Detector configurations to target particular astrophysical
                  sources. Shown are the optimal tunings of the Advanced LIGO interferometer
                 for (NS/NS) neutron star binary inspirals, (BH/BH) for intermediate mass
                black hole binary inspirals, and (Pulsars) for narrowband sources (such as pulsars)
                emitting gravitational radiation around 1~kHz. The Broadband configuration
                has the best overall sensitivity and is expected to be the easiest to operate.}
   \label{fig:aLIGO_cases}
\end{figure}

The dynamic tuning capability of signal recycled interferometers (e.g. Fig.~\ref{fig:DRFPMI}) 
is a powerful one and makes
these instruments qualitatively different from their predecessors. By adjusting the length
of the signal recycling cavity by fractions of a wavelength, the coupled resonance between the
arm cavities and the signal mirror can be fine-tuned to match the frequency content of
astrophysical sources as shown in Fig.~\ref{fig:aLIGO_cases}. The low frequency
response is due to the radiation pressure induced optical spring (cf. Sec.~\ref{sec:Quantum})
and can also be tuned by adjusting the laser power. Both the signal mirror position and the laser
power can be adjusted remotely to arrive at a new configuration within minutes, in principle.

The baseline configuration of the Virgo and KAGRA
interferometers will have the signal recycling cavity slightly detuned from resonance
in order to maximize the sensitivity to a specific astrophysical source: the
inspiral of a binary neutron star system~($M_1 = M_2 = 1.4\,M_{\odot}$). The Advanced LIGO 
and GEO-HF interferometers will begin operation in a broadband resonant sideband
extraction configuration.

\paragraph{Mode Healing}
In addition to the ability to tune the response of the interferometer to the spacetime strain, the
interferometers with signal recycling cavities also exhibit the phenomenon known as
\textit{mode healing}~\cite{Mee1988, StMe1991, Gerhard:1998, DMC:healing93}. Without this mirror,
differences in the spot size or wavefront curvature of the beams from the two Michelson arms
result in an imperfect destructive interference at the Michelson anti-symmetric port (where the GW signal
is recorded). This extra light produces no signal but contributes to extra shot noise as well as
introducing technical difficulties with the interferometer control system~\cite{Nic:2011}. The signal recycling
cavity can be designed to be anti-resonant for this 'junk light' so as to preferentially keep it
from getting to the detection port while allowing the signal light to 
pass~\cite{YiPan:SRC, Brett:2003}. With different storage times for each higher-order spatial
mode, some of the energy which is initially scattered out from the fundamental mode can
come back into this mode due to the mode mixing which occurs at each perturbed optical
surface. Depending on the details of mirror roughness, signal cavity tuning, and g-factors of the
arms and signal cavity, there can be either mode healing or mode harming for the fundamental
mode.

% \begin{itemize}
% \item makes phase shift on the HOM
% \item some HOM get suppressed
% \item some HOM get enhanced
% \item contrast can be improved
% \end{itemize}

% Second Gen comparison tables
\begin{table*}[ht]
    \begin{centering}
      \begin{tabular*}{0.85\textwidth}[t]{@{\extracolsep{\fill}}l | p{30mm}      p{30mm}        p{30mm}           p{30mm} }
\hline
Detector                         & KAGRA              & GEO-HF              & Adv. Virgo      & Adv. LIGO \\
\hline
Arm Length [m]               &  3000             &    600                    & 3000               &  3995        \\
Mirror Mass [kg]              &    27               &   5.5                       & 40                   &    40     \\
Beam Spot Size [cm]         &   3.5               &   2.4                     & 6                    & 5.9 \\
\# of seismic stages       &   1 + 5           & 1 + 3 + 3               &  1 + 6              & 1 + 2 + 4 \\   
Stored Power [kW]           &   400             &   10                       &  760                  &    800        \\
Strain Noise [$10^{-23}/\sqrt{\rm Hz}$]  &  0.3    & 3.5          & 0.3                    & 0.3 \\
Sensitive Band [kHz]       & 0.02--3           & 0.1--5                  & 0.02--3            & 0.01--5 \\
Location                         & Japan           & Germany                & Italy                  & USA \\
\hline
    \end{tabular*}
    \caption{Comparison of 2$^{\rm nd}$ generation interferometers (KAGRA\,\cite{Somiya:2011tb},
                   GEO-HF\,\cite{Harald:2010un}, Advanced Virgo\,\cite{aVir2009,aVirgo:TDR}, and Advanced 
                   LIGO\,\cite{Gregg:aLIGO2010, aLIGO:web}).
                  The numbers in the seismic row refer to the number of external active, internal passive, and 
                  pendulum suspension stages, respectively.}
    \label{t:2G}
  \end{centering}
\end{table*}

\subsection{High-Power Opto-Mechanics}
\label{sec:Lasers}
\begin{figure}[h]
   \centering
   \includegraphics[width=\columnwidth, trim=0 0 -3cm 0, clip]{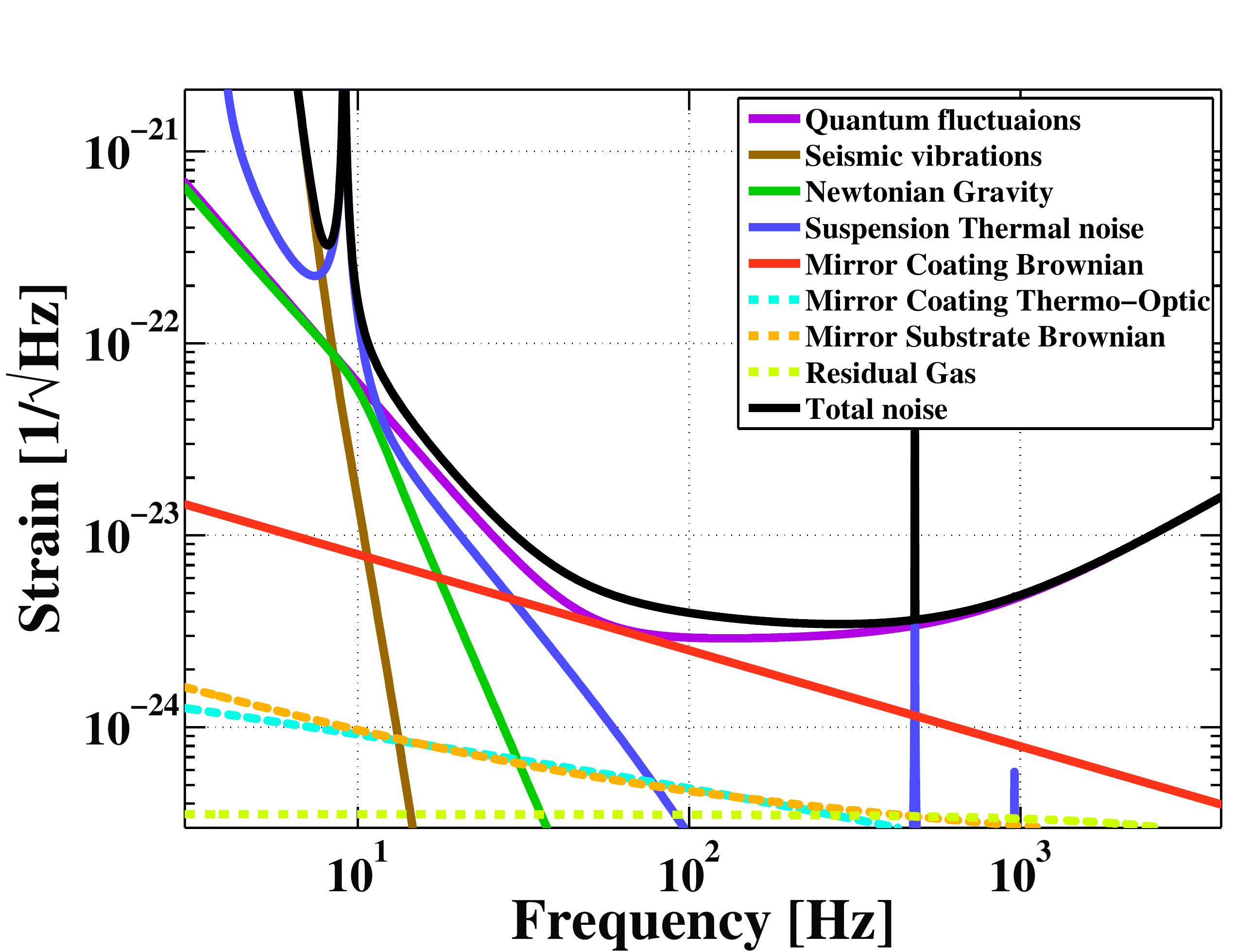} 
   \caption{(Color online) Noise budget of the Advanced LIGO interferometers operating
                  in a broadband configuration with the parameters of Table~\ref{t:2G}.}
   \label{fig:aLIGO_gwinc}
\end{figure}

These new high-quality optics make it possible to use massively higher power levels. The designs of
the Advanced interferometers call for storing 0.5\,--\,1\,MW in the arm cavities in order to improve the
shot-noise limited sensitivity (shown in Fig.~\ref{fig:aLIGO_gwinc}); this is a factor of 10\,--\,50 higher than the previous generation. 

While some differences exist among the laser designs for LIGO, Virgo~\cite{Greverie:10}, and KAGRA, 
they share a set of common themes. First, a low noise master oscillator ($\sim$1--2\,W) is 
amplified with one or two amplifier stages. The light is then passed through a low finesse
cavity in order to filter angular fluctuations and to provide filtering of amplitude noise at RF
frequencies.

The LIGO design has a 2\,W Innolight Non-Planar Ring Oscillator (NPRO) 
as the master oscillator, followed by a single pass power amplifier
with an output of 35\,W. This configuration was used as the laser for the Enhanced 
LIGO~\cite{Tobin:DC}. This 35\,W system has now been augmented by a high power stage to 
produce 200\,W of single mode 
light at 1064\,nm~\cite{Benno:2008}.

\begin{figure}[h]
   \centering
   \includegraphics[width=\columnwidth, trim=0 0 -2cm 0, clip]{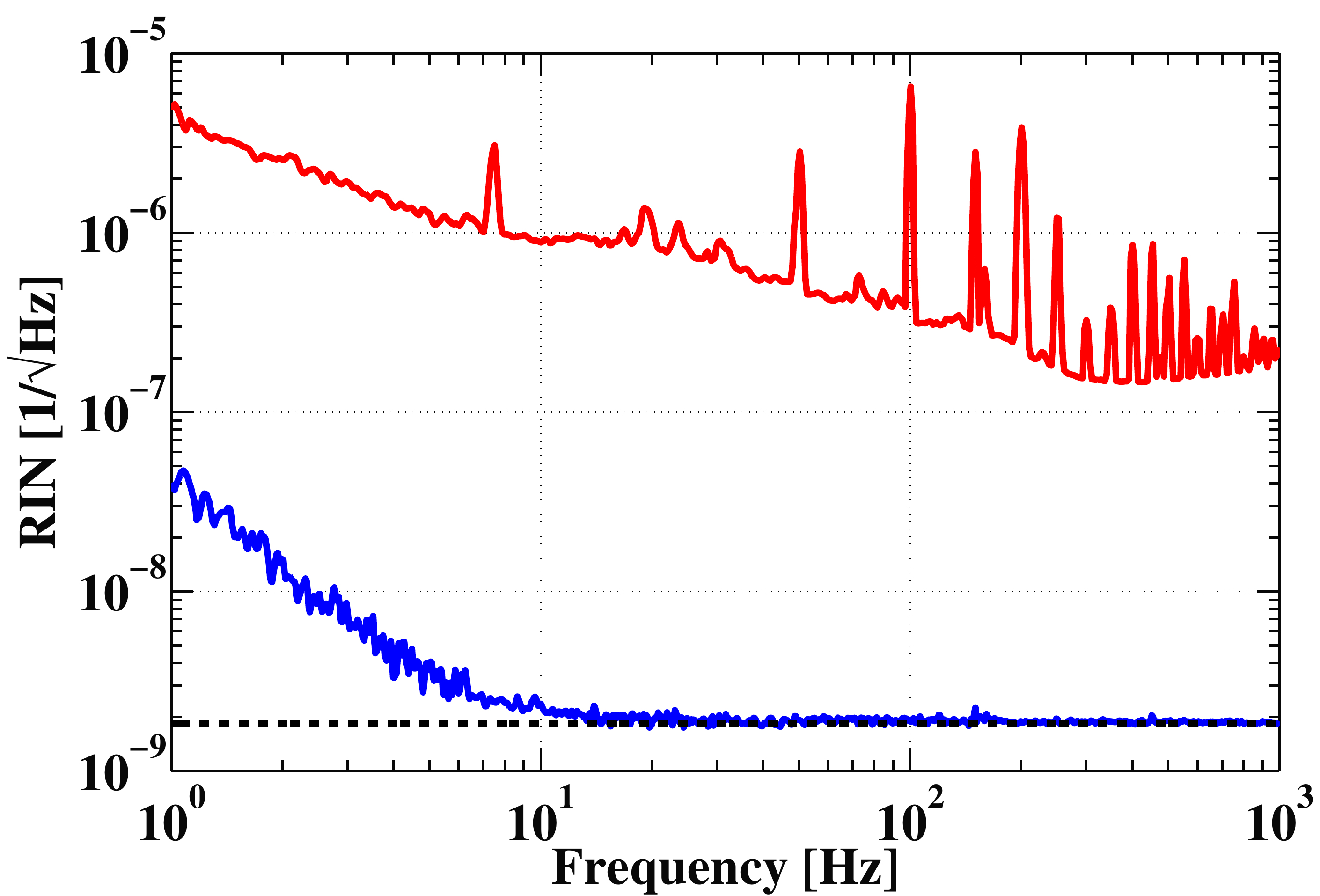}
   \caption{(Color online) Relative power fluctuations after stabilization of a prototype laser system:
                  (RED) free running laser noise, (BLUE) stabilized level (out of loop), (BLACK) shot noise limit.
                   The goal for Advanced LIGO is $2~\times~10^{-9}~/\sqrt{\rm Hz}$~\cite{Kwee:09}.}
   \label{fig:IntensityNoise}
\end{figure}

With the direct (homodyne) readout scheme adopted for LIGO, Virgo, GEO, and KAGRA, the laser 
power fluctuations show up directly in the readout signal. At high power levels, the dominant coupling path 
for laser power fluctuations is not so direct. The classical radiation pressure 
from the laser power fluctuations pushes the mirrors directly and couples to the 
anti-symmetric port through the imbalance in the finesse of the arm cavities.
To mitigate this somewhat, multi-stage active stabilization is used to suppress the raw 
laser noise by several orders of magnitude.
In the end, the relative power stability of the light (shown in Fig.~\ref{fig:IntensityNoise}) entering the interferometer is 
$\lesssim 10^{-8} /\sqrt{\rm Hz}$ in the GW band~\cite{Kwee:09}. 

The sensitivity to laser frequency noise is expected to be no greater than it was for
the first generation detectors. Therefore, the same strategy of using a multi-stage
active stabilization scheme~\cite{Mavalvala:2001vn, PF:RPP2009, Virgo:fnoise2009} 
is expected to be sufficient.

\subsubsection{Angular Instabilities}
\label{sec:SiggSidles}
\begin{figure}[h]
   \centering
   \includegraphics[width=\columnwidth, trim=0 0 0cm 0, clip]{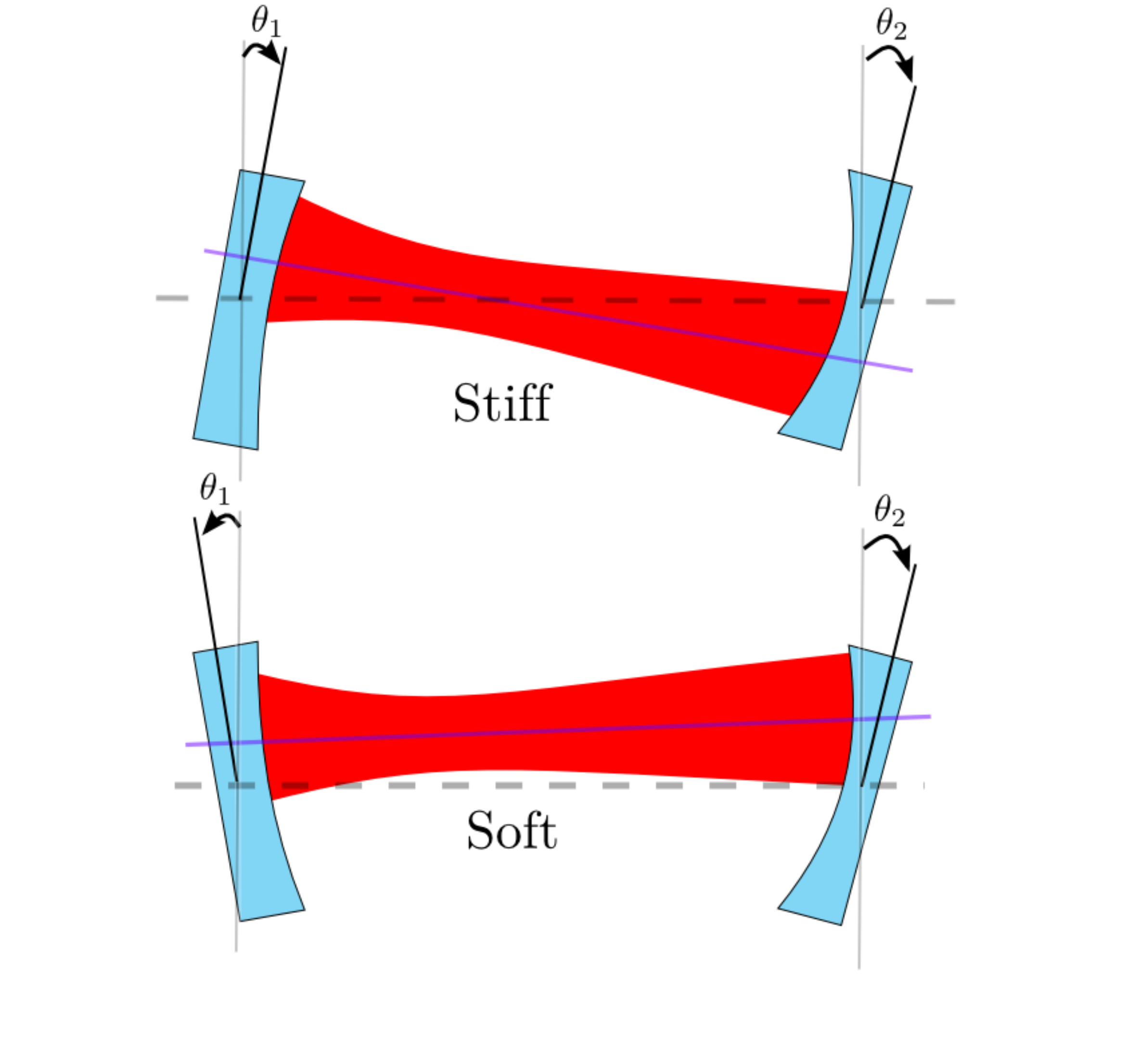}
   \caption{(Color online) The common and differential angular modes of the Fabry-Perot cavity 
                  mirrors are softened (bottom)
                 and stiffened (top) by the radiation pressure torque.}
   \label{fig:SiggSidles}
\end{figure}

In 2002, the LIGO interferometers were beset by weak angular instabilities as
the stored powers in the arm cavities exceeded $\sim$\,1\,kW. Sidles and Sigg 
pointed out~\cite{Sidles:2006un} that these
instabilities must be due to radiation pressure overwhelming the mechanical restoring torques 
of the mirror suspensions.

The mechanism behind this 'Sigg-Sidles' instability is illustrated in Fig.~\ref{fig:SiggSidles}. In this
picture the radiation pressure couples the suspended optics at either end of the cavity. Including this optical
torque, the two mirror system can now be seen as having a 'soft' and 'stiff' mode. With enough stored
optical power, the radiation pressure torque can statically de-stabilize the cavity in the soft mode.

In the individual mirror angle basis, we can define an optical torsional stiffness matrix:
\begin{equation}
\kappa_{RP} = \frac{2 P}{c} \frac{L}{1 - g_1 g_2} 
     \begin{pmatrix}
       -g_2 & 1 \\
          1   & -g_1
     \end{pmatrix}
\label{eq:SiggSidles}
\end{equation}
where $P$ is the cavity power, $L$ is the cavity length, and the cavity g-factors for each mirror
are defined as $g_i = 1 - L/R_i$, where $R_i$ is the radius of curvature of the $i^{th}$ mirror. 
The cavity instability occurs when the eigenvalue from this torsional matrix 
corresponding to the 'soft' mode exceeds
the mechanical torsional stiffness of the mirror suspension.

As described in Sec.~\ref{sec:mirror_thermal}, the cavity beam sizes are maximized to reduce the impact
of the mirror's thermal noise. This has the unfortunate side-effect of amplifying these optical
torsional stiffnesses. The large beam sizes can be realized by utilizing either a plane-parallel
or concentric cavity design~\cite{Siegman:Lasers}. As can be seen from Eq.~\ref{eq:SiggSidles}, the
concentric design (which has negative g-factors) causes the dominant mode to have a positive sign
and thereby contribute to the 'stiff', self-aligning mode. The plane-parallel design, on the other
hand, has positive g-factors. In this case the denominator of Eq.~\ref{eq:SiggSidles} blows up
as the g-factors approach unity (as they must to increase the spot sizes). For this reason,
the concentric design has been adopted for all modern GW detectors.

This 'Sigg-Sidles' effect was first characterized for the initial LIGO detectors~\cite{Hirose:10}
and then subsequently in the Enhanced LIGO where a modal control approach was used to
stabilize it~\cite{Dooley:thesis}. This modal approach seems to be sufficient to control
the instability~\cite{aLIGO:ASC} but the noise from the control system is likely to be comparable to the
more fundamental limits (e.g., suspension thermal noise).

\subsubsection{Parametric Instabilities}
With high circulating powers in the arm cavities, a parametric instability can occur involving
the high-Q mechanical modes of the mirrors and higher-order transverse optical modes of the
Fabry-Perot cavity~\cite{BSV2001, BSV2002, StVy2007}. Although not observed in 
the first generation detectors, similar instabilities have been observed in toroidal 
microcavities~\cite{Vahala:PRL2005} and in short, kilogram-scale Fabry-Perot cavities~\cite{Nergis:PI2005}.

\begin{figure}[h]
   \centering
   \includegraphics[width=\columnwidth, trim=0 0 0cm 0, clip]{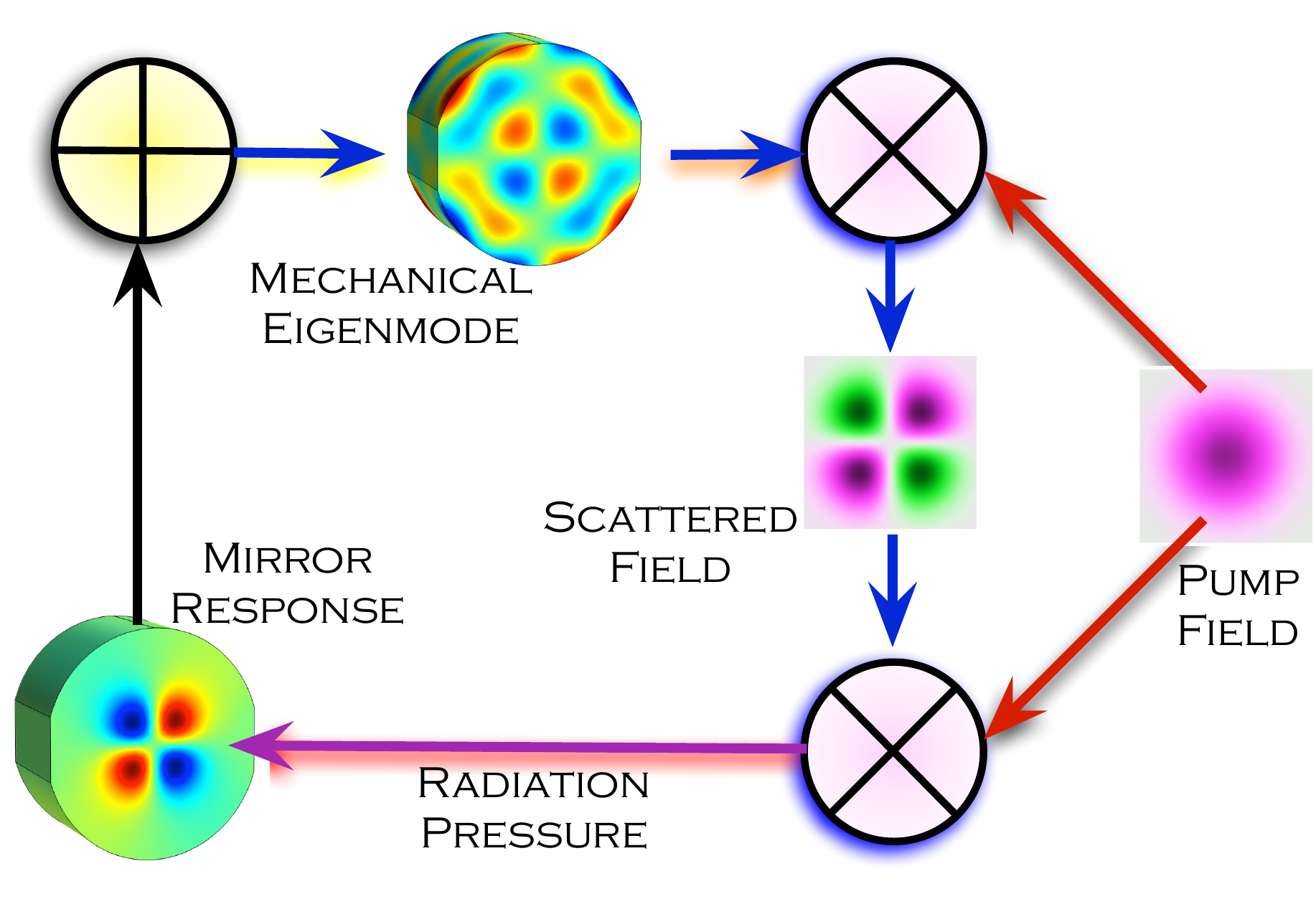}
   \caption{(Color online) Feedback loop diagram of the parametric instability process. 
           The oscillation of one of the mirror's mechanical eigenmodes scatters the resonant
           cavity mode into a higher order transverse mode which resonates partially in the coupled optical cavities
            of the interferometer and returns to excite the mirror via radiation pressure.}
   \label{fig:PI}
\end{figure}

Following~\cite{Matt:PI}, we can write the round-trip parametric gain for the $m^{th}$ mechanical
mode as:
\begin{equation}
R_m = \frac{4 \pi Q_m P}{M \omega_{m}^2 c \lambda} \sum_{n=0}^{\infty} \mathfrak{R}\{G_n\} B_{m,n}^2
\label{eq:PI}
\end{equation}
where $Q_m$ is the mechanical Q of the mode, $P$ is the arm cavity stored power, $M$ is
the mirror mass, $\omega_m$ is the mechanical eigenfrequency, $\lambda$ is the laser wavelength, $B_{m,n}$
is the overlap coefficient between the mechanical mode and the optical mode, and $G_n$ is the
round trip gain for the scattered field within the entire interferometer. This process is shown schematically
in Fig.~\ref{fig:PI}.

Even considering the optical resonance of the full interferometer, predicting the impact of
parametric instabilities is difficult. The details of the surface figure for each of the mirrors
shifts the resonant frequency for the higher order optical modes by a significant
fraction of the cavity linewidth. Small differences in dimensions of mirrors and long term
drifts in the laboratory temperatures can make order of magnitude changes in the round trip
gain by reducing the frequency overlap between the mechanical and optical modes.

A Monte Carlo analysis~\cite{Matt:PI} indicates that there is likely to be $\sim$several
unstable modes in a full power Advanced LIGO interferometer. As the masses, Q's, and power
levels are similar, most likely the same problems will afflict the Advanced Virgo and KAGRA
interferometers.

Several mitigation strategies have been proposed to suppress these instabilities: adding
passive damping films to the mirror 'barrel'~\cite{Gras:PI2009}, attaching a resonant electro-mechanical
damper, active feedback via the existing mirror actuators~\cite{Miller:ESD}
or using the radiation pressure of an external laser~\cite{Blair:PIStrategy}. While there are challenges
to be overcome with all of these techniques, it seems likely that a combination of them will be able to suppress
the instabilities down to the nuisance level. Future interferometers should be able to scale
the mirror mass directly with laser power and thereby stay at a nearly invariant instability level.

\subsection{Low Frequency Seismic Isolation}
\begin{figure}[h]
   \centering
   \includegraphics[width=\columnwidth, trim=0 0 0cm 0, clip]{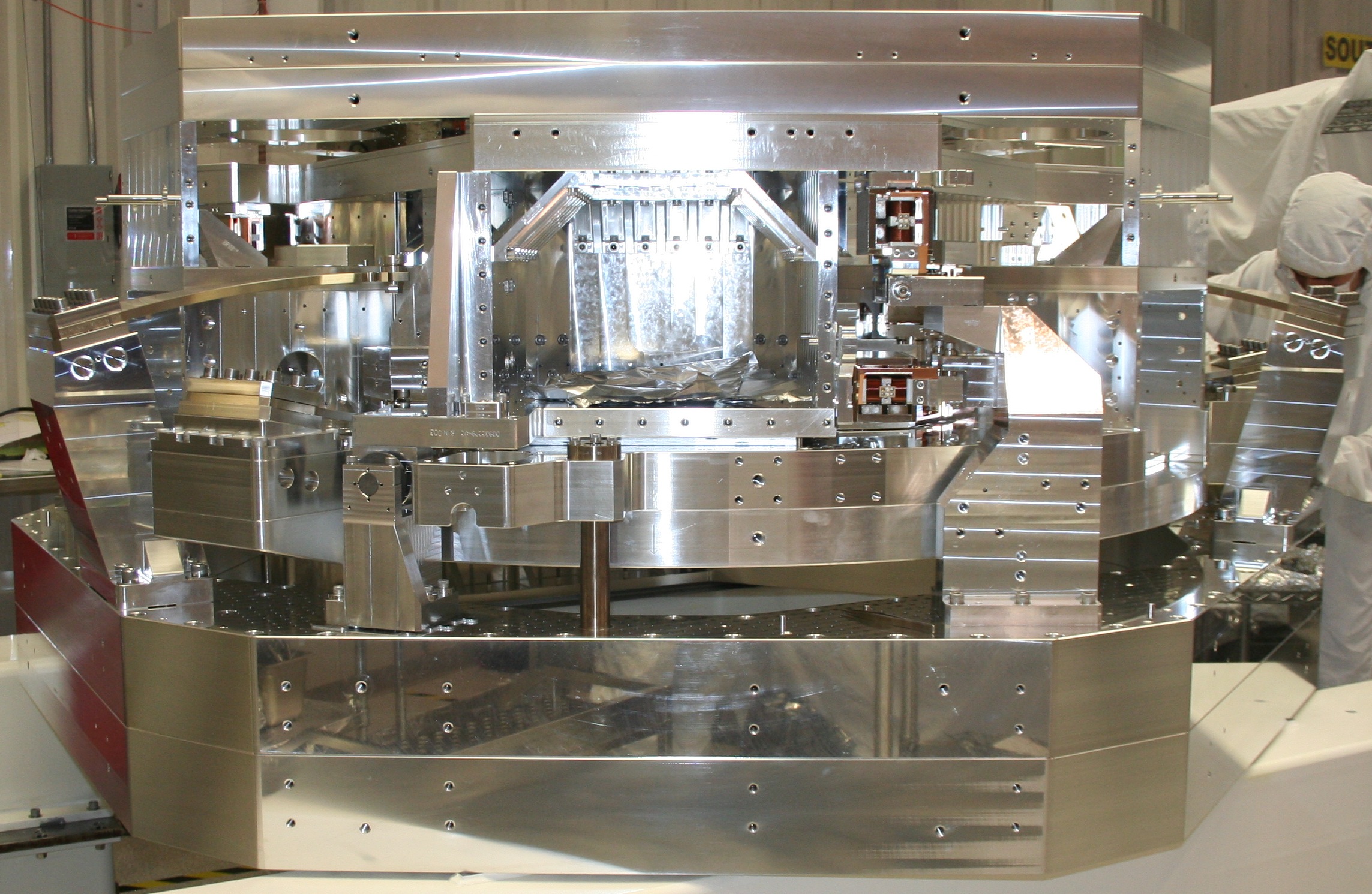}
   \caption{(Color online) Advanced LIGO Vibration Isolation platform: this double stage, 
       in-vacuum platform provides active isolation from 0.5 - 30\,Hz and passive 
       isolation above $\sim$1\,Hz. The leaf springs around the outer edge of the image
       provide the vertical compliance. The copper coils near the center of the image
       are part of the coil-magnet actuators used in the active feedback. Inertial
       sensors in sealed pods are attached to the platforms to provide the readback signals
       in the isolation servos.}
   \label{fig:BSCISI}
\end{figure}

The experience with the initial interferometers highlighted the multitudinous ways in which
large, low-frequency seismic motions can produce noise in the GW signal band through
nonlinear upconversion (see Sec.~\ref{sec:lowfmotion} above). As a result, all of the 
2$^{\rm nd}$ generation
vibration isolation systems seek to reduce motions not only in the GW band, but also in
the 0.01--10~Hz band.

The Advanced LIGO system is a 3-stage hybrid, active-passive platform~\cite{aLIGO:Seismic2002}.
There is a hydraulic pre-isolator to provide coarse positioning and coarse active
vibration control. This is followed by 2 compliant platforms (shown in Fig.~\ref{fig:BSCISI})
which provide passive isolation above $\sim$1~Hz and active isolation from 
0.1 to 30~Hz. An array of seismometers placed
near each mirror will be used to reduce the fluctuations in the low frequency, global
interferometric lengths~\cite{Ryan:FFW2012} that arise from the microseismic peaks~\cite{Ed:MSFF}.

A comparison of the vibration isolation performance of all ground based GW detectors 
is shown in Fig.~\ref{fig:VibrationIsolation}.

% ===============================================================
\section{Third Generation Detectors}
   \label{sec:3G}
Even conservative estimates of astrophysical event rates~\cite{Sterl:Minimal, LSC:rates, CuTh2002} 
predict many detections per year for the second generation detectors. Once the first detections are 
well established, one would like to move on to using the waveforms to make tests of astrophysical 
models, use 'standard' sirens for high precision cosmography, and make
tests of fundamental physics~\cite{3G:Science}. In order to
pursue this type of science, the sensitivity must be pushed beyond what the 
second generation detectors are capable of.

\begin{figure}[h]
   \centering
   \includegraphics[width=\columnwidth, trim=0 0 -3cm 0, clip]{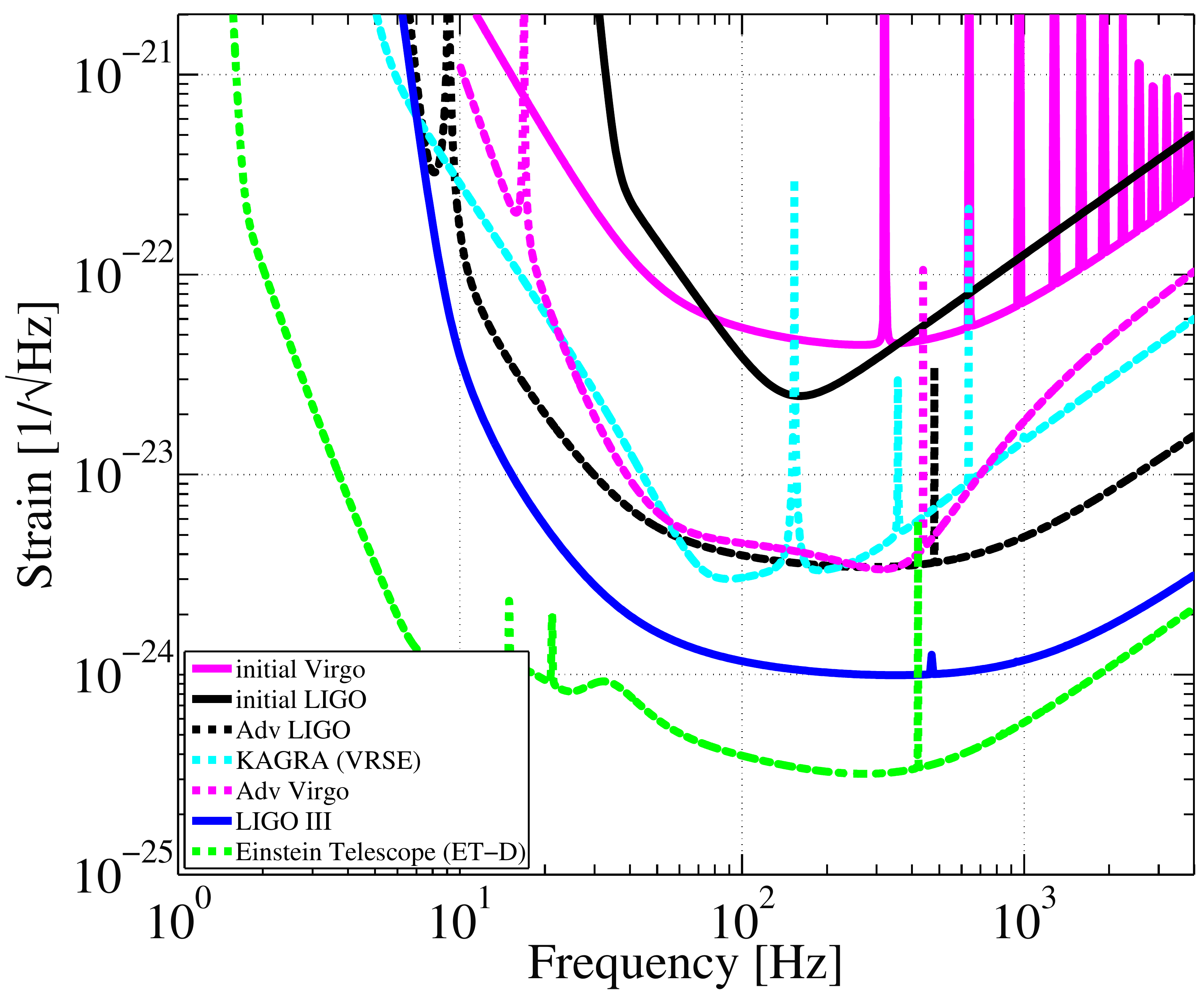} 
   \caption{(Color online) Comparison of strain noise estimates for
                  the ground based detectors.
                  The 'LIGO-III' trace refers to an upgrade of the Advanced LIGO detector
          including several of the ideas mentioned in Sec.~\ref{sec:3G}.}
   \label{fig:Gen2_noise}
\end{figure}

A combination of astrophysical motivations and technical developments has driven the 
European design of the Einstein Telescope~\cite{ET2011,Sathyaprakash:2012jt}. The 
Einstein Telescope (ET) is foreseen to be an underground, triangular, 10~km interferometer 
array operating at cryogenic temperatures. The goal is
to improve upon the broadband sensitivity by an order of magnitude over the 2$^{\rm nd}$ 
generation instruments and to lower the low frequency cutoff by a factor of 2--3. The most 
recent estimates of the ET sensitivity goal are shown in Fig.~\ref{fig:Gen2_noise}.

The LIGO Scientific Collaboration is currently studying the possibility of a complementary
3$^{\rm rd}$ generation detector network. As the LIGO detectors have yet to reach the fundamental
limits of the existing facilities, the study focuses on implementing the new interferometers in the
existing vacuum system.

\begin{figure}[h]
   \includegraphics[width=\columnwidth, trim=0 0 -2.75cm 0, clip]{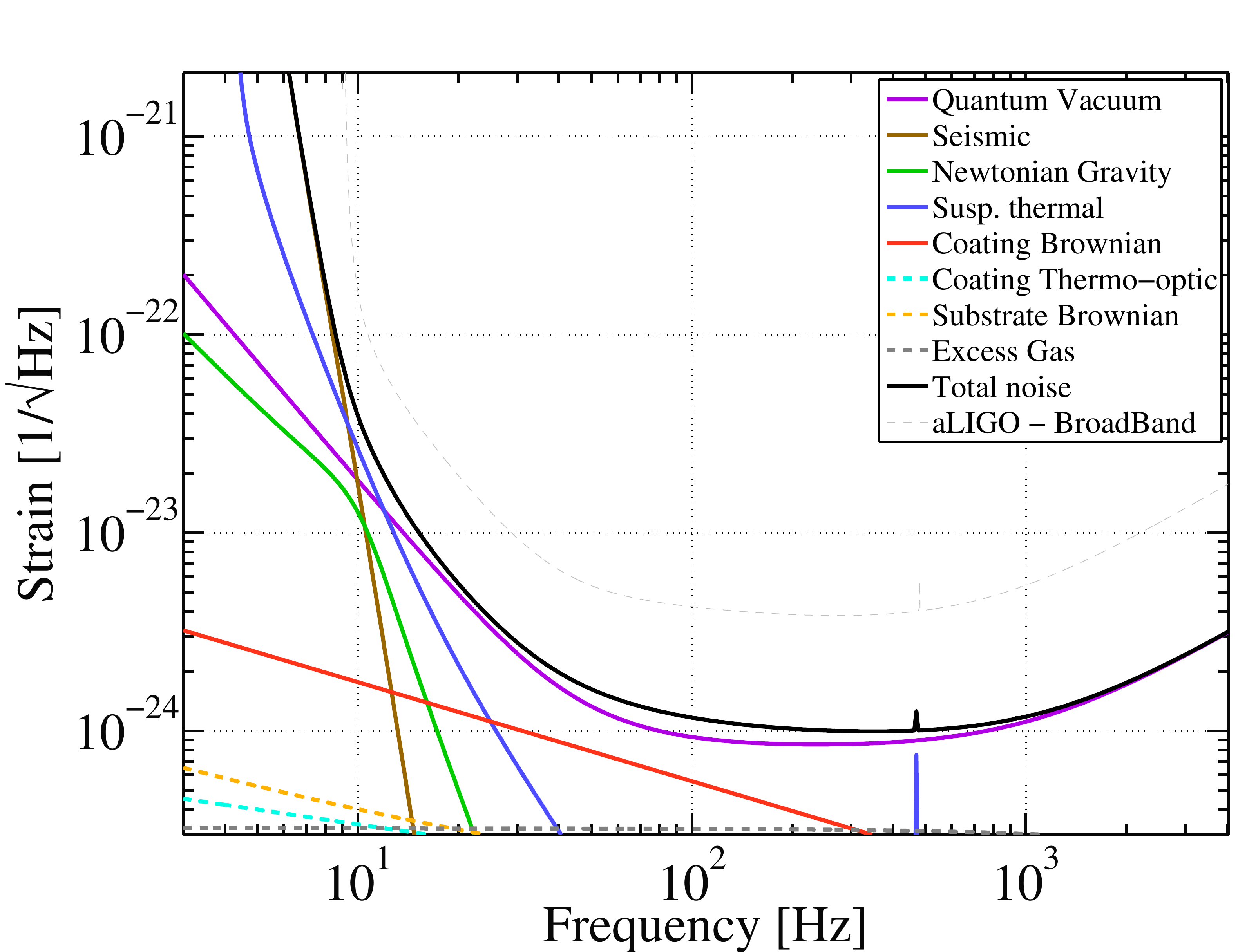} 
   \caption{(Color online) Limiting noise sources for a potential 3$^{\rm rd}$
     generation LIGO detector with 3~MW of arm cavity power, 10~dB of frequency
     dependent squeezed light
     injection, 140~kg Si mirrors with GaAs coatings operating cryogenically at 120~K, 
     and 30$\times$ subtraction of Newtonian gravity noise.}
   \label{fig:LIGO3Blue}
\end{figure}

In the following sections, several of the key techniques to making this 
improvement (summarized in Fig.~\ref{fig:LIGO3Blue} and Table~\ref{t:3G}) are described.

\subsection{Quantum Non-Demolition}
\label{sec:QND}
Most of the noise limits for the large interferometers have already been reduced to below the
usual quantum limits (cf.\,Fig.~\ref{fig:aLIGO_gwinc}). Improving the quantum limits will give a
larger scientific payoff than any other technical improvement. Correspondingly, there has 
been an explosion of research into QND readout schemes for GW interferometers 
in the 21$^{\rm st}$~century. Here we will just describe several of the most promising 
ideas~\cite{BrKh1996a}.

Recent reviews of the state of the art in QND for GW detectors describes well some
of the more promising techniques~\cite{Corbitt:Review2004, Chen:2010da, 
McClelland:2011fe, Yanbei:MQMreview, Schnabel:2010review}. In the past decade,
there has been a number of theoretical
and experimental advances which have led to better estimates of what is possible.
We can categorize the basic optical topologies in the following way:

\subsubsection{Frequency Dependent Squeezed State Injection}
The injection of squeezed light has long been seen as a panacea for the quantum noise limits of
GW detection. However, the direct injection of squeezed light can only reduce the noise
in the quadrature which has been squeezed~\cite{CaSc1985}; phase-squeezed light 
would improve the shot
noise limited region, but add, \textit{at least}, a corresponding amount of radiation pressure noise.
Work by \cite{KLMTV2001} and later by \cite{Harms:2003wv} showed that kilometer sized cavities
could be used to apply a frequency dependent phase shift to the squeezed fields. This phase
shift can be tuned to provide amplitude squeezing in the band where the radiation pressure is dominant
and phase squeezing where the shot noise dominates. For the broadband (tuned-RSE) configuration
of Advanced LIGO, this can be accomplished with a single cavity. For the detuned-RSE configurations
chosen by Virgo and KAGRA, two cavities are required to optimally match the squeeze quadrature to
the interferometer's opto-mechanical response.

Following early work on producing squeezed states at high frequencies~\cite{Kimble:1986}, the GW community
pushed the technology to produce high levels of squeezing at audio 
frequencies~\cite{Kirk:2004, Henning:PRL2006, KirksThesis, 
ANU:BackSqueeze2011, Stefszky:Balanced2012, Vahlbruch:2007da} on table-top prototypes. In the last
few years, moderate levels of noise improvement have been observed
from injecting squeezed light
into a suspended prototype~\cite{Go:40m} as well as the GEO600~\cite{GEO:Squeezing} and
Enhanced LIGO~\cite{H1:Squeezing, Dwyer:Thesis} detectors.

With the confidence gained from these demonstrations and the imminent prototyping of 
quadrature rotating filter cavities, it is very likely that the 2$^{\rm nd}$ generation 
detectors can be upgraded with
the injection of squeezed states of light before the end of the decade.

\begin{table*}[ht]
    \begin{centering}
      \begin{tabular*}{0.95\textwidth}[t]{@{\extracolsep{\fill}} l|c r c | l | c r c}
\hline
Parameter                       & Symbol   & Value & Units  &          Parameter                                     & Symbol   & Value & Units \\
\hline\hline
Light Wavelength            & $\lambda$ & 1064 & nm   &  Substrate Young's Modulus                  & $Y_{\rm sub}$ &185  & GPa\\
Arm Cavity Mirror Mass  & $m$           & 145   & kg    &  Suspension Ribbon Young's Modulus   & $Y_{si}$    & 130 & GPa \\ 
Arm Cavity Length          &  $L$           & 4000 & m     & Suspension Ribbon Thickness               & $h_{\rm sus}$ & 0.2 & mm \\
Arm Cavity Finesse         & $F$           &  550  & -      & Suspension Ribbon Width                       & $d_{\rm sus}$ & 2                  & mm\\
Arm Cavity Power           & $P_{\rm cav}$   & 3000  & kW   &  Substrate Loss Angle           & $\phi_{\rm sub}$  & $3 \times 10^{-9}$ & rad \\
Beam Radius                  & $\omega$  & 5.8  & cm  &  Coating Loss Angle              & $\phi_{\rm coat}$ & $2 \times 10^{-5}$ & rad \\
Detection Efficiency        & $\eta$       &  0.95    & -      & Mirror Coating                                      & -            & GaAs:AlAs & - \\
Squeeze Factor              & $R$            & 10     & dB   &                                                                & & \\
Filter Cavity Length         & $L_{\rm fc}$     & 100    &  m    & Newtonian Noise Subtraction Factor & $\aleph_{\rm NN}$    &     30       &     -         \\
Filter Cavity Loss     & $A_{\rm fc}$     &  33      & ppm  & Mirror / Suspension Temperature                 & $T$   &   120 & K \\
\hline
    \end{tabular*}
    \caption{Nominal values of some LIGO-III interferometer parameters used for Fig.~\ref{fig:LIGO3Blue}}
    \label{t:3G}
  \end{centering}
\end{table*}

\subsubsection{Frequency Dependent Readout Quadrature}
The quantum correlations built up in the signal recycled interferometers make it possible
to surpass the Standard Quantum Limit in a narrow band~\cite{VyMa1996a, BuCh2002}.
At high power levels, the vacuum fields in the amplitude quadrature drive the mirror
and produce signals in the phase quadrature as well. By choosing the appropriate
combination of homodyne readout quadratures after the photodetection, the amplitude
noise can be partially cancelled. 

The addition of one of the long quadrature rotation cavities can allow one to rotate
the \textit{readout} quadrature as a function of frequency and cancel the sensitivity to
the radiation pressure noise~\cite{KLMTV2001, Kha2007}. The radiation pressure noise itself
has not been cancelled; the mirrors are still moving. Rather, we have just chosen to adjust the
phase of our optical readout so as to ignore such perturbations. This technique is often referred 
to as the \textit{variational readout} technique. However, this delicate cancellation
by tuning of the readout quadrature has its problems: optical losses in the rotation cavity
degrade this scheme faster than the squeezed light injection scheme above. Since the
cancellation also subtracts much of the signal, it becomes more sensitive to any degradation
of the internal squeezing due to losses.

\subsubsection{QND Observable Readout}
In the Heisenberg picture, the increase in our positional resolution comes at the expense
of increased momentum perturbations since the position and momentum do not commute.
The momentum perturbations influence the time evolution of the mirror position and
spoil the low frequency sensitivity. An alternative to this approach is to read out
some observable which carries the gravitational wave information and also commutes
with itself at later times. In this way, we would have a true QND observable readout and
not have to worry about the quantum back-action effects~\cite{BrKh1999a}. 
One such observable is the mirror's 
momentum (or speed)~\cite{BGKT2000, PuCh2002, Che2003}.
Practically, this can be done by adding one of the long filter cavities into the interferometer in
such a way so as to differentiate the usual positional signal. By taking the differences
between successive position measurements, the readout variable closely approximates
momentum and so this type of interferometer is often referred to as
a \textit{speedmeter}.

\begin{figure}[h]
   \includegraphics[width=\columnwidth, trim=0 0 -3cm 0, clip]{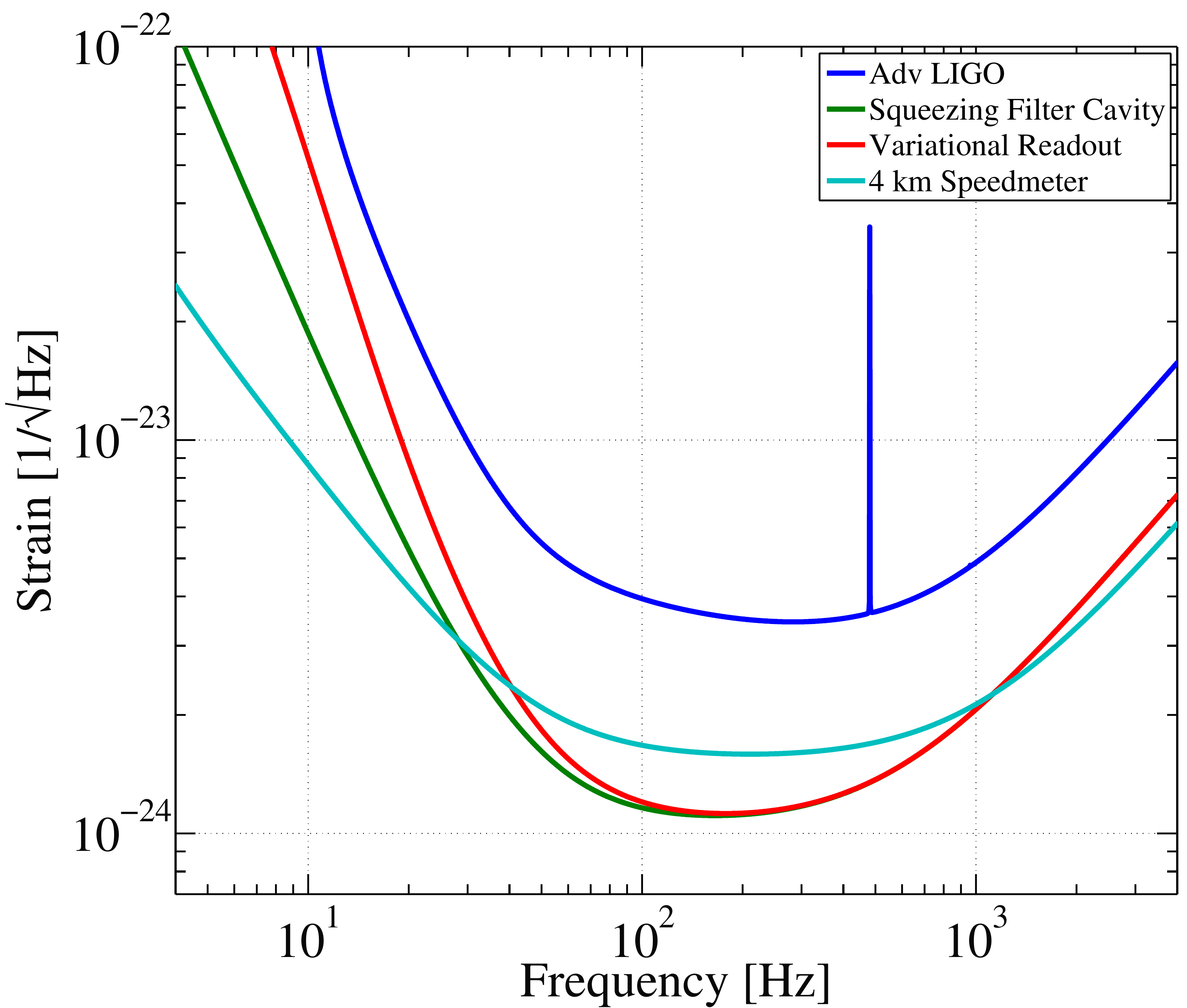} 
   \caption{(Color online) Comparison of the equivalent strain noise levels 
                 of various quantum non-demolition schemes implemented
                 on this 'LIGO-III' concept. 10~dB squeezed light is injected in all cases.
                 For the 'Filter Cavity' trace, squeezed light is filtered with a 100~m cavity. For the
                 'Variational'  trace, the output of the interferometer is filtered by a 100~m filter cavity,
               and for the speedmeter case, the 'Filter Cavity' configuration is modified by adding
               a 4~km speedmeter cavity. }
   \label{fig:AIC_compare}
\end{figure}

\subsubsection{Optical Losses and QND}
\label{sec:QNDlosses}
In addition to the squeezing input, the variational readout, and the speedmeter, there are a host of
other possibilities for QND upgrades: optical 'levers'~\cite{Kha2002}, 
multi-wavelength optical springs~\cite{Yanbei:DoubleSpring}, 
and multi-wavelength 'xylophones'~\cite{Yanbei:Local}, etc. The issues with most of the previous
inter-comparisons is that they do not include losses in a realistic way. In addition to the optical
losses due to scattering within the interferometer, losses in the readout chain, and finite
quantum efficiency of photodetectors, one must also include the losses in the quadrature
rotating filter cavities. To include these losses in a realistic way, it is important to remember
that the true loss will scale with the beam size (cf. Sec.~\ref{sec:mirrors}) and 
thereby the cavity length.

A numerical comparison that incorporates realistic losses within the framework of the 'LIGO-III' 
design has been carried out by~\cite{Haixing:CC2013} and is shown in Fig.~\ref{fig:AIC_compare}.
Here it has been assumed that the round-trip losses in the filter cavity are 33~ppm (consistent with
the past experience about large cavities).

\subsection{Circumventing Mirror Thermal Noise}
\label{sec:3Gcoating}
The relatively large mechanical dissipation in the mirror coatings and the shallow frequency 
dependence of the resulting mirror surface fluctuations makes the coating thermal noise one 
of the most serious limits for future detectors. Broadly speaking, two approaches are being 
pursued to avoid this limit: new coatings with higher mechanical Qs and alternative optical
cavity mode shapes that can partially reject the noise.

\subsubsection{Non-Gaussian Beam Shapes}
\label{sec:NGbeams}
A straightforward approach to reducing the effects of coating thermal noise is to increase
the beam size (cf.\,Sec.~\ref{sec:mirror_thermal}). However, as described in 
Section~\ref{sec:SiggSidles}, this can exacerbate the radiation pressure induced angular 
instability. Even if this can be compensated
by an exceptionally sophisticated feedback control system, it is unlikely that beam size alone
will offer more than a factor of two improvement in the long run.

A more effective approach might be using higher order spatial modes of the cavity.
The field inside of a Fabry-Perot cavity with spherical mirrors can be well approximated with the
set of orthonormal Laguerre-Gaussian functions~\cite{Siegman:Lasers}. Within the Virgo 
project~\cite{Vinet:LG}, it has been proposed to use axially symmetric 
Laguerre-Gaussian modes (chiefly the LG$_{3,3}$ and the LG$_{5,5}$ modes).

In addition to the technical difficulties associated with generation~\cite{Rob:LG} and 
control~\cite{Andreas:LG2009} of such beams, a stability analysis of the cavities~\cite{Ting:LG}
shows that the cavity field is strongly distorted when taking into account
the realistic surface imperfections (see Fig.~\ref{fig:aLIGO_mirror}) of the best available mirrors.
The several-fold degeneracy of these higher order modes is weakly split by the surface deformations
and all of the degenerate modes are partially resonant. In this perturbed state, the fields from the
two arm cavities are no longer well-matched and this degrades the interference at the anti-symmetric port of
the Michelson. Consequently, the ability to make a low phase noise optical readout is compromised.

An even more complicated option is to use a particular linear combination of
Laguerre-Gaussian modes. The so-called 
'Mesa beams'~\cite{Mesa:Erika, JM:Mesa2007, JM:Mesa2008} are one such combination. 
Simulations~\cite{Ting:LG} show that they are not much 
worse than TEM$_{0,0}$ Gaussian modes in their susceptibility to angular instabilities 
or modal degeneracy. Unfortunately, it is not yet straightforward to produce the 
non-spherical mirrors required for Mesa beams.

The theoretical maximum improvement from any of the above beam shaping techniques is
$\sim\,70\%$. To make any further improvements it will be necessary to either have radical
improvements in the mechanical loss of mirror coatings or 
build a much longer interferometer.

\subsubsection{Heteroepitaxial Bragg Mirrors}
\label{sec:mbe}
As described in Sec.~\ref{sec:mirror_thermal}, the thermal fluctuations of the mirror surface 
are dominated by the Langevin thermal forces generated in the high-reflectivity 
dielectric coatings. It has been shown \cite{Pohl:RMP, Phillips:1987} that the mechanical 
dissipation (and consequently the thermal noise) of nearly all amorphous, thin-film 
materials is higher than that of crystalline materials. The cause of the
dissipation, almost universally, is known to be due to the presence of a set of low 
energy modes (which are not 'frozen-out'). Tunneling into this vast sea of available 
modes leads to the observed mechanical dissipation.
Of all amorphous solids, fused silica seems to be singular in its extremely low dissipation at room
temperature and above~\cite{Penn:2004}. Unfortunately, this high Q of the bulk material
does not translate into high Q for the silica thin films used in the optics industry. 

One strategy in avoiding this thermal noise source is to eschew coatings altogether and to
use corner reflectors~\cite{Braginsky:2004fp, Cella:2006ki} or total internal reflection
\cite{Schiller:92, Gossler:2007uu}. Although these approaches introduce new technical problems,
there is, so far, no known fundamental reason why they cannot be used to supplant coatings. A
rigorous theoretical treatment followed by a direct thermal noise experiment is required.

Another approach to avoiding mechanically lossy coatings is to pattern the surface of the substrate
in order to produce grating based waveguide reflectors~\cite{Roman:Gratings2006}. Work in this
area has resulted in impressive performance in recent 
years~\cite{Gratings:2008, Gratings:2011, Kroker:11}
approaching power reflectivities of up to 99.9\%. Incorporating gratings into interferometers for
gravitational-wave detection will require substantial hurdles to be overcome: the coupling of mirror
alignment fluctuations~\cite{Freise:2007ed, kroker2013coupled} and transverse mirror 
motions~\cite{Wise:2005, brown2013invariance} into
longitudinal phase noise, the control of micro-roughness to reduce the diffuse scattered 
light~\cite{Grating:Scatter94, Josh:scatter2012}, the control of the large scale flatness to control 
the mirror figure error, 
and reducing the transmission losses by another factor of 10 ($R = 99.999\%$).

A less exotic option is to search more widely for lower mechanical loss materials which can produce
Bragg reflectors in the same manner as is done with the standard dielectric coatings.
The poor mechanical Q of amorphous materials leads one towards crystalline coatings. Epitaxial 
deposition techniques (e.g., chemical vapor deposition, molecular beam epitaxy, atomic layer 
deposition) have advanced dramatically over the past several decades to support the 
development of electronic circuits and opto-electronics. 

A promising set of prospects are trinary AlGaAs layers grown on GaAs substrates and then attached
to silica or silicon mirrors via epitaxial liftoff (ELO)~\cite{ELO}.
These structures have been grown on GaAs substrates and the resulting mechanical Q is $\sim30\times$
larger than the best amorphous high reflectivity coatings~\cite{cole:261108}. Another
possibility is to grow AlGaP:GaP~\cite{Angie:GaP2011} layers directly onto silicon substrates where the
lattice matching is quite good. The matching may largely mitigate the thermal stresses and allow
operation of the interferometer at cryogenic temperatures where the thermal noise is further
reduced. If the mechanical dissipation can be maintained at such low levels after ELO and the
absorption can be reduced to $\lesssim$\,5~ppm, these epitaxial coatings have the promise of expanding
the astrophysical reach of the detectors by a factor of 3--10 in the most critical frequency band.

\subsection{Newtonian Gravity Noise Subtraction}
\label{sec:NN_sub}
As described in Sec.~\ref{sec:NN}, the fluctuations in the local gravitational field will limit 
any further progress below $\sim$\,20\,Hz due to the inability to shield the mirrors from 
local gravitational perturbations. 
The seismic noise shown in Fig.~\ref{fig:SeismicNoise} indicates that the situation is 
largely the same for LIGO and Virgo, while the Newtonian noise may be as much as an 
order of magnitude smaller for KAGRA.
Underground detectors such as KAGRA and the Einstein Telescope should be designed 
to have a high degree of symmetry in the shape of the caverns around each test mass. 
The symmetry of these caverns can then passively cancel much of the 
Newtonian gravity noise~\cite{Harms:2009he, Cella:2006vd}.
Although it is not possible to significantly reduce the ambient vibrations,
it is possible, in principle, to subtract this gravitational noise either by applying canceling 
forces on the mirrors or by regressing it from the data stream offline.

Clearly the major impediment to subtracting out noise sources, in general, is to determine what part of
the interferometer output is noise and what part is signal. If this was straightforward, then all of the
important noises could be removed in this manner. The distinguishing feature of the Newtonian noise,
however, is that the source terms are readily measured. As we can see from Fig.~\ref{fig:aLIGO_NN},
the dominant component comes from the ambient ground motion in the vicinity of the test masses.
Of the various modes of the ground, the chief contributors to the gravitational noise are the Rayleigh
waves~\cite{HuTh1998, Cella:1998} on the surface. The body waves in the ground 
produce only small density
perturbations and are at least 10$\times$ smaller in their gravitational impact.

\begin{figure}[h]
   \includegraphics[width=\columnwidth, trim=0 0 -1cm 0, clip]{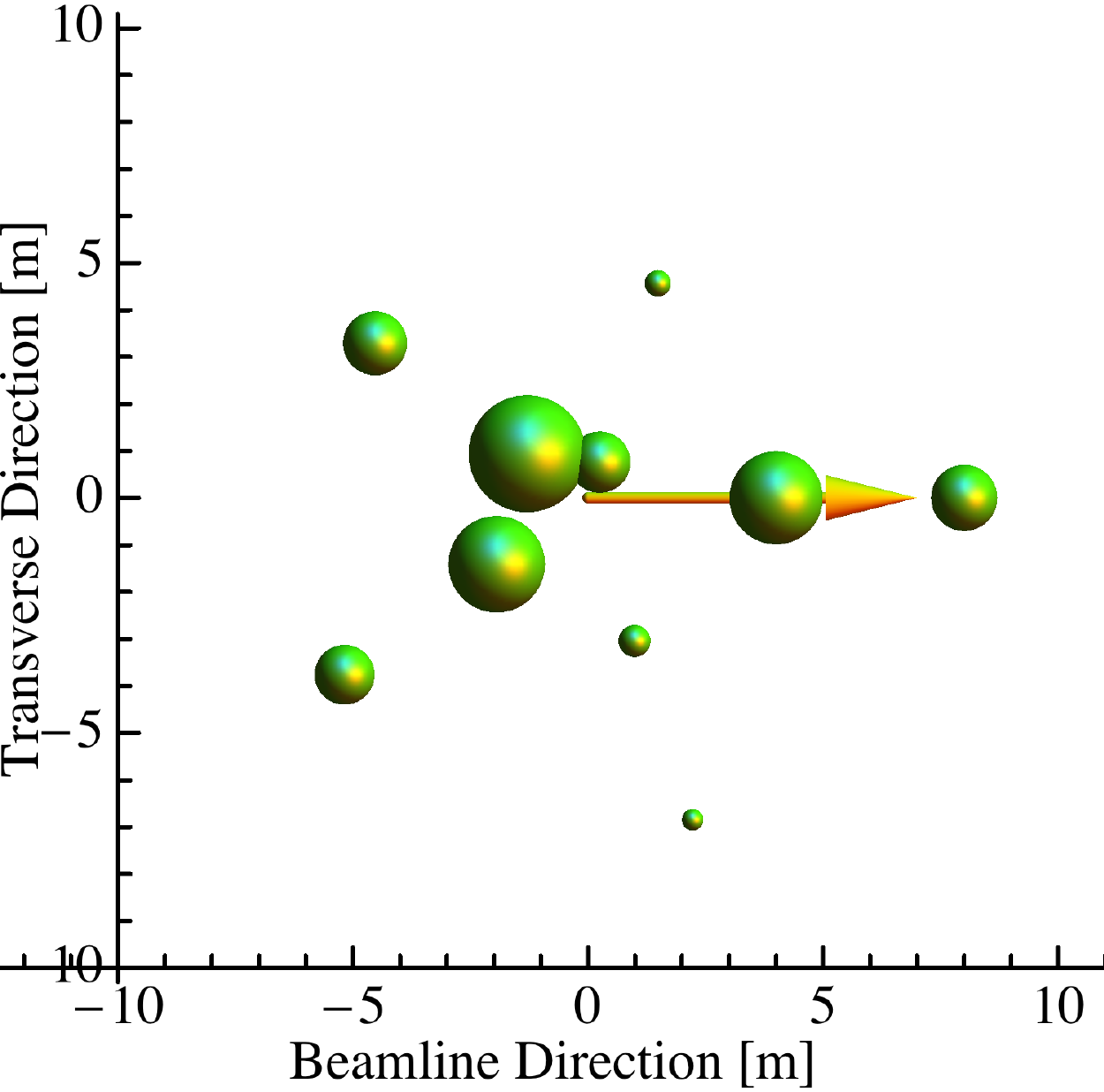} 
   \caption{(Color online) An example spiral array optimized for subtracting Newtonian noise due to
                 surface waves~\cite{NN:subtract2012} for
                 the 'LIGO-III' concept show in in Fig.~\ref{fig:LIGO3Blue}. The center of each
                 sphere indicates the location of one seismometer and the size of the sphere
                 is proportional to the coherence between the seismometer and the Newtonian
                 gravitational perturbations on the test mass. In this plot, the test mass is at
                 (0, 0) and the laser beam direction is indicated by the arrow.}
   \label{fig:NN_array}
\end{figure}

In principle, an array of seismometers near each mirror could measure this surface wave 
contribution. Given the time series of seismic noise, the remaining step is to then determine
the Green's function that relates the motion of each sensor to the mirror motion. Given sufficient
knowledge about the ground and the surrounding laboratory environment this could possibly give
some moderate subtraction quality, but would require significant effort to perform the characterization
and construct such an elaborate model with any accuracy.

A more promising approach is to use adaptive noise cancellation algorithms to 'learn' the
Green's function and apply the resulting digital filters to the data 
stream~\cite{Haykin:Adaptive, Say2003, Huang:MIMO, Beker:2011}. This approach has
proven to be successful in the laboratory~\cite{NIST:Wiener, Till:2013cav, Driggers:2012fl} 
in subtracting the direct
seismic influences from fixed cavities and suspended interferometers using an array of 
several low noise seismometers
and accelerometers. Moreover, this technique was employed in the recent LIGO 
Science run (S6) to remove the seismic influence from several of the interferometric 
degrees of freedom~\cite{Ryan:FFW2012} as well as magnetic field fluctuations at the mains 
frequencies~\cite{Tobin:DC}.

Early estimates of the noise at the LIGO sites and simulations of the subtraction systems
indicate that an array (shown in Fig.~\ref{fig:NN_array}) of $\sim$10--20 sensors per test 
mass will be sufficient to subtract 90\% of the noise
in the 5--20 Hz band~\cite{NN:subtract2012}. Experience with Advanced LIGO should allow for
making improvements to this depending upon the complexity of the seismic fields.

At the low frequencies where Newtonian noise is dominant, the main sources of gravitational
waves are expected to be the mergers of intermediate mass ($M \simeq 10-1000\,M_{\odot}$) 
black holes and the early part of the
inspiral for solar mass compact binaries. The implementation of a Newtonian noise subtraction
system should eventually allow for localizing the solar mass binaries to within a reasonable window
in the sky and allow electromagnetic telescopes to point to the source well ahead of the merger.

\subsection{Beyond Silica Suspensions}
\label{sec:cryo_sus}
In the 3$^{\rm rd}$ generation mirror suspensions, the frontier is not in improving the
vibration isolation, but rather it is in reducing the thermal noise due to the suspension
fiber. As discussed in Sec.~\ref{sec:monolithic}, the present limit to the mechanical dissipation
comes from the residual defects in the surface layer of the silica fibers. From 
Eqs.~\ref{eq:FDT} and \ref{eq:dsho} we can see that progress may be made on 
two fronts: reducing the temperature and reducing the loss.

\subsubsection{Silicon Suspensions and Cryogenics}
To reduce the loss it will be necessary to use a material with a very high mechanical
Q (e.g., sapphire, silicon, niobium, diamond~\cite{Diamond:Q})
as well as an extremely high quality surface. Fortunately, surface treatments of monocrystalline
silicon have advanced dramatically over the past decade. Present-day technologies can already
produce silicon with 10$\times$ less surface loss than fused silica~\cite{Ronny:Surface2010} 
and new etching and passivation methods pioneered by the opto-mechanics 
community~\cite{Oskar:SurfaceChem} may surpass this limit by another order of 
magnitude in the coming decade.

A seemingly straightforward option is to simply operate the interferometer at cryogenic 
temperatures, thereby winning in the thermal noise as $T^{-1/2}$. In addition, cryogenic 
silicon has many other excellent low temperature properties: the thermal expansion 
coefficient goes through zero at 18 and 120\,K, the thermal conductivity is at least 300$\times$ 
higher than that of silica (thus reducing thermal gradients and distortions), and, as is the case with
many crystalline substances, the mechanical Q increases with decreasing temperature.

With MW level laser power in the Fabry-Perot cavities, extracting heat from the mirrors becomes
an issue for low temperature operation. The high thermal conductivity of
materials such as silicon and sapphire may make it possible to extract $\sim$\,10\,mW of
heat through the suspension fibers~\cite{LCGT:heatextract} before the thickness of the fiber
compromises the thermal noise benefits. 

%In order to extract a few Watts of power via
%conduction through the fibers (or ribbons), the diameter

At the higher zero crossing temperature of 120\,K, the radiative cooling power of a
large mirror can exceed $\sim$\,10\,W. This should make it possible to cool the 
mirrors entirely by radiation using a cold shield around the suspension. This
essentially noiseless approach should permit the use of 10$\times$ higher circulating power
in the interferometer while maintaining the thermal noise benefits of low temperature
operation. It remains to be seen if the surfaces can be treated in a way so as to have a
high emissivity while not spoiling the mechanical Q too much.

%Blades. Nonlinear TE cancellation.

%Surface to volume effects. Increasing mass decreases the surface to volume
%ratio. Plot? Function?

%\R{Flaw in anti-spring ideas...where to go?}

\subsubsection{Electro-Magnetic Suspensions}
A natural route to explore is that of eschewing the fiber altogether and using purely magnetic
suspension forces~\cite{Drever:Magnetic, Jayawant:1981}. In principle,
the lack of any mechanical support element will eliminate the suspension thermal noise
contribution to the interferometer displacement noise. Attaching magnets directly to
the mirror is likely to lead to too much mechanical dissipation in the attachments and in
the magnets themselves. Another possibility is to find a paramagnetic mirror material
and to levitate it with strong permanent magnets~\cite{Drever:ParaQ} although it seems
problematic to simultaneously have a large magnetic susceptibility and high mechanical
Q. In either case, the mirror must be well shielded from the ambient magnetic field fluctuations
and even then, the Barkhausen noise in the permanent magnets could well introduce an
insurmountable noise floor. Even superconducting magnetic suspensions may have 
dissipation~\cite{Hebard:RSI, Giles:SuperTorsion}
due to nearby eddy currents or small normal regions of the material.

Rather than directly levitating the mirror, the magnetic suspension could be used to support
an upper stage of a multi-stage
suspension system~\cite{Monica:Magnetic}. The mirror could
then be supported from this magnetically levitated platform by a passive mechanical suspension.
Although this approach would not avoid the fiber's thermal noise, it could allow for a
very low frequency suspension and concomitant improvement in filtering of seismic noise.

Another option is to instead use electrostatic~\cite{Giazotto:ElectroSUS, onera:2000} 
suspensions. This would seemingly avoid the problems due to coupling from ambient
magnetic fields. In any case, the passive stability of any such system is forbidden
by Earnshaw's theorem, and some kind of active feedback must be used to stabilize
at least one degree of freedom. The sensitivity of such a sensor limits the ultimate
low frequency performance of such a suspension/levitation system, but it may be very
useful as an intermediate stage in a compound pendulum system.

% Monica    http://www.sciencedirect.com/science/article/pii/S0927650504000180
% Drever  proceedings of the moriond 1996
% Drever para http://link.aip.org/link/doi/10.1063/1.1291877
% Jayawant  http://dx.doi.org/10.1088/0034-4885/44/4/002

% blog      http://blogs.vassar.edu/magnes/author/joandrade/

% Giaz   http://www.sciencedirect.com/science/article/pii/S0375960198004113
% Willemenot  http://link.aip.org/link/doi/10.1063/1.1150198

% Hebard http://link.aip.org/link/doi/10.1063/1.1686149
% Giles  http://link.aip.org/link/doi/10.1063/1.1651631

%\paragraph{Electro-Mechanical Suspension}
% addition of a EM spring can shift the real part of the spring constant but not the
% imaginary part.

% ===============================================================
\section{Low Frequency Detectors}
\label{sec:LF}

% The gravitational wave spectrum is shown in Fig.~\ref{fig:GWspec}. All of the proposed
% terrestrial detectors operate in the 'indigo' regime; 16 rich decades of lower frequency
% space remain inaudible from the earth. 

The gravitational wave spectrum spans twenty decades in frequency:
at the lowest frequencies, corresponding to the age of the universe, the polarization of the 
cosmic microwave background should contain signals from the primordial gravitational waves 
due to cosmic inflation~\cite{Hu:AR2002}.
The nano- to micro-Hertz band is covered by timing
of pulsars and artificial satellites (cf. Sec.~\ref{sec:PulsarTiming}). Between the timing
measurements and the ground based detectors, the wide
10$^{-5}$ to 1\,Hz band will be pursued with space-based interferometers in the near
future.

\subsection{Interferometers in Space}
\label{sec:space}
Space detectors have tremendous advantages over ground based detectors
below $\sim$\,5~Hz. Direct seismic vibrations and Newtonian gravitational
fluctuations are almost completely absent. All
of the proposed space missions, therefore, are designed to focus on
sub-Hz frequencies. 
\begin{figure}[h]
   \includegraphics[width=\columnwidth, trim=0 0 -25mm 0, clip]{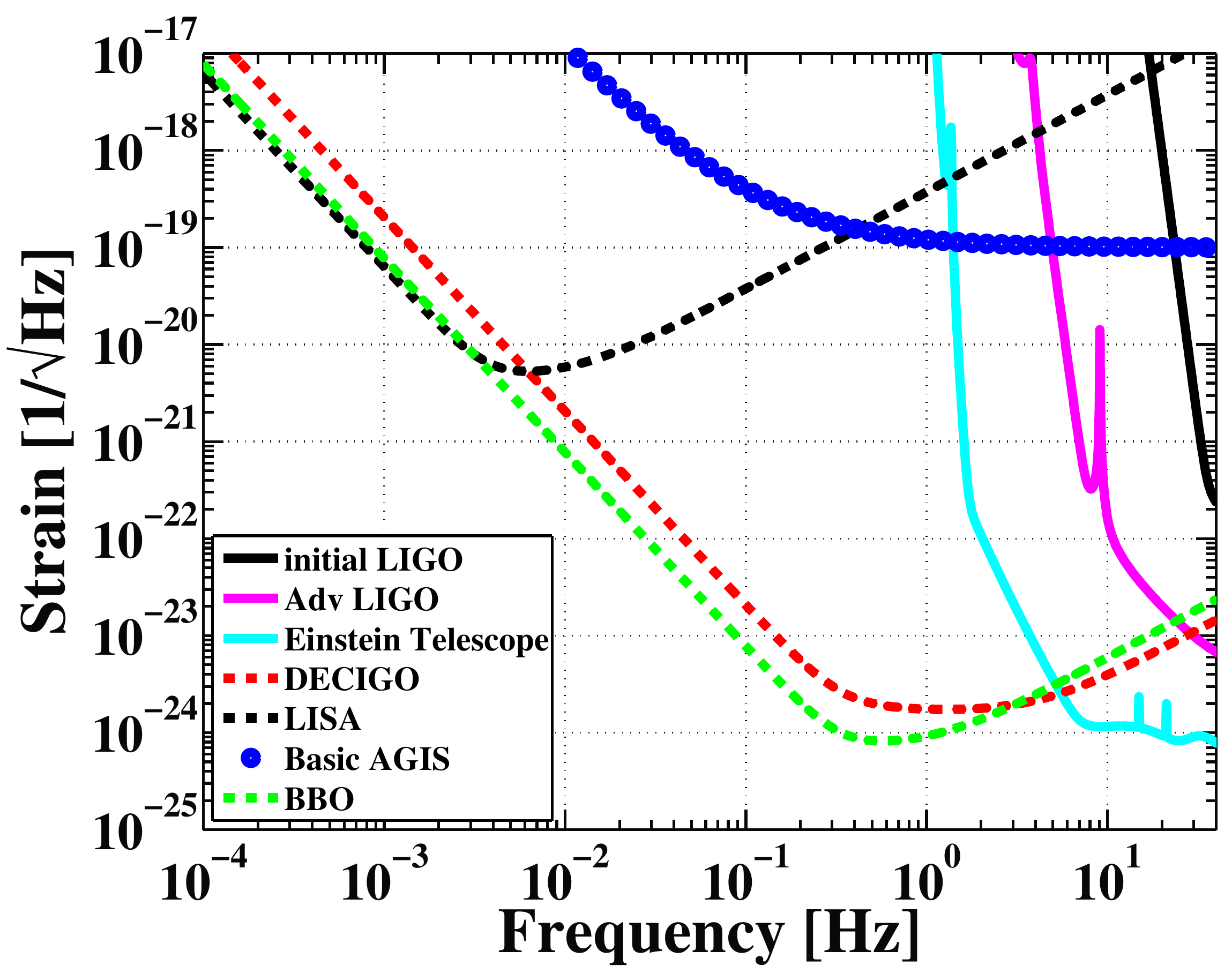} 
   \caption{(Color online) Comparison of strain noise estimates for
                  future detectors: LISA, DECIGO, BBO, Basic AGIS, and ET~(D).
                  The LIGO sensitivity curves are included for reference.}
   \label{fig:Gen3_noise}
\end{figure}

% clean this up to make sure its all LISA or eLISA or whatever...
\subsubsection{eLISA}
\label{sec:LISA}
eLISA (previously LISA: the Laser Interferometer Space Antenna) is a proposed
European Space Agency variant~\cite{eLISA:WhitePaper}
of LISA~\cite{Faller1989107, PrEA2002, Danzmann:2003, LISA:2009} 
slated for launch in the early part of the 21$^{\rm st}$ century.
The design has three spacecraft flying in a near-equilateral 
triangle formation in orbit around the sun, trailing the earth by $\sim\,20^{\circ}$. Whereas
the original LISA mission had links between each of the spacecraft, the new eLISA concept has
two interferometric links.
From the central satellite, a laser beam is sent to the others using a large beam expanding telescope. Due
to the large distances, most of the light is lost through diffraction, yielding very little power
for detection at each end. The local lasers at each receiving satellite are then phase locked to the 
incoming light. The local laser light is then sent to the central satellite similar to 
standard transponder methods. The phase differences between
the lasers contain the GW strain signal as well as various technical noise sources which
can be removed by the techniques of Time-Delay 
Interferometry~\cite{Armstrong:1999hp, Glenn:TDI2010, lrr-2005-4}. The low
power levels which are received at each satellite result in the interferometer being shot noise
limited above a few mHz. 
Although the expected displacement
sensitivity is 'only' $\sim\,10^{-11}\,\rm m/\sqrt{Hz}$, the impressive strain 
sensitivity is achieved by having arm lengths of $\sim\,10^6\,\rm km$. In contrast 
to the ground-based detectors, eLISA will
operate in the limit of having many high SNR signals enabling it to do extremely precise
tests of astrophysical models and general relativity. In fact, the high sensitivity
is expected to lead to a so-called 'confusion noise' limit~\cite{KTV2004} where the low 
frequency end of the
spectrum is dominated by a large foreground of gravitational radiation from 
galactic and extragalactic compact binaries. In order to reach the sensitivity shown in 
Fig.~\ref{fig:Gen3_noise},
sophisticated subtraction algorithms will have to be used in post-processing~\cite{Cornish:2007}.
The rotation of the eLISA constellation and its orbit around the sun will produce phase and 
amplitude modulations of the detected signals. These modulations in turn will allow the 
analysis to reconstruct the angular position of the sources with orders-of-magnitude 
better resolution than the ground based detectors. Details of the mission technology and
science goals can be found in the eLISA Yellow Book~\cite{NGO:YellowBook}.

\subsubsection{DECIGO and BBO}
\label{sec:DECIGO}
Of all of the proposed sources of gravitational radiation, the most exciting one for cosmologists
is perhaps the early universe (cf.\,Sec.~\ref{sec:SGWB}). Due to the weak coupling of gravitational
waves with matter, a detection of a primordial stochastic background would allow us to peer
back into the time when the age of the universe was less than $\sim\,10^{-20}$ seconds. For
a scale invariant spectrum of radiation, we have the best chance of detection at low frequencies.
Unfortunately, the astrophysical foreground of gravitational waves in the $10^{-9}-10^{-1}$~Hz
band makes the detection of an inflationary background nearly hopeless. 
Nearly all of the white dwarf binaries have merged before their orbital frequencies have increased 
to 0.1\,Hz~\cite{FaPh2003} and so only the relatively
small number of binaries containing neutron stars and black holes remain in the
0.1--1\,Hz band.

\begin{figure}[h]
   \includegraphics[width=\columnwidth, trim=0 0 -2mm 0, clip]{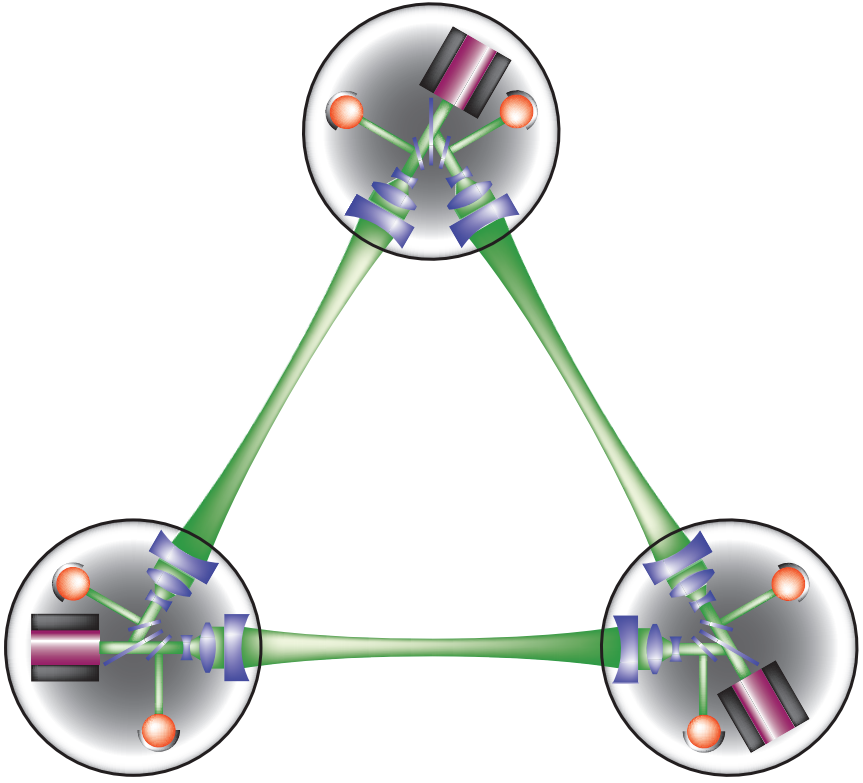} 
   \caption{DECIGO constellation concept~\cite{Sato:DECIGO2009}}
   \label{fig:DECIGO}
\end{figure}

Two space missions are being studied to probe this frequency band: the 
Japanese Deci-Hertz Gravitational-wave Observatory (DECIGO)~\cite{Ando:DECIGO,DECIGO} and
the international Big Bang Observer (BBO)~\cite{Phi2003,CuHa2006}. In addition to the eventual
detection of cosmological backgrounds, there is a wealth of wonderful astrophysical science
which can be extracted during the foreground removal of these 
detectors~\cite{Cutler:2009, Yagi:DECIGOalt}. Unfortunately, it is unlikely that either of these
missions will fly within the next decade due to budgetary constraints. A three constellation
concept which is common to DECIGO, BBO, and the early versions of LISA/eLISA is shown in
Fig.~\ref{fig:DECIGO}.

\subsection{Low Frequency Terrestrial Detectors}
The natural way to avoid terrestrial disturbances is to make an extra-terrestrial detector. However,
recent advanced in our understanding of Newtonian gravitational noise have made it reasonable
to reexamine this issue. In particular, the relatively higher abundance of sources in the 0.01--10\,Hz
band and their long duration, make it possible to have astrophysically interesting detectors
even if their noise floors are higher by a factor of $10^5$ than that of the km-scale, ground based
detectors~\cite{MANGO:2013}.

\subsubsection{Torsion Bar Antenna}
\label{sec:TOBA}
Recently, a novel arrangement of torsion bars has been proposed to readout sub-Hz gravitational
waves~\cite{Ando:TOBA, TOBA:2011}. The tidal force from an incoming wave will twist the crossed
torsion bars differentially. A high sensitivity interferometric sensor is used to read out the differential
torsion angle. Early estimates project the strain sensitivity to be near 10$^{-19}/\sqrt{\rm Hz}$ 
above $\sim$\,0.1~Hz using 10~m bars. With such a sensitivity
it should be possible to observe the mergers of intermediate mass black holes out to cosmological 
distances, search for the merger of galactic white dwarfs~\cite{FaPh2003}, and serve as an early 
warning system for extra-galactic compact object inspirals for the ground based detectors.

\subsubsection{Atom Interferometers}
An alternative to standard laser interferometry is to use clouds of atoms instead of 
mirrors~\cite{Dim:Atom2008, Holger:GRG2011}. This method uses pulses of light to change the momentum
states of some of the atoms in the clouds. These clouds then take different free fall paths. A
final pulse is used to synchronize the momentum states of the atoms and the interference of the
atomic clouds is used to read out the GW signal.

The advantages of these atomic techniques are many: the clouds have a very high immunity to
radiation pressure noise, very low thermal noise, and no suspension noise. The common launch
for the atomic clouds makes the influence of seismic noise nearly zero. However,
the Newtonian noise is a problem for the atom interferometers just as it is for 
laser interferometers. A spaced based detector, the Atomic 
Gravitational wave Interferometric Sensor (AGIS), has also been proposed to
circumvent these terrestrial limits~\cite{AGIS:2011}.

Bender has highlighted~\cite{Bender:Atom2011, Bender:PRD2011} several 
additional complications (including wavefront aberration and beam jitter) 
with the light-pulse atom interferometers which limit 
significantly the achievable sensitivity; these issues are being addressed by 
the atomic community~\cite{Dim:PRD2011} It remains to be seen if this type 
of atom interferometry can be made to be competitive with other 
technologies (such as DECIGO).

%\subsubsection{Requisite Technologies}

% ===============================================================
\section{Conclusion}
Many of the most interesting objects in the universe remain invisible so far to those of us
on the earth. Our understanding of astrophysics and cosmology has been transformed
in the past millenium by observations of electromagnetic radiation, looking into new
wavelengths, looking farther back into the early universe, and looking more deeply at
our local neighborhood.

We have yet to witness the same revolution through our observations of gravitational radiation,
and yet the promise for discovery and revolution remains as profound as before.

\begin{figure}[h]
   \includegraphics[width=\columnwidth, trim=0 0 -25mm 0, clip]{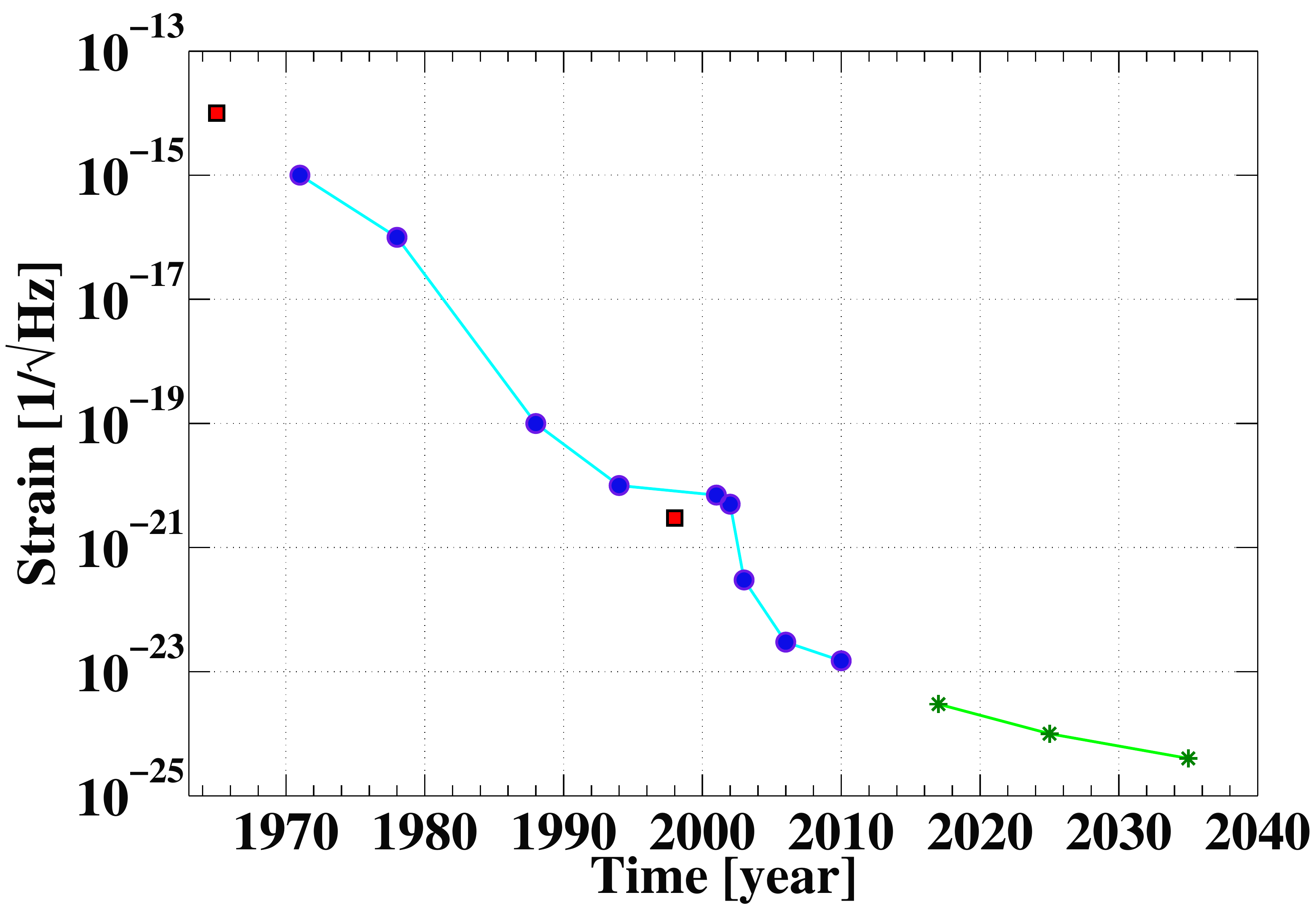} 
   \caption{(Color online) Evolution in gravitational wave detector sensitivity from 1965 into
            the near future. On the y-axis is plotted the \textit{minimum} of the
            strain noise spectral density for the given detectors. The acoustic
            bar detectors are shown as (RED) squares and the laser interferometers
            as (BLUE) circles. Estimates for future detectors are (GREEN) stars.}
   \label{fig:SensEvo}
\end{figure}

The recent progress in numerical relativity, wide area astronomical surveys, and gravitational wave
detector technology (shown in Fig.~\ref{fig:SensEvo}) all point to a wonderful convergence of science that will ineluctably lead
to another series of revolutions in our understanding of the universe.

The upcoming crop of ground based detectors is almost guaranteed to make detection in the next
few years and the laboratory research of today promises to turn the gravitational wave
astronomy of the future into a precision science. Buoyed by the likely detections of signals 
by pulsar timing and terrestrial interferometers, the space
missions should complete our coverage of the gravitational wave spectrum. 
The sources of gravitational waves may often be dark but the future is bright.

% ===============================================================
\begin{acknowledgments}
I must primarily thank the international community of scientists involved in the search 
for gravitational waves; I have learned much from them in the last decade.

In particular, I thank H.~Grote, H.~L\"{u}ck, D.~Tatsumi, K.~Arai, M.~Barsuglia, 
L.~Barsotti, J.~Marque, G.~Vajente, and 
V.~Frolov for detailed information on the first generation interferometers and their noise 
sources.
I am grateful to S.~Hild, Y.~Aso, M.~Ando, G.~Losurdo, and P.~Fritschel for detailed
information on the second generation interferometers. 
I must also thank J.~Harms, M.~Coughlin, and J.~C.~Driggers for assembling the updated
seismic information from the worldwide network and producing the recent estimates of
Newtonian gravity noise, to P.~Kwee for the information on the AEI 200~W laser, C.~Vorvick
for images of the LIGO optic scattering, T.~Akutsu for information on DECIGO, W.~W.~Johnson
for elucidation on the sensitivity of bar detectors, S.~E.~Whitcomb for discussions about
excess gas, and to E.~K.~Gustafson and K.~Arai for a careful reading of the manuscript.
I gratefully acknowledge the support of the National Science Foundation under 
grant PHY-0555406.
\end{acknowledgments}

% ===============================================================

%\bibliographystyle{apsrmp4-1}
\bibliographystyle{apsrmp}
\bibliography{GWreferences}

\begin{thebibliography}{402}
\expandafter\ifx\csname natexlab\endcsname\relax\def\natexlab#1{#1}\fi
\expandafter\ifx\csname bibnamefont\endcsname\relax
  \def\bibnamefont#1{#1}\fi
\expandafter\ifx\csname bibfnamefont\endcsname\relax
  \def\bibfnamefont#1{#1}\fi
\expandafter\ifx\csname citenamefont\endcsname\relax
  \def\citenamefont#1{#1}\fi
\expandafter\ifx\csname url\endcsname\relax
  \def\url#1{\texttt{#1}}\fi
\expandafter\ifx\csname urlprefix\endcsname\relax\def\urlprefix{URL }\fi
\providecommand{\bibinfo}[2]{#2}
\providecommand{\eprint}[2][]{\url{#2}}

\bibitem[{\citenamefont{Aasi} \emph{et~al.}(2013)\citenamefont{Aasi, Abadie,
  Abbott, Abbott, Abbott, Abernathy, Adams, Adams, Addesso, Adhikari, Affeldt,
  Aguiar} \emph{et~al.}}]{H1:Squeezing}
\bibinfo{author}{\bibnamefont{Aasi}, \bibfnamefont{J.}},
  \bibinfo{author}{\bibfnamefont{J.}~\bibnamefont{Abadie}},
  \bibinfo{author}{\bibfnamefont{B.~P.} \bibnamefont{Abbott}},
  \bibinfo{author}{\bibfnamefont{R.}~\bibnamefont{Abbott}},
  \bibinfo{author}{\bibfnamefont{T.~D.} \bibnamefont{Abbott}},
  \bibinfo{author}{\bibfnamefont{M.~R.} \bibnamefont{Abernathy}},
  \bibinfo{author}{\bibfnamefont{C.}~\bibnamefont{Adams}},
  \bibinfo{author}{\bibfnamefont{T.}~\bibnamefont{Adams}},
  \bibinfo{author}{\bibfnamefont{P.}~\bibnamefont{Addesso}},
  \bibinfo{author}{\bibfnamefont{R.~X.} \bibnamefont{Adhikari}},
  \bibinfo{author}{\bibfnamefont{C.}~\bibnamefont{Affeldt}},
  \bibinfo{author}{\bibfnamefont{O.~D.} \bibnamefont{Aguiar}}, \emph{et~al.},
  \bibinfo{year}{2013}, \bibinfo{journal}{Nat Photon}
  \textbf{\bibinfo{volume}{7}}(\bibinfo{number}{8}), \bibinfo{pages}{613}, ISSN
  \bibinfo{issn}{1749-4885},
  \urlprefix\url{http://dx.doi.org/10.1038/nphoton.2013.177}.

\bibitem[{\citenamefont{Abadie} \emph{et~al.}(2010)\citenamefont{Abadie,
  Abbott, Abbott, Abernathy, Accadia, Acernese, Adams, Adhikari, Ajith, Allen,
  Allen, Ceron} \emph{et~al.}}]{LSC:rates}
\bibinfo{author}{\bibnamefont{Abadie}, \bibfnamefont{J.}},
  \bibinfo{author}{\bibfnamefont{B.~P.} \bibnamefont{Abbott}},
  \bibinfo{author}{\bibfnamefont{R.}~\bibnamefont{Abbott}},
  \bibinfo{author}{\bibfnamefont{M.}~\bibnamefont{Abernathy}},
  \bibinfo{author}{\bibfnamefont{T.}~\bibnamefont{Accadia}},
  \bibinfo{author}{\bibfnamefont{F.}~\bibnamefont{Acernese}},
  \bibinfo{author}{\bibfnamefont{C.}~\bibnamefont{Adams}},
  \bibinfo{author}{\bibfnamefont{R.}~\bibnamefont{Adhikari}},
  \bibinfo{author}{\bibfnamefont{P.}~\bibnamefont{Ajith}},
  \bibinfo{author}{\bibfnamefont{B.}~\bibnamefont{Allen}},
  \bibinfo{author}{\bibfnamefont{G.}~\bibnamefont{Allen}},
  \bibinfo{author}{\bibfnamefont{E.~A.} \bibnamefont{Ceron}}, \emph{et~al.},
  \bibinfo{year}{2010}, \bibinfo{journal}{Classical and Quantum Gravity}
  \textbf{\bibinfo{volume}{27}}(\bibinfo{number}{17}), \bibinfo{pages}{173001}.

\bibitem[{\citenamefont{Abbott}
  \emph{et~al.}(2004{\natexlab{a}})\citenamefont{Abbott, Abbott, Adhikari,
  Ageev, Allen, Amin, Anderson, Anderson, Araya, Armandula}
  \emph{et~al.}}]{DHS:Detector2004}
\bibinfo{author}{\bibnamefont{Abbott}, \bibfnamefont{B.}},
  \bibinfo{author}{\bibfnamefont{R.}~\bibnamefont{Abbott}},
  \bibinfo{author}{\bibfnamefont{R.}~\bibnamefont{Adhikari}},
  \bibinfo{author}{\bibfnamefont{A.}~\bibnamefont{Ageev}},
  \bibinfo{author}{\bibfnamefont{B.}~\bibnamefont{Allen}},
  \bibinfo{author}{\bibfnamefont{R.}~\bibnamefont{Amin}},
  \bibinfo{author}{\bibfnamefont{S.}~\bibnamefont{Anderson}},
  \bibinfo{author}{\bibfnamefont{W.}~\bibnamefont{Anderson}},
  \bibinfo{author}{\bibfnamefont{M.}~\bibnamefont{Araya}},
  \bibinfo{author}{\bibfnamefont{H.}~\bibnamefont{Armandula}}, \emph{et~al.},
  \bibinfo{year}{2004}{\natexlab{a}}, \bibinfo{journal}{Nuclear Instruments and
  Methods in Physics Research Section A: Accelerators, Spectrometers, Detectors
  and Associated Equipment}
  \textbf{\bibinfo{volume}{517}}(\bibinfo{number}{1}), \bibinfo{pages}{154}.

\bibitem[{\citenamefont{Abbott} \emph{et~al.}(2008)\citenamefont{Abbott,
  Abbott, Adhikari, Ajith, Allen, Allen, Amin, Anderson, Anderson, Arain,
  Araya, Armandula} \emph{et~al.}}]{LIGO:SGR}
\bibinfo{author}{\bibnamefont{Abbott}, \bibfnamefont{B.}},
  \bibinfo{author}{\bibfnamefont{R.}~\bibnamefont{Abbott}},
  \bibinfo{author}{\bibfnamefont{R.}~\bibnamefont{Adhikari}},
  \bibinfo{author}{\bibfnamefont{P.}~\bibnamefont{Ajith}},
  \bibinfo{author}{\bibfnamefont{B.}~\bibnamefont{Allen}},
  \bibinfo{author}{\bibfnamefont{G.}~\bibnamefont{Allen}},
  \bibinfo{author}{\bibfnamefont{R.}~\bibnamefont{Amin}},
  \bibinfo{author}{\bibfnamefont{S.~B.} \bibnamefont{Anderson}},
  \bibinfo{author}{\bibfnamefont{W.~G.} \bibnamefont{Anderson}},
  \bibinfo{author}{\bibfnamefont{M.~A.} \bibnamefont{Arain}},
  \bibinfo{author}{\bibfnamefont{M.}~\bibnamefont{Araya}},
  \bibinfo{author}{\bibfnamefont{H.}~\bibnamefont{Armandula}}, \emph{et~al.}
  (\bibinfo{collaboration}{LIGO Scientific Collaboration}),
  \bibinfo{year}{2008}, \bibinfo{journal}{Phys. Rev. Lett.}
  \textbf{\bibinfo{volume}{101}}, \bibinfo{pages}{211102},
  \urlprefix\url{http://link.aps.org/doi/10.1103/PhysRevLett.101.211102}.

\bibitem[{\citenamefont{{Abbott}} \emph{et~al.}(2009)\citenamefont{{Abbott},
  {Abbott}, {Acernese}, {Adhikari}, {Ajith}, {Allen}, {Allen}, {Alshourbagy},
  {Amin}, {Anderson}, and et~al.}}]{LIGO:SGWB2009}
\bibinfo{author}{\bibnamefont{{Abbott}}, \bibfnamefont{B.~P.}},
  \bibinfo{author}{\bibfnamefont{R.}~\bibnamefont{{Abbott}}},
  \bibinfo{author}{\bibfnamefont{F.}~\bibnamefont{{Acernese}}},
  \bibinfo{author}{\bibfnamefont{R.}~\bibnamefont{{Adhikari}}},
  \bibinfo{author}{\bibfnamefont{P.}~\bibnamefont{{Ajith}}},
  \bibinfo{author}{\bibfnamefont{B.}~\bibnamefont{{Allen}}},
  \bibinfo{author}{\bibfnamefont{G.}~\bibnamefont{{Allen}}},
  \bibinfo{author}{\bibfnamefont{M.}~\bibnamefont{{Alshourbagy}}},
  \bibinfo{author}{\bibfnamefont{R.~S.} \bibnamefont{{Amin}}},
  \bibinfo{author}{\bibfnamefont{S.~B.} \bibnamefont{{Anderson}}}, and
  \bibinfo{author}{\bibnamefont{et~al.}}, \bibinfo{year}{2009},
  \bibinfo{journal}{Nature} \textbf{\bibinfo{volume}{460}},
  \bibinfo{pages}{990}.

\bibitem[{\citenamefont{Abbott}
  \emph{et~al.}(2009{\natexlab{a}})\citenamefont{Abbott, Abbott, Adhikari,
  Ajith, Allen, Allen, Amin, Anderson, Anderson, Arain, Araya, Armandula}
  \emph{et~al.}}]{S5:Pulsars}
\bibinfo{author}{\bibnamefont{Abbott}, \bibfnamefont{B.~P.}},
  \bibinfo{author}{\bibfnamefont{R.}~\bibnamefont{Abbott}},
  \bibinfo{author}{\bibfnamefont{R.}~\bibnamefont{Adhikari}},
  \bibinfo{author}{\bibfnamefont{P.}~\bibnamefont{Ajith}},
  \bibinfo{author}{\bibfnamefont{B.}~\bibnamefont{Allen}},
  \bibinfo{author}{\bibfnamefont{G.}~\bibnamefont{Allen}},
  \bibinfo{author}{\bibfnamefont{R.~S.} \bibnamefont{Amin}},
  \bibinfo{author}{\bibfnamefont{S.~B.} \bibnamefont{Anderson}},
  \bibinfo{author}{\bibfnamefont{W.~G.} \bibnamefont{Anderson}},
  \bibinfo{author}{\bibfnamefont{M.~A.} \bibnamefont{Arain}},
  \bibinfo{author}{\bibfnamefont{M.}~\bibnamefont{Araya}},
  \bibinfo{author}{\bibfnamefont{H.}~\bibnamefont{Armandula}}, \emph{et~al.},
  \bibinfo{year}{2009}{\natexlab{a}}, \bibinfo{journal}{Phys. Rev. Lett.}
  \textbf{\bibinfo{volume}{102}}, \bibinfo{pages}{111102}.

\bibitem[{\citenamefont{Abbott}
  \emph{et~al.}(2009{\natexlab{b}})\citenamefont{Abbott, Abbott, Adhikari,
  Ajith, Allen, Allen, Amin, Anderson, Anderson, Arain, Araya, Armandula}
  \emph{et~al.}}]{PF:RPP2009}
\bibinfo{author}{\bibnamefont{Abbott}, \bibfnamefont{B.~P.}},
  \bibinfo{author}{\bibfnamefont{R.}~\bibnamefont{Abbott}},
  \bibinfo{author}{\bibfnamefont{R.}~\bibnamefont{Adhikari}},
  \bibinfo{author}{\bibfnamefont{P.}~\bibnamefont{Ajith}},
  \bibinfo{author}{\bibfnamefont{B.}~\bibnamefont{Allen}},
  \bibinfo{author}{\bibfnamefont{G.}~\bibnamefont{Allen}},
  \bibinfo{author}{\bibfnamefont{R.~S.} \bibnamefont{Amin}},
  \bibinfo{author}{\bibfnamefont{S.~B.} \bibnamefont{Anderson}},
  \bibinfo{author}{\bibfnamefont{W.~G.} \bibnamefont{Anderson}},
  \bibinfo{author}{\bibfnamefont{M.~A.} \bibnamefont{Arain}},
  \bibinfo{author}{\bibfnamefont{M.}~\bibnamefont{Araya}},
  \bibinfo{author}{\bibfnamefont{H.}~\bibnamefont{Armandula}}, \emph{et~al.},
  \bibinfo{year}{2009}{\natexlab{b}}, \bibinfo{journal}{Reports on Progress in
  Physics} \textbf{\bibinfo{volume}{72}}(\bibinfo{number}{7}),
  \bibinfo{pages}{076901+}, ISSN \bibinfo{issn}{0034-4885},
  \urlprefix\url{http://dx.doi.org/10.1088/0034-4885/72/7/076901}.

\bibitem[{\citenamefont{Abbott and Wise}(1984)}]{Wise:GW}
\bibinfo{author}{\bibnamefont{Abbott}, \bibfnamefont{L.}}, and
  \bibinfo{author}{\bibfnamefont{M.~B.} \bibnamefont{Wise}},
  \bibinfo{year}{1984}, \bibinfo{journal}{Nuclear Physics B}
  \textbf{\bibinfo{volume}{244}}(\bibinfo{number}{2}), \bibinfo{pages}{541 },
  ISSN \bibinfo{issn}{0550-3213},
  \urlprefix\url{http://www.sciencedirect.com/science/article/pii/0550321384903298}.

\bibitem[{\citenamefont{Abbott}
  \emph{et~al.}(2004{\natexlab{b}})\citenamefont{Abbott, Adhikari, Allen,
  Baglino, Campbell, Coyne, Daw, DeBra, Faludi, Fritschel}
  \emph{et~al.}}]{abbott2004seismic}
\bibinfo{author}{\bibnamefont{Abbott}, \bibfnamefont{R.}},
  \bibinfo{author}{\bibfnamefont{R.}~\bibnamefont{Adhikari}},
  \bibinfo{author}{\bibfnamefont{G.}~\bibnamefont{Allen}},
  \bibinfo{author}{\bibfnamefont{D.}~\bibnamefont{Baglino}},
  \bibinfo{author}{\bibfnamefont{C.}~\bibnamefont{Campbell}},
  \bibinfo{author}{\bibfnamefont{D.}~\bibnamefont{Coyne}},
  \bibinfo{author}{\bibfnamefont{E.}~\bibnamefont{Daw}},
  \bibinfo{author}{\bibfnamefont{D.}~\bibnamefont{DeBra}},
  \bibinfo{author}{\bibfnamefont{J.}~\bibnamefont{Faludi}},
  \bibinfo{author}{\bibfnamefont{P.}~\bibnamefont{Fritschel}}, \emph{et~al.},
  \bibinfo{year}{2004}{\natexlab{b}}, \bibinfo{journal}{Classical and Quantum
  Gravity} \textbf{\bibinfo{volume}{21}}(\bibinfo{number}{5}),
  \bibinfo{pages}{S915}.

\bibitem[{\citenamefont{Abbott} \emph{et~al.}(2002)\citenamefont{Abbott,
  Adhikari, Allen, Cowley, Daw, DeBra, Giaime, Hammond, Hammond, Hardham, How,
  Hua} \emph{et~al.}}]{aLIGO:Seismic2002}
\bibinfo{author}{\bibnamefont{Abbott}, \bibfnamefont{R.}},
  \bibinfo{author}{\bibfnamefont{R.}~\bibnamefont{Adhikari}},
  \bibinfo{author}{\bibfnamefont{G.}~\bibnamefont{Allen}},
  \bibinfo{author}{\bibfnamefont{S.}~\bibnamefont{Cowley}},
  \bibinfo{author}{\bibfnamefont{E.}~\bibnamefont{Daw}},
  \bibinfo{author}{\bibfnamefont{D.}~\bibnamefont{DeBra}},
  \bibinfo{author}{\bibfnamefont{J.}~\bibnamefont{Giaime}},
  \bibinfo{author}{\bibfnamefont{G.}~\bibnamefont{Hammond}},
  \bibinfo{author}{\bibfnamefont{M.}~\bibnamefont{Hammond}},
  \bibinfo{author}{\bibfnamefont{C.}~\bibnamefont{Hardham}},
  \bibinfo{author}{\bibfnamefont{J.}~\bibnamefont{How}},
  \bibinfo{author}{\bibfnamefont{W.}~\bibnamefont{Hua}}, \emph{et~al.},
  \bibinfo{year}{2002}, \bibinfo{journal}{Classical and Quantum Gravity}
  \textbf{\bibinfo{volume}{19}}(\bibinfo{number}{7}), \bibinfo{pages}{1591}.

\bibitem[{\citenamefont{Abramovici}
  \emph{et~al.}(1992)\citenamefont{Abramovici, Althouse, Drever, G{\"u}rsel,
  Kawamura, Raab, Shoemaker, Sievers, Spero, Thorne, Vogt, Weiss}
  \emph{et~al.}}]{LIGO:Science}
\bibinfo{author}{\bibnamefont{Abramovici}, \bibfnamefont{A.}},
  \bibinfo{author}{\bibfnamefont{W.~E.} \bibnamefont{Althouse}},
  \bibinfo{author}{\bibfnamefont{R.~W.~P.} \bibnamefont{Drever}},
  \bibinfo{author}{\bibfnamefont{Y.}~\bibnamefont{G{\"u}rsel}},
  \bibinfo{author}{\bibfnamefont{S.}~\bibnamefont{Kawamura}},
  \bibinfo{author}{\bibfnamefont{F.~J.} \bibnamefont{Raab}},
  \bibinfo{author}{\bibfnamefont{D.}~\bibnamefont{Shoemaker}},
  \bibinfo{author}{\bibfnamefont{L.}~\bibnamefont{Sievers}},
  \bibinfo{author}{\bibfnamefont{R.~E.} \bibnamefont{Spero}},
  \bibinfo{author}{\bibfnamefont{K.~S.} \bibnamefont{Thorne}},
  \bibinfo{author}{\bibfnamefont{R.~E.} \bibnamefont{Vogt}},
  \bibinfo{author}{\bibfnamefont{R.}~\bibnamefont{Weiss}}, \emph{et~al.},
  \bibinfo{year}{1992}, \bibinfo{journal}{Science}
  \textbf{\bibinfo{volume}{256}}(\bibinfo{number}{5055}), \bibinfo{pages}{325}.

\bibitem[{\citenamefont{{Accadia}} \emph{et~al.}(2012)\citenamefont{{Accadia},
  {Acernese}, {Alshourbagy}, {Amico}, {Antonucci}, {Aoudia}, {Arnaud},
  {Arnault}, {Arun}, {Astone}, and et~al.}}]{Virgo:Review2012}
\bibinfo{author}{\bibnamefont{{Accadia}}, \bibfnamefont{T.}},
  \bibinfo{author}{\bibfnamefont{F.}~\bibnamefont{{Acernese}}},
  \bibinfo{author}{\bibfnamefont{M.}~\bibnamefont{{Alshourbagy}}},
  \bibinfo{author}{\bibfnamefont{P.}~\bibnamefont{{Amico}}},
  \bibinfo{author}{\bibfnamefont{F.}~\bibnamefont{{Antonucci}}},
  \bibinfo{author}{\bibfnamefont{S.}~\bibnamefont{{Aoudia}}},
  \bibinfo{author}{\bibfnamefont{N.}~\bibnamefont{{Arnaud}}},
  \bibinfo{author}{\bibfnamefont{C.}~\bibnamefont{{Arnault}}},
  \bibinfo{author}{\bibfnamefont{K.~G.} \bibnamefont{{Arun}}},
  \bibinfo{author}{\bibfnamefont{P.}~\bibnamefont{{Astone}}}, and
  \bibinfo{author}{\bibnamefont{et~al.}}, \bibinfo{year}{2012},
  \bibinfo{journal}{Journal of Instrumentation} \textbf{\bibinfo{volume}{7}},
  \bibinfo{pages}{3012}.

\bibitem[{\citenamefont{Accadia}
  \emph{et~al.}(2010{\natexlab{a}})\citenamefont{Accadia, Acernese, Antonucci,
  Astone, Ballardin, Barone, Barsuglia, Basti, Bauer, and
  Beker}}]{Virgo:ASC2010}
\bibinfo{author}{\bibnamefont{Accadia}, \bibfnamefont{T.}},
  \bibinfo{author}{\bibfnamefont{F.}~\bibnamefont{Acernese}},
  \bibinfo{author}{\bibfnamefont{F.}~\bibnamefont{Antonucci}},
  \bibinfo{author}{\bibfnamefont{P.}~\bibnamefont{Astone}},
  \bibinfo{author}{\bibfnamefont{G.}~\bibnamefont{Ballardin}},
  \bibinfo{author}{\bibfnamefont{F.}~\bibnamefont{Barone}},
  \bibinfo{author}{\bibfnamefont{M.}~\bibnamefont{Barsuglia}},
  \bibinfo{author}{\bibfnamefont{A.}~\bibnamefont{Basti}},
  \bibinfo{author}{\bibfnamefont{T.}~\bibnamefont{Bauer}}, and
  \bibinfo{author}{\bibfnamefont{M.~G.} \bibnamefont{Beker}},
  \bibinfo{year}{2010}{\natexlab{a}}, \bibinfo{journal}{Astroparticle Physics}
  ISSN \bibinfo{issn}{09276505},
  \urlprefix\url{http://dx.doi.org/10.1016/j.astropartphys.2010.10.005}.

\bibitem[{\citenamefont{{Accadia}} \emph{et~al.}(2011)\citenamefont{{Accadia},
  {Acernese}, {Antonucci}, {Astone}, {Ballardin}, {Barone}, {Barsuglia},
  {Basti}, {Bauer}, {Beker}, {Belletoile}, {Birindelli}}
  \emph{et~al.}}]{Virgo:LSC2011}
\bibinfo{author}{\bibnamefont{{Accadia}}, \bibfnamefont{T.}},
  \bibinfo{author}{\bibfnamefont{F.}~\bibnamefont{{Acernese}}},
  \bibinfo{author}{\bibfnamefont{F.}~\bibnamefont{{Antonucci}}},
  \bibinfo{author}{\bibfnamefont{P.}~\bibnamefont{{Astone}}},
  \bibinfo{author}{\bibfnamefont{G.}~\bibnamefont{{Ballardin}}},
  \bibinfo{author}{\bibfnamefont{F.}~\bibnamefont{{Barone}}},
  \bibinfo{author}{\bibfnamefont{M.}~\bibnamefont{{Barsuglia}}},
  \bibinfo{author}{\bibfnamefont{A.}~\bibnamefont{{Basti}}},
  \bibinfo{author}{\bibfnamefont{T.~S.} \bibnamefont{{Bauer}}},
  \bibinfo{author}{\bibfnamefont{M.~G.} \bibnamefont{{Beker}}},
  \bibinfo{author}{\bibfnamefont{A.}~\bibnamefont{{Belletoile}}},
  \bibinfo{author}{\bibfnamefont{S.}~\bibnamefont{{Birindelli}}},
  \emph{et~al.}, \bibinfo{year}{2011}, \bibinfo{journal}{Astroparticle Physics}
  \textbf{\bibinfo{volume}{34}}, \bibinfo{pages}{521}.

\bibitem[{\citenamefont{Accadia} \emph{et~al.}(2011)\citenamefont{Accadia,
  Acernese, Antonucci, Astone, Ballardin, Barone, Barsuglia, Bauer, Beker,
  Belletoile, Birindelli, Bitossi} \emph{et~al.}}]{Accadia:2011jh}
\bibinfo{author}{\bibnamefont{Accadia}, \bibfnamefont{T.}},
  \bibinfo{author}{\bibfnamefont{F.}~\bibnamefont{Acernese}},
  \bibinfo{author}{\bibfnamefont{F.}~\bibnamefont{Antonucci}},
  \bibinfo{author}{\bibfnamefont{P.}~\bibnamefont{Astone}},
  \bibinfo{author}{\bibfnamefont{G.}~\bibnamefont{Ballardin}},
  \bibinfo{author}{\bibfnamefont{F.}~\bibnamefont{Barone}},
  \bibinfo{author}{\bibfnamefont{M.}~\bibnamefont{Barsuglia}},
  \bibinfo{author}{\bibfnamefont{T.~S.} \bibnamefont{Bauer}},
  \bibinfo{author}{\bibfnamefont{M.~G.} \bibnamefont{Beker}},
  \bibinfo{author}{\bibfnamefont{A.}~\bibnamefont{Belletoile}},
  \bibinfo{author}{\bibfnamefont{S.}~\bibnamefont{Birindelli}},
  \bibinfo{author}{\bibfnamefont{M.}~\bibnamefont{Bitossi}}, \emph{et~al.},
  \bibinfo{year}{2011}, \bibinfo{journal}{Low Frequency Noise, Vibration and
  Active Control} \textbf{\bibinfo{volume}{30}}(\bibinfo{number}{1}),
  \bibinfo{pages}{63}.

\bibitem[{\citenamefont{Accadia} \emph{et~al.}(2012)\citenamefont{Accadia,
  Acernese, Astone, Ballardin, Barone, Barsuglia, Basti, Bauer, Bebronne,
  Beker, Belletoile, Bitossi} \emph{et~al.}}]{Accadia:2011ge}
\bibinfo{author}{\bibnamefont{Accadia}, \bibfnamefont{T.}},
  \bibinfo{author}{\bibfnamefont{F.}~\bibnamefont{Acernese}},
  \bibinfo{author}{\bibfnamefont{P.}~\bibnamefont{Astone}},
  \bibinfo{author}{\bibfnamefont{G.}~\bibnamefont{Ballardin}},
  \bibinfo{author}{\bibfnamefont{F.}~\bibnamefont{Barone}},
  \bibinfo{author}{\bibfnamefont{M.}~\bibnamefont{Barsuglia}},
  \bibinfo{author}{\bibfnamefont{A.}~\bibnamefont{Basti}},
  \bibinfo{author}{\bibfnamefont{T.~S.} \bibnamefont{Bauer}},
  \bibinfo{author}{\bibfnamefont{M.}~\bibnamefont{Bebronne}},
  \bibinfo{author}{\bibfnamefont{M.~G.} \bibnamefont{Beker}},
  \bibinfo{author}{\bibfnamefont{A.}~\bibnamefont{Belletoile}},
  \bibinfo{author}{\bibfnamefont{M.}~\bibnamefont{Bitossi}}, \emph{et~al.},
  \bibinfo{year}{2012}, \bibinfo{journal}{Classical and Quantum Gravity}
  \textbf{\bibinfo{volume}{29}}(\bibinfo{number}{2}), \bibinfo{pages}{025005},
  \urlprefix\url{http://stacks.iop.org/0264-9381/29/i=2/a=025005}.

\bibitem[{\citenamefont{Accadia}
  \emph{et~al.}(2010{\natexlab{b}})\citenamefont{Accadia, Swinkels, and
  Collaboration}}]{Bas:Virgo2010}
\bibinfo{author}{\bibnamefont{Accadia}, \bibfnamefont{T.}},
  \bibinfo{author}{\bibfnamefont{B.~L.} \bibnamefont{Swinkels}}, and
  \bibinfo{author}{\bibfnamefont{V.}~\bibnamefont{Collaboration}},
  \bibinfo{year}{2010}{\natexlab{b}}, \bibinfo{journal}{Classical and Quantum
  Gravity} \textbf{\bibinfo{volume}{27}}(\bibinfo{number}{8}),
  \bibinfo{pages}{4002}.

\bibitem[{\citenamefont{{Acernese}}
  \emph{et~al.}(2008)\citenamefont{{Acernese}, {Alshourbagy}, {Amico},
  {Antonucci}, {Aoudia}, {Arun}, {Astone}, {Avino}, {Baggio}, {Ballardin},
  {Barone}, {Barsotti}} \emph{et~al.}}]{VIR2008}
\bibinfo{author}{\bibnamefont{{Acernese}}, \bibfnamefont{F.}},
  \bibinfo{author}{\bibfnamefont{M.}~\bibnamefont{{Alshourbagy}}},
  \bibinfo{author}{\bibfnamefont{P.}~\bibnamefont{{Amico}}},
  \bibinfo{author}{\bibfnamefont{F.}~\bibnamefont{{Antonucci}}},
  \bibinfo{author}{\bibfnamefont{S.}~\bibnamefont{{Aoudia}}},
  \bibinfo{author}{\bibfnamefont{K.~G.} \bibnamefont{{Arun}}},
  \bibinfo{author}{\bibfnamefont{P.}~\bibnamefont{{Astone}}},
  \bibinfo{author}{\bibfnamefont{S.}~\bibnamefont{{Avino}}},
  \bibinfo{author}{\bibfnamefont{L.}~\bibnamefont{{Baggio}}},
  \bibinfo{author}{\bibfnamefont{G.}~\bibnamefont{{Ballardin}}},
  \bibinfo{author}{\bibfnamefont{F.}~\bibnamefont{{Barone}}},
  \bibinfo{author}{\bibfnamefont{L.}~\bibnamefont{{Barsotti}}}, \emph{et~al.},
  \bibinfo{year}{2008}, \bibinfo{journal}{Classical and Quantum Gravity}
  \textbf{\bibinfo{volume}{25}}(\bibinfo{number}{18}), \bibinfo{pages}{184001}.

\bibitem[{\citenamefont{Acernese}
  \emph{et~al.}(2010{\natexlab{a}})\citenamefont{Acernese, Alshourbagy,
  Antonucci, Aoudia, Arun, Astone, Ballardin, Barone, Barsuglia, Bauer}
  \emph{et~al.}}]{Virgo:ASC}
\bibinfo{author}{\bibnamefont{Acernese}, \bibfnamefont{F.}},
  \bibinfo{author}{\bibfnamefont{M.}~\bibnamefont{Alshourbagy}},
  \bibinfo{author}{\bibfnamefont{F.}~\bibnamefont{Antonucci}},
  \bibinfo{author}{\bibfnamefont{S.}~\bibnamefont{Aoudia}},
  \bibinfo{author}{\bibfnamefont{K.}~\bibnamefont{Arun}},
  \bibinfo{author}{\bibfnamefont{P.}~\bibnamefont{Astone}},
  \bibinfo{author}{\bibfnamefont{G.}~\bibnamefont{Ballardin}},
  \bibinfo{author}{\bibfnamefont{F.}~\bibnamefont{Barone}},
  \bibinfo{author}{\bibfnamefont{M.}~\bibnamefont{Barsuglia}},
  \bibinfo{author}{\bibfnamefont{T.}~\bibnamefont{Bauer}}, \emph{et~al.},
  \bibinfo{year}{2010}{\natexlab{a}}, \bibinfo{journal}{Astroparticle Physics}
  \textbf{\bibinfo{volume}{33}}(\bibinfo{number}{3}), \bibinfo{pages}{131}.

\bibitem[{\citenamefont{Acernese} \emph{et~al.}(2009)\citenamefont{Acernese,
  Alshourbagy, Antonucci, Aoudia, Arun, Astone, Ballardin, Barone, Barsotti,
  Barsuglia, Bauer, Bigotta} \emph{et~al.}}]{Virgo:fnoise2009}
\bibinfo{author}{\bibnamefont{Acernese}, \bibfnamefont{F.}},
  \bibinfo{author}{\bibfnamefont{M.}~\bibnamefont{Alshourbagy}},
  \bibinfo{author}{\bibfnamefont{F.}~\bibnamefont{Antonucci}},
  \bibinfo{author}{\bibfnamefont{S.}~\bibnamefont{Aoudia}},
  \bibinfo{author}{\bibfnamefont{K.~G.} \bibnamefont{Arun}},
  \bibinfo{author}{\bibfnamefont{P.}~\bibnamefont{Astone}},
  \bibinfo{author}{\bibfnamefont{G.}~\bibnamefont{Ballardin}},
  \bibinfo{author}{\bibfnamefont{F.}~\bibnamefont{Barone}},
  \bibinfo{author}{\bibfnamefont{L.}~\bibnamefont{Barsotti}},
  \bibinfo{author}{\bibfnamefont{M.}~\bibnamefont{Barsuglia}},
  \bibinfo{author}{\bibfnamefont{T.~S.} \bibnamefont{Bauer}},
  \bibinfo{author}{\bibfnamefont{S.}~\bibnamefont{Bigotta}}, \emph{et~al.},
  \bibinfo{year}{2009}, \bibinfo{journal}{Phys. Rev. A}
  \textbf{\bibinfo{volume}{79}}, \bibinfo{pages}{053824},
  \urlprefix\url{http://link.aps.org/doi/10.1103/PhysRevA.79.053824}.

\bibitem[{\citenamefont{Acernese} \emph{et~al.}(2006)\citenamefont{Acernese,
  Amico, Al-Shourbagy, Aoudia, Avino, Babusci, Ballardin, Barille, Barone,
  Barsotti, Barsuglia, Beauville} \emph{et~al.}}]{Acernese:2006gm}
\bibinfo{author}{\bibnamefont{Acernese}, \bibfnamefont{F.}},
  \bibinfo{author}{\bibfnamefont{P.}~\bibnamefont{Amico}},
  \bibinfo{author}{\bibfnamefont{M.}~\bibnamefont{Al-Shourbagy}},
  \bibinfo{author}{\bibfnamefont{S.}~\bibnamefont{Aoudia}},
  \bibinfo{author}{\bibfnamefont{S.}~\bibnamefont{Avino}},
  \bibinfo{author}{\bibfnamefont{D.}~\bibnamefont{Babusci}},
  \bibinfo{author}{\bibfnamefont{G.}~\bibnamefont{Ballardin}},
  \bibinfo{author}{\bibfnamefont{R.}~\bibnamefont{Barille}},
  \bibinfo{author}{\bibfnamefont{F.}~\bibnamefont{Barone}},
  \bibinfo{author}{\bibfnamefont{L.}~\bibnamefont{Barsotti}},
  \bibinfo{author}{\bibfnamefont{M.}~\bibnamefont{Barsuglia}},
  \bibinfo{author}{\bibfnamefont{F.}~\bibnamefont{Beauville}}, \emph{et~al.},
  \bibinfo{year}{2006}, \bibinfo{journal}{IEEE Transactions on Instrumentation
  and Measurement} \textbf{\bibinfo{volume}{55}}(\bibinfo{number}{6}),
  \bibinfo{pages}{1985}.

\bibitem[{\citenamefont{{Acernese}}
  \emph{et~al.}(2006)\citenamefont{{Acernese}, {Amico}, {Al-Shourbagy},
  {Aoudia}, {Avino}, {Babusci}, {Ballardin}, {Barill{\'e}}, {Barone},
  {Barsotti}, {Barsuglia}, {Beauville}} \emph{et~al.}}]{Virgo:ASC2006}
\bibinfo{author}{\bibnamefont{{Acernese}}, \bibfnamefont{F.}},
  \bibinfo{author}{\bibfnamefont{P.}~\bibnamefont{{Amico}}},
  \bibinfo{author}{\bibfnamefont{M.}~\bibnamefont{{Al-Shourbagy}}},
  \bibinfo{author}{\bibfnamefont{S.}~\bibnamefont{{Aoudia}}},
  \bibinfo{author}{\bibfnamefont{S.}~\bibnamefont{{Avino}}},
  \bibinfo{author}{\bibfnamefont{D.}~\bibnamefont{{Babusci}}},
  \bibinfo{author}{\bibfnamefont{G.}~\bibnamefont{{Ballardin}}},
  \bibinfo{author}{\bibfnamefont{R.}~\bibnamefont{{Barill{\'e}}}},
  \bibinfo{author}{\bibfnamefont{F.}~\bibnamefont{{Barone}}},
  \bibinfo{author}{\bibfnamefont{L.}~\bibnamefont{{Barsotti}}},
  \bibinfo{author}{\bibfnamefont{M.}~\bibnamefont{{Barsuglia}}},
  \bibinfo{author}{\bibfnamefont{F.}~\bibnamefont{{Beauville}}}, \emph{et~al.},
  \bibinfo{year}{2006}, \bibinfo{journal}{Classical and Quantum Gravity}
  \textbf{\bibinfo{volume}{23}}, \bibinfo{pages}{91}.

\bibitem[{\citenamefont{Acernese} \emph{et~al.}(2007)\citenamefont{Acernese,
  Amico, Alshourbagy, Antonucci, Aoudia, Astone, Avino, Babusci, Ballardin,
  Barone} \emph{et~al.}}]{Virgo:Optical}
\bibinfo{author}{\bibnamefont{Acernese}, \bibfnamefont{F.}},
  \bibinfo{author}{\bibfnamefont{P.}~\bibnamefont{Amico}},
  \bibinfo{author}{\bibfnamefont{M.}~\bibnamefont{Alshourbagy}},
  \bibinfo{author}{\bibfnamefont{F.}~\bibnamefont{Antonucci}},
  \bibinfo{author}{\bibfnamefont{S.}~\bibnamefont{Aoudia}},
  \bibinfo{author}{\bibfnamefont{P.}~\bibnamefont{Astone}},
  \bibinfo{author}{\bibfnamefont{S.}~\bibnamefont{Avino}},
  \bibinfo{author}{\bibfnamefont{D.}~\bibnamefont{Babusci}},
  \bibinfo{author}{\bibfnamefont{G.}~\bibnamefont{Ballardin}},
  \bibinfo{author}{\bibfnamefont{F.}~\bibnamefont{Barone}}, \emph{et~al.},
  \bibinfo{year}{2007}, \bibinfo{journal}{Applied Optics}
  \textbf{\bibinfo{volume}{46}}, \bibinfo{pages}{3466}.

\bibitem[{\citenamefont{{Acernese}}
  \emph{et~al.}(2004)\citenamefont{{Acernese}, {Amico}, {Arnaud}, {Babusci},
  {Barill{\'e}}, {Barone}, {Barsotti}, {Barsuglia}, {Beauville}, {Bizouard},
  {Boccara}, {Bondu}} \emph{et~al.}}]{VirgoSeis:2004}
\bibinfo{author}{\bibnamefont{{Acernese}}, \bibfnamefont{F.}},
  \bibinfo{author}{\bibfnamefont{P.}~\bibnamefont{{Amico}}},
  \bibinfo{author}{\bibfnamefont{N.}~\bibnamefont{{Arnaud}}},
  \bibinfo{author}{\bibfnamefont{D.}~\bibnamefont{{Babusci}}},
  \bibinfo{author}{\bibfnamefont{R.}~\bibnamefont{{Barill{\'e}}}},
  \bibinfo{author}{\bibfnamefont{F.}~\bibnamefont{{Barone}}},
  \bibinfo{author}{\bibfnamefont{L.}~\bibnamefont{{Barsotti}}},
  \bibinfo{author}{\bibfnamefont{M.}~\bibnamefont{{Barsuglia}}},
  \bibinfo{author}{\bibfnamefont{F.}~\bibnamefont{{Beauville}}},
  \bibinfo{author}{\bibfnamefont{M.~A.} \bibnamefont{{Bizouard}}},
  \bibinfo{author}{\bibfnamefont{C.}~\bibnamefont{{Boccara}}},
  \bibinfo{author}{\bibfnamefont{F.}~\bibnamefont{{Bondu}}}, \emph{et~al.},
  \bibinfo{year}{2004}, \bibinfo{journal}{Classical and Quantum Gravity}
  \textbf{\bibinfo{volume}{21}}, \bibinfo{pages}{433}.

\bibitem[{\citenamefont{Acernese}
  \emph{et~al.}(2010{\natexlab{b}})\citenamefont{Acernese, Antonucci, Aoudia,
  Arun, Astone, Ballardin, Barone, Barsuglia, Bauer, Beker, Bigotta,
  Birindelli} \emph{et~al.}}]{Virgo:SA2010}
\bibinfo{author}{\bibnamefont{Acernese}, \bibfnamefont{F.}},
  \bibinfo{author}{\bibfnamefont{F.}~\bibnamefont{Antonucci}},
  \bibinfo{author}{\bibfnamefont{S.}~\bibnamefont{Aoudia}},
  \bibinfo{author}{\bibfnamefont{K.}~\bibnamefont{Arun}},
  \bibinfo{author}{\bibfnamefont{P.}~\bibnamefont{Astone}},
  \bibinfo{author}{\bibfnamefont{G.}~\bibnamefont{Ballardin}},
  \bibinfo{author}{\bibfnamefont{F.}~\bibnamefont{Barone}},
  \bibinfo{author}{\bibfnamefont{M.}~\bibnamefont{Barsuglia}},
  \bibinfo{author}{\bibfnamefont{T.}~\bibnamefont{Bauer}},
  \bibinfo{author}{\bibfnamefont{M.}~\bibnamefont{Beker}},
  \bibinfo{author}{\bibfnamefont{S.}~\bibnamefont{Bigotta}},
  \bibinfo{author}{\bibfnamefont{S.}~\bibnamefont{Birindelli}}, \emph{et~al.},
  \bibinfo{year}{2010}{\natexlab{b}}, \bibinfo{journal}{Astroparticle Physics}
  \textbf{\bibinfo{volume}{33}}(\bibinfo{number}{3}), \bibinfo{pages}{182 },
  ISSN \bibinfo{issn}{0927-6505}.

\bibitem[{\citenamefont{{Acernese}}
  \emph{et~al.}(2010)\citenamefont{{Acernese}, {Antonucci}, {Aoudia}, {Arun},
  {Astone}, {Ballardin}, {Barone}, {Barsuglia}, {Bauer}, {Beker}, {Bigotta},
  {Birindelli}} \emph{et~al.}}]{Virgo:LSC2010}
\bibinfo{author}{\bibnamefont{{Acernese}}, \bibfnamefont{F.}},
  \bibinfo{author}{\bibfnamefont{F.}~\bibnamefont{{Antonucci}}},
  \bibinfo{author}{\bibfnamefont{S.}~\bibnamefont{{Aoudia}}},
  \bibinfo{author}{\bibfnamefont{K.~G.} \bibnamefont{{Arun}}},
  \bibinfo{author}{\bibfnamefont{P.}~\bibnamefont{{Astone}}},
  \bibinfo{author}{\bibfnamefont{G.}~\bibnamefont{{Ballardin}}},
  \bibinfo{author}{\bibfnamefont{F.}~\bibnamefont{{Barone}}},
  \bibinfo{author}{\bibfnamefont{M.}~\bibnamefont{{Barsuglia}}},
  \bibinfo{author}{\bibfnamefont{T.~S.} \bibnamefont{{Bauer}}},
  \bibinfo{author}{\bibfnamefont{M.~G.} \bibnamefont{{Beker}}},
  \bibinfo{author}{\bibfnamefont{S.}~\bibnamefont{{Bigotta}}},
  \bibinfo{author}{\bibfnamefont{S.}~\bibnamefont{{Birindelli}}},
  \emph{et~al.}, \bibinfo{year}{2010}, \bibinfo{journal}{Astroparticle Physics}
  \textbf{\bibinfo{volume}{33}}, \bibinfo{pages}{75}.

\bibitem[{\citenamefont{Adhikari}(2004)}]{Rana:PhD}
\bibinfo{author}{\bibnamefont{Adhikari}, \bibfnamefont{R.}},
  \bibinfo{year}{2004}, \emph{\bibinfo{title}{Sensitivity and Noise Analysis of
  4 km Laser Interferometric Gravitational Wave Antennae}}, Ph.D. thesis,
  \bibinfo{school}{{Massachusetts Institute of Technology}},
  \urlprefix\url{http://hdl.handle.net/1721.1/28646}.

\bibitem[{\citenamefont{Ageev} \emph{et~al.}(2004)\citenamefont{Ageev, Palmer,
  Felice, Penn, and Saulson}}]{Penn:2004}
\bibinfo{author}{\bibnamefont{Ageev}, \bibfnamefont{A.}},
  \bibinfo{author}{\bibfnamefont{B.~C.} \bibnamefont{Palmer}},
  \bibinfo{author}{\bibfnamefont{A.~D.} \bibnamefont{Felice}},
  \bibinfo{author}{\bibfnamefont{S.~D.} \bibnamefont{Penn}}, and
  \bibinfo{author}{\bibfnamefont{P.~R.} \bibnamefont{Saulson}},
  \bibinfo{year}{2004}, \bibinfo{journal}{Classical and Quantum Gravity}
  \textbf{\bibinfo{volume}{21}}(\bibinfo{number}{16}), \bibinfo{pages}{3887}.

\bibitem[{\citenamefont{Ajith} \emph{et~al.}(2012)\citenamefont{Ajith, Boyle,
  Brown, Br{\"u}gmann, Buchman, Cadonati, Campanelli, Chu, Etienne, Fairhurst,
  Hannam, Healy} \emph{et~al.}}]{ninja2}
\bibinfo{author}{\bibnamefont{Ajith}, \bibfnamefont{P.}},
  \bibinfo{author}{\bibfnamefont{M.}~\bibnamefont{Boyle}},
  \bibinfo{author}{\bibfnamefont{D.~A.} \bibnamefont{Brown}},
  \bibinfo{author}{\bibfnamefont{B.}~\bibnamefont{Br{\"u}gmann}},
  \bibinfo{author}{\bibfnamefont{L.~T.} \bibnamefont{Buchman}},
  \bibinfo{author}{\bibfnamefont{L.}~\bibnamefont{Cadonati}},
  \bibinfo{author}{\bibfnamefont{M.}~\bibnamefont{Campanelli}},
  \bibinfo{author}{\bibfnamefont{T.}~\bibnamefont{Chu}},
  \bibinfo{author}{\bibfnamefont{Z.~B.} \bibnamefont{Etienne}},
  \bibinfo{author}{\bibfnamefont{S.}~\bibnamefont{Fairhurst}},
  \bibinfo{author}{\bibfnamefont{M.}~\bibnamefont{Hannam}},
  \bibinfo{author}{\bibfnamefont{J.}~\bibnamefont{Healy}}, \emph{et~al.},
  \bibinfo{year}{2012}, \bibinfo{journal}{Classical and Quantum Gravity}
  \textbf{\bibinfo{volume}{29}}(\bibinfo{number}{12}), \bibinfo{pages}{124001},
  \urlprefix\url{http://stacks.iop.org/0264-9381/29/i=12/a=124001}.

\bibitem[{\citenamefont{Aki and Richards}(2009)}]{AkRi2009}
\bibinfo{author}{\bibnamefont{Aki}, \bibfnamefont{K.}}, and
  \bibinfo{author}{\bibfnamefont{P.~G.} \bibnamefont{Richards}},
  \bibinfo{year}{2009}, \emph{\bibinfo{title}{{Quantitative Seismology, 2nd
  edition}}} (\bibinfo{publisher}{University Science Books}).

\bibitem[{\citenamefont{Allen}(1988)}]{Bruce:1988}
\bibinfo{author}{\bibnamefont{Allen}, \bibfnamefont{B.}}, \bibinfo{year}{1988},
  \bibinfo{journal}{Phys. Rev. D} \textbf{\bibinfo{volume}{37}},
  \bibinfo{pages}{2078},
  \urlprefix\url{http://link.aps.org/doi/10.1103/PhysRevD.37.2078}.

\bibitem[{\citenamefont{{Allen}}(1997)}]{All1996}
\bibinfo{author}{\bibnamefont{{Allen}}, \bibfnamefont{B.}},
  \bibinfo{year}{1997}, in \emph{\bibinfo{booktitle}{Relativistic Gravitation
  and Gravitational Radiation}}, edited by
  \bibinfo{editor}{\bibnamefont{{J.-A.~Marck \& J.-P.~Lasota}}}, p.
  \bibinfo{pages}{373}, \eprint{arXiv:gr-qc/9604033}.

\bibitem[{\citenamefont{Allen} \emph{et~al.}(2012)\citenamefont{Allen,
  Anderson, Brady, Brown, and Creighton}}]{FINDCHIRP}
\bibinfo{author}{\bibnamefont{Allen}, \bibfnamefont{B.}},
  \bibinfo{author}{\bibfnamefont{W.~G.} \bibnamefont{Anderson}},
  \bibinfo{author}{\bibfnamefont{P.~R.} \bibnamefont{Brady}},
  \bibinfo{author}{\bibfnamefont{D.~A.} \bibnamefont{Brown}}, and
  \bibinfo{author}{\bibfnamefont{J.~D.~E.} \bibnamefont{Creighton}},
  \bibinfo{year}{2012}, \bibinfo{journal}{Phys. Rev. D}
  \textbf{\bibinfo{volume}{85}}, \bibinfo{pages}{122006},
  \urlprefix\url{http://link.aps.org/doi/10.1103/PhysRevD.85.122006}.

\bibitem[{\citenamefont{Ando}
  \emph{et~al.}(2010{\natexlab{a}})\citenamefont{Ando, Ishidoshiro, Yamamoto,
  Yagi, Kokuyama, Tsubono, and Takamori}}]{Ando:TOBA}
\bibinfo{author}{\bibnamefont{Ando}, \bibfnamefont{M.}},
  \bibinfo{author}{\bibfnamefont{K.}~\bibnamefont{Ishidoshiro}},
  \bibinfo{author}{\bibfnamefont{K.}~\bibnamefont{Yamamoto}},
  \bibinfo{author}{\bibfnamefont{K.}~\bibnamefont{Yagi}},
  \bibinfo{author}{\bibfnamefont{W.}~\bibnamefont{Kokuyama}},
  \bibinfo{author}{\bibfnamefont{K.}~\bibnamefont{Tsubono}}, and
  \bibinfo{author}{\bibfnamefont{A.}~\bibnamefont{Takamori}},
  \bibinfo{year}{2010}{\natexlab{a}}, \bibinfo{journal}{Phys. Rev. Lett.}
  \textbf{\bibinfo{volume}{105}}, \bibinfo{pages}{161101}.

\bibitem[{\citenamefont{Ando}
  \emph{et~al.}(2010{\natexlab{b}})\citenamefont{Ando, Kawamura, Seto, Sato,
  Nakamura, Tsubono, Takashima, Funaki, Numata, Kanda, Tanaka, Ioka}
  \emph{et~al.}}]{Ando:DECIGO}
\bibinfo{author}{\bibnamefont{Ando}, \bibfnamefont{M.}},
  \bibinfo{author}{\bibfnamefont{S.}~\bibnamefont{Kawamura}},
  \bibinfo{author}{\bibfnamefont{N.}~\bibnamefont{Seto}},
  \bibinfo{author}{\bibfnamefont{S.}~\bibnamefont{Sato}},
  \bibinfo{author}{\bibfnamefont{T.}~\bibnamefont{Nakamura}},
  \bibinfo{author}{\bibfnamefont{K.}~\bibnamefont{Tsubono}},
  \bibinfo{author}{\bibfnamefont{T.}~\bibnamefont{Takashima}},
  \bibinfo{author}{\bibfnamefont{I.}~\bibnamefont{Funaki}},
  \bibinfo{author}{\bibfnamefont{K.}~\bibnamefont{Numata}},
  \bibinfo{author}{\bibfnamefont{N.}~\bibnamefont{Kanda}},
  \bibinfo{author}{\bibfnamefont{T.}~\bibnamefont{Tanaka}},
  \bibinfo{author}{\bibfnamefont{K.}~\bibnamefont{Ioka}}, \emph{et~al.},
  \bibinfo{year}{2010}{\natexlab{b}}, \bibinfo{journal}{Classical and Quantum
  Gravity} \textbf{\bibinfo{volume}{27}}(\bibinfo{number}{8}),
  \bibinfo{pages}{084010}.

\bibitem[{\citenamefont{Ando} \emph{et~al.}(2012)\citenamefont{Ando, Baret,
  Bouhou, Chassande-Motting, Kouchner, Moscoso, Van~Elewyck, Bartos, Marka,
  Marka, Corsi, Di~Palma} \emph{et~al.}}]{GWHENreview}
\bibinfo{author}{\bibnamefont{Ando}, \bibfnamefont{S.}},
  \bibinfo{author}{\bibfnamefont{B.}~\bibnamefont{Baret}},
  \bibinfo{author}{\bibfnamefont{B.}~\bibnamefont{Bouhou}},
  \bibinfo{author}{\bibfnamefont{E.}~\bibnamefont{Chassande-Motting}},
  \bibinfo{author}{\bibfnamefont{A.}~\bibnamefont{Kouchner}},
  \bibinfo{author}{\bibfnamefont{L.}~\bibnamefont{Moscoso}},
  \bibinfo{author}{\bibfnamefont{V.}~\bibnamefont{Van~Elewyck}},
  \bibinfo{author}{\bibfnamefont{I.}~\bibnamefont{Bartos}},
  \bibinfo{author}{\bibfnamefont{S.}~\bibnamefont{Marka}},
  \bibinfo{author}{\bibfnamefont{Z.}~\bibnamefont{Marka}},
  \bibinfo{author}{\bibfnamefont{A.}~\bibnamefont{Corsi}},
  \bibinfo{author}{\bibfnamefont{I.}~\bibnamefont{Di~Palma}}, \emph{et~al.},
  \bibinfo{year}{2012}, \bibinfo{journal}{submitted to RMP}
  \bibinfo{note}{(LIGO-P1100194)}.

\bibitem[{\citenamefont{Anholm} \emph{et~al.}(2009)\citenamefont{Anholm,
  Ballmer, Creighton, Price, and Siemens}}]{Larry:PulsarTiming2009}
\bibinfo{author}{\bibnamefont{Anholm}, \bibfnamefont{M.}},
  \bibinfo{author}{\bibfnamefont{S.}~\bibnamefont{Ballmer}},
  \bibinfo{author}{\bibfnamefont{J.~D.~E.} \bibnamefont{Creighton}},
  \bibinfo{author}{\bibfnamefont{L.~R.} \bibnamefont{Price}}, and
  \bibinfo{author}{\bibfnamefont{X.}~\bibnamefont{Siemens}},
  \bibinfo{year}{2009}, \bibinfo{journal}{Phys. Rev. D}
  \textbf{\bibinfo{volume}{79}}, \bibinfo{pages}{084030},
  \urlprefix\url{http://link.aps.org/doi/10.1103/PhysRevD.79.084030}.

\bibitem[{\citenamefont{Arai} \emph{et~al.}(2009)\citenamefont{Arai, Takahashi,
  Tatsumi, Izumi, Wakabayashi, Ishizaki, Fukushima, Yamazaki, Fujimoto,
  Takamori, Tsubono, DeSalvo} \emph{et~al.}}]{TAMA:2009}
\bibinfo{author}{\bibnamefont{Arai}, \bibfnamefont{K.}},
  \bibinfo{author}{\bibfnamefont{R.}~\bibnamefont{Takahashi}},
  \bibinfo{author}{\bibfnamefont{D.}~\bibnamefont{Tatsumi}},
  \bibinfo{author}{\bibfnamefont{K.}~\bibnamefont{Izumi}},
  \bibinfo{author}{\bibfnamefont{Y.}~\bibnamefont{Wakabayashi}},
  \bibinfo{author}{\bibfnamefont{H.}~\bibnamefont{Ishizaki}},
  \bibinfo{author}{\bibfnamefont{M.}~\bibnamefont{Fukushima}},
  \bibinfo{author}{\bibfnamefont{T.}~\bibnamefont{Yamazaki}},
  \bibinfo{author}{\bibfnamefont{M.-K.} \bibnamefont{Fujimoto}},
  \bibinfo{author}{\bibfnamefont{A.}~\bibnamefont{Takamori}},
  \bibinfo{author}{\bibfnamefont{K.}~\bibnamefont{Tsubono}},
  \bibinfo{author}{\bibfnamefont{R.}~\bibnamefont{DeSalvo}}, \emph{et~al.},
  \bibinfo{year}{2009}, \bibinfo{journal}{Classical and Quantum Gravity}
  \textbf{\bibinfo{volume}{26}}(\bibinfo{number}{20}), \bibinfo{pages}{204020}.

\bibitem[{\citenamefont{{Arai} and {TAMA Collaboration}}(2002)}]{TAMA:LSC}
\bibinfo{author}{\bibnamefont{{Arai}}, \bibfnamefont{K.}}, and
  \bibinfo{author}{\bibnamefont{{TAMA Collaboration}}}, \bibinfo{year}{2002},
  \bibinfo{journal}{Classical and Quantum Gravity}
  \textbf{\bibinfo{volume}{19}}, \bibinfo{pages}{1843}.

\bibitem[{\citenamefont{Arain and Mueller}(2008)}]{Arain:2008fu}
\bibinfo{author}{\bibnamefont{Arain}, \bibfnamefont{M.~A.}}, and
  \bibinfo{author}{\bibfnamefont{G.}~\bibnamefont{Mueller}},
  \bibinfo{year}{2008}, \bibinfo{journal}{Optics Express}
  \textbf{\bibinfo{volume}{16}}(\bibinfo{number}{14}), \bibinfo{pages}{10018}.

\bibitem[{\citenamefont{Armstrong}(2006)}]{Armstrong:LRR}
\bibinfo{author}{\bibnamefont{Armstrong}, \bibfnamefont{J.~W.}},
  \bibinfo{year}{2006}, \bibinfo{journal}{Living Reviews in Relativity}
  \textbf{\bibinfo{volume}{9}}(\bibinfo{number}{1}),
  \urlprefix\url{http://www.livingreviews.org/lrr-2006-1}.

\bibitem[{\citenamefont{Armstrong} \emph{et~al.}(1999)\citenamefont{Armstrong,
  Estabrook, and Tinto}}]{Armstrong:1999hp}
\bibinfo{author}{\bibnamefont{Armstrong}, \bibfnamefont{J.~W.}},
  \bibinfo{author}{\bibfnamefont{F.~B.} \bibnamefont{Estabrook}}, and
  \bibinfo{author}{\bibfnamefont{M.}~\bibnamefont{Tinto}},
  \bibinfo{year}{1999}, \bibinfo{journal}{The Astrophysical Journal}
  \textbf{\bibinfo{volume}{527}}(\bibinfo{number}{2}), \bibinfo{pages}{814}.

\bibitem[{\citenamefont{Armstrong} \emph{et~al.}(2003)\citenamefont{Armstrong,
  Iess, Tortora, and Bertotti}}]{Armstrong:SGW}
\bibinfo{author}{\bibnamefont{Armstrong}, \bibfnamefont{J.~W.}},
  \bibinfo{author}{\bibfnamefont{L.}~\bibnamefont{Iess}},
  \bibinfo{author}{\bibfnamefont{P.}~\bibnamefont{Tortora}}, and
  \bibinfo{author}{\bibfnamefont{B.}~\bibnamefont{Bertotti}},
  \bibinfo{year}{2003}, \bibinfo{journal}{Astrophysical Journal} .

\bibitem[{\citenamefont{{Asmar}} \emph{et~al.}(2005)\citenamefont{{Asmar},
  {Armstrong}, {Iess}, and {Tortora}}}]{Armstrong:2005}
\bibinfo{author}{\bibnamefont{{Asmar}}, \bibfnamefont{S.~W.}},
  \bibinfo{author}{\bibfnamefont{J.~W.} \bibnamefont{{Armstrong}}},
  \bibinfo{author}{\bibfnamefont{L.}~\bibnamefont{{Iess}}}, and
  \bibinfo{author}{\bibfnamefont{P.}~\bibnamefont{{Tortora}}},
  \bibinfo{year}{2005}, \bibinfo{journal}{Radio Science}
  \textbf{\bibinfo{volume}{40}}, \bibinfo{eid}{RS2001}.

\bibitem[{\citenamefont{Aso and Araya}(2012)}]{Aso:Seismic}
\bibinfo{author}{\bibnamefont{Aso}, \bibfnamefont{Y.}}, and
  \bibinfo{author}{\bibfnamefont{A.}~\bibnamefont{Araya}},
  \bibinfo{year}{2012}, \bibinfo{title}{Kamioka seismic data},
  \bibinfo{howpublished}{personal communication}.

\bibitem[{\citenamefont{Aston} \emph{et~al.}(2012)\citenamefont{Aston, Barton,
  Bell, Beveridge, Bland, Brummitt, Cagnoli, Cantley, Carbone, Cumming,
  Cunningham, Cutler} \emph{et~al.}}]{Aston:2012}
\bibinfo{author}{\bibnamefont{Aston}, \bibfnamefont{S.~M.}},
  \bibinfo{author}{\bibfnamefont{M.~A.} \bibnamefont{Barton}},
  \bibinfo{author}{\bibfnamefont{A.~S.} \bibnamefont{Bell}},
  \bibinfo{author}{\bibfnamefont{N.}~\bibnamefont{Beveridge}},
  \bibinfo{author}{\bibfnamefont{B.}~\bibnamefont{Bland}},
  \bibinfo{author}{\bibfnamefont{A.~J.} \bibnamefont{Brummitt}},
  \bibinfo{author}{\bibfnamefont{G.}~\bibnamefont{Cagnoli}},
  \bibinfo{author}{\bibfnamefont{C.~A.} \bibnamefont{Cantley}},
  \bibinfo{author}{\bibfnamefont{L.}~\bibnamefont{Carbone}},
  \bibinfo{author}{\bibfnamefont{A.~V.} \bibnamefont{Cumming}},
  \bibinfo{author}{\bibfnamefont{L.}~\bibnamefont{Cunningham}},
  \bibinfo{author}{\bibfnamefont{R.~M.} \bibnamefont{Cutler}}, \emph{et~al.},
  \bibinfo{year}{2012}, \bibinfo{journal}{Classical and Quantum Gravity}
  \textbf{\bibinfo{volume}{29}}(\bibinfo{number}{23}), \bibinfo{pages}{235004},
  \urlprefix\url{http://stacks.iop.org/0264-9381/29/i=23/a=235004}.

\bibitem[{\citenamefont{Aufmuth and Danzmann}(2005)}]{Aufmuth:2005vv}
\bibinfo{author}{\bibnamefont{Aufmuth}, \bibfnamefont{P.}}, and
  \bibinfo{author}{\bibfnamefont{K.}~\bibnamefont{Danzmann}},
  \bibinfo{year}{2005}, \bibinfo{journal}{New Journal of Physics}
  \textbf{\bibinfo{volume}{7}}, \bibinfo{pages}{252}.

\bibitem[{\citenamefont{Augst and Drever}(2000)}]{Drever:ParaQ}
\bibinfo{author}{\bibnamefont{Augst}, \bibfnamefont{S.~J.}}, and
  \bibinfo{author}{\bibfnamefont{R.~W.~P.} \bibnamefont{Drever}},
  \bibinfo{year}{2000}, \bibinfo{journal}{AIP Conference Proceedings}
  \textbf{\bibinfo{volume}{523}}(\bibinfo{number}{1}), \bibinfo{pages}{338},
  \urlprefix\url{http://link.aip.org/link/?APC/523/338/1}.

\bibitem[{\citenamefont{Ballardin} \emph{et~al.}(2001)\citenamefont{Ballardin,
  Bracci, Braccini, Bradaschia, Casciano, Calamai, Cavalieri, Cecchi, Cella,
  Cuoco, D'Ambrosio, Dattilo} \emph{et~al.}}]{Stefano:2001}
\bibinfo{author}{\bibnamefont{Ballardin}, \bibfnamefont{G.}},
  \bibinfo{author}{\bibfnamefont{L.}~\bibnamefont{Bracci}},
  \bibinfo{author}{\bibfnamefont{S.}~\bibnamefont{Braccini}},
  \bibinfo{author}{\bibfnamefont{C.}~\bibnamefont{Bradaschia}},
  \bibinfo{author}{\bibfnamefont{C.}~\bibnamefont{Casciano}},
  \bibinfo{author}{\bibfnamefont{G.}~\bibnamefont{Calamai}},
  \bibinfo{author}{\bibfnamefont{R.}~\bibnamefont{Cavalieri}},
  \bibinfo{author}{\bibfnamefont{R.}~\bibnamefont{Cecchi}},
  \bibinfo{author}{\bibfnamefont{G.}~\bibnamefont{Cella}},
  \bibinfo{author}{\bibfnamefont{E.}~\bibnamefont{Cuoco}},
  \bibinfo{author}{\bibfnamefont{E.}~\bibnamefont{D'Ambrosio}},
  \bibinfo{author}{\bibfnamefont{V.}~\bibnamefont{Dattilo}}, \emph{et~al.},
  \bibinfo{year}{2001}, \bibinfo{journal}{Review of Scientific Instruments}
  \textbf{\bibinfo{volume}{72}}(\bibinfo{number}{9}), \bibinfo{pages}{3643}.

\bibitem[{\citenamefont{{Ballmer}}(2006)}]{Stefan:Thesis}
\bibinfo{author}{\bibnamefont{{Ballmer}}, \bibfnamefont{S.~W.}},
  \bibinfo{year}{2006}, \emph{\bibinfo{title}{{LIGO interferometer operating at
  design sensitivity with application to gravitational radiometry}}}, Ph.D.
  thesis, \bibinfo{school}{Massachusetts Institute of Technology,
  Massachusetts, USA}.

\bibitem[{\citenamefont{{Balser} and {Wagner}}(1960)}]{Balser:1960}
\bibinfo{author}{\bibnamefont{{Balser}}, \bibfnamefont{M.}}, and
  \bibinfo{author}{\bibfnamefont{C.~A.} \bibnamefont{{Wagner}}},
  \bibinfo{year}{1960}, \bibinfo{journal}{Nature}
  \textbf{\bibinfo{volume}{188}}, \bibinfo{pages}{638}.

\bibitem[{\citenamefont{Barish and Weiss}(1999)}]{BaWe:PhysToday}
\bibinfo{author}{\bibnamefont{Barish}, \bibfnamefont{B.}}, and
  \bibinfo{author}{\bibfnamefont{R.}~\bibnamefont{Weiss}},
  \bibinfo{year}{1999}, \bibinfo{journal}{Physics Today}
  \textbf{\bibinfo{volume}{52}}(\bibinfo{number}{10}), \bibinfo{pages}{44}.

\bibitem[{\citenamefont{Barkhausen}(1919)}]{barkhausen1919rauschen}
\bibinfo{author}{\bibnamefont{Barkhausen}, \bibfnamefont{H.}},
  \bibinfo{year}{1919}, \bibinfo{journal}{Phys. Zeitschrift}
  \textbf{\bibinfo{volume}{20}}, \bibinfo{pages}{401}.

\bibitem[{\citenamefont{Barsotti}(2006)}]{Lisa:PhD}
\bibinfo{author}{\bibnamefont{Barsotti}, \bibfnamefont{L.}},
  \bibinfo{year}{2006}, \emph{\bibinfo{title}{The control of the Virgo
  interferometer for gravitational wave detection}}, Ph.D. thesis,
  \bibinfo{school}{University of Pisa}.

\bibitem[{\citenamefont{Barsotti} \emph{et~al.}(2010)\citenamefont{Barsotti,
  Evans, and Fritschel}}]{aLIGO:ASC}
\bibinfo{author}{\bibnamefont{Barsotti}, \bibfnamefont{L.}},
  \bibinfo{author}{\bibfnamefont{M.}~\bibnamefont{Evans}}, and
  \bibinfo{author}{\bibfnamefont{P.}~\bibnamefont{Fritschel}},
  \bibinfo{year}{2010}, \bibinfo{journal}{Classical and Quantum Gravity}
  \textbf{\bibinfo{volume}{27}}(\bibinfo{number}{8}), \bibinfo{pages}{084026},
  \urlprefix\url{http://stacks.iop.org/0264-9381/27/i=8/a=084026}.

\bibitem[{\citenamefont{Bass and Mahajan}(2009)}]{bass2009handbook}
\bibinfo{author}{\bibnamefont{Bass}, \bibfnamefont{M.}}, and
  \bibinfo{author}{\bibfnamefont{V.}~\bibnamefont{Mahajan}},
  \bibinfo{year}{2009}, \emph{\bibinfo{title}{Handbook of Optics: Geometrical
  and Physical Optics, Polarized Light, Components and Instruments}},
  \bibinfo{number}{v. 1} (\bibinfo{publisher}{McGraw-Hill}), ISBN
  \bibinfo{isbn}{9780071498890},
  \urlprefix\url{http://books.google.com/books?id=xmDntwAACAAJ}.

\bibitem[{\citenamefont{Bassiri} \emph{et~al.}(2012)\citenamefont{Bassiri,
  Evans, Borisenko, Fejer, Hough, MacLaren, Martin, Route, and
  Rowan}}]{Bassiri2012}
\bibinfo{author}{\bibnamefont{Bassiri}, \bibfnamefont{R.}},
  \bibinfo{author}{\bibfnamefont{K.}~\bibnamefont{Evans}},
  \bibinfo{author}{\bibfnamefont{K.}~\bibnamefont{Borisenko}},
  \bibinfo{author}{\bibfnamefont{M.}~\bibnamefont{Fejer}},
  \bibinfo{author}{\bibfnamefont{J.}~\bibnamefont{Hough}},
  \bibinfo{author}{\bibfnamefont{I.}~\bibnamefont{MacLaren}},
  \bibinfo{author}{\bibfnamefont{I.}~\bibnamefont{Martin}},
  \bibinfo{author}{\bibfnamefont{R.}~\bibnamefont{Route}}, and
  \bibinfo{author}{\bibfnamefont{S.}~\bibnamefont{Rowan}},
  \bibinfo{year}{2012}, \bibinfo{journal}{Acta Materialia}
  (\bibinfo{number}{0}), , ISSN \bibinfo{issn}{1359-6454},
  \urlprefix\url{http://www.sciencedirect.com/science/article/pii/S1359645412007343}.

\bibitem[{\citenamefont{{Baumann}} \emph{et~al.}(2009)\citenamefont{{Baumann},
  {Cooray}, {Dodelson}, {Dunkley}, {Fraisse}, {Jackson}, {Kogut}, {Krauss},
  {Smith}, and {Zaldarriaga}}}]{CMBpol:2009}
\bibinfo{author}{\bibnamefont{{Baumann}}, \bibfnamefont{D.}},
  \bibinfo{author}{\bibfnamefont{A.}~\bibnamefont{{Cooray}}},
  \bibinfo{author}{\bibfnamefont{S.}~\bibnamefont{{Dodelson}}},
  \bibinfo{author}{\bibfnamefont{J.}~\bibnamefont{{Dunkley}}},
  \bibinfo{author}{\bibfnamefont{A.~A.} \bibnamefont{{Fraisse}}},
  \bibinfo{author}{\bibfnamefont{M.~G.} \bibnamefont{{Jackson}}},
  \bibinfo{author}{\bibfnamefont{A.}~\bibnamefont{{Kogut}}},
  \bibinfo{author}{\bibfnamefont{L.~M.} \bibnamefont{{Krauss}}},
  \bibinfo{author}{\bibfnamefont{K.~M.} \bibnamefont{{Smith}}}, and
  \bibinfo{author}{\bibfnamefont{M.}~\bibnamefont{{Zaldarriaga}}},
  \bibinfo{year}{2009}, in \emph{\bibinfo{booktitle}{American Institute of
  Physics Conference Series}}, edited by
  \bibinfo{editor}{\bibfnamefont{S.}~\bibnamefont{{Dodelson}}},
  \bibinfo{editor}{\bibfnamefont{D.}~\bibnamefont{{Baumann}}},
  \bibinfo{editor}{\bibfnamefont{A.}~\bibnamefont{{Cooray}}},
  \bibinfo{editor}{\bibfnamefont{J.}~\bibnamefont{{Dunkley}}},
  \bibinfo{editor}{\bibfnamefont{A.}~\bibnamefont{{Fraisse}}},
  \bibinfo{editor}{\bibfnamefont{M.~G.} \bibnamefont{{Jackson}}},
  \bibinfo{editor}{\bibfnamefont{A.}~\bibnamefont{{Kogut}}},
  \bibinfo{editor}{\bibfnamefont{L.}~\bibnamefont{{Krauss}}},
  \bibinfo{editor}{\bibfnamefont{M.}~\bibnamefont{{Zaldarriaga}}}, and
  \bibinfo{editor}{\bibfnamefont{K.}~\bibnamefont{{Smith}}}, volume
  \bibinfo{volume}{1141} of \emph{\bibinfo{series}{American Institute of
  Physics Conference Series}}, pp. \bibinfo{pages}{3--9}, \eprint{0811.3911}.

\bibitem[{\citenamefont{Beccaria}
  \emph{et~al.}(1998{\natexlab{a}})\citenamefont{Beccaria, Bernardini,
  Braccini, Bradaschia, Bozzi, Casciano, Cella, Ciampa, Cuoco, Curci,
  D'Ambrosio, Dattilo} \emph{et~al.}}]{BeEA1998}
\bibinfo{author}{\bibnamefont{Beccaria}, \bibfnamefont{M.}},
  \bibinfo{author}{\bibfnamefont{M.}~\bibnamefont{Bernardini}},
  \bibinfo{author}{\bibfnamefont{S.}~\bibnamefont{Braccini}},
  \bibinfo{author}{\bibfnamefont{C.}~\bibnamefont{Bradaschia}},
  \bibinfo{author}{\bibfnamefont{A.}~\bibnamefont{Bozzi}},
  \bibinfo{author}{\bibfnamefont{C.}~\bibnamefont{Casciano}},
  \bibinfo{author}{\bibfnamefont{G.}~\bibnamefont{Cella}},
  \bibinfo{author}{\bibfnamefont{A.}~\bibnamefont{Ciampa}},
  \bibinfo{author}{\bibfnamefont{E.}~\bibnamefont{Cuoco}},
  \bibinfo{author}{\bibfnamefont{G.}~\bibnamefont{Curci}},
  \bibinfo{author}{\bibfnamefont{E.}~\bibnamefont{D'Ambrosio}},
  \bibinfo{author}{\bibfnamefont{V.}~\bibnamefont{Dattilo}}, \emph{et~al.},
  \bibinfo{year}{1998}{\natexlab{a}}, \bibinfo{journal}{Classical and Quantum
  Gravity} \textbf{\bibinfo{volume}{15}}(\bibinfo{number}{11}),
  \bibinfo{pages}{3339}.

\bibitem[{\citenamefont{Beccaria}
  \emph{et~al.}(1998{\natexlab{b}})\citenamefont{Beccaria, Bernardini,
  Braccini, Bradaschia, Bozzi, Casciano, Cella, Ciampa, Cuoco, Curci,
  D'Ambrosio, Dattilo} \emph{et~al.}}]{Cella:1998}
\bibinfo{author}{\bibnamefont{Beccaria}, \bibfnamefont{M.}},
  \bibinfo{author}{\bibfnamefont{M.}~\bibnamefont{Bernardini}},
  \bibinfo{author}{\bibfnamefont{S.}~\bibnamefont{Braccini}},
  \bibinfo{author}{\bibfnamefont{C.}~\bibnamefont{Bradaschia}},
  \bibinfo{author}{\bibfnamefont{A.}~\bibnamefont{Bozzi}},
  \bibinfo{author}{\bibfnamefont{C.}~\bibnamefont{Casciano}},
  \bibinfo{author}{\bibfnamefont{G.}~\bibnamefont{Cella}},
  \bibinfo{author}{\bibfnamefont{A.}~\bibnamefont{Ciampa}},
  \bibinfo{author}{\bibfnamefont{E.}~\bibnamefont{Cuoco}},
  \bibinfo{author}{\bibfnamefont{G.}~\bibnamefont{Curci}},
  \bibinfo{author}{\bibfnamefont{E.}~\bibnamefont{D'Ambrosio}},
  \bibinfo{author}{\bibfnamefont{V.}~\bibnamefont{Dattilo}}, \emph{et~al.},
  \bibinfo{year}{1998}{\natexlab{b}}, \bibinfo{journal}{Classical and Quantum
  Gravity} \textbf{\bibinfo{volume}{15}}(\bibinfo{number}{11}),
  \bibinfo{pages}{3339},
  \urlprefix\url{http://stacks.iop.org/0264-9381/15/i=11/a=004}.

\bibitem[{\citenamefont{{Beker}} \emph{et~al.}(2011)\citenamefont{{Beker},
  {Cella}, {Desalvo}, {Doets}, {Grote}, {Harms}, {Hennes}, {Mandic},
  {Rabeling}, {van den Brand}, and {van Leeuwen}}}]{Beker:2011}
\bibinfo{author}{\bibnamefont{{Beker}}, \bibfnamefont{M.~G.}},
  \bibinfo{author}{\bibfnamefont{G.}~\bibnamefont{{Cella}}},
  \bibinfo{author}{\bibfnamefont{R.}~\bibnamefont{{Desalvo}}},
  \bibinfo{author}{\bibfnamefont{M.}~\bibnamefont{{Doets}}},
  \bibinfo{author}{\bibfnamefont{H.}~\bibnamefont{{Grote}}},
  \bibinfo{author}{\bibfnamefont{J.}~\bibnamefont{{Harms}}},
  \bibinfo{author}{\bibfnamefont{E.}~\bibnamefont{{Hennes}}},
  \bibinfo{author}{\bibfnamefont{V.}~\bibnamefont{{Mandic}}},
  \bibinfo{author}{\bibfnamefont{D.~S.} \bibnamefont{{Rabeling}}},
  \bibinfo{author}{\bibfnamefont{J.~F.~J.} \bibnamefont{{van den Brand}}}, and
  \bibinfo{author}{\bibfnamefont{C.~M.} \bibnamefont{{van Leeuwen}}},
  \bibinfo{year}{2011}, \bibinfo{journal}{General Relativity and Gravitation}
  \textbf{\bibinfo{volume}{43}}, \bibinfo{pages}{623}.

\bibitem[{\citenamefont{{Belczynski}}
  \emph{et~al.}(2002)\citenamefont{{Belczynski}, {Kalogera}, and
  {Bulik}}}]{BKB2002}
\bibinfo{author}{\bibnamefont{{Belczynski}}, \bibfnamefont{K.}},
  \bibinfo{author}{\bibfnamefont{V.}~\bibnamefont{{Kalogera}}}, and
  \bibinfo{author}{\bibfnamefont{T.}~\bibnamefont{{Bulik}}},
  \bibinfo{year}{2002}, \bibinfo{journal}{The Astrophysical Journal}
  \textbf{\bibinfo{volume}{572}}, \bibinfo{pages}{407}.

\bibitem[{\citenamefont{Bender}(2011)}]{Bender:PRD2011}
\bibinfo{author}{\bibnamefont{Bender}, \bibfnamefont{P.~L.}},
  \bibinfo{year}{2011}, \bibinfo{journal}{Phys. Rev. D}
  \textbf{\bibinfo{volume}{84}}, \bibinfo{pages}{028101},
  \urlprefix\url{http://link.aps.org/doi/10.1103/PhysRevD.84.028101}.

\bibitem[{\citenamefont{{Bender}}(2012)}]{Bender:Atom2011}
\bibinfo{author}{\bibnamefont{{Bender}}, \bibfnamefont{P.~L.}},
  \bibinfo{year}{2012}, \bibinfo{journal}{General Relativity and Gravitation}
  \textbf{\bibinfo{volume}{44}}, \bibinfo{pages}{711}.

\bibitem[{\citenamefont{{Bennett}} \emph{et~al.}(2012)\citenamefont{{Bennett},
  {Larson}, {Weiland}, {Jarosik}, {Hinshaw}, {Odegard}, {Smith}, {Hill},
  {Gold}, {Halpern}, {Komatsu}, {Nolta}} \emph{et~al.}}]{WMAP:9year}
\bibinfo{author}{\bibnamefont{{Bennett}}, \bibfnamefont{C.~L.}},
  \bibinfo{author}{\bibfnamefont{D.}~\bibnamefont{{Larson}}},
  \bibinfo{author}{\bibfnamefont{J.~L.} \bibnamefont{{Weiland}}},
  \bibinfo{author}{\bibfnamefont{N.}~\bibnamefont{{Jarosik}}},
  \bibinfo{author}{\bibfnamefont{G.}~\bibnamefont{{Hinshaw}}},
  \bibinfo{author}{\bibfnamefont{N.}~\bibnamefont{{Odegard}}},
  \bibinfo{author}{\bibfnamefont{K.~M.} \bibnamefont{{Smith}}},
  \bibinfo{author}{\bibfnamefont{R.~S.} \bibnamefont{{Hill}}},
  \bibinfo{author}{\bibfnamefont{B.}~\bibnamefont{{Gold}}},
  \bibinfo{author}{\bibfnamefont{M.}~\bibnamefont{{Halpern}}},
  \bibinfo{author}{\bibfnamefont{E.}~\bibnamefont{{Komatsu}}},
  \bibinfo{author}{\bibfnamefont{M.~R.} \bibnamefont{{Nolta}}}, \emph{et~al.},
  \bibinfo{year}{2012}, \bibinfo{journal}{ArXiv e-prints} \eprint{1212.5225}.

\bibitem[{\citenamefont{Bernard and Callen}(1959)}]{Callen:1959}
\bibinfo{author}{\bibnamefont{Bernard}, \bibfnamefont{W.}}, and
  \bibinfo{author}{\bibfnamefont{H.~B.} \bibnamefont{Callen}},
  \bibinfo{year}{1959}, \bibinfo{journal}{Rev. Mod. Phys.}
  \textbf{\bibinfo{volume}{31}}, \bibinfo{pages}{1017},
  \urlprefix\url{http://link.aps.org/doi/10.1103/RevModPhys.31.1017}.

\bibitem[{\citenamefont{Bertotti}(1998)}]{bertotti1998hysteresis}
\bibinfo{author}{\bibnamefont{Bertotti}, \bibfnamefont{G.}},
  \bibinfo{year}{1998}, \emph{\bibinfo{title}{Hysteresis in Magnetism: For
  Physicists, Materials Scientists, and Engineers}}, Electromagnetism Series
  (\bibinfo{publisher}{Academic Press}), ISBN \bibinfo{isbn}{9780120932702},
  \urlprefix\url{http://books.google.com/books?id=B7xZctzCPfgC}.

\bibitem[{\citenamefont{Beyersdorf}
  \emph{et~al.}(2000)\citenamefont{Beyersdorf, Byer, and Fejer}}]{Byer:Sagnac}
\bibinfo{author}{\bibnamefont{Beyersdorf}, \bibfnamefont{P.~T.}},
  \bibinfo{author}{\bibfnamefont{R.~L.} \bibnamefont{Byer}}, and
  \bibinfo{author}{\bibfnamefont{M.~M.} \bibnamefont{Fejer}},
  \bibinfo{year}{2000}, \bibinfo{journal}{AIP Conference Proceedings}
  \textbf{\bibinfo{volume}{523}}(\bibinfo{number}{1}), \bibinfo{pages}{200}.

\bibitem[{\citenamefont{Bildsten}(1998)}]{Bildsten:1998}
\bibinfo{author}{\bibnamefont{Bildsten}, \bibfnamefont{L.}},
  \bibinfo{year}{1998}, \bibinfo{journal}{The Astrophysical Journal Letters}
  \textbf{\bibinfo{volume}{501}}(\bibinfo{number}{1}), \bibinfo{pages}{L89}.

\bibitem[{\citenamefont{{Bittel}}(1969)}]{Bittel:1969}
\bibinfo{author}{\bibnamefont{{Bittel}}, \bibfnamefont{H.}},
  \bibinfo{year}{1969}, \bibinfo{journal}{IEEE Transactions on Magnetics}
  \textbf{\bibinfo{volume}{5}}, \bibinfo{pages}{359}.

\bibitem[{\citenamefont{Blair} \emph{et~al.}(2012)\citenamefont{Blair, Howell,
  Ju, and Zhao}}]{blair2012advanced}
\bibinfo{author}{\bibnamefont{Blair}, \bibfnamefont{D.}},
  \bibinfo{author}{\bibfnamefont{E.}~\bibnamefont{Howell}},
  \bibinfo{author}{\bibfnamefont{L.}~\bibnamefont{Ju}}, and
  \bibinfo{author}{\bibfnamefont{C.}~\bibnamefont{Zhao}}, \bibinfo{year}{2012},
  \emph{\bibinfo{title}{Advanced Gravitational Wave Detectors}}
  (\bibinfo{publisher}{Cambridge University Press}), ISBN
  \bibinfo{isbn}{9780521874298},
  \urlprefix\url{http://books.google.com/books?id=mvVBkgQZrecC}.

\bibitem[{\citenamefont{Bochner}(2003)}]{Brett:2003}
\bibinfo{author}{\bibnamefont{Bochner}, \bibfnamefont{B.}},
  \bibinfo{year}{2003}, \bibinfo{journal}{General Relativity and Gravitation}
  \textbf{\bibinfo{volume}{35}}, \bibinfo{pages}{1029}, ISSN
  \bibinfo{issn}{0001-7701}, \bibinfo{note}{10.1023/A:1024016917669}.

\bibitem[{\citenamefont{{Borselli}}
  \emph{et~al.}(2006)\citenamefont{{Borselli}, {Johnson}, and
  {Painter}}}]{Oskar:SurfaceChem}
\bibinfo{author}{\bibnamefont{{Borselli}}, \bibfnamefont{M.}},
  \bibinfo{author}{\bibfnamefont{T.~J.} \bibnamefont{{Johnson}}}, and
  \bibinfo{author}{\bibfnamefont{O.}~\bibnamefont{{Painter}}},
  \bibinfo{year}{2006}, \bibinfo{journal}{Applied Physics Letters}
  \textbf{\bibinfo{volume}{88}}(\bibinfo{number}{13}), \bibinfo{eid}{131114}.

\bibitem[{\citenamefont{Brady} \emph{et~al.}(1998)\citenamefont{Brady,
  Creighton, Cutler, and Schutz}}]{BCCS1998}
\bibinfo{author}{\bibnamefont{Brady}, \bibfnamefont{P.~R.}},
  \bibinfo{author}{\bibfnamefont{T.}~\bibnamefont{Creighton}},
  \bibinfo{author}{\bibfnamefont{C.}~\bibnamefont{Cutler}}, and
  \bibinfo{author}{\bibfnamefont{B.~F.} \bibnamefont{Schutz}},
  \bibinfo{year}{1998}, \bibinfo{journal}{Phys.~Rev.~D}
  \textbf{\bibinfo{volume}{57}}, \bibinfo{pages}{2101}.

\bibitem[{\citenamefont{Braginsky}(2008)}]{Braginsky:2008ea}
\bibinfo{author}{\bibnamefont{Braginsky}, \bibfnamefont{V.~B.}},
  \bibinfo{year}{2008}, \bibinfo{journal}{Astronomy Letters}
  \textbf{\bibinfo{volume}{34}}(\bibinfo{number}{8}), \bibinfo{pages}{558}.

\bibitem[{\citenamefont{Braginsky} \emph{et~al.}(2000)\citenamefont{Braginsky,
  Gorodetsky, Khalili, and Thorne}}]{BGKT2000}
\bibinfo{author}{\bibnamefont{Braginsky}, \bibfnamefont{V.~B.}},
  \bibinfo{author}{\bibfnamefont{M.~L.} \bibnamefont{Gorodetsky}},
  \bibinfo{author}{\bibfnamefont{F.~Y.} \bibnamefont{Khalili}}, and
  \bibinfo{author}{\bibfnamefont{K.~S.} \bibnamefont{Thorne}},
  \bibinfo{year}{2000}, \bibinfo{journal}{Phys.~Rev.~D}
  \textbf{\bibinfo{volume}{61}}, \bibinfo{pages}{044002}.

\bibitem[{\citenamefont{Braginsky and Khalili}(1996)}]{BrKh1996a}
\bibinfo{author}{\bibnamefont{Braginsky}, \bibfnamefont{V.~B.}}, and
  \bibinfo{author}{\bibfnamefont{F.~Y.} \bibnamefont{Khalili}},
  \bibinfo{year}{1996}, \bibinfo{journal}{Rev.~Mod.~Phys.}
  \textbf{\bibinfo{volume}{68}}, \bibinfo{pages}{1}.

\bibitem[{\citenamefont{Braginsky and Khalili}(1999)}]{BrKh1999a}
\bibinfo{author}{\bibnamefont{Braginsky}, \bibfnamefont{V.~B.}}, and
  \bibinfo{author}{\bibfnamefont{F.~Y.} \bibnamefont{Khalili}},
  \bibinfo{year}{1999}, \emph{\bibinfo{title}{Quantum Measurement}}
  (\bibinfo{publisher}{Cambridge University Press}).

\bibitem[{\citenamefont{Braginsky} \emph{et~al.}(2006)\citenamefont{Braginsky,
  Ryazhskaya, and Vyatchanin}}]{Braginsky:2006ip}
\bibinfo{author}{\bibnamefont{Braginsky}, \bibfnamefont{V.~B.}},
  \bibinfo{author}{\bibfnamefont{O.~G.} \bibnamefont{Ryazhskaya}}, and
  \bibinfo{author}{\bibfnamefont{S.~P.} \bibnamefont{Vyatchanin}},
  \bibinfo{year}{2006}, \bibinfo{journal}{Physics Letters A}
  \textbf{\bibinfo{volume}{359}}(\bibinfo{number}{2}), \bibinfo{pages}{86}.

\bibitem[{\citenamefont{{Braginsky}}
  \emph{et~al.}(2001)\citenamefont{{Braginsky}, {Strigin}, and
  {Vyatchanin}}}]{BSV2001}
\bibinfo{author}{\bibnamefont{{Braginsky}}, \bibfnamefont{V.~B.}},
  \bibinfo{author}{\bibfnamefont{S.~E.} \bibnamefont{{Strigin}}}, and
  \bibinfo{author}{\bibfnamefont{S.~P.} \bibnamefont{{Vyatchanin}}},
  \bibinfo{year}{2001}, \bibinfo{journal}{Physics Letters A}
  \textbf{\bibinfo{volume}{287}}, \bibinfo{pages}{331}.

\bibitem[{\citenamefont{{Braginsky}}
  \emph{et~al.}(2002)\citenamefont{{Braginsky}, {Strigin}, and
  {Vyatchanin}}}]{BSV2002}
\bibinfo{author}{\bibnamefont{{Braginsky}}, \bibfnamefont{V.~B.}},
  \bibinfo{author}{\bibfnamefont{S.~E.} \bibnamefont{{Strigin}}}, and
  \bibinfo{author}{\bibfnamefont{S.~P.} \bibnamefont{{Vyatchanin}}},
  \bibinfo{year}{2002}, \bibinfo{journal}{Physics Letters A}
  \textbf{\bibinfo{volume}{305}}, \bibinfo{pages}{111}.

\bibitem[{\citenamefont{Braginsky and Vyatchanin}(2003)}]{BrVy2003}
\bibinfo{author}{\bibnamefont{Braginsky}, \bibfnamefont{V.~B.}}, and
  \bibinfo{author}{\bibfnamefont{S.~P.} \bibnamefont{Vyatchanin}},
  \bibinfo{year}{2003}, \bibinfo{journal}{Phys.~Lett.~A}
  \textbf{\bibinfo{volume}{312}}, \bibinfo{pages}{244}.

\bibitem[{\citenamefont{Braginsky and Vyatchanin}(2004)}]{Braginsky:2004fp}
\bibinfo{author}{\bibnamefont{Braginsky}, \bibfnamefont{V.~B.}}, and
  \bibinfo{author}{\bibfnamefont{S.~P.} \bibnamefont{Vyatchanin}},
  \bibinfo{year}{2004}, \bibinfo{journal}{Physics Letters A}
  \textbf{\bibinfo{volume}{324}}(\bibinfo{number}{5-6}), \bibinfo{pages}{345}.

\bibitem[{\citenamefont{{Brooks}} \emph{et~al.}(2009)\citenamefont{{Brooks},
  {Hosken}, {Munch}, {Veitch}, {Yan}, {Zhao}, {Fan}, {Ju}, {Blair}, {Willems},
  {Slagmolen}, and {Degallaix}}}]{Aidan:2009}
\bibinfo{author}{\bibnamefont{{Brooks}}, \bibfnamefont{A.~F.}},
  \bibinfo{author}{\bibfnamefont{D.}~\bibnamefont{{Hosken}}},
  \bibinfo{author}{\bibfnamefont{J.}~\bibnamefont{{Munch}}},
  \bibinfo{author}{\bibfnamefont{P.~J.} \bibnamefont{{Veitch}}},
  \bibinfo{author}{\bibfnamefont{Z.}~\bibnamefont{{Yan}}},
  \bibinfo{author}{\bibfnamefont{C.}~\bibnamefont{{Zhao}}},
  \bibinfo{author}{\bibfnamefont{Y.}~\bibnamefont{{Fan}}},
  \bibinfo{author}{\bibfnamefont{L.}~\bibnamefont{{Ju}}},
  \bibinfo{author}{\bibfnamefont{D.}~\bibnamefont{{Blair}}},
  \bibinfo{author}{\bibfnamefont{P.}~\bibnamefont{{Willems}}},
  \bibinfo{author}{\bibfnamefont{B.}~\bibnamefont{{Slagmolen}}}, and
  \bibinfo{author}{\bibfnamefont{J.}~\bibnamefont{{Degallaix}}},
  \bibinfo{year}{2009}, \bibinfo{journal}{\ao} \textbf{\bibinfo{volume}{48}},
  \bibinfo{pages}{355}.

\bibitem[{\citenamefont{Brown} \emph{et~al.}(1982)\citenamefont{Brown, Mills,
  and Tyson}}]{Tyson:1982}
\bibinfo{author}{\bibnamefont{Brown}, \bibfnamefont{B.~L.}},
  \bibinfo{author}{\bibfnamefont{A.~P.} \bibnamefont{Mills}}, and
  \bibinfo{author}{\bibfnamefont{J.~A.} \bibnamefont{Tyson}},
  \bibinfo{year}{1982}, \bibinfo{journal}{Phys. Rev. D}
  \textbf{\bibinfo{volume}{26}}, \bibinfo{pages}{1209}.

\bibitem[{\citenamefont{Brown} \emph{et~al.}(2013)\citenamefont{Brown,
  Friedrich, Br{\"u}ckner, Carbone, Schnabel, and
  Freise}}]{brown2013invariance}
\bibinfo{author}{\bibnamefont{Brown}, \bibfnamefont{D.}},
  \bibinfo{author}{\bibfnamefont{D.}~\bibnamefont{Friedrich}},
  \bibinfo{author}{\bibfnamefont{F.}~\bibnamefont{Br{\"u}ckner}},
  \bibinfo{author}{\bibfnamefont{L.}~\bibnamefont{Carbone}},
  \bibinfo{author}{\bibfnamefont{R.}~\bibnamefont{Schnabel}}, and
  \bibinfo{author}{\bibfnamefont{A.}~\bibnamefont{Freise}},
  \bibinfo{year}{2013}, \bibinfo{journal}{Optics letters}
  \textbf{\bibinfo{volume}{38}}(\bibinfo{number}{11}), \bibinfo{pages}{1844}.

\bibitem[{\citenamefont{{Br{\"u}ckner}}
  \emph{et~al.}(2008)\citenamefont{{Br{\"u}ckner}, {Clausnitzer}, {Burmeister},
  {Friedrich}, {Kley}, {Danzmann}, {T{\"u}nnermann}, and
  {Schnabel}}}]{Gratings:2008}
\bibinfo{author}{\bibnamefont{{Br{\"u}ckner}}, \bibfnamefont{F.}},
  \bibinfo{author}{\bibfnamefont{T.}~\bibnamefont{{Clausnitzer}}},
  \bibinfo{author}{\bibfnamefont{O.}~\bibnamefont{{Burmeister}}},
  \bibinfo{author}{\bibfnamefont{D.}~\bibnamefont{{Friedrich}}},
  \bibinfo{author}{\bibfnamefont{E.-B.} \bibnamefont{{Kley}}},
  \bibinfo{author}{\bibfnamefont{K.}~\bibnamefont{{Danzmann}}},
  \bibinfo{author}{\bibfnamefont{A.}~\bibnamefont{{T{\"u}nnermann}}}, and
  \bibinfo{author}{\bibfnamefont{R.}~\bibnamefont{{Schnabel}}},
  \bibinfo{year}{2008}, \bibinfo{journal}{Optics Letters}
  \textbf{\bibinfo{volume}{33}}, \bibinfo{pages}{264}.

\bibitem[{\citenamefont{{Bunkowski}}
  \emph{et~al.}(2006)\citenamefont{{Bunkowski}, {Burmeister}, {Friedrich},
  {Danzmann}, and {Schnabel}}}]{Roman:Gratings2006}
\bibinfo{author}{\bibnamefont{{Bunkowski}}, \bibfnamefont{A.}},
  \bibinfo{author}{\bibfnamefont{O.}~\bibnamefont{{Burmeister}}},
  \bibinfo{author}{\bibfnamefont{D.}~\bibnamefont{{Friedrich}}},
  \bibinfo{author}{\bibfnamefont{K.}~\bibnamefont{{Danzmann}}}, and
  \bibinfo{author}{\bibfnamefont{R.}~\bibnamefont{{Schnabel}}},
  \bibinfo{year}{2006}, \bibinfo{journal}{Classical and Quantum Gravity}
  \textbf{\bibinfo{volume}{23}}, \bibinfo{pages}{7297}.

\bibitem[{\citenamefont{Buonanno and Chen}(2001)}]{BuCh2001}
\bibinfo{author}{\bibnamefont{Buonanno}, \bibfnamefont{A.}}, and
  \bibinfo{author}{\bibfnamefont{Y.}~\bibnamefont{Chen}}, \bibinfo{year}{2001},
  \bibinfo{journal}{Phys.~Rev.~D} \textbf{\bibinfo{volume}{64}},
  \bibinfo{pages}{042006}.

\bibitem[{\citenamefont{Buonanno and Chen}(2002)}]{BuCh2002}
\bibinfo{author}{\bibnamefont{Buonanno}, \bibfnamefont{A.}}, and
  \bibinfo{author}{\bibfnamefont{Y.}~\bibnamefont{Chen}}, \bibinfo{year}{2002},
  \bibinfo{journal}{Phys. Rev. D} \textbf{\bibinfo{volume}{65}},
  \bibinfo{pages}{042001},
  \urlprefix\url{http://link.aps.org/doi/10.1103/PhysRevD.65.042001}.

\bibitem[{\citenamefont{{Cagnoli}} \emph{et~al.}(1999)\citenamefont{{Cagnoli},
  {Gammaitoni}, {Kovalik}, {Marchesoni}, and {Punturo}}}]{Cagnoli:1999ud}
\bibinfo{author}{\bibnamefont{{Cagnoli}}, \bibfnamefont{G.}},
  \bibinfo{author}{\bibfnamefont{L.}~\bibnamefont{{Gammaitoni}}},
  \bibinfo{author}{\bibfnamefont{J.}~\bibnamefont{{Kovalik}}},
  \bibinfo{author}{\bibfnamefont{F.}~\bibnamefont{{Marchesoni}}}, and
  \bibinfo{author}{\bibfnamefont{M.}~\bibnamefont{{Punturo}}},
  \bibinfo{year}{1999}, \bibinfo{journal}{Physics Letters A}
  \textbf{\bibinfo{volume}{255}}, \bibinfo{pages}{230}.

\bibitem[{\citenamefont{Cagnoli} \emph{et~al.}(2000)\citenamefont{Cagnoli,
  Hough, DeBra, Fejer, Gustafson, Rowan, and Mitrofanov}}]{Geppo:2000}
\bibinfo{author}{\bibnamefont{Cagnoli}, \bibfnamefont{G.}},
  \bibinfo{author}{\bibfnamefont{J.}~\bibnamefont{Hough}},
  \bibinfo{author}{\bibfnamefont{D.}~\bibnamefont{DeBra}},
  \bibinfo{author}{\bibfnamefont{M.}~\bibnamefont{Fejer}},
  \bibinfo{author}{\bibfnamefont{E.}~\bibnamefont{Gustafson}},
  \bibinfo{author}{\bibfnamefont{S.}~\bibnamefont{Rowan}}, and
  \bibinfo{author}{\bibfnamefont{V.}~\bibnamefont{Mitrofanov}},
  \bibinfo{year}{2000}, \bibinfo{journal}{Physics Letters A}
  \textbf{\bibinfo{volume}{272}}(\bibinfo{number}{1-2}), \bibinfo{pages}{39 },
  ISSN \bibinfo{issn}{0375-9601}.

\bibitem[{\citenamefont{Cagnoli and Willems}(2002)}]{Phil:Nonlinear}
\bibinfo{author}{\bibnamefont{Cagnoli}, \bibfnamefont{G.}}, and
  \bibinfo{author}{\bibfnamefont{P.~A.} \bibnamefont{Willems}},
  \bibinfo{year}{2002}, \bibinfo{journal}{Phys. Rev. B}
  \textbf{\bibinfo{volume}{65}}, \bibinfo{pages}{174111},
  \urlprefix\url{http://link.aps.org/doi/10.1103/PhysRevB.65.174111}.

\bibitem[{\citenamefont{Caldwell} \emph{et~al.}(1998)\citenamefont{Caldwell,
  Kamionkowski, and Wadley}}]{Caldwell:1998dq}
\bibinfo{author}{\bibnamefont{Caldwell}, \bibfnamefont{R.}},
  \bibinfo{author}{\bibfnamefont{M.}~\bibnamefont{Kamionkowski}}, and
  \bibinfo{author}{\bibfnamefont{L.}~\bibnamefont{Wadley}},
  \bibinfo{year}{1998}, \bibinfo{journal}{Physical Review D}
  \textbf{\bibinfo{volume}{59}}(\bibinfo{number}{2}), \bibinfo{pages}{027101}.

\bibitem[{\citenamefont{Callen and Welton}(1951)}]{CaWe1951}
\bibinfo{author}{\bibnamefont{Callen}, \bibfnamefont{H.~B.}}, and
  \bibinfo{author}{\bibfnamefont{T.~A.} \bibnamefont{Welton}},
  \bibinfo{year}{1951}, \bibinfo{journal}{Phys.~Rev.}
  \textbf{\bibinfo{volume}{83}}, \bibinfo{pages}{34}.

\bibitem[{\citenamefont{{Camp}} \emph{et~al.}(2000)\citenamefont{{Camp},
  {Yamamoto}, {Whitcomb}, and {McClelland}}}]{Jordan:Light}
\bibinfo{author}{\bibnamefont{{Camp}}, \bibfnamefont{J.~B.}},
  \bibinfo{author}{\bibfnamefont{H.}~\bibnamefont{{Yamamoto}}},
  \bibinfo{author}{\bibfnamefont{S.~E.} \bibnamefont{{Whitcomb}}}, and
  \bibinfo{author}{\bibfnamefont{D.~E.} \bibnamefont{{McClelland}}},
  \bibinfo{year}{2000}, \bibinfo{journal}{Journal of the Optical Society of
  America A} \textbf{\bibinfo{volume}{17}}, \bibinfo{pages}{120}.

\bibitem[{\citenamefont{{Campbell}}(1965)}]{Campbell:1965}
\bibinfo{author}{\bibnamefont{{Campbell}}, \bibfnamefont{W.~H.}},
  \bibinfo{year}{1965}, in \emph{\bibinfo{booktitle}{Report on Equatorial
  Aeronomy}}, edited by \bibinfo{editor}{\bibfnamefont{F.}~\bibnamefont{{de
  Mendon{\c c}a}}}, p. \bibinfo{pages}{495}.

\bibitem[{\citenamefont{Caves}(1981)}]{Cav1980}
\bibinfo{author}{\bibnamefont{Caves}, \bibfnamefont{C.~M.}},
  \bibinfo{year}{1981}, \bibinfo{journal}{Phys. Rev. D}
  \textbf{\bibinfo{volume}{23}}, \bibinfo{pages}{1693},
  \urlprefix\url{http://link.aps.org/doi/10.1103/PhysRevD.23.1693}.

\bibitem[{\citenamefont{Caves and Schumaker}(1985)}]{CaSc1985}
\bibinfo{author}{\bibnamefont{Caves}, \bibfnamefont{C.~M.}}, and
  \bibinfo{author}{\bibfnamefont{B.~L.} \bibnamefont{Schumaker}},
  \bibinfo{year}{1985}, \bibinfo{journal}{Phys.~Rev.~A}
  \textbf{\bibinfo{volume}{31}}, \bibinfo{pages}{3068 \& 3093}.

\bibitem[{\citenamefont{Cella}(2006)}]{Cella:2006vd}
\bibinfo{author}{\bibnamefont{Cella}, \bibfnamefont{G.}}, \bibinfo{year}{2006},
  \bibinfo{title}{{Underground Reduction of Gravity Gradient Noise}},
  \bibinfo{note}{{Gravitational Wave Advanced Detector Workshop}},
  \urlprefix\url{https://dcc.ligo.org/LIGO-G060311/public}.

\bibitem[{\citenamefont{Cella and Giazotto}(2006)}]{Cella:2006ki}
\bibinfo{author}{\bibnamefont{Cella}, \bibfnamefont{G.}}, and
  \bibinfo{author}{\bibfnamefont{A.}~\bibnamefont{Giazotto}},
  \bibinfo{year}{2006}, \bibinfo{journal}{Phys. Rev. D}
  \textbf{\bibinfo{volume}{74}}, \bibinfo{pages}{042001},
  \urlprefix\url{http://link.aps.org/doi/10.1103/PhysRevD.74.042001}.

\bibitem[{\citenamefont{Cella and Giazotto}(2011)}]{Cella:2011}
\bibinfo{author}{\bibnamefont{Cella}, \bibfnamefont{G.}}, and
  \bibinfo{author}{\bibfnamefont{A.}~\bibnamefont{Giazotto}},
  \bibinfo{year}{2011}, \bibinfo{journal}{Review of Scientific Instruments}
  \textbf{\bibinfo{volume}{82}}(\bibinfo{number}{10}), \bibinfo{eid}{101101}
  (pages~\bibinfo{numpages}{44}).

\bibitem[{\citenamefont{Chakrabarty}
  \emph{et~al.}(2003)\citenamefont{Chakrabarty, Morgan, Muno, Galloway,
  Wijnands, van~der Klis, and Markwardt}}]{Deepto:2003}
\bibinfo{author}{\bibnamefont{Chakrabarty}, \bibfnamefont{D.}},
  \bibinfo{author}{\bibfnamefont{E.~H.} \bibnamefont{Morgan}},
  \bibinfo{author}{\bibfnamefont{M.~P.} \bibnamefont{Muno}},
  \bibinfo{author}{\bibfnamefont{D.~K.} \bibnamefont{Galloway}},
  \bibinfo{author}{\bibfnamefont{R.}~\bibnamefont{Wijnands}},
  \bibinfo{author}{\bibfnamefont{M.}~\bibnamefont{van~der Klis}}, and
  \bibinfo{author}{\bibfnamefont{C.~B.} \bibnamefont{Markwardt}},
  \bibinfo{year}{2003}, \bibinfo{journal}{Nature}
  \textbf{\bibinfo{volume}{424}}(\bibinfo{number}{6944}), \bibinfo{pages}{42}.

\bibitem[{\citenamefont{Chapline} \emph{et~al.}(1974)\citenamefont{Chapline,
  Nuckolls, and Wood}}]{Nukes:1974}
\bibinfo{author}{\bibnamefont{Chapline}, \bibfnamefont{G.~F.}},
  \bibinfo{author}{\bibfnamefont{J.}~\bibnamefont{Nuckolls}}, and
  \bibinfo{author}{\bibfnamefont{L.~L.} \bibnamefont{Wood}},
  \bibinfo{year}{1974}, \bibinfo{journal}{Phys. Rev. D}
  \textbf{\bibinfo{volume}{10}}, \bibinfo{pages}{1064}.

\bibitem[{\citenamefont{Chelkowski}
  \emph{et~al.}(2009)\citenamefont{Chelkowski, Hild, and
  Freise}}]{Andreas:LG2009}
\bibinfo{author}{\bibnamefont{Chelkowski}, \bibfnamefont{S.}},
  \bibinfo{author}{\bibfnamefont{S.}~\bibnamefont{Hild}}, and
  \bibinfo{author}{\bibfnamefont{A.}~\bibnamefont{Freise}},
  \bibinfo{year}{2009}, \bibinfo{journal}{Phys. Rev. D}
  \textbf{\bibinfo{volume}{79}}, \bibinfo{pages}{122002}.

\bibitem[{\citenamefont{Chen}(2003)}]{Che2003}
\bibinfo{author}{\bibnamefont{Chen}, \bibfnamefont{Y.}}, \bibinfo{year}{2003},
  \bibinfo{journal}{Phys. Rev. D} \textbf{\bibinfo{volume}{67}},
  \bibinfo{pages}{122004},
  \urlprefix\url{http://link.aps.org/doi/10.1103/PhysRevD.67.122004}.

\bibitem[{\citenamefont{{Chen}}(2013)}]{Yanbei:MQMreview}
\bibinfo{author}{\bibnamefont{{Chen}}, \bibfnamefont{Y.}},
  \bibinfo{year}{2013}, \bibinfo{journal}{ArXiv e-prints} \eprint{1302.1924}.

\bibitem[{\citenamefont{Chen} \emph{et~al.}(2010)\citenamefont{Chen,
  Danilishin, Khalili, and M{\"u}ller-Ebhardt}}]{Chen:2010da}
\bibinfo{author}{\bibnamefont{Chen}, \bibfnamefont{Y.}},
  \bibinfo{author}{\bibfnamefont{S.~L.} \bibnamefont{Danilishin}},
  \bibinfo{author}{\bibfnamefont{F.~Y.} \bibnamefont{Khalili}}, and
  \bibinfo{author}{\bibfnamefont{H.}~\bibnamefont{M{\"u}ller-Ebhardt}},
  \bibinfo{year}{2010}, \bibinfo{journal}{General Relativity and Gravitation}
  \textbf{\bibinfo{volume}{43}}(\bibinfo{number}{2}), \bibinfo{pages}{671}.

\bibitem[{\citenamefont{Christensen}(1992)}]{Nelson:PRD92}
\bibinfo{author}{\bibnamefont{Christensen}, \bibfnamefont{N.}},
  \bibinfo{year}{1992}, \bibinfo{journal}{Phys. Rev. D}
  \textbf{\bibinfo{volume}{46}}, \bibinfo{pages}{5250},
  \urlprefix\url{http://link.aps.org/doi/10.1103/PhysRevD.46.5250}.

\bibitem[{\citenamefont{{Chua}} \emph{et~al.}(2011)\citenamefont{{Chua},
  {Stefszky}, {Mow-Lowry}, {Buchler}, {Dwyer}, {Shaddock}, {Lam}, and
  {McClelland}}}]{ANU:BackSqueeze2011}
\bibinfo{author}{\bibnamefont{{Chua}}, \bibfnamefont{S.~S.~Y.}},
  \bibinfo{author}{\bibfnamefont{M.~S.} \bibnamefont{{Stefszky}}},
  \bibinfo{author}{\bibfnamefont{C.~M.} \bibnamefont{{Mow-Lowry}}},
  \bibinfo{author}{\bibfnamefont{B.~C.} \bibnamefont{{Buchler}}},
  \bibinfo{author}{\bibfnamefont{S.}~\bibnamefont{{Dwyer}}},
  \bibinfo{author}{\bibfnamefont{D.~A.} \bibnamefont{{Shaddock}}},
  \bibinfo{author}{\bibfnamefont{P.~K.} \bibnamefont{{Lam}}}, and
  \bibinfo{author}{\bibfnamefont{D.~E.} \bibnamefont{{McClelland}}},
  \bibinfo{year}{2011}, \bibinfo{journal}{Optics Letters}
  \textbf{\bibinfo{volume}{36}}, \bibinfo{pages}{4680}.

\bibitem[{\citenamefont{Cole} \emph{et~al.}(2008)\citenamefont{Cole,
  Groblacher, Gugler, Gigan, and Aspelmeyer}}]{cole:261108}
\bibinfo{author}{\bibnamefont{Cole}, \bibfnamefont{G.~D.}},
  \bibinfo{author}{\bibfnamefont{S.}~\bibnamefont{Groblacher}},
  \bibinfo{author}{\bibfnamefont{K.}~\bibnamefont{Gugler}},
  \bibinfo{author}{\bibfnamefont{S.}~\bibnamefont{Gigan}}, and
  \bibinfo{author}{\bibfnamefont{M.}~\bibnamefont{Aspelmeyer}},
  \bibinfo{year}{2008}, \bibinfo{journal}{Applied Physics Letters}
  \textbf{\bibinfo{volume}{92}}(\bibinfo{number}{26}), \bibinfo{eid}{261108}
  (pages~\bibinfo{numpages}{3}).

\bibitem[{\citenamefont{Collaboration}(2011)}]{aVirgo:web}
\bibinfo{author}{\bibnamefont{Collaboration}, \bibfnamefont{V.}},
  \bibinfo{year}{2011},
  \urlprefix\url{https://wwwcascina.virgo.infn.it/advirgo/}.

\bibitem[{\citenamefont{Corbitt and Mavalvala}(2004)}]{Corbitt:Review2004}
\bibinfo{author}{\bibnamefont{Corbitt}, \bibfnamefont{T.}}, and
  \bibinfo{author}{\bibfnamefont{N.}~\bibnamefont{Mavalvala}},
  \bibinfo{year}{2004}, \bibinfo{journal}{Journal of Optics B: Quantum and
  Semiclassical Optics} \textbf{\bibinfo{volume}{6}}(\bibinfo{number}{8}),
  \bibinfo{pages}{S675}.

\bibitem[{\citenamefont{Corbitt} \emph{et~al.}(2006)\citenamefont{Corbitt,
  Ottaway, Innerhofer, Pelc, and Mavalvala}}]{Nergis:PI2005}
\bibinfo{author}{\bibnamefont{Corbitt}, \bibfnamefont{T.}},
  \bibinfo{author}{\bibfnamefont{D.}~\bibnamefont{Ottaway}},
  \bibinfo{author}{\bibfnamefont{E.}~\bibnamefont{Innerhofer}},
  \bibinfo{author}{\bibfnamefont{J.}~\bibnamefont{Pelc}}, and
  \bibinfo{author}{\bibfnamefont{N.}~\bibnamefont{Mavalvala}},
  \bibinfo{year}{2006}, \bibinfo{journal}{Phys. Rev. A}
  \textbf{\bibinfo{volume}{74}}, \bibinfo{pages}{021802},
  \urlprefix\url{http://link.aps.org/doi/10.1103/PhysRevA.74.021802}.

\bibitem[{\citenamefont{Cornish and Porter}(2007)}]{Cornish:2007}
\bibinfo{author}{\bibnamefont{Cornish}, \bibfnamefont{N.~J.}}, and
  \bibinfo{author}{\bibfnamefont{E.~K.} \bibnamefont{Porter}},
  \bibinfo{year}{2007}, \bibinfo{journal}{Phys. Rev. D}
  \textbf{\bibinfo{volume}{75}}, \bibinfo{pages}{021301}.

\bibitem[{\citenamefont{Creighton}(2008)}]{Cre2008}
\bibinfo{author}{\bibnamefont{Creighton}, \bibfnamefont{T.}},
  \bibinfo{year}{2008}, \bibinfo{journal}{Classical and Quantum Gravity}
  \textbf{\bibinfo{volume}{25}}(\bibinfo{number}{12}), \bibinfo{pages}{125011}.

\bibitem[{\citenamefont{Cumming} \emph{et~al.}(2009)\citenamefont{Cumming,
  Heptonstall, Kumar, Cunningham, Torrie, Barton, Strain, Hough, and
  Rowan}}]{SUS:FEA2009}
\bibinfo{author}{\bibnamefont{Cumming}, \bibfnamefont{A.}},
  \bibinfo{author}{\bibfnamefont{A.}~\bibnamefont{Heptonstall}},
  \bibinfo{author}{\bibfnamefont{R.}~\bibnamefont{Kumar}},
  \bibinfo{author}{\bibfnamefont{W.}~\bibnamefont{Cunningham}},
  \bibinfo{author}{\bibfnamefont{C.}~\bibnamefont{Torrie}},
  \bibinfo{author}{\bibfnamefont{M.}~\bibnamefont{Barton}},
  \bibinfo{author}{\bibfnamefont{K.~A.} \bibnamefont{Strain}},
  \bibinfo{author}{\bibfnamefont{J.}~\bibnamefont{Hough}}, and
  \bibinfo{author}{\bibfnamefont{S.}~\bibnamefont{Rowan}},
  \bibinfo{year}{2009}, \bibinfo{journal}{Classical and Quantum Gravity}
  \textbf{\bibinfo{volume}{26}}(\bibinfo{number}{21}), \bibinfo{pages}{215012},
  \urlprefix\url{http://stacks.iop.org/0264-9381/26/i=21/a=215012}.

\bibitem[{\citenamefont{Cumming} \emph{et~al.}(2012)\citenamefont{Cumming,
  Bell, Barsotti, Barton, Cagnoli, Cook, Cunningham, Evans, Hammond, Harry,
  Heptonstall, Hough} \emph{et~al.}}]{SUS:2012}
\bibinfo{author}{\bibnamefont{Cumming}, \bibfnamefont{A.~V.}},
  \bibinfo{author}{\bibfnamefont{A.~S.} \bibnamefont{Bell}},
  \bibinfo{author}{\bibfnamefont{L.}~\bibnamefont{Barsotti}},
  \bibinfo{author}{\bibfnamefont{M.~A.} \bibnamefont{Barton}},
  \bibinfo{author}{\bibfnamefont{G.}~\bibnamefont{Cagnoli}},
  \bibinfo{author}{\bibfnamefont{D.}~\bibnamefont{Cook}},
  \bibinfo{author}{\bibfnamefont{L.}~\bibnamefont{Cunningham}},
  \bibinfo{author}{\bibfnamefont{M.}~\bibnamefont{Evans}},
  \bibinfo{author}{\bibfnamefont{G.~D.} \bibnamefont{Hammond}},
  \bibinfo{author}{\bibfnamefont{G.~M.} \bibnamefont{Harry}},
  \bibinfo{author}{\bibfnamefont{A.}~\bibnamefont{Heptonstall}},
  \bibinfo{author}{\bibfnamefont{J.}~\bibnamefont{Hough}}, \emph{et~al.},
  \bibinfo{year}{2012}, \bibinfo{journal}{Classical and Quantum Gravity}
  \textbf{\bibinfo{volume}{29}}(\bibinfo{number}{3}), \bibinfo{pages}{035003},
  \urlprefix\url{http://stacks.iop.org/0264-9381/29/i=3/a=035003}.

\bibitem[{\citenamefont{Cutler and Harms}(2006)}]{CuHa2006}
\bibinfo{author}{\bibnamefont{Cutler}, \bibfnamefont{C.}}, and
  \bibinfo{author}{\bibfnamefont{J.}~\bibnamefont{Harms}},
  \bibinfo{year}{2006}, \bibinfo{journal}{Phys.~Rev.~D}
  \textbf{\bibinfo{volume}{73}}, \bibinfo{pages}{042001}.

\bibitem[{\citenamefont{Cutler and Holz}(2009)}]{Cutler:2009}
\bibinfo{author}{\bibnamefont{Cutler}, \bibfnamefont{C.}}, and
  \bibinfo{author}{\bibfnamefont{D.~E.} \bibnamefont{Holz}},
  \bibinfo{year}{2009}, \bibinfo{journal}{Phys. Rev. D}
  \textbf{\bibinfo{volume}{80}}, \bibinfo{pages}{104009},
  \urlprefix\url{http://link.aps.org/doi/10.1103/PhysRevD.80.104009}.

\bibitem[{\citenamefont{{Cutler} and {Thorne}}(2002)}]{CuTh2002}
\bibinfo{author}{\bibnamefont{{Cutler}}, \bibfnamefont{C.}}, and
  \bibinfo{author}{\bibfnamefont{K.~S.} \bibnamefont{{Thorne}}},
  \bibinfo{year}{2002}, \bibinfo{journal}{ArXiv General Relativity and Quantum
  Cosmology e-prints} \eprint{arXiv:gr-qc/0204090}.

\bibitem[{\citenamefont{D'Ambrosio}
  \emph{et~al.}(2004)\citenamefont{D'Ambrosio, O'Shaugnessy, Thorne, Willems,
  Strigin, and Vyatchanin}}]{Mesa:Erika}
\bibinfo{author}{\bibnamefont{D'Ambrosio}, \bibfnamefont{E.}},
  \bibinfo{author}{\bibfnamefont{R.}~\bibnamefont{O'Shaugnessy}},
  \bibinfo{author}{\bibfnamefont{K.}~\bibnamefont{Thorne}},
  \bibinfo{author}{\bibfnamefont{P.}~\bibnamefont{Willems}},
  \bibinfo{author}{\bibfnamefont{S.}~\bibnamefont{Strigin}}, and
  \bibinfo{author}{\bibfnamefont{S.}~\bibnamefont{Vyatchanin}},
  \bibinfo{year}{2004}, \bibinfo{journal}{Classical and Quantum Gravity}
  \textbf{\bibinfo{volume}{21}}(\bibinfo{number}{5}), \bibinfo{pages}{S867}.

\bibitem[{\citenamefont{Danzmann}(2013)}]{eLISA:WhitePaper}
\bibinfo{author}{\bibnamefont{Danzmann}, \bibfnamefont{K.}},
  \bibinfo{year}{2013}, \emph{\bibinfo{title}{The Gravitational Universe}},
  \bibinfo{type}{Technical Report}, \bibinfo{institution}{Albert Einstein
  Institute Hannover},
  \urlprefix\url{https://www.elisascience.org/whitepaper/}.

\bibitem[{\citenamefont{Danzmann and R{\"u}diger}(2003)}]{Danzmann:2003}
\bibinfo{author}{\bibnamefont{Danzmann}, \bibfnamefont{K.}}, and
  \bibinfo{author}{\bibfnamefont{A.}~\bibnamefont{R{\"u}diger}},
  \bibinfo{year}{2003}, \bibinfo{journal}{Classical and Quantum Gravity}
  \textbf{\bibinfo{volume}{20}}(\bibinfo{number}{10}), \bibinfo{pages}{S1}.

\bibitem[{\citenamefont{{Daw}} \emph{et~al.}(2004)\citenamefont{{Daw},
  {Giaime}, {Lormand}, {Lubi{\'n}ski}, and {Zweizig}}}]{Daw:2004}
\bibinfo{author}{\bibnamefont{{Daw}}, \bibfnamefont{E.~J.}},
  \bibinfo{author}{\bibfnamefont{J.~A.} \bibnamefont{{Giaime}}},
  \bibinfo{author}{\bibfnamefont{D.}~\bibnamefont{{Lormand}}},
  \bibinfo{author}{\bibfnamefont{M.}~\bibnamefont{{Lubi{\'n}ski}}}, and
  \bibinfo{author}{\bibfnamefont{J.}~\bibnamefont{{Zweizig}}},
  \bibinfo{year}{2004}, \bibinfo{journal}{Classical and Quantum Gravity}
  \textbf{\bibinfo{volume}{21}}, \bibinfo{pages}{2255}.

\bibitem[{\citenamefont{{DECIGO}}(2011)}]{DECIGO}
\bibinfo{author}{\bibnamefont{{DECIGO}}}, \bibinfo{year}{2011},
  \bibinfo{title}{{DECIGO} homepage},
  \urlprefix\url{http://tamago.mtk.nao.ac.jp/decigo/}.

\bibitem[{\citenamefont{Demeester} \emph{et~al.}(1993)\citenamefont{Demeester,
  Pollentier, Dobbelaere, Brys, and Daele}}]{ELO}
\bibinfo{author}{\bibnamefont{Demeester}, \bibfnamefont{P.}},
  \bibinfo{author}{\bibfnamefont{I.}~\bibnamefont{Pollentier}},
  \bibinfo{author}{\bibfnamefont{P.~D.} \bibnamefont{Dobbelaere}},
  \bibinfo{author}{\bibfnamefont{C.}~\bibnamefont{Brys}}, and
  \bibinfo{author}{\bibfnamefont{P.~V.} \bibnamefont{Daele}},
  \bibinfo{year}{1993}, \bibinfo{journal}{Semiconductor Science and Technology}
  \textbf{\bibinfo{volume}{8}}(\bibinfo{number}{6}), \bibinfo{pages}{1124},
  \urlprefix\url{http://stacks.iop.org/0268-1242/8/i=6/a=021}.

\bibitem[{\citenamefont{DeRosa} \emph{et~al.}(2012)\citenamefont{DeRosa,
  Driggers, Atkinson, Miao, Frolov, Landry, Giaime, and
  Adhikari}}]{Ryan:FFW2012}
\bibinfo{author}{\bibnamefont{DeRosa}, \bibfnamefont{R.}},
  \bibinfo{author}{\bibfnamefont{J.~C.} \bibnamefont{Driggers}},
  \bibinfo{author}{\bibfnamefont{D.}~\bibnamefont{Atkinson}},
  \bibinfo{author}{\bibfnamefont{H.}~\bibnamefont{Miao}},
  \bibinfo{author}{\bibfnamefont{V.}~\bibnamefont{Frolov}},
  \bibinfo{author}{\bibfnamefont{M.}~\bibnamefont{Landry}},
  \bibinfo{author}{\bibfnamefont{J.~A.} \bibnamefont{Giaime}}, and
  \bibinfo{author}{\bibfnamefont{R.~X.} \bibnamefont{Adhikari}},
  \bibinfo{year}{2012}, \bibinfo{journal}{Classical and Quantum Gravity}
  \textbf{\bibinfo{volume}{29}}(\bibinfo{number}{21}), \bibinfo{pages}{215008},
  \urlprefix\url{http://stacks.iop.org/0264-9381/29/i=21/a=215008}.

\bibitem[{\citenamefont{Detweiler}(1979)}]{Detweiler:PTA}
\bibinfo{author}{\bibnamefont{Detweiler}, \bibfnamefont{S.}},
  \bibinfo{year}{1979}, \bibinfo{journal}{The Astrophysical Journal}
  \textbf{\bibinfo{volume}{234}}.

\bibitem[{\citenamefont{Dimopoulos}
  \emph{et~al.}(2008)\citenamefont{Dimopoulos, Graham, Hogan, Kasevich, and
  Rajendran}}]{Dim:Atom2008}
\bibinfo{author}{\bibnamefont{Dimopoulos}, \bibfnamefont{S.}},
  \bibinfo{author}{\bibfnamefont{P.~W.} \bibnamefont{Graham}},
  \bibinfo{author}{\bibfnamefont{J.~M.} \bibnamefont{Hogan}},
  \bibinfo{author}{\bibfnamefont{M.~A.} \bibnamefont{Kasevich}}, and
  \bibinfo{author}{\bibfnamefont{S.}~\bibnamefont{Rajendran}},
  \bibinfo{year}{2008}, \bibinfo{journal}{Phys. Rev. D}
  \textbf{\bibinfo{volume}{78}}, \bibinfo{pages}{122002}.

\bibitem[{\citenamefont{Dimopoulos}
  \emph{et~al.}(2011)\citenamefont{Dimopoulos, Graham, Hogan, Kasevich, and
  Rajendran}}]{Dim:PRD2011}
\bibinfo{author}{\bibnamefont{Dimopoulos}, \bibfnamefont{S.}},
  \bibinfo{author}{\bibfnamefont{P.~W.} \bibnamefont{Graham}},
  \bibinfo{author}{\bibfnamefont{J.~M.} \bibnamefont{Hogan}},
  \bibinfo{author}{\bibfnamefont{M.~A.} \bibnamefont{Kasevich}}, and
  \bibinfo{author}{\bibfnamefont{S.}~\bibnamefont{Rajendran}},
  \bibinfo{year}{2011}, \bibinfo{journal}{Phys. Rev. D}
  \textbf{\bibinfo{volume}{84}}, \bibinfo{pages}{028102},
  \urlprefix\url{http://link.aps.org/doi/10.1103/PhysRevD.84.028102}.

\bibitem[{\citenamefont{Dooley}(2011)}]{Dooley:thesis}
\bibinfo{author}{\bibnamefont{Dooley}, \bibfnamefont{K.}},
  \bibinfo{year}{2011}, \emph{\bibinfo{title}{Design and Performance of High
  Laser Power Interferometers for Gravitational-wave Detection}}, Ph.D. thesis,
  \bibinfo{school}{University of Florida}.

\bibitem[{\citenamefont{Douglass} \emph{et~al.}(1975)\citenamefont{Douglass,
  Gram, Tyson, and Lee}}]{Tyson:Nothing}
\bibinfo{author}{\bibnamefont{Douglass}, \bibfnamefont{D.~H.}},
  \bibinfo{author}{\bibfnamefont{R.~Q.} \bibnamefont{Gram}},
  \bibinfo{author}{\bibfnamefont{J.~A.} \bibnamefont{Tyson}}, and
  \bibinfo{author}{\bibfnamefont{R.~W.} \bibnamefont{Lee}},
  \bibinfo{year}{1975}, \bibinfo{journal}{Phys. Rev. Lett.}
  \textbf{\bibinfo{volume}{35}}, \bibinfo{pages}{480}.

\bibitem[{\citenamefont{{Drever}}(1983)}]{Ron:GR1983}
\bibinfo{author}{\bibnamefont{{Drever}}, \bibfnamefont{R.~W.~P.}},
  \bibinfo{year}{1983}, in \emph{\bibinfo{booktitle}{Lecture Notes in
  Physics}}, volume \bibinfo{volume}{124} of \emph{\bibinfo{series}{Berlin
  Springer Verlag}}, pp. \bibinfo{pages}{321--338}.

\bibitem[{\citenamefont{Drever}(1983)}]{Drever:PR}
\bibinfo{author}{\bibnamefont{Drever}, \bibfnamefont{R.~W.~P.}},
  \bibinfo{year}{1983}, \bibinfo{journal}{AIP Conference Proceedings}
  \textbf{\bibinfo{volume}{96}}(\bibinfo{number}{1}), \bibinfo{pages}{336}.

\bibitem[{\citenamefont{{Drever}}(1991)}]{Drever:FP1991}
\bibinfo{author}{\bibnamefont{{Drever}}, \bibfnamefont{R.~W.~P.}},
  \bibinfo{year}{1991}, in \emph{\bibinfo{booktitle}{The Detection of
  Gravitational Waves}}, edited by
  \bibinfo{editor}{\bibnamefont{{D.~G.~Blair}}}, p. \bibinfo{pages}{306}.

\bibitem[{\citenamefont{{Drever}}(1996)}]{Drever:Magnetic}
\bibinfo{author}{\bibnamefont{{Drever}}, \bibfnamefont{R.~W.~P.}},
  \bibinfo{year}{1996}, in \emph{\bibinfo{booktitle}{Dark Matter in Cosmology
  Quantam Measurements Experimental Gravitation}}, edited by
  \bibinfo{editor}{\bibfnamefont{R.}~\bibnamefont{{Ansari}}},
  \bibinfo{editor}{\bibfnamefont{Y.}~\bibnamefont{{Giraud-Heraud}}}, and
  \bibinfo{editor}{\bibfnamefont{J.}~\bibnamefont{{Tran Thanh Van}}}, p.
  \bibinfo{pages}{375}.

\bibitem[{\citenamefont{{Drever}} \emph{et~al.}(1983)\citenamefont{{Drever},
  {Hall}, {Kowalski}, {Hough}, {Ford}, {Munley}, and {Ward}}}]{PDH:1983}
\bibinfo{author}{\bibnamefont{{Drever}}, \bibfnamefont{R.~W.~P.}},
  \bibinfo{author}{\bibfnamefont{J.~L.} \bibnamefont{{Hall}}},
  \bibinfo{author}{\bibfnamefont{F.~V.} \bibnamefont{{Kowalski}}},
  \bibinfo{author}{\bibfnamefont{J.}~\bibnamefont{{Hough}}},
  \bibinfo{author}{\bibfnamefont{G.~M.} \bibnamefont{{Ford}}},
  \bibinfo{author}{\bibfnamefont{A.~J.} \bibnamefont{{Munley}}}, and
  \bibinfo{author}{\bibfnamefont{H.}~\bibnamefont{{Ward}}},
  \bibinfo{year}{1983}, \bibinfo{journal}{Applied Physics B: Lasers and Optics}
  \textbf{\bibinfo{volume}{31}}, \bibinfo{pages}{97}.

\bibitem[{\citenamefont{Driggers}
  \emph{et~al.}(2012{\natexlab{a}})\citenamefont{Driggers, Evans, Pepper, and
  Adhikari}}]{Driggers:2012fl}
\bibinfo{author}{\bibnamefont{Driggers}, \bibfnamefont{J.~C.}},
  \bibinfo{author}{\bibfnamefont{M.}~\bibnamefont{Evans}},
  \bibinfo{author}{\bibfnamefont{K.}~\bibnamefont{Pepper}}, and
  \bibinfo{author}{\bibfnamefont{R.}~\bibnamefont{Adhikari}},
  \bibinfo{year}{2012}{\natexlab{a}}, \bibinfo{journal}{Review of Scientific
  Instruments} \textbf{\bibinfo{volume}{83}}(\bibinfo{number}{2}),
  \bibinfo{pages}{024501}.

\bibitem[{\citenamefont{Driggers and Harms}(2011)}]{DrHa2011}
\bibinfo{author}{\bibnamefont{Driggers}, \bibfnamefont{J.~C.}}, and
  \bibinfo{author}{\bibfnamefont{J.}~\bibnamefont{Harms}},
  \bibinfo{year}{2011}, \emph{\bibinfo{title}{Results of Phase 1 Newtonian
  Noise Measurements at the LIGO Sites}}, \bibinfo{type}{Technical Report}
  \bibinfo{number}{T1100237}, \bibinfo{institution}{LIGO},
  \urlprefix\url{https://dcc.ligo.org/LIGO-T1100237/public}.

\bibitem[{\citenamefont{Driggers}
  \emph{et~al.}(2012{\natexlab{b}})\citenamefont{Driggers, Harms, and
  Adhikari}}]{NN:subtract2012}
\bibinfo{author}{\bibnamefont{Driggers}, \bibfnamefont{J.~C.}},
  \bibinfo{author}{\bibfnamefont{J.}~\bibnamefont{Harms}}, and
  \bibinfo{author}{\bibfnamefont{R.~X.} \bibnamefont{Adhikari}},
  \bibinfo{year}{2012}{\natexlab{b}}, \bibinfo{journal}{Phys. Rev. D}
  \textbf{\bibinfo{volume}{86}}, \bibinfo{pages}{102001},
  \urlprefix\url{http://link.aps.org/doi/10.1103/PhysRevD.86.102001}.

\bibitem[{\citenamefont{Durin and Zapperi}(2004)}]{Durin:2004}
\bibinfo{author}{\bibnamefont{Durin}, \bibfnamefont{G.}}, and
  \bibinfo{author}{\bibfnamefont{S.}~\bibnamefont{Zapperi}},
  \bibinfo{year}{2004}, \eprint{cond-mat/0404512},
  \urlprefix\url{http://arxiv.org/abs/cond-mat/0404512}.

\bibitem[{\citenamefont{Dwyer}(2013)}]{Dwyer:Thesis}
\bibinfo{author}{\bibnamefont{Dwyer}, \bibfnamefont{S.}}, \bibinfo{year}{2013},
  \emph{\bibinfo{title}{Quantum noise reduction using squeezed states in
  {LIGO}}}, Ph.D. thesis, \bibinfo{school}{{Massachusetts Institute of
  Technology}}, \urlprefix\url{http://hdl.handle.net/1721.1/79427}.

\bibitem[{\citenamefont{Einstein}(1916)}]{Einstein:1916b}
\bibinfo{author}{\bibnamefont{Einstein}, \bibfnamefont{A.}},
  \bibinfo{year}{1916}, \bibinfo{journal}{Sitzungsberichte der K{\"o}niglich
  Preu{\ss}ischen Akademie der Wissenschaften (Berlin)}
  \textbf{\bibinfo{volume}{33}}, \bibinfo{pages}{688}.

\bibitem[{\citenamefont{{Einstein}}(1918)}]{Einstein:1918a}
\bibinfo{author}{\bibnamefont{{Einstein}}, \bibfnamefont{A.}},
  \bibinfo{year}{1918}, \bibinfo{journal}{Sitzungsberichte der K{\"o}niglich
  Preu{\ss}ischen Akademie der Wissenschaften (Berlin)}
  \textbf{\bibinfo{volume}{8}}, \bibinfo{pages}{154}.

\bibitem[{\citenamefont{Einstein and Engel}(1997)}]{einstein1997collected}
\bibinfo{author}{\bibnamefont{Einstein}, \bibfnamefont{A.}}, and
  \bibinfo{author}{\bibfnamefont{A.}~\bibnamefont{Engel}},
  \bibinfo{year}{1997}, \emph{\bibinfo{title}{The Collected Papers of Albert
  Einstein: The Berlin Years: Writings, 1914-1917}}, The Collected Papers of
  Albert Einstein (\bibinfo{publisher}{Princeton University Press}), ISBN
  \bibinfo{isbn}{9780691017341},
  \urlprefix\url{http://books.google.com/books?id=AfMclQEACAAJ}.

\bibitem[{\citenamefont{{Einstein @ Home}}(2012)}]{EatHweb}
\bibinfo{author}{\bibnamefont{{Einstein @ Home}}}, \bibinfo{year}{2012},
  \urlprefix\url{http://einstein.phys.uwm.edu/}.

\bibitem[{\citenamefont{{Einstein Telescope Science Team}}(2011)}]{ET2011}
\bibinfo{author}{\bibnamefont{{Einstein Telescope Science Team}}},
  \bibinfo{year}{2011}, \bibinfo{title}{{Einstein gravitational wave Telescope
  conceptual design study}}, \urlprefix\url{http://www.et-gw.eu/}.

\bibitem[{\citenamefont{{Estabrook} and {Wahlquist}}(1975)}]{EsWa1975}
\bibinfo{author}{\bibnamefont{{Estabrook}}, \bibfnamefont{F.~B.}}, and
  \bibinfo{author}{\bibfnamefont{H.~D.} \bibnamefont{{Wahlquist}}},
  \bibinfo{year}{1975}, \bibinfo{journal}{General Relativity and Gravitation}
  \textbf{\bibinfo{volume}{6}}, \bibinfo{pages}{439}.

\bibitem[{\citenamefont{Evans} \emph{et~al.}(2012)\citenamefont{Evans, Bassiri,
  Maclaren, Rowan, Martin, Hough, and Borisenko}}]{evans2012reduced}
\bibinfo{author}{\bibnamefont{Evans}, \bibfnamefont{K.}},
  \bibinfo{author}{\bibfnamefont{R.}~\bibnamefont{Bassiri}},
  \bibinfo{author}{\bibfnamefont{I.}~\bibnamefont{Maclaren}},
  \bibinfo{author}{\bibfnamefont{S.}~\bibnamefont{Rowan}},
  \bibinfo{author}{\bibfnamefont{I.}~\bibnamefont{Martin}},
  \bibinfo{author}{\bibfnamefont{J.}~\bibnamefont{Hough}}, and
  \bibinfo{author}{\bibfnamefont{K.}~\bibnamefont{Borisenko}},
  \bibinfo{year}{2012}, in \emph{\bibinfo{booktitle}{Journal of Physics:
  Conference Series}} (\bibinfo{organization}{IOP Publishing}), volume
  \bibinfo{volume}{371}, p. \bibinfo{pages}{012058}.

\bibitem[{\citenamefont{Evans} \emph{et~al.}(2008)\citenamefont{Evans, Ballmer,
  Fejer, Fritschel, Harry, and Ogin}}]{Matt:TOnoise}
\bibinfo{author}{\bibnamefont{Evans}, \bibfnamefont{M.}},
  \bibinfo{author}{\bibfnamefont{S.}~\bibnamefont{Ballmer}},
  \bibinfo{author}{\bibfnamefont{M.}~\bibnamefont{Fejer}},
  \bibinfo{author}{\bibfnamefont{P.}~\bibnamefont{Fritschel}},
  \bibinfo{author}{\bibfnamefont{G.}~\bibnamefont{Harry}}, and
  \bibinfo{author}{\bibfnamefont{G.}~\bibnamefont{Ogin}}, \bibinfo{year}{2008},
  \bibinfo{journal}{Phys. Rev. D} \textbf{\bibinfo{volume}{78}},
  \bibinfo{pages}{102003}.

\bibitem[{\citenamefont{Evans} \emph{et~al.}(2010)\citenamefont{Evans,
  Barsotti, and Fritschel}}]{Matt:PI}
\bibinfo{author}{\bibnamefont{Evans}, \bibfnamefont{M.}},
  \bibinfo{author}{\bibfnamefont{L.}~\bibnamefont{Barsotti}}, and
  \bibinfo{author}{\bibfnamefont{P.}~\bibnamefont{Fritschel}},
  \bibinfo{year}{2010}, \bibinfo{journal}{Physics Letters A}
  \textbf{\bibinfo{volume}{374}}(\bibinfo{number}{4}), \bibinfo{pages}{665 },
  ISSN \bibinfo{issn}{0375-9601}.

\bibitem[{\citenamefont{Faller} \emph{et~al.}(1989)\citenamefont{Faller,
  Bender, Hall, Hils, Stebbins, and Vincent}}]{Faller1989107}
\bibinfo{author}{\bibnamefont{Faller}, \bibfnamefont{J.}},
  \bibinfo{author}{\bibfnamefont{P.}~\bibnamefont{Bender}},
  \bibinfo{author}{\bibfnamefont{J.}~\bibnamefont{Hall}},
  \bibinfo{author}{\bibfnamefont{D.}~\bibnamefont{Hils}},
  \bibinfo{author}{\bibfnamefont{R.}~\bibnamefont{Stebbins}}, and
  \bibinfo{author}{\bibfnamefont{M.}~\bibnamefont{Vincent}},
  \bibinfo{year}{1989}, \bibinfo{journal}{Advances in Space Research}
  \textbf{\bibinfo{volume}{9}}(\bibinfo{number}{9}), \bibinfo{pages}{107 },
  ISSN \bibinfo{issn}{0273-1177}.

\bibitem[{\citenamefont{Farmer and Phinney}(2003)}]{FaPh2003}
\bibinfo{author}{\bibnamefont{Farmer}, \bibfnamefont{A.~J.}}, and
  \bibinfo{author}{\bibfnamefont{E.~S.} \bibnamefont{Phinney}},
  \bibinfo{year}{2003}, \bibinfo{journal}{Monthly Notices of the Royal
  Astronomical Society} \textbf{\bibinfo{volume}{346}}(\bibinfo{number}{4}),
  \bibinfo{pages}{1197}, ISSN \bibinfo{issn}{1365-2966}.

\bibitem[{\citenamefont{Faye} \emph{et~al.}(2012)\citenamefont{Faye, Marsat,
  Blanchet, and Iyer}}]{Bala:7half}
\bibinfo{author}{\bibnamefont{Faye}, \bibfnamefont{G.}},
  \bibinfo{author}{\bibfnamefont{S.}~\bibnamefont{Marsat}},
  \bibinfo{author}{\bibfnamefont{L.}~\bibnamefont{Blanchet}}, and
  \bibinfo{author}{\bibfnamefont{B.~R.} \bibnamefont{Iyer}},
  \bibinfo{year}{2012}, \bibinfo{journal}{Classical and Quantum Gravity}
  \textbf{\bibinfo{volume}{29}}(\bibinfo{number}{17}), \bibinfo{pages}{175004},
  \urlprefix\url{http://stacks.iop.org/0264-9381/29/i=17/a=175004}.

\bibitem[{\citenamefont{{Feynman}}(1965)}]{Feynman:III}
\bibinfo{author}{\bibnamefont{{Feynman}}, \bibfnamefont{R.~P.}},
  \bibinfo{year}{1965}, \emph{\bibinfo{title}{{Feynman lectures on physics.
  Volume 3: Quantum mechanics}}} (\bibinfo{publisher}{Addison-Wesley}).

\bibitem[{\citenamefont{Flaminio} \emph{et~al.}(2010)\citenamefont{Flaminio,
  Franc, Michel, Morgado, Pinard, and Sassolas}}]{Flaminio:2010uk}
\bibinfo{author}{\bibnamefont{Flaminio}, \bibfnamefont{R.}},
  \bibinfo{author}{\bibfnamefont{J.}~\bibnamefont{Franc}},
  \bibinfo{author}{\bibfnamefont{C.}~\bibnamefont{Michel}},
  \bibinfo{author}{\bibfnamefont{N.}~\bibnamefont{Morgado}},
  \bibinfo{author}{\bibfnamefont{L.}~\bibnamefont{Pinard}}, and
  \bibinfo{author}{\bibfnamefont{B.}~\bibnamefont{Sassolas}},
  \bibinfo{year}{2010}, \bibinfo{journal}{Classical and Quantum Gravity}
  \textbf{\bibinfo{volume}{27}}(\bibinfo{number}{8}), \bibinfo{pages}{084030}.

\bibitem[{\citenamefont{Flanagan and Thorne}(1995)}]{Kip:Scatter95}
\bibinfo{author}{\bibnamefont{Flanagan}, \bibfnamefont{E.~E.}}, and
  \bibinfo{author}{\bibfnamefont{K.~S.} \bibnamefont{Thorne}},
  \bibinfo{year}{1995}, \emph{\bibinfo{title}{Scattered-Light Noise for LIGO}},
  \bibinfo{type}{Technical Report} \bibinfo{number}{T950102},
  \bibinfo{institution}{LIGO},
  \urlprefix\url{https://dcc.ligo.org/LIGO-T950102/public}.

\bibitem[{\citenamefont{{Fraisse}} \emph{et~al.}(2009)\citenamefont{{Fraisse},
  {Brown}, {Dobler}, {Dotson}, {Draine}, {Frisch}, {Haverkorn}, {Hirata},
  {Jansson}, {Lazarian}, {Magalha\~{}Es}, {Waelkens}}
  \emph{et~al.}}]{CMBpol:foreground}
\bibinfo{author}{\bibnamefont{{Fraisse}}, \bibfnamefont{A.~A.}},
  \bibinfo{author}{\bibfnamefont{J.-A.~C.} \bibnamefont{{Brown}}},
  \bibinfo{author}{\bibfnamefont{G.}~\bibnamefont{{Dobler}}},
  \bibinfo{author}{\bibfnamefont{J.~L.} \bibnamefont{{Dotson}}},
  \bibinfo{author}{\bibfnamefont{B.~T.} \bibnamefont{{Draine}}},
  \bibinfo{author}{\bibfnamefont{P.~C.} \bibnamefont{{Frisch}}},
  \bibinfo{author}{\bibfnamefont{M.}~\bibnamefont{{Haverkorn}}},
  \bibinfo{author}{\bibfnamefont{C.~M.} \bibnamefont{{Hirata}}},
  \bibinfo{author}{\bibfnamefont{R.}~\bibnamefont{{Jansson}}},
  \bibinfo{author}{\bibfnamefont{A.}~\bibnamefont{{Lazarian}}},
  \bibinfo{author}{\bibfnamefont{A.~M.} \bibnamefont{{Magalha\~{}Es}}},
  \bibinfo{author}{\bibfnamefont{A.}~\bibnamefont{{Waelkens}}}, \emph{et~al.},
  \bibinfo{year}{2009}, in \emph{\bibinfo{booktitle}{American Institute of
  Physics Conference Series}}, edited by
  \bibinfo{editor}{\bibfnamefont{S.}~\bibnamefont{{Dodelson}}},
  \bibinfo{editor}{\bibfnamefont{D.}~\bibnamefont{{Baumann}}},
  \bibinfo{editor}{\bibfnamefont{A.}~\bibnamefont{{Cooray}}},
  \bibinfo{editor}{\bibfnamefont{J.}~\bibnamefont{{Dunkley}}},
  \bibinfo{editor}{\bibfnamefont{A.}~\bibnamefont{{Fraisse}}},
  \bibinfo{editor}{\bibfnamefont{M.~G.} \bibnamefont{{Jackson}}},
  \bibinfo{editor}{\bibfnamefont{A.}~\bibnamefont{{Kogut}}},
  \bibinfo{editor}{\bibfnamefont{L.}~\bibnamefont{{Krauss}}},
  \bibinfo{editor}{\bibfnamefont{M.}~\bibnamefont{{Zaldarriaga}}}, and
  \bibinfo{editor}{\bibfnamefont{K.}~\bibnamefont{{Smith}}}, volume
  \bibinfo{volume}{1141} of \emph{\bibinfo{series}{American Institute of
  Physics Conference Series}}, pp. \bibinfo{pages}{265--310},
  \eprint{0811.3920}.

\bibitem[{\citenamefont{Freise} \emph{et~al.}(2007)\citenamefont{Freise,
  Bunkowski, and Schnabel}}]{Freise:2007ed}
\bibinfo{author}{\bibnamefont{Freise}, \bibfnamefont{A.}},
  \bibinfo{author}{\bibfnamefont{A.}~\bibnamefont{Bunkowski}}, and
  \bibinfo{author}{\bibfnamefont{R.}~\bibnamefont{Schnabel}},
  \bibinfo{year}{2007}, \bibinfo{journal}{New Journal of Physics}
  \textbf{\bibinfo{volume}{9}}(\bibinfo{number}{12}), \bibinfo{pages}{433},
  \urlprefix\url{http://stacks.iop.org/1367-2630/9/i=12/a=433}.

\bibitem[{\citenamefont{Freise and Strain}(2010)}]{lrr-2010-1}
\bibinfo{author}{\bibnamefont{Freise}, \bibfnamefont{A.}}, and
  \bibinfo{author}{\bibfnamefont{K.~A.} \bibnamefont{Strain}},
  \bibinfo{year}{2010}, \bibinfo{journal}{Living Reviews in Relativity}
  \textbf{\bibinfo{volume}{13}}(\bibinfo{number}{1}),
  \urlprefix\url{http://www.livingreviews.org/lrr-2010-1}.

\bibitem[{\citenamefont{Fricke} \emph{et~al.}(2012)\citenamefont{Fricke,
  Smith-Lefebvre, Abbott, Adhikari, Dooley, Evans, Fritschel, Frolov, Kawabe,
  Kissel, Slagmolen, and Waldman}}]{Tobin:DC}
\bibinfo{author}{\bibnamefont{Fricke}, \bibfnamefont{T.~T.}},
  \bibinfo{author}{\bibfnamefont{N.~D.} \bibnamefont{Smith-Lefebvre}},
  \bibinfo{author}{\bibfnamefont{R.}~\bibnamefont{Abbott}},
  \bibinfo{author}{\bibfnamefont{R.}~\bibnamefont{Adhikari}},
  \bibinfo{author}{\bibfnamefont{K.~L.} \bibnamefont{Dooley}},
  \bibinfo{author}{\bibfnamefont{M.}~\bibnamefont{Evans}},
  \bibinfo{author}{\bibfnamefont{P.}~\bibnamefont{Fritschel}},
  \bibinfo{author}{\bibfnamefont{V.~V.} \bibnamefont{Frolov}},
  \bibinfo{author}{\bibfnamefont{K.}~\bibnamefont{Kawabe}},
  \bibinfo{author}{\bibfnamefont{J.~S.} \bibnamefont{Kissel}},
  \bibinfo{author}{\bibfnamefont{B.~J.~J.} \bibnamefont{Slagmolen}}, and
  \bibinfo{author}{\bibfnamefont{S.~J.} \bibnamefont{Waldman}},
  \bibinfo{year}{2012}, \bibinfo{journal}{Classical and Quantum Gravity}
  \textbf{\bibinfo{volume}{29}}(\bibinfo{number}{6}), \bibinfo{pages}{065005},
  \urlprefix\url{http://stacks.iop.org/CQG/29/065005}.

\bibitem[{\citenamefont{{Friedrich}}
  \emph{et~al.}(2011)\citenamefont{{Friedrich}, {Barr}, {Br{\"u}ckner}, {Hild},
  {Nelson}, {MacArthur}, {Plissi}, {Edgar}, {Huttner}, {Sorazu}, {Kroker},
  {Britzger}} \emph{et~al.}}]{Gratings:2011}
\bibinfo{author}{\bibnamefont{{Friedrich}}, \bibfnamefont{D.}},
  \bibinfo{author}{\bibfnamefont{B.~W.} \bibnamefont{{Barr}}},
  \bibinfo{author}{\bibfnamefont{F.}~\bibnamefont{{Br{\"u}ckner}}},
  \bibinfo{author}{\bibfnamefont{S.}~\bibnamefont{{Hild}}},
  \bibinfo{author}{\bibfnamefont{J.}~\bibnamefont{{Nelson}}},
  \bibinfo{author}{\bibfnamefont{J.}~\bibnamefont{{MacArthur}}},
  \bibinfo{author}{\bibfnamefont{M.~V.} \bibnamefont{{Plissi}}},
  \bibinfo{author}{\bibfnamefont{M.~P.} \bibnamefont{{Edgar}}},
  \bibinfo{author}{\bibfnamefont{S.~H.} \bibnamefont{{Huttner}}},
  \bibinfo{author}{\bibfnamefont{B.}~\bibnamefont{{Sorazu}}},
  \bibinfo{author}{\bibfnamefont{S.}~\bibnamefont{{Kroker}}},
  \bibinfo{author}{\bibfnamefont{M.}~\bibnamefont{{Britzger}}}, \emph{et~al.},
  \bibinfo{year}{2011}, \bibinfo{journal}{Optics Express}
  \textbf{\bibinfo{volume}{19}}, \bibinfo{pages}{14955}.

\bibitem[{\citenamefont{Fritschel} \emph{et~al.}(2001)\citenamefont{Fritschel,
  Bork, Gonz\'{a}lez, Mavalvala, Ouimette, Rong, Sigg, and
  Zucker}}]{Mavalvala:2001vn}
\bibinfo{author}{\bibnamefont{Fritschel}, \bibfnamefont{P.}},
  \bibinfo{author}{\bibfnamefont{R.}~\bibnamefont{Bork}},
  \bibinfo{author}{\bibfnamefont{G.}~\bibnamefont{Gonz\'{a}lez}},
  \bibinfo{author}{\bibfnamefont{N.}~\bibnamefont{Mavalvala}},
  \bibinfo{author}{\bibfnamefont{D.}~\bibnamefont{Ouimette}},
  \bibinfo{author}{\bibfnamefont{H.}~\bibnamefont{Rong}},
  \bibinfo{author}{\bibfnamefont{D.}~\bibnamefont{Sigg}}, and
  \bibinfo{author}{\bibfnamefont{M.}~\bibnamefont{Zucker}},
  \bibinfo{year}{2001}, \bibinfo{journal}{Appl. Opt.}
  \textbf{\bibinfo{volume}{40}}(\bibinfo{number}{28}), \bibinfo{pages}{4988},
  \urlprefix\url{http://dx.doi.org/10.1364/AO.40.004988}.

\bibitem[{\citenamefont{Fritschel} \emph{et~al.}(1998)\citenamefont{Fritschel,
  Mavalvala, Shoemaker, Sigg, Zucker, and Gonz\'{a}lez}}]{Mavalvala:1998ur}
\bibinfo{author}{\bibnamefont{Fritschel}, \bibfnamefont{P.}},
  \bibinfo{author}{\bibfnamefont{N.}~\bibnamefont{Mavalvala}},
  \bibinfo{author}{\bibfnamefont{D.}~\bibnamefont{Shoemaker}},
  \bibinfo{author}{\bibfnamefont{D.}~\bibnamefont{Sigg}},
  \bibinfo{author}{\bibfnamefont{M.}~\bibnamefont{Zucker}}, and
  \bibinfo{author}{\bibfnamefont{G.}~\bibnamefont{Gonz\'{a}lez}},
  \bibinfo{year}{1998}, \bibinfo{journal}{Appl. Opt.}
  \textbf{\bibinfo{volume}{37}}(\bibinfo{number}{28}), \bibinfo{pages}{6734},
  \urlprefix\url{http://dx.doi.org/10.1364/ao.37.006734}.

\bibitem[{\citenamefont{Fritschel} \emph{et~al.}(1992)\citenamefont{Fritschel,
  Shoemaker, and Weiss}}]{Fritschel:1992jk}
\bibinfo{author}{\bibnamefont{Fritschel}, \bibfnamefont{P.}},
  \bibinfo{author}{\bibfnamefont{D.}~\bibnamefont{Shoemaker}}, and
  \bibinfo{author}{\bibfnamefont{R.}~\bibnamefont{Weiss}},
  \bibinfo{year}{1992}, \bibinfo{journal}{Appl. Opt}
  \textbf{\bibinfo{volume}{31}}(\bibinfo{number}{10}), \bibinfo{pages}{1412}.

\bibitem[{\citenamefont{{Gaidarzhy}}
  \emph{et~al.}(2007)\citenamefont{{Gaidarzhy}, {Imboden}, {Mohanty}, {Rankin},
  and {Sheldon}}}]{Diamond:Q}
\bibinfo{author}{\bibnamefont{{Gaidarzhy}}, \bibfnamefont{A.}},
  \bibinfo{author}{\bibfnamefont{M.}~\bibnamefont{{Imboden}}},
  \bibinfo{author}{\bibfnamefont{P.}~\bibnamefont{{Mohanty}}},
  \bibinfo{author}{\bibfnamefont{J.}~\bibnamefont{{Rankin}}}, and
  \bibinfo{author}{\bibfnamefont{B.~W.} \bibnamefont{{Sheldon}}},
  \bibinfo{year}{2007}, \bibinfo{journal}{Applied Physics Letters}
  \textbf{\bibinfo{volume}{91}}(\bibinfo{number}{20}), \bibinfo{eid}{203503}.

\bibitem[{\citenamefont{Giaime} \emph{et~al.}(1996)\citenamefont{Giaime, Saha,
  Shoemaker, and Sievers}}]{Giaime:1996}
\bibinfo{author}{\bibnamefont{Giaime}, \bibfnamefont{J.}},
  \bibinfo{author}{\bibfnamefont{P.}~\bibnamefont{Saha}},
  \bibinfo{author}{\bibfnamefont{D.}~\bibnamefont{Shoemaker}}, and
  \bibinfo{author}{\bibfnamefont{L.}~\bibnamefont{Sievers}},
  \bibinfo{year}{1996}, \bibinfo{journal}{Review of Scientific Instruments}
  \textbf{\bibinfo{volume}{67}}(\bibinfo{number}{1}), \bibinfo{pages}{208}.

\bibitem[{\citenamefont{Giaime} \emph{et~al.}(2003)\citenamefont{Giaime, Daw,
  Weitz, Adhikari, Fritschel, Abbott, Bork, and Heefner}}]{Ed:MSFF}
\bibinfo{author}{\bibnamefont{Giaime}, \bibfnamefont{J.~A.}},
  \bibinfo{author}{\bibfnamefont{E.~J.} \bibnamefont{Daw}},
  \bibinfo{author}{\bibfnamefont{M.}~\bibnamefont{Weitz}},
  \bibinfo{author}{\bibfnamefont{R.}~\bibnamefont{Adhikari}},
  \bibinfo{author}{\bibfnamefont{P.}~\bibnamefont{Fritschel}},
  \bibinfo{author}{\bibfnamefont{R.}~\bibnamefont{Abbott}},
  \bibinfo{author}{\bibfnamefont{R.}~\bibnamefont{Bork}}, and
  \bibinfo{author}{\bibfnamefont{J.}~\bibnamefont{Heefner}},
  \bibinfo{year}{2003}, \bibinfo{journal}{Review of Scientific Instruments}
  \textbf{\bibinfo{volume}{74}}(\bibinfo{number}{1}), \bibinfo{pages}{218},
  \urlprefix\url{http://link.aip.org/link/?RSI/74/218/1}.

\bibitem[{\citenamefont{Giazotto}(1989)}]{Giazotto:1989dr}
\bibinfo{author}{\bibnamefont{Giazotto}, \bibfnamefont{A.}},
  \bibinfo{year}{1989}, \bibinfo{journal}{Physics Reports}
  \textbf{\bibinfo{volume}{182}}(\bibinfo{number}{6}), \bibinfo{pages}{365}.

\bibitem[{\citenamefont{Giazotto}(1998)}]{Giazotto:ElectroSUS}
\bibinfo{author}{\bibnamefont{Giazotto}, \bibfnamefont{A.}},
  \bibinfo{year}{1998}, \bibinfo{journal}{Physics Letters A}
  \textbf{\bibinfo{volume}{245}}(\bibinfo{number}{3--4}), \bibinfo{pages}{203
  }, ISSN \bibinfo{issn}{0375-9601},
  \urlprefix\url{http://www.sciencedirect.com/science/article/pii/S0375960198004113}.

\bibitem[{\citenamefont{Giazotto}(2011)}]{Giazotto:2011ux}
\bibinfo{author}{\bibnamefont{Giazotto}, \bibfnamefont{A.}},
  \bibinfo{year}{2011}, \bibinfo{journal}{Physics Letters A} .

\bibitem[{\citenamefont{Gillespie and Raab}(1995)}]{Gillespie:1995}
\bibinfo{author}{\bibnamefont{Gillespie}, \bibfnamefont{A.}}, and
  \bibinfo{author}{\bibfnamefont{F.}~\bibnamefont{Raab}}, \bibinfo{year}{1995},
  \bibinfo{journal}{Phys. Rev. D} \textbf{\bibinfo{volume}{52}},
  \bibinfo{pages}{577}.

\bibitem[{\citenamefont{{Goda}} \emph{et~al.}(2008)\citenamefont{{Goda},
  {Miyakawa}, {Mikhailov}, {Saraf}, {Adhikari}, {McKenzie}, {Ward}, {Vass},
  {Weinstein}, and {Mavalvala}}}]{Go:40m}
\bibinfo{author}{\bibnamefont{{Goda}}, \bibfnamefont{K.}},
  \bibinfo{author}{\bibfnamefont{O.}~\bibnamefont{{Miyakawa}}},
  \bibinfo{author}{\bibfnamefont{E.~E.} \bibnamefont{{Mikhailov}}},
  \bibinfo{author}{\bibfnamefont{S.}~\bibnamefont{{Saraf}}},
  \bibinfo{author}{\bibfnamefont{R.}~\bibnamefont{{Adhikari}}},
  \bibinfo{author}{\bibfnamefont{K.}~\bibnamefont{{McKenzie}}},
  \bibinfo{author}{\bibfnamefont{R.}~\bibnamefont{{Ward}}},
  \bibinfo{author}{\bibfnamefont{S.}~\bibnamefont{{Vass}}},
  \bibinfo{author}{\bibfnamefont{A.~J.} \bibnamefont{{Weinstein}}}, and
  \bibinfo{author}{\bibfnamefont{N.}~\bibnamefont{{Mavalvala}}},
  \bibinfo{year}{2008}, \bibinfo{journal}{Nature Physics}
  \textbf{\bibinfo{volume}{4}}, \bibinfo{pages}{472}.

\bibitem[{\citenamefont{Gonz{\'a}lez}(2000)}]{Gonzalez:2000vw}
\bibinfo{author}{\bibnamefont{Gonz{\'a}lez}, \bibfnamefont{G.}},
  \bibinfo{year}{2000}, \bibinfo{journal}{Classical and Quantum Gravity} .

\bibitem[{\citenamefont{Gonz{\'a}lez and Saulson}(1994)}]{gonzalez:207}
\bibinfo{author}{\bibnamefont{Gonz{\'a}lez}, \bibfnamefont{G.~I.}}, and
  \bibinfo{author}{\bibfnamefont{P.~R.} \bibnamefont{Saulson}},
  \bibinfo{year}{1994}, \bibinfo{journal}{The Journal of the Acoustical Society
  of America} \textbf{\bibinfo{volume}{96}}(\bibinfo{number}{1}),
  \bibinfo{pages}{207}.

\bibitem[{\citenamefont{Go\ss{}ler}
  \emph{et~al.}(2007)\citenamefont{Go\ss{}ler, Cumpston, McKenzie, Mow-Lowry,
  Gray, and McClelland}}]{Gossler:2007uu}
\bibinfo{author}{\bibnamefont{Go\ss{}ler}, \bibfnamefont{S.}},
  \bibinfo{author}{\bibfnamefont{J.}~\bibnamefont{Cumpston}},
  \bibinfo{author}{\bibfnamefont{K.}~\bibnamefont{McKenzie}},
  \bibinfo{author}{\bibfnamefont{C.~M.} \bibnamefont{Mow-Lowry}},
  \bibinfo{author}{\bibfnamefont{M.~B.} \bibnamefont{Gray}}, and
  \bibinfo{author}{\bibfnamefont{D.~E.} \bibnamefont{McClelland}},
  \bibinfo{year}{2007}, \bibinfo{journal}{Phys. Rev. A}
  \textbf{\bibinfo{volume}{76}}, \bibinfo{pages}{053810},
  \urlprefix\url{http://link.aps.org/doi/10.1103/PhysRevA.76.053810}.

\bibitem[{\citenamefont{Granata}
  \emph{et~al.}(2010{\natexlab{a}})\citenamefont{Granata, Barsuglia, Flaminio,
  Freise, Hild, and Marque}}]{Granata:2010}
\bibinfo{author}{\bibnamefont{Granata}, \bibfnamefont{M.}},
  \bibinfo{author}{\bibfnamefont{M.}~\bibnamefont{Barsuglia}},
  \bibinfo{author}{\bibfnamefont{R.}~\bibnamefont{Flaminio}},
  \bibinfo{author}{\bibfnamefont{A.}~\bibnamefont{Freise}},
  \bibinfo{author}{\bibfnamefont{S.}~\bibnamefont{Hild}}, and
  \bibinfo{author}{\bibfnamefont{J.}~\bibnamefont{Marque}},
  \bibinfo{year}{2010}{\natexlab{a}}, \bibinfo{journal}{Journal of Physics:
  Conference Series} \textbf{\bibinfo{volume}{228}}(\bibinfo{number}{1}),
  \bibinfo{pages}{012016}.

\bibitem[{\citenamefont{Granata}
  \emph{et~al.}(2010{\natexlab{b}})\citenamefont{Granata, Buy, Ward, and
  Barsuglia}}]{Rob:LG}
\bibinfo{author}{\bibnamefont{Granata}, \bibfnamefont{M.}},
  \bibinfo{author}{\bibfnamefont{C.}~\bibnamefont{Buy}},
  \bibinfo{author}{\bibfnamefont{R.}~\bibnamefont{Ward}}, and
  \bibinfo{author}{\bibfnamefont{M.}~\bibnamefont{Barsuglia}},
  \bibinfo{year}{2010}{\natexlab{b}}, \bibinfo{journal}{Phys. Rev. Lett.}
  \textbf{\bibinfo{volume}{105}}, \bibinfo{pages}{231102}.

\bibitem[{\citenamefont{Gras} \emph{et~al.}(2009)\citenamefont{Gras, Blair, and
  Zhao}}]{Gras:PI2009}
\bibinfo{author}{\bibnamefont{Gras}, \bibfnamefont{S.}},
  \bibinfo{author}{\bibfnamefont{D.~G.} \bibnamefont{Blair}}, and
  \bibinfo{author}{\bibfnamefont{C.}~\bibnamefont{Zhao}}, \bibinfo{year}{2009},
  \bibinfo{journal}{Classical and Quantum Gravity}
  \textbf{\bibinfo{volume}{26}}(\bibinfo{number}{13}), \bibinfo{pages}{135012},
  \urlprefix\url{http://stacks.iop.org/0264-9381/26/i=13/a=135012}.

\bibitem[{\citenamefont{Gretarsson and
  Harry}(1999)}]{gretarsson1999dissipation}
\bibinfo{author}{\bibnamefont{Gretarsson}, \bibfnamefont{A.}}, and
  \bibinfo{author}{\bibfnamefont{G.}~\bibnamefont{Harry}},
  \bibinfo{year}{1999}, \bibinfo{journal}{Review of scientific instruments}
  \textbf{\bibinfo{volume}{70}}(\bibinfo{number}{10}), \bibinfo{pages}{4081}.

\bibitem[{\citenamefont{Gretarsson}
  \emph{et~al.}(2007)\citenamefont{Gretarsson, D'Ambrosio, Frolov, O'Reilly,
  and Fritschel}}]{Andri:PRCoffset}
\bibinfo{author}{\bibnamefont{Gretarsson}, \bibfnamefont{A.~M.}},
  \bibinfo{author}{\bibfnamefont{E.}~\bibnamefont{D'Ambrosio}},
  \bibinfo{author}{\bibfnamefont{V.}~\bibnamefont{Frolov}},
  \bibinfo{author}{\bibfnamefont{B.}~\bibnamefont{O'Reilly}}, and
  \bibinfo{author}{\bibfnamefont{P.~K.} \bibnamefont{Fritschel}},
  \bibinfo{year}{2007}, \bibinfo{journal}{J. Opt. Soc. Am. B}
  \textbf{\bibinfo{volume}{24}}(\bibinfo{number}{11}), \bibinfo{pages}{2821},
  \urlprefix\url{http://dx.doi.org/10.1364/JOSAB.24.002821}.

\bibitem[{\citenamefont{{Gretarsson}}
  \emph{et~al.}(2000)\citenamefont{{Gretarsson}, {Harry}, {Penn}, {Saulson},
  {Startin}, {Rowan}, {Cagnoli}, and {Hough}}}]{Andri:Surface}
\bibinfo{author}{\bibnamefont{{Gretarsson}}, \bibfnamefont{A.~M.}},
  \bibinfo{author}{\bibfnamefont{G.~M.} \bibnamefont{{Harry}}},
  \bibinfo{author}{\bibfnamefont{S.~D.} \bibnamefont{{Penn}}},
  \bibinfo{author}{\bibfnamefont{P.~R.} \bibnamefont{{Saulson}}},
  \bibinfo{author}{\bibfnamefont{W.~J.} \bibnamefont{{Startin}}},
  \bibinfo{author}{\bibfnamefont{S.}~\bibnamefont{{Rowan}}},
  \bibinfo{author}{\bibfnamefont{G.}~\bibnamefont{{Cagnoli}}}, and
  \bibinfo{author}{\bibfnamefont{J.}~\bibnamefont{{Hough}}},
  \bibinfo{year}{2000}, \bibinfo{journal}{Physics Letters A}
  \textbf{\bibinfo{volume}{270}}, \bibinfo{pages}{108}.

\bibitem[{\citenamefont{Gr\'{e}verie}
  \emph{et~al.}(2010)\citenamefont{Gr\'{e}verie, Brillet, Man, Chaibi, Coulon,
  and Feliksik}}]{Greverie:10}
\bibinfo{author}{\bibnamefont{Gr\'{e}verie}, \bibfnamefont{C.}},
  \bibinfo{author}{\bibfnamefont{A.}~\bibnamefont{Brillet}},
  \bibinfo{author}{\bibfnamefont{C.~N.} \bibnamefont{Man}},
  \bibinfo{author}{\bibfnamefont{W.}~\bibnamefont{Chaibi}},
  \bibinfo{author}{\bibfnamefont{J.~P.} \bibnamefont{Coulon}}, and
  \bibinfo{author}{\bibfnamefont{K.}~\bibnamefont{Feliksik}},
  \bibinfo{year}{2010}, in \emph{\bibinfo{booktitle}{Conference on Lasers and
  Electro-Optics}} (\bibinfo{publisher}{Optical Society of America}), p.
  \bibinfo{pages}{JTuD36},
  \urlprefix\url{http://www.opticsinfobase.org/abstract.cfm?URI=CLEO-2010-JTuD36}.

\bibitem[{\citenamefont{Grote}(2003)}]{Hartmut:PhD}
\bibinfo{author}{\bibnamefont{Grote}, \bibfnamefont{H.}}, \bibinfo{year}{2003},
  \emph{\bibinfo{title}{Making it Work: Second Generation Interferometry in GEO
  600!}}, Ph.D. thesis, \bibinfo{school}{{Universit\"at Hannover}}.

\bibitem[{\citenamefont{Grote}(2008)}]{grote2008status}
\bibinfo{author}{\bibnamefont{Grote}, \bibfnamefont{H.}}, \bibinfo{year}{2008},
  \bibinfo{journal}{Classical and Quantum Gravity}
  \textbf{\bibinfo{volume}{25}}(\bibinfo{number}{11}), \bibinfo{pages}{114043}.

\bibitem[{\citenamefont{Grote}(2010)}]{Hartmut:2010}
\bibinfo{author}{\bibnamefont{Grote}, \bibfnamefont{H.}}, \bibinfo{year}{2010},
  \bibinfo{journal}{Classical and Quantum Gravity}
  \textbf{\bibinfo{volume}{27}}(\bibinfo{number}{8}), \bibinfo{pages}{084003}.

\bibitem[{\citenamefont{{Grote}} \emph{et~al.}(2002)\citenamefont{{Grote},
  {Heinzel}, {Freise}, {Gossler}, {Willke}, {L{\"u}ck}, {Ward}, {Casey},
  {Strain}, {Robertson}, {Hough}, and {Danzmann}}}]{Grote:2002wl}
\bibinfo{author}{\bibnamefont{{Grote}}, \bibfnamefont{H.}},
  \bibinfo{author}{\bibfnamefont{G.}~\bibnamefont{{Heinzel}}},
  \bibinfo{author}{\bibfnamefont{A.}~\bibnamefont{{Freise}}},
  \bibinfo{author}{\bibfnamefont{S.}~\bibnamefont{{Gossler}}},
  \bibinfo{author}{\bibfnamefont{B.}~\bibnamefont{{Willke}}},
  \bibinfo{author}{\bibfnamefont{H.}~\bibnamefont{{L{\"u}ck}}},
  \bibinfo{author}{\bibfnamefont{H.}~\bibnamefont{{Ward}}},
  \bibinfo{author}{\bibfnamefont{M.}~\bibnamefont{{Casey}}},
  \bibinfo{author}{\bibfnamefont{K.~A.} \bibnamefont{{Strain}}},
  \bibinfo{author}{\bibfnamefont{D.}~\bibnamefont{{Robertson}}},
  \bibinfo{author}{\bibfnamefont{J.}~\bibnamefont{{Hough}}}, and
  \bibinfo{author}{\bibfnamefont{K.}~\bibnamefont{{Danzmann}}},
  \bibinfo{year}{2002}, \bibinfo{journal}{Classical and Quantum Gravity}
  \textbf{\bibinfo{volume}{19}}, \bibinfo{pages}{1849}.

\bibitem[{\citenamefont{{Hammond}} \emph{et~al.}(2012)\citenamefont{{Hammond},
  {Cumming}, {Hough}, {Kumar}, {Tokmakov}, {Reid}, and
  {Rowan}}}]{GIles:SUS2012}
\bibinfo{author}{\bibnamefont{{Hammond}}, \bibfnamefont{G.~D.}},
  \bibinfo{author}{\bibfnamefont{A.~V.} \bibnamefont{{Cumming}}},
  \bibinfo{author}{\bibfnamefont{J.}~\bibnamefont{{Hough}}},
  \bibinfo{author}{\bibfnamefont{R.}~\bibnamefont{{Kumar}}},
  \bibinfo{author}{\bibfnamefont{K.}~\bibnamefont{{Tokmakov}}},
  \bibinfo{author}{\bibfnamefont{S.}~\bibnamefont{{Reid}}}, and
  \bibinfo{author}{\bibfnamefont{S.}~\bibnamefont{{Rowan}}},
  \bibinfo{year}{2012}, \bibinfo{journal}{Classical and Quantum Gravity}
  \textbf{\bibinfo{volume}{29}}(\bibinfo{number}{12}), \bibinfo{pages}{124009}.

\bibitem[{\citenamefont{{Hammond}} \emph{et~al.}(2004)\citenamefont{{Hammond},
  {Pulido-Paton}, {Speake}, and {Trenkel}}}]{Giles:SuperTorsion}
\bibinfo{author}{\bibnamefont{{Hammond}}, \bibfnamefont{G.~D.}},
  \bibinfo{author}{\bibfnamefont{A.}~\bibnamefont{{Pulido-Paton}}},
  \bibinfo{author}{\bibfnamefont{C.~C.} \bibnamefont{{Speake}}}, and
  \bibinfo{author}{\bibfnamefont{C.}~\bibnamefont{{Trenkel}}},
  \bibinfo{year}{2004}, \bibinfo{journal}{Review of Scientific Instruments}
  \textbf{\bibinfo{volume}{75}}, \bibinfo{pages}{955}.

\bibitem[{\citenamefont{Hardham} \emph{et~al.}(2004)\citenamefont{Hardham,
  Abbott, Abbott, Allen, Bork, Campbell, Carter, Coyne, DeBra, Evans}
  \emph{et~al.}}]{hardham2004multi}
\bibinfo{author}{\bibnamefont{Hardham}, \bibfnamefont{C.}},
  \bibinfo{author}{\bibfnamefont{B.}~\bibnamefont{Abbott}},
  \bibinfo{author}{\bibfnamefont{R.}~\bibnamefont{Abbott}},
  \bibinfo{author}{\bibfnamefont{G.}~\bibnamefont{Allen}},
  \bibinfo{author}{\bibfnamefont{R.}~\bibnamefont{Bork}},
  \bibinfo{author}{\bibfnamefont{C.}~\bibnamefont{Campbell}},
  \bibinfo{author}{\bibfnamefont{K.}~\bibnamefont{Carter}},
  \bibinfo{author}{\bibfnamefont{D.}~\bibnamefont{Coyne}},
  \bibinfo{author}{\bibfnamefont{D.}~\bibnamefont{DeBra}},
  \bibinfo{author}{\bibfnamefont{T.}~\bibnamefont{Evans}}, \emph{et~al.},
  \bibinfo{year}{2004}, in \emph{\bibinfo{booktitle}{in Proceedings of ASPE
  Spring Topical Meeting on Control of Precision Systems}}, pp.
  \bibinfo{pages}{127--132}.

\bibitem[{\citenamefont{Harms} \emph{et~al.}(2003)\citenamefont{Harms, Chen,
  Chelkowski, Franzen, Vahlbruch, Danzmann, and Schnabel}}]{Harms:2003wv}
\bibinfo{author}{\bibnamefont{Harms}, \bibfnamefont{J.}},
  \bibinfo{author}{\bibfnamefont{Y.}~\bibnamefont{Chen}},
  \bibinfo{author}{\bibfnamefont{S.}~\bibnamefont{Chelkowski}},
  \bibinfo{author}{\bibfnamefont{A.}~\bibnamefont{Franzen}},
  \bibinfo{author}{\bibfnamefont{H.}~\bibnamefont{Vahlbruch}},
  \bibinfo{author}{\bibfnamefont{K.}~\bibnamefont{Danzmann}}, and
  \bibinfo{author}{\bibfnamefont{R.}~\bibnamefont{Schnabel}},
  \bibinfo{year}{2003}, \bibinfo{journal}{Physical Review D}
  \textbf{\bibinfo{volume}{68}}(\bibinfo{number}{4}), \bibinfo{pages}{042001}.

\bibitem[{\citenamefont{Harms} \emph{et~al.}(2009)\citenamefont{Harms, DeSalvo,
  Dorsher, and Mandic}}]{Harms:2009he}
\bibinfo{author}{\bibnamefont{Harms}, \bibfnamefont{J.}},
  \bibinfo{author}{\bibfnamefont{R.}~\bibnamefont{DeSalvo}},
  \bibinfo{author}{\bibfnamefont{S.}~\bibnamefont{Dorsher}}, and
  \bibinfo{author}{\bibfnamefont{V.}~\bibnamefont{Mandic}},
  \bibinfo{year}{2009}, \bibinfo{journal}{Phys. Rev. D}
  \textbf{\bibinfo{volume}{80}}, \bibinfo{pages}{122001},
  \urlprefix\url{http://link.aps.org/doi/10.1103/PhysRevD.80.122001}.

\bibitem[{\citenamefont{{Harms}} \emph{et~al.}(2013)\citenamefont{{Harms},
  {Slagmolen}, {Adhikari}, {Miller}, {Evans}, {Chen}, {M{\"u}ller}, and
  {Ando}}}]{MANGO:2013}
\bibinfo{author}{\bibnamefont{{Harms}}, \bibfnamefont{J.}},
  \bibinfo{author}{\bibfnamefont{B.~J.~J.} \bibnamefont{{Slagmolen}}},
  \bibinfo{author}{\bibfnamefont{R.~X.} \bibnamefont{{Adhikari}}},
  \bibinfo{author}{\bibfnamefont{M.~C.} \bibnamefont{{Miller}}},
  \bibinfo{author}{\bibfnamefont{M.}~\bibnamefont{{Evans}}},
  \bibinfo{author}{\bibfnamefont{Y.}~\bibnamefont{{Chen}}},
  \bibinfo{author}{\bibfnamefont{H.}~\bibnamefont{{M{\"u}ller}}}, and
  \bibinfo{author}{\bibfnamefont{M.}~\bibnamefont{{Ando}}},
  \bibinfo{year}{2013}, \bibinfo{journal}{ArXiv e-prints} \eprint{1308.2074}.

\bibitem[{\citenamefont{Harry} \emph{et~al.}(2012)\citenamefont{Harry, Bodiya,
  and DeSalvo}}]{harry2012optical}
\bibinfo{author}{\bibnamefont{Harry}, \bibfnamefont{G.}},
  \bibinfo{author}{\bibfnamefont{T.}~\bibnamefont{Bodiya}}, and
  \bibinfo{author}{\bibfnamefont{R.}~\bibnamefont{DeSalvo}},
  \bibinfo{year}{2012}, \emph{\bibinfo{title}{Optical Coatings and Thermal
  Noise in Precision Measurement}}, Optical Coatings and Thermal Noise in
  Precision Measurement (\bibinfo{publisher}{Cambridge University Press}), ISBN
  \bibinfo{isbn}{9781107003385},
  \urlprefix\url{http://books.google.com/books?id=770n4QLmBrMC}.

\bibitem[{\citenamefont{Harry}(2010)}]{Gregg:aLIGO2010}
\bibinfo{author}{\bibnamefont{Harry}, \bibfnamefont{G.~M.}},
  \bibinfo{year}{2010}, \bibinfo{journal}{Classical and Quantum Gravity}
  \textbf{\bibinfo{volume}{27}}(\bibinfo{number}{8}), \bibinfo{pages}{084006}.

\bibitem[{\citenamefont{Harry} \emph{et~al.}(2007)\citenamefont{Harry,
  Abernathy, Becerra-Toledo, Armandula, Black, Dooley, Eichenfield, Nwabugwu,
  Villar, Crooks, Cagnoli, Hough} \emph{et~al.}}]{Harry:CQG2007}
\bibinfo{author}{\bibnamefont{Harry}, \bibfnamefont{G.~M.}},
  \bibinfo{author}{\bibfnamefont{M.~R.} \bibnamefont{Abernathy}},
  \bibinfo{author}{\bibfnamefont{A.~E.} \bibnamefont{Becerra-Toledo}},
  \bibinfo{author}{\bibfnamefont{H.}~\bibnamefont{Armandula}},
  \bibinfo{author}{\bibfnamefont{E.}~\bibnamefont{Black}},
  \bibinfo{author}{\bibfnamefont{K.}~\bibnamefont{Dooley}},
  \bibinfo{author}{\bibfnamefont{M.}~\bibnamefont{Eichenfield}},
  \bibinfo{author}{\bibfnamefont{C.}~\bibnamefont{Nwabugwu}},
  \bibinfo{author}{\bibfnamefont{A.}~\bibnamefont{Villar}},
  \bibinfo{author}{\bibfnamefont{D.~R.~M.} \bibnamefont{Crooks}},
  \bibinfo{author}{\bibfnamefont{G.}~\bibnamefont{Cagnoli}},
  \bibinfo{author}{\bibfnamefont{J.}~\bibnamefont{Hough}}, \emph{et~al.},
  \bibinfo{year}{2007}, \bibinfo{journal}{Classical and Quantum Gravity}
  \textbf{\bibinfo{volume}{24}}(\bibinfo{number}{2}), \bibinfo{pages}{405}.

\bibitem[{\citenamefont{Harry} \emph{et~al.}(2006)\citenamefont{Harry,
  Armandula, Black, Crooks, Cagnoli, Hough, Murray, Reid, Rowan, Sneddon,
  Fejer, Route} \emph{et~al.}}]{Harry:AO2006}
\bibinfo{author}{\bibnamefont{Harry}, \bibfnamefont{G.~M.}},
  \bibinfo{author}{\bibfnamefont{H.}~\bibnamefont{Armandula}},
  \bibinfo{author}{\bibfnamefont{E.}~\bibnamefont{Black}},
  \bibinfo{author}{\bibfnamefont{D.~R.~M.} \bibnamefont{Crooks}},
  \bibinfo{author}{\bibfnamefont{G.}~\bibnamefont{Cagnoli}},
  \bibinfo{author}{\bibfnamefont{J.}~\bibnamefont{Hough}},
  \bibinfo{author}{\bibfnamefont{P.}~\bibnamefont{Murray}},
  \bibinfo{author}{\bibfnamefont{S.}~\bibnamefont{Reid}},
  \bibinfo{author}{\bibfnamefont{S.}~\bibnamefont{Rowan}},
  \bibinfo{author}{\bibfnamefont{P.}~\bibnamefont{Sneddon}},
  \bibinfo{author}{\bibfnamefont{M.~M.} \bibnamefont{Fejer}},
  \bibinfo{author}{\bibfnamefont{R.}~\bibnamefont{Route}}, \emph{et~al.},
  \bibinfo{year}{2006}, \bibinfo{journal}{Appl. Opt.}
  \textbf{\bibinfo{volume}{45}}(\bibinfo{number}{7}), \bibinfo{pages}{1569}.

\bibitem[{\citenamefont{Harry} \emph{et~al.}(2002)\citenamefont{Harry,
  Gretarsson, Saulson, Kittelberger, Penn, Startin, Rowan, Fejer, Crooks,
  Cagnoli, Hough, and Nakagawa}}]{Harry:CQG2002}
\bibinfo{author}{\bibnamefont{Harry}, \bibfnamefont{G.~M.}},
  \bibinfo{author}{\bibfnamefont{A.~M.} \bibnamefont{Gretarsson}},
  \bibinfo{author}{\bibfnamefont{P.~R.} \bibnamefont{Saulson}},
  \bibinfo{author}{\bibfnamefont{S.~E.} \bibnamefont{Kittelberger}},
  \bibinfo{author}{\bibfnamefont{S.~D.} \bibnamefont{Penn}},
  \bibinfo{author}{\bibfnamefont{W.~J.} \bibnamefont{Startin}},
  \bibinfo{author}{\bibfnamefont{S.}~\bibnamefont{Rowan}},
  \bibinfo{author}{\bibfnamefont{M.~M.} \bibnamefont{Fejer}},
  \bibinfo{author}{\bibfnamefont{D.~R.~M.} \bibnamefont{Crooks}},
  \bibinfo{author}{\bibfnamefont{G.}~\bibnamefont{Cagnoli}},
  \bibinfo{author}{\bibfnamefont{J.}~\bibnamefont{Hough}}, and
  \bibinfo{author}{\bibfnamefont{N.}~\bibnamefont{Nakagawa}},
  \bibinfo{year}{2002}, \bibinfo{journal}{Classical and Quantum Gravity}
  \textbf{\bibinfo{volume}{19}}(\bibinfo{number}{5}), \bibinfo{pages}{897}.

\bibitem[{\citenamefont{Hawking and Israel}(1989)}]{300:years}
\bibinfo{editor}{\bibnamefont{Hawking}, \bibfnamefont{S.}}, and
  \bibinfo{editor}{\bibfnamefont{W.}~\bibnamefont{Israel}} (eds.),
  \bibinfo{year}{1989}, \emph{\bibinfo{title}{{Three Hundred Years of
  Gravitation}}}, Philosophiae Naturalis, Principia Mathematica
  (\bibinfo{publisher}{Cambridge University Press}), ISBN
  \bibinfo{isbn}{9780521379762}.

\bibitem[{\citenamefont{Haykin}(2002)}]{Haykin:Adaptive}
\bibinfo{author}{\bibnamefont{Haykin}, \bibfnamefont{S.}},
  \bibinfo{year}{2002}, \emph{\bibinfo{title}{Adaptive Filter Theory}},
  Prentice-Hall information and system sciences series
  (\bibinfo{publisher}{Prentice Hall}), ISBN \bibinfo{isbn}{9780130901262}.

\bibitem[{\citenamefont{{Hebard}}(1973)}]{Hebard:RSI}
\bibinfo{author}{\bibnamefont{{Hebard}}, \bibfnamefont{A.~F.}},
  \bibinfo{year}{1973}, \bibinfo{journal}{Review of Scientific Instruments}
  \textbf{\bibinfo{volume}{44}}, \bibinfo{pages}{425}.

\bibitem[{\citenamefont{Heinzel} \emph{et~al.}(1998)\citenamefont{Heinzel,
  Strain, Mizuno, Skeldon, Willke, Winkler, Schilling, R\"udiger, and
  Danzmann}}]{Gerhard:1998}
\bibinfo{author}{\bibnamefont{Heinzel}, \bibfnamefont{G.}},
  \bibinfo{author}{\bibfnamefont{K.~A.} \bibnamefont{Strain}},
  \bibinfo{author}{\bibfnamefont{J.}~\bibnamefont{Mizuno}},
  \bibinfo{author}{\bibfnamefont{K.~D.} \bibnamefont{Skeldon}},
  \bibinfo{author}{\bibfnamefont{B.}~\bibnamefont{Willke}},
  \bibinfo{author}{\bibfnamefont{W.}~\bibnamefont{Winkler}},
  \bibinfo{author}{\bibfnamefont{R.}~\bibnamefont{Schilling}},
  \bibinfo{author}{\bibfnamefont{A.}~\bibnamefont{R\"udiger}}, and
  \bibinfo{author}{\bibfnamefont{K.}~\bibnamefont{Danzmann}},
  \bibinfo{year}{1998}, \bibinfo{journal}{Phys. Rev. Lett.}
  \textbf{\bibinfo{volume}{81}}, \bibinfo{pages}{5493}.

\bibitem[{\citenamefont{{Hellings} and {Downs}}(1983)}]{Hellings:PTA}
\bibinfo{author}{\bibnamefont{{Hellings}}, \bibfnamefont{R.~W.}}, and
  \bibinfo{author}{\bibfnamefont{G.~S.} \bibnamefont{{Downs}}},
  \bibinfo{year}{1983}, \bibinfo{journal}{The Astrophysical Journal}
  \textbf{\bibinfo{volume}{265}}, \bibinfo{pages}{L39}.

\bibitem[{\citenamefont{Hello and Vinet}(1990)}]{hello1990analytical}
\bibinfo{author}{\bibnamefont{Hello}, \bibfnamefont{P.}}, and
  \bibinfo{author}{\bibfnamefont{J.}~\bibnamefont{Vinet}},
  \bibinfo{year}{1990}, \bibinfo{journal}{Journal de Physique}
  \textbf{\bibinfo{volume}{51}}(\bibinfo{number}{12}), \bibinfo{pages}{1267}.

\bibitem[{\citenamefont{Hello and Vinet}(1993)}]{Hello:1993}
\bibinfo{author}{\bibnamefont{Hello}, \bibfnamefont{P.}}, and
  \bibinfo{author}{\bibfnamefont{J.-Y.} \bibnamefont{Vinet}},
  \bibinfo{year}{1993}, \bibinfo{journal}{Physics Letters A}
  \textbf{\bibinfo{volume}{178}}(\bibinfo{number}{5}), \bibinfo{pages}{351}.

\bibitem[{\citenamefont{Hensley} \emph{et~al.}(1999)\citenamefont{Hensley,
  Peters, and Chu}}]{Chu:1999}
\bibinfo{author}{\bibnamefont{Hensley}, \bibfnamefont{J.~M.}},
  \bibinfo{author}{\bibfnamefont{A.}~\bibnamefont{Peters}}, and
  \bibinfo{author}{\bibfnamefont{S.}~\bibnamefont{Chu}}, \bibinfo{year}{1999},
  \bibinfo{journal}{Review of Scientific Instruments}
  \textbf{\bibinfo{volume}{70}}(\bibinfo{number}{6}), \bibinfo{pages}{2735}.

\bibitem[{\citenamefont{Heptonstall}
  \emph{et~al.}(2010)\citenamefont{Heptonstall, Barton, Cantley, Cumming,
  Cagnoli, Hough, Jones, Kumar, Martin, Rowan}
  \emph{et~al.}}]{heptonstall2010investigation}
\bibinfo{author}{\bibnamefont{Heptonstall}, \bibfnamefont{A.}},
  \bibinfo{author}{\bibfnamefont{M.}~\bibnamefont{Barton}},
  \bibinfo{author}{\bibfnamefont{C.}~\bibnamefont{Cantley}},
  \bibinfo{author}{\bibfnamefont{A.}~\bibnamefont{Cumming}},
  \bibinfo{author}{\bibfnamefont{G.}~\bibnamefont{Cagnoli}},
  \bibinfo{author}{\bibfnamefont{J.}~\bibnamefont{Hough}},
  \bibinfo{author}{\bibfnamefont{R.}~\bibnamefont{Jones}},
  \bibinfo{author}{\bibfnamefont{R.}~\bibnamefont{Kumar}},
  \bibinfo{author}{\bibfnamefont{I.}~\bibnamefont{Martin}},
  \bibinfo{author}{\bibfnamefont{S.}~\bibnamefont{Rowan}}, \emph{et~al.},
  \bibinfo{year}{2010}, \bibinfo{journal}{Classical and Quantum Gravity}
  \textbf{\bibinfo{volume}{27}}(\bibinfo{number}{3}), \bibinfo{pages}{035013}.

\bibitem[{\citenamefont{Herriott and Schulte}(1965)}]{Herriott:65}
\bibinfo{author}{\bibnamefont{Herriott}, \bibfnamefont{D.~R.}}, and
  \bibinfo{author}{\bibfnamefont{H.~J.} \bibnamefont{Schulte}},
  \bibinfo{year}{1965}, \bibinfo{journal}{Appl. Opt.}
  \textbf{\bibinfo{volume}{4}}(\bibinfo{number}{8}), \bibinfo{pages}{883}.

\bibitem[{\citenamefont{Hewish} \emph{et~al.}(1968)\citenamefont{Hewish, Bell,
  Pilkington, Scott, and Collins}}]{Bell:1968}
\bibinfo{author}{\bibnamefont{Hewish}, \bibfnamefont{A.}},
  \bibinfo{author}{\bibfnamefont{S.~J.} \bibnamefont{Bell}},
  \bibinfo{author}{\bibfnamefont{J.~D.~H.} \bibnamefont{Pilkington}},
  \bibinfo{author}{\bibfnamefont{P.~F.} \bibnamefont{Scott}}, and
  \bibinfo{author}{\bibfnamefont{R.~A.} \bibnamefont{Collins}},
  \bibinfo{year}{1968}, \bibinfo{journal}{Nature}
  \textbf{\bibinfo{volume}{217}}(\bibinfo{number}{5130}), \bibinfo{pages}{709}.

\bibitem[{\citenamefont{{Hild}} \emph{et~al.}(2006)\citenamefont{{Hild},
  {L{\"u}ck}, {Winkler}, {Strain}, {Grote}, {Smith}, {Malec}, {Hewitson},
  {Willke}, {Hough}, and {Danzmann}}}]{GEO:Absorption}
\bibinfo{author}{\bibnamefont{{Hild}}, \bibfnamefont{S.}},
  \bibinfo{author}{\bibfnamefont{H.}~\bibnamefont{{L{\"u}ck}}},
  \bibinfo{author}{\bibfnamefont{W.}~\bibnamefont{{Winkler}}},
  \bibinfo{author}{\bibfnamefont{K.}~\bibnamefont{{Strain}}},
  \bibinfo{author}{\bibfnamefont{H.}~\bibnamefont{{Grote}}},
  \bibinfo{author}{\bibfnamefont{J.}~\bibnamefont{{Smith}}},
  \bibinfo{author}{\bibfnamefont{M.}~\bibnamefont{{Malec}}},
  \bibinfo{author}{\bibfnamefont{M.}~\bibnamefont{{Hewitson}}},
  \bibinfo{author}{\bibfnamefont{B.}~\bibnamefont{{Willke}}},
  \bibinfo{author}{\bibfnamefont{J.}~\bibnamefont{{Hough}}}, and
  \bibinfo{author}{\bibfnamefont{K.}~\bibnamefont{{Danzmann}}},
  \bibinfo{year}{2006}, \bibinfo{journal}{\ao} \textbf{\bibinfo{volume}{45}},
  \bibinfo{pages}{7269}.

\bibitem[{\citenamefont{Hirose} \emph{et~al.}(2010)\citenamefont{Hirose,
  Kawabe, Sigg, Adhikari, and Saulson}}]{Hirose:10}
\bibinfo{author}{\bibnamefont{Hirose}, \bibfnamefont{E.}},
  \bibinfo{author}{\bibfnamefont{K.}~\bibnamefont{Kawabe}},
  \bibinfo{author}{\bibfnamefont{D.}~\bibnamefont{Sigg}},
  \bibinfo{author}{\bibfnamefont{R.}~\bibnamefont{Adhikari}}, and
  \bibinfo{author}{\bibfnamefont{P.~R.} \bibnamefont{Saulson}},
  \bibinfo{year}{2010}, \bibinfo{journal}{Appl. Opt.}
  \textbf{\bibinfo{volume}{49}}(\bibinfo{number}{18}), \bibinfo{pages}{3474}.

\bibitem[{\citenamefont{{Hogan}} \emph{et~al.}(2011)\citenamefont{{Hogan},
  {Johnson}, {Dickerson}, {Kovachy}, {Sugarbaker}, {Chiow}, {Graham},
  {Kasevich}, {Saif}, {Rajendran}, {Bouyer}, {Seery}}
  \emph{et~al.}}]{AGIS:2011}
\bibinfo{author}{\bibnamefont{{Hogan}}, \bibfnamefont{J.~M.}},
  \bibinfo{author}{\bibfnamefont{D.~M.~S.} \bibnamefont{{Johnson}}},
  \bibinfo{author}{\bibfnamefont{S.}~\bibnamefont{{Dickerson}}},
  \bibinfo{author}{\bibfnamefont{T.}~\bibnamefont{{Kovachy}}},
  \bibinfo{author}{\bibfnamefont{A.}~\bibnamefont{{Sugarbaker}}},
  \bibinfo{author}{\bibfnamefont{S.-W.} \bibnamefont{{Chiow}}},
  \bibinfo{author}{\bibfnamefont{P.~W.} \bibnamefont{{Graham}}},
  \bibinfo{author}{\bibfnamefont{M.~A.} \bibnamefont{{Kasevich}}},
  \bibinfo{author}{\bibfnamefont{B.}~\bibnamefont{{Saif}}},
  \bibinfo{author}{\bibfnamefont{S.}~\bibnamefont{{Rajendran}}},
  \bibinfo{author}{\bibfnamefont{P.}~\bibnamefont{{Bouyer}}},
  \bibinfo{author}{\bibfnamefont{B.~D.} \bibnamefont{{Seery}}}, \emph{et~al.},
  \bibinfo{year}{2011}, \bibinfo{journal}{General Relativity and Gravitation}
  \textbf{\bibinfo{volume}{43}}, \bibinfo{pages}{1953}.

\bibitem[{\citenamefont{Hohensee} \emph{et~al.}(2011)\citenamefont{Hohensee,
  Lan, Houtz, Chan, Estey, Kim, Kuan, and M{\"u}ller}}]{Holger:GRG2011}
\bibinfo{author}{\bibnamefont{Hohensee}, \bibfnamefont{M.}},
  \bibinfo{author}{\bibfnamefont{S.-Y.} \bibnamefont{Lan}},
  \bibinfo{author}{\bibfnamefont{R.}~\bibnamefont{Houtz}},
  \bibinfo{author}{\bibfnamefont{C.}~\bibnamefont{Chan}},
  \bibinfo{author}{\bibfnamefont{B.}~\bibnamefont{Estey}},
  \bibinfo{author}{\bibfnamefont{G.}~\bibnamefont{Kim}},
  \bibinfo{author}{\bibfnamefont{P.-C.} \bibnamefont{Kuan}}, and
  \bibinfo{author}{\bibfnamefont{H.}~\bibnamefont{M{\"u}ller}},
  \bibinfo{year}{2011}, \bibinfo{journal}{General Relativity and Gravitation}
  \textbf{\bibinfo{volume}{43}}, \bibinfo{pages}{1905}, ISSN
  \bibinfo{issn}{0001-7701}, \bibinfo{note}{10.1007/s10714-010-1118-x}.

\bibitem[{\citenamefont{Hong} \emph{et~al.}(2011)\citenamefont{Hong, Miller,
  Yamamoto, Chen, and Adhikari}}]{Ting:LG}
\bibinfo{author}{\bibnamefont{Hong}, \bibfnamefont{T.}},
  \bibinfo{author}{\bibfnamefont{J.}~\bibnamefont{Miller}},
  \bibinfo{author}{\bibfnamefont{H.}~\bibnamefont{Yamamoto}},
  \bibinfo{author}{\bibfnamefont{Y.}~\bibnamefont{Chen}}, and
  \bibinfo{author}{\bibfnamefont{R.}~\bibnamefont{Adhikari}},
  \bibinfo{year}{2011}, \bibinfo{journal}{Phys. Rev. D}
  \textbf{\bibinfo{volume}{84}}, \bibinfo{pages}{102001}.

\bibitem[{\citenamefont{Hong} \emph{et~al.}(2012)\citenamefont{Hong, Yang,
  Gustafson, Adhikari, and Chen}}]{Ting:Brownian2012}
\bibinfo{author}{\bibnamefont{Hong}, \bibfnamefont{T.}},
  \bibinfo{author}{\bibfnamefont{H.}~\bibnamefont{Yang}},
  \bibinfo{author}{\bibfnamefont{E.}~\bibnamefont{Gustafson}},
  \bibinfo{author}{\bibfnamefont{R.}~\bibnamefont{Adhikari}}, and
  \bibinfo{author}{\bibfnamefont{Y.}~\bibnamefont{Chen}}, \bibinfo{year}{2012},
  \bibinfo{journal}{arXiv preprint arXiv:1207.6145} .

\bibitem[{\citenamefont{{Hu} and {Dodelson}}(2002)}]{Hu:AR2002}
\bibinfo{author}{\bibnamefont{{Hu}}, \bibfnamefont{W.}}, and
  \bibinfo{author}{\bibfnamefont{S.}~\bibnamefont{{Dodelson}}},
  \bibinfo{year}{2002}, \bibinfo{journal}{Annual Review of Astronomy and
  Astrophysics} \textbf{\bibinfo{volume}{40}}, \bibinfo{pages}{171}.

\bibitem[{\citenamefont{Hu and White}(1997)}]{Hu:1997vp}
\bibinfo{author}{\bibnamefont{Hu}, \bibfnamefont{W.}}, and
  \bibinfo{author}{\bibfnamefont{M.}~\bibnamefont{White}},
  \bibinfo{year}{1997}, \bibinfo{journal}{New Astronomy}
  \textbf{\bibinfo{volume}{2}}(\bibinfo{number}{4}), \bibinfo{pages}{323 },
  ISSN \bibinfo{issn}{1384-1076},
  \urlprefix\url{http://www.sciencedirect.com/science/article/pii/S1384107697000225}.

\bibitem[{\citenamefont{Huang} \emph{et~al.}(2006)\citenamefont{Huang, Benesty,
  and Chen}}]{Huang:MIMO}
\bibinfo{author}{\bibnamefont{Huang}, \bibfnamefont{Y.}},
  \bibinfo{author}{\bibfnamefont{J.}~\bibnamefont{Benesty}}, and
  \bibinfo{author}{\bibfnamefont{J.}~\bibnamefont{Chen}}, \bibinfo{year}{2006},
  \emph{\bibinfo{title}{Acoustic MIMO signal processing}}, Signals and
  communication technology (\bibinfo{publisher}{Springer}), ISBN
  \bibinfo{isbn}{9783540376309}.

\bibitem[{\citenamefont{Hughes and Thorne}(1998)}]{HuTh1998}
\bibinfo{author}{\bibnamefont{Hughes}, \bibfnamefont{S.~A.}}, and
  \bibinfo{author}{\bibfnamefont{K.~S.} \bibnamefont{Thorne}},
  \bibinfo{year}{1998}, \bibinfo{journal}{Phys. Rev. D}
  \textbf{\bibinfo{volume}{58}}, \bibinfo{pages}{122002}.

\bibitem[{\citenamefont{{Ipser}}(1971)}]{Ipser:1971}
\bibinfo{author}{\bibnamefont{{Ipser}}, \bibfnamefont{J.~R.}},
  \bibinfo{year}{1971}, \bibinfo{journal}{\apj} \textbf{\bibinfo{volume}{166}},
  \bibinfo{pages}{175}.

\bibitem[{\citenamefont{Ishidoshiro}
  \emph{et~al.}(2011)\citenamefont{Ishidoshiro, Ando, Takamori, Takahashi,
  Okada, Matsumoto, Kokuyama, Kanda, Aso, and Tsubono}}]{TOBA:2011}
\bibinfo{author}{\bibnamefont{Ishidoshiro}, \bibfnamefont{K.}},
  \bibinfo{author}{\bibfnamefont{M.}~\bibnamefont{Ando}},
  \bibinfo{author}{\bibfnamefont{A.}~\bibnamefont{Takamori}},
  \bibinfo{author}{\bibfnamefont{H.}~\bibnamefont{Takahashi}},
  \bibinfo{author}{\bibfnamefont{K.}~\bibnamefont{Okada}},
  \bibinfo{author}{\bibfnamefont{N.}~\bibnamefont{Matsumoto}},
  \bibinfo{author}{\bibfnamefont{W.}~\bibnamefont{Kokuyama}},
  \bibinfo{author}{\bibfnamefont{N.}~\bibnamefont{Kanda}},
  \bibinfo{author}{\bibfnamefont{Y.}~\bibnamefont{Aso}}, and
  \bibinfo{author}{\bibfnamefont{K.}~\bibnamefont{Tsubono}},
  \bibinfo{year}{2011}, \bibinfo{journal}{Phys. Rev. Lett.}
  \textbf{\bibinfo{volume}{106}}, \bibinfo{pages}{161101}.

\bibitem[{\citenamefont{Jayawant}(1981)}]{Jayawant:1981}
\bibinfo{author}{\bibnamefont{Jayawant}, \bibfnamefont{B.~V.}},
  \bibinfo{year}{1981}, \bibinfo{journal}{Reports on Progress in Physics}
  \textbf{\bibinfo{volume}{44}}(\bibinfo{number}{4}), \bibinfo{pages}{411},
  \urlprefix\url{http://stacks.iop.org/0034-4885/44/i=4/a=002}.

\bibitem[{\citenamefont{Ju} \emph{et~al.}(2000)\citenamefont{Ju, Blair, and
  Zhao}}]{Blair:RPP2000}
\bibinfo{author}{\bibnamefont{Ju}, \bibfnamefont{L.}},
  \bibinfo{author}{\bibfnamefont{D.~G.} \bibnamefont{Blair}}, and
  \bibinfo{author}{\bibfnamefont{C.}~\bibnamefont{Zhao}}, \bibinfo{year}{2000},
  \bibinfo{journal}{Reports on Progress in Physics}
  \textbf{\bibinfo{volume}{63}}(\bibinfo{number}{9}), \bibinfo{pages}{1317}.

\bibitem[{\citenamefont{Ju} \emph{et~al.}(2009)\citenamefont{Ju, Blair, Zhao,
  Gras, Zhang, Barriga, Miao, Fan, and Merrill}}]{Blair:PIStrategy}
\bibinfo{author}{\bibnamefont{Ju}, \bibfnamefont{L.}},
  \bibinfo{author}{\bibfnamefont{D.~G.} \bibnamefont{Blair}},
  \bibinfo{author}{\bibfnamefont{C.}~\bibnamefont{Zhao}},
  \bibinfo{author}{\bibfnamefont{S.}~\bibnamefont{Gras}},
  \bibinfo{author}{\bibfnamefont{Z.}~\bibnamefont{Zhang}},
  \bibinfo{author}{\bibfnamefont{P.}~\bibnamefont{Barriga}},
  \bibinfo{author}{\bibfnamefont{H.}~\bibnamefont{Miao}},
  \bibinfo{author}{\bibfnamefont{Y.}~\bibnamefont{Fan}}, and
  \bibinfo{author}{\bibfnamefont{L.}~\bibnamefont{Merrill}},
  \bibinfo{year}{2009}, \bibinfo{journal}{Classical and Quantum Gravity}
  \textbf{\bibinfo{volume}{26}}(\bibinfo{number}{1}), \bibinfo{pages}{015002},
  \urlprefix\url{http://stacks.iop.org/0264-9381/26/i=1/a=015002}.

\bibitem[{\citenamefont{Kafka and Schnupp}(1978)}]{Kafka:1978}
\bibinfo{author}{\bibnamefont{Kafka}, \bibfnamefont{R.}}, and
  \bibinfo{author}{\bibfnamefont{L.}~\bibnamefont{Schnupp}},
  \bibinfo{year}{1978}, \bibinfo{journal}{Astronomy and Astrophysics}
  \textbf{\bibinfo{volume}{70}}, \bibinfo{pages}{97}.

\bibitem[{\citenamefont{KAGRA}(2011)}]{LCGT:web}
\bibinfo{author}{\bibnamefont{KAGRA}}, \bibinfo{year}{2011},
  \bibinfo{title}{{KAGRA} home page},
  \urlprefix\url{http://gwcenter.icrr.u-tokyo.ac.jp/en/}.

\bibitem[{\citenamefont{Khalili}(2002)}]{Kha2002}
\bibinfo{author}{\bibnamefont{Khalili}, \bibfnamefont{F.~Y.}},
  \bibinfo{year}{2002}, \bibinfo{journal}{Phys.~Lett.~A}
  \textbf{\bibinfo{volume}{298}}, \bibinfo{pages}{308}.

\bibitem[{\citenamefont{Khalili}(2007)}]{Kha2007}
\bibinfo{author}{\bibnamefont{Khalili}, \bibfnamefont{F.~Y.}},
  \bibinfo{year}{2007}, \bibinfo{journal}{Phys.~Rev.~D}
  \textbf{\bibinfo{volume}{76}}, \bibinfo{pages}{102002}.

\bibitem[{\citenamefont{Kimble} \emph{et~al.}(2001)\citenamefont{Kimble, Levin,
  Matsko, Thorne, and Vyatchanin}}]{KLMTV2001}
\bibinfo{author}{\bibnamefont{Kimble}, \bibfnamefont{H.~J.}},
  \bibinfo{author}{\bibfnamefont{Y.}~\bibnamefont{Levin}},
  \bibinfo{author}{\bibfnamefont{A.~B.} \bibnamefont{Matsko}},
  \bibinfo{author}{\bibfnamefont{K.~S.} \bibnamefont{Thorne}}, and
  \bibinfo{author}{\bibfnamefont{S.~P.} \bibnamefont{Vyatchanin}},
  \bibinfo{year}{2001}, \bibinfo{journal}{Phys.~Rev.~D}
  \textbf{\bibinfo{volume}{65}}, \bibinfo{pages}{022002}.

\bibitem[{\citenamefont{Kippenberg}
  \emph{et~al.}(2005)\citenamefont{Kippenberg, Rokhsari, Carmon, Scherer, and
  Vahala}}]{Vahala:PRL2005}
\bibinfo{author}{\bibnamefont{Kippenberg}, \bibfnamefont{T.~J.}},
  \bibinfo{author}{\bibfnamefont{H.}~\bibnamefont{Rokhsari}},
  \bibinfo{author}{\bibfnamefont{T.}~\bibnamefont{Carmon}},
  \bibinfo{author}{\bibfnamefont{A.}~\bibnamefont{Scherer}}, and
  \bibinfo{author}{\bibfnamefont{K.~J.} \bibnamefont{Vahala}},
  \bibinfo{year}{2005}, \bibinfo{journal}{Phys. Rev. Lett.}
  \textbf{\bibinfo{volume}{95}}, \bibinfo{pages}{033901},
  \urlprefix\url{http://link.aps.org/doi/10.1103/PhysRevLett.95.033901}.

\bibitem[{\citenamefont{Knispel} \emph{et~al.}(2011)\citenamefont{Knispel,
  Lazarus, Allen, Anderson, Aulbert, Bhat, Bock, Bogdanov, Brazier, Camilo,
  Chatterjee, Cordes} \emph{et~al.}}]{EatH:2011}
\bibinfo{author}{\bibnamefont{Knispel}, \bibfnamefont{B.}},
  \bibinfo{author}{\bibfnamefont{P.}~\bibnamefont{Lazarus}},
  \bibinfo{author}{\bibfnamefont{B.}~\bibnamefont{Allen}},
  \bibinfo{author}{\bibfnamefont{D.}~\bibnamefont{Anderson}},
  \bibinfo{author}{\bibfnamefont{C.}~\bibnamefont{Aulbert}},
  \bibinfo{author}{\bibfnamefont{N.~D.~R.} \bibnamefont{Bhat}},
  \bibinfo{author}{\bibfnamefont{O.}~\bibnamefont{Bock}},
  \bibinfo{author}{\bibfnamefont{S.}~\bibnamefont{Bogdanov}},
  \bibinfo{author}{\bibfnamefont{A.}~\bibnamefont{Brazier}},
  \bibinfo{author}{\bibfnamefont{F.}~\bibnamefont{Camilo}},
  \bibinfo{author}{\bibfnamefont{S.}~\bibnamefont{Chatterjee}},
  \bibinfo{author}{\bibfnamefont{J.~M.} \bibnamefont{Cordes}}, \emph{et~al.},
  \bibinfo{year}{2011}, \bibinfo{journal}{The Astrophysical Journal Letters}
  \textbf{\bibinfo{volume}{732}}(\bibinfo{number}{1}), \bibinfo{pages}{L1},
  \urlprefix\url{http://stacks.iop.org/2041-8205/732/i=1/a=L1}.

\bibitem[{\citenamefont{Kondratiev}
  \emph{et~al.}(2011)\citenamefont{Kondratiev, Gurkovsky, and
  Gorodetsky}}]{Goro:2011}
\bibinfo{author}{\bibnamefont{Kondratiev}, \bibfnamefont{N.~M.}},
  \bibinfo{author}{\bibfnamefont{A.~G.} \bibnamefont{Gurkovsky}}, and
  \bibinfo{author}{\bibfnamefont{M.~L.} \bibnamefont{Gorodetsky}},
  \bibinfo{year}{2011}, \bibinfo{journal}{Phys. Rev. D}
  \textbf{\bibinfo{volume}{84}}, \bibinfo{pages}{022001}.

\bibitem[{\citenamefont{Kovalik and Saulson}(1993)}]{Kovalik:1993}
\bibinfo{author}{\bibnamefont{Kovalik}, \bibfnamefont{J.}}, and
  \bibinfo{author}{\bibfnamefont{P.~R.} \bibnamefont{Saulson}},
  \bibinfo{year}{1993}, \bibinfo{journal}{Review of Scientific Instruments}
  \textbf{\bibinfo{volume}{64}}(\bibinfo{number}{10}), \bibinfo{pages}{2942},
  \urlprefix\url{http://link.aip.org/link/?RSI/64/2942/1}.

\bibitem[{\citenamefont{Kroker} \emph{et~al.}(2011)\citenamefont{Kroker,
  K\"{a}sebier, Br\"{u}ckner, Fuchs, Kley, and T\"{u}nnermann}}]{Kroker:11}
\bibinfo{author}{\bibnamefont{Kroker}, \bibfnamefont{S.}},
  \bibinfo{author}{\bibfnamefont{T.}~\bibnamefont{K\"{a}sebier}},
  \bibinfo{author}{\bibfnamefont{F.}~\bibnamefont{Br\"{u}ckner}},
  \bibinfo{author}{\bibfnamefont{F.}~\bibnamefont{Fuchs}},
  \bibinfo{author}{\bibfnamefont{E.-B.} \bibnamefont{Kley}}, and
  \bibinfo{author}{\bibfnamefont{A.}~\bibnamefont{T\"{u}nnermann}},
  \bibinfo{year}{2011}, \bibinfo{journal}{Opt. Express}
  \textbf{\bibinfo{volume}{19}}(\bibinfo{number}{17}), \bibinfo{pages}{16466},
  \urlprefix\url{http://www.opticsexpress.org/abstract.cfm?URI=oe-19-17-16466}.

\bibitem[{\citenamefont{Kroker} \emph{et~al.}(2013)\citenamefont{Kroker,
  K{\"a}sebier, Kley, and T{\"u}nnermann}}]{kroker2013coupled}
\bibinfo{author}{\bibnamefont{Kroker}, \bibfnamefont{S.}},
  \bibinfo{author}{\bibfnamefont{T.}~\bibnamefont{K{\"a}sebier}},
  \bibinfo{author}{\bibfnamefont{E.-B.} \bibnamefont{Kley}}, and
  \bibinfo{author}{\bibfnamefont{A.}~\bibnamefont{T{\"u}nnermann}},
  \bibinfo{year}{2013}, \bibinfo{journal}{Optics Letters}
  \textbf{\bibinfo{volume}{38}}(\bibinfo{number}{17}), \bibinfo{pages}{3336}.

\bibitem[{\citenamefont{Kr{\'o}lak}
  \emph{et~al.}(2004)\citenamefont{Kr{\'o}lak, Tinto, and
  Vallisneri}}]{KTV2004}
\bibinfo{author}{\bibnamefont{Kr{\'o}lak}, \bibfnamefont{A.}},
  \bibinfo{author}{\bibfnamefont{M.}~\bibnamefont{Tinto}}, and
  \bibinfo{author}{\bibfnamefont{M.}~\bibnamefont{Vallisneri}},
  \bibinfo{year}{2004}, \bibinfo{journal}{Phys.~Rev.~D}
  \textbf{\bibinfo{volume}{70}}, \bibinfo{pages}{022003}.

\bibitem[{\citenamefont{{Kubo}}(1966)}]{Kubo:FDT}
\bibinfo{author}{\bibnamefont{{Kubo}}, \bibfnamefont{R.}},
  \bibinfo{year}{1966}, \bibinfo{journal}{Reports on Progress in Physics}
  \textbf{\bibinfo{volume}{29}}, \bibinfo{pages}{255}.

\bibitem[{\citenamefont{Kwee} \emph{et~al.}(2009)\citenamefont{Kwee, Willke,
  and Danzmann}}]{Kwee:09}
\bibinfo{author}{\bibnamefont{Kwee}, \bibfnamefont{P.}},
  \bibinfo{author}{\bibfnamefont{B.}~\bibnamefont{Willke}}, and
  \bibinfo{author}{\bibfnamefont{K.}~\bibnamefont{Danzmann}},
  \bibinfo{year}{2009}, \bibinfo{journal}{Opt. Lett.}
  \textbf{\bibinfo{volume}{34}}(\bibinfo{number}{19}), \bibinfo{pages}{2912},
  \urlprefix\url{http://ol.osa.org/abstract.cfm?URI=ol-34-19-2912}.

\bibitem[{\citenamefont{Lada}(2006)}]{Lada:2006}
\bibinfo{author}{\bibnamefont{Lada}, \bibfnamefont{C.~J.}},
  \bibinfo{year}{2006}, \bibinfo{journal}{The Astrophysical Journal Letters}
  \textbf{\bibinfo{volume}{640}}(\bibinfo{number}{1}), \bibinfo{pages}{L63},
  \urlprefix\url{http://stacks.iop.org/1538-4357/640/i=1/a=L63}.

\bibitem[{\citenamefont{Lantz} \emph{et~al.}(2009)\citenamefont{Lantz,
  Schofield, O'Reilly, Clark, and DeBra}}]{Lantz:2009dw}
\bibinfo{author}{\bibnamefont{Lantz}, \bibfnamefont{B.}},
  \bibinfo{author}{\bibfnamefont{R.}~\bibnamefont{Schofield}},
  \bibinfo{author}{\bibfnamefont{B.}~\bibnamefont{O'Reilly}},
  \bibinfo{author}{\bibfnamefont{D.~E.} \bibnamefont{Clark}}, and
  \bibinfo{author}{\bibfnamefont{D.}~\bibnamefont{DeBra}},
  \bibinfo{year}{2009}, \bibinfo{journal}{Bulletin of the Seismological Society
  of America} \textbf{\bibinfo{volume}{99}}(\bibinfo{number}{2B}),
  \bibinfo{pages}{980}.

\bibitem[{\citenamefont{Lawrence} \emph{et~al.}(2004)\citenamefont{Lawrence,
  Ottaway, Zucker, and Fritschel}}]{Ryan:TCS}
\bibinfo{author}{\bibnamefont{Lawrence}, \bibfnamefont{R.}},
  \bibinfo{author}{\bibfnamefont{D.}~\bibnamefont{Ottaway}},
  \bibinfo{author}{\bibfnamefont{M.}~\bibnamefont{Zucker}}, and
  \bibinfo{author}{\bibfnamefont{P.}~\bibnamefont{Fritschel}},
  \bibinfo{year}{2004}, \bibinfo{journal}{Optics Letters}
  \textbf{\bibinfo{volume}{29}}(\bibinfo{number}{22}), \bibinfo{pages}{2635}.

\bibitem[{\citenamefont{{Leibrandt}}
  \emph{et~al.}(2013)\citenamefont{{Leibrandt}, {Bergquist}, and
  {Rosenband}}}]{Till:2013cav}
\bibinfo{author}{\bibnamefont{{Leibrandt}}, \bibfnamefont{D.~R.}},
  \bibinfo{author}{\bibfnamefont{J.~C.} \bibnamefont{{Bergquist}}}, and
  \bibinfo{author}{\bibfnamefont{T.}~\bibnamefont{{Rosenband}}},
  \bibinfo{year}{2013}, \bibinfo{journal}{ArXiv e-prints} \eprint{1301.0022}.

\bibitem[{\citenamefont{Levin}(1998)}]{Levin:Direct}
\bibinfo{author}{\bibnamefont{Levin}, \bibfnamefont{Y.}}, \bibinfo{year}{1998},
  \bibinfo{journal}{Phys. Rev. D} \textbf{\bibinfo{volume}{57}},
  \bibinfo{pages}{659},
  \urlprefix\url{http://link.aps.org/doi/10.1103/PhysRevD.57.659}.

\bibitem[{\citenamefont{LIGO}(2011)}]{aLIGO:web}
\bibinfo{author}{\bibnamefont{LIGO}}, \bibinfo{year}{2011},
  \urlprefix\url{https://www.advancedligo.mit.edu/}.

\bibitem[{\citenamefont{{LIGO Scientific Collaboration}}
  \emph{et~al.}(2011)\citenamefont{{LIGO Scientific Collaboration}, {Abadie},
  {Abbott}, {Abbott}, {Abbott}, {Abernathy}, {Adams}, {Adhikari}, {Affeldt},
  {Allen}, and et~al.}}]{GEO:Squeezing}
\bibinfo{author}{\bibnamefont{{LIGO Scientific Collaboration}}},
  \bibinfo{author}{\bibfnamefont{J.}~\bibnamefont{{Abadie}}},
  \bibinfo{author}{\bibfnamefont{B.~P.} \bibnamefont{{Abbott}}},
  \bibinfo{author}{\bibfnamefont{R.}~\bibnamefont{{Abbott}}},
  \bibinfo{author}{\bibfnamefont{T.~D.} \bibnamefont{{Abbott}}},
  \bibinfo{author}{\bibfnamefont{M.}~\bibnamefont{{Abernathy}}},
  \bibinfo{author}{\bibfnamefont{C.}~\bibnamefont{{Adams}}},
  \bibinfo{author}{\bibfnamefont{R.}~\bibnamefont{{Adhikari}}},
  \bibinfo{author}{\bibfnamefont{C.}~\bibnamefont{{Affeldt}}},
  \bibinfo{author}{\bibfnamefont{B.}~\bibnamefont{{Allen}}}, and
  \bibinfo{author}{\bibnamefont{et~al.}}, \bibinfo{year}{2011},
  \bibinfo{journal}{Nature Physics} \textbf{\bibinfo{volume}{7}},
  \bibinfo{pages}{962}.

\bibitem[{\citenamefont{{Lin}} \emph{et~al.}(2011)\citenamefont{{Lin},
  {Harris}, and {Fejer}}}]{Angie:GaP2011}
\bibinfo{author}{\bibnamefont{{Lin}}, \bibfnamefont{A.~C.}},
  \bibinfo{author}{\bibfnamefont{J.~S.} \bibnamefont{{Harris}}}, and
  \bibinfo{author}{\bibfnamefont{M.~M.} \bibnamefont{{Fejer}}},
  \bibinfo{year}{2011}, \bibinfo{journal}{Journal of Vacuum Science Technology
  B: Microelectronics and Nanometer Structures}
  \textbf{\bibinfo{volume}{29}}(\bibinfo{number}{3}), \bibinfo{pages}{030000}.

\bibitem[{\citenamefont{{Liu} and {Lindblom}}(2001)}]{Lee:NS2001}
\bibinfo{author}{\bibnamefont{{Liu}}, \bibfnamefont{Y.~T.}}, and
  \bibinfo{author}{\bibfnamefont{L.}~\bibnamefont{{Lindblom}}},
  \bibinfo{year}{2001}, \bibinfo{journal}{Monthly Notices of the Royal
  Astronomical Society} \textbf{\bibinfo{volume}{324}}, \bibinfo{pages}{1063}.

\bibitem[{\citenamefont{Logan} \emph{et~al.}(1993)\citenamefont{Logan, Hough,
  and Robertson}}]{Logan:1993}
\bibinfo{author}{\bibnamefont{Logan}, \bibfnamefont{J.~E.}},
  \bibinfo{author}{\bibfnamefont{J.}~\bibnamefont{Hough}}, and
  \bibinfo{author}{\bibfnamefont{N.~A.} \bibnamefont{Robertson}},
  \bibinfo{year}{1993}, \bibinfo{journal}{Physics Letters A} .

\bibitem[{\citenamefont{{Lor{\'e}n-Aguilar}}
  \emph{et~al.}(2005)\citenamefont{{Lor{\'e}n-Aguilar}, {Guerrero}, {Isern},
  {Lobo}, and {Garc{\'{\i}}a-Berro}}}]{Lobo:WD2005}
\bibinfo{author}{\bibnamefont{{Lor{\'e}n-Aguilar}}, \bibfnamefont{P.}},
  \bibinfo{author}{\bibfnamefont{J.}~\bibnamefont{{Guerrero}}},
  \bibinfo{author}{\bibfnamefont{J.}~\bibnamefont{{Isern}}},
  \bibinfo{author}{\bibfnamefont{J.~A.} \bibnamefont{{Lobo}}}, and
  \bibinfo{author}{\bibfnamefont{E.}~\bibnamefont{{Garc{\'{\i}}a-Berro}}},
  \bibinfo{year}{2005}, \bibinfo{journal}{Monthly Notices of the Royal
  Astronomical Society} \textbf{\bibinfo{volume}{356}}, \bibinfo{pages}{627}.

\bibitem[{\citenamefont{Lorimer}(2008)}]{Lorimer:LRR}
\bibinfo{author}{\bibnamefont{Lorimer}, \bibfnamefont{D.~R.}},
  \bibinfo{year}{2008}, \bibinfo{journal}{Living Reviews in Relativity}
  \textbf{\bibinfo{volume}{11}}(\bibinfo{number}{8}).

\bibitem[{\citenamefont{Loudon}(1981)}]{Lou1981}
\bibinfo{author}{\bibnamefont{Loudon}, \bibfnamefont{R.}},
  \bibinfo{year}{1981}, \bibinfo{journal}{Phys.~Rev.~Lett.}
  \textbf{\bibinfo{volume}{47}}, \bibinfo{pages}{815}.

\bibitem[{\citenamefont{L\"{u}ck} \emph{et~al.}(2010)\citenamefont{L\"{u}ck,
  Affeldt, Degallaix, Freise, Grote, Hewitson, Hild, Leong, Prijatelj, Strain,
  Willke, Wittel} \emph{et~al.}}]{Harald:2010un}
\bibinfo{author}{\bibnamefont{L\"{u}ck}, \bibfnamefont{H.}},
  \bibinfo{author}{\bibfnamefont{C.}~\bibnamefont{Affeldt}},
  \bibinfo{author}{\bibfnamefont{J.}~\bibnamefont{Degallaix}},
  \bibinfo{author}{\bibfnamefont{A.}~\bibnamefont{Freise}},
  \bibinfo{author}{\bibfnamefont{H.}~\bibnamefont{Grote}},
  \bibinfo{author}{\bibfnamefont{M.}~\bibnamefont{Hewitson}},
  \bibinfo{author}{\bibfnamefont{S.}~\bibnamefont{Hild}},
  \bibinfo{author}{\bibfnamefont{J.}~\bibnamefont{Leong}},
  \bibinfo{author}{\bibfnamefont{M.}~\bibnamefont{Prijatelj}},
  \bibinfo{author}{\bibfnamefont{K.~A.} \bibnamefont{Strain}},
  \bibinfo{author}{\bibfnamefont{B.}~\bibnamefont{Willke}},
  \bibinfo{author}{\bibfnamefont{H.}~\bibnamefont{Wittel}}, \emph{et~al.},
  \bibinfo{year}{2010}, \bibinfo{journal}{Journal of Physics: Conference
  Series} \textbf{\bibinfo{volume}{228}}(\bibinfo{number}{1}),
  \bibinfo{pages}{012012+}, ISSN \bibinfo{issn}{1742-6596},
  \urlprefix\url{http://dx.doi.org/10.1088/1742-6596/228/1/012012}.

\bibitem[{\citenamefont{{L{\"u}ck}}
  \emph{et~al.}(2008)\citenamefont{{L{\"u}ck}, {Degallaix}, {Grote},
  {Hewitson}, {Hild}, {Willke}, and {Danzmann}}}]{GEO:scatter2008}
\bibinfo{author}{\bibnamefont{{L{\"u}ck}}, \bibfnamefont{H.}},
  \bibinfo{author}{\bibfnamefont{J.}~\bibnamefont{{Degallaix}}},
  \bibinfo{author}{\bibfnamefont{H.}~\bibnamefont{{Grote}}},
  \bibinfo{author}{\bibfnamefont{M.}~\bibnamefont{{Hewitson}}},
  \bibinfo{author}{\bibfnamefont{S.}~\bibnamefont{{Hild}}},
  \bibinfo{author}{\bibfnamefont{B.}~\bibnamefont{{Willke}}}, and
  \bibinfo{author}{\bibfnamefont{K.}~\bibnamefont{{Danzmann}}},
  \bibinfo{year}{2008}, \bibinfo{journal}{Journal of Optics A: Pure and Applied
  Optics} \textbf{\bibinfo{volume}{10}}(\bibinfo{number}{8}),
  \bibinfo{pages}{085004}.

\bibitem[{\citenamefont{L{\"u}ck} \emph{et~al.}(2004)\citenamefont{L{\"u}ck,
  Freise, Go{\ss}ler, Hild, Kawabe, and Danzmann}}]{Hild:TCS}
\bibinfo{author}{\bibnamefont{L{\"u}ck}, \bibfnamefont{H.}},
  \bibinfo{author}{\bibfnamefont{A.}~\bibnamefont{Freise}},
  \bibinfo{author}{\bibfnamefont{S.}~\bibnamefont{Go{\ss}ler}},
  \bibinfo{author}{\bibfnamefont{S.}~\bibnamefont{Hild}},
  \bibinfo{author}{\bibfnamefont{K.}~\bibnamefont{Kawabe}}, and
  \bibinfo{author}{\bibfnamefont{K.}~\bibnamefont{Danzmann}},
  \bibinfo{year}{2004}, \bibinfo{journal}{Classical and Quantum Gravity}
  \textbf{\bibinfo{volume}{21}}(\bibinfo{number}{5}), \bibinfo{pages}{985}.

\bibitem[{\citenamefont{{L{\"u}ck}}
  \emph{et~al.}(2006)\citenamefont{{L{\"u}ck}, {Hewitson}, {Ajith}, {Allen},
  {Aufmuth}, {Aulbert}, {Babak}, {Balasubramanian}, {Barr}, {Berukoff},
  {Bunkowski}, {Cagnoli}} \emph{et~al.}}]{LuEA2006}
\bibinfo{author}{\bibnamefont{{L{\"u}ck}}, \bibfnamefont{H.}},
  \bibinfo{author}{\bibfnamefont{M.}~\bibnamefont{{Hewitson}}},
  \bibinfo{author}{\bibfnamefont{P.}~\bibnamefont{{Ajith}}},
  \bibinfo{author}{\bibfnamefont{B.}~\bibnamefont{{Allen}}},
  \bibinfo{author}{\bibfnamefont{P.}~\bibnamefont{{Aufmuth}}},
  \bibinfo{author}{\bibfnamefont{C.}~\bibnamefont{{Aulbert}}},
  \bibinfo{author}{\bibfnamefont{S.}~\bibnamefont{{Babak}}},
  \bibinfo{author}{\bibfnamefont{R.}~\bibnamefont{{Balasubramanian}}},
  \bibinfo{author}{\bibfnamefont{B.~W.} \bibnamefont{{Barr}}},
  \bibinfo{author}{\bibfnamefont{S.}~\bibnamefont{{Berukoff}}},
  \bibinfo{author}{\bibfnamefont{A.}~\bibnamefont{{Bunkowski}}},
  \bibinfo{author}{\bibfnamefont{G.}~\bibnamefont{{Cagnoli}}}, \emph{et~al.},
  \bibinfo{year}{2006}, \bibinfo{journal}{Classical and Quantum Gravity}
  \textbf{\bibinfo{volume}{23}}, \bibinfo{pages}{71}.

\bibitem[{\citenamefont{{Maga{\~n}a-Sandoval}}
  \emph{et~al.}(2012)\citenamefont{{Maga{\~n}a-Sandoval}, {Adhikari}, {Frolov},
  {Harms}, {Lee}, {Sankar}, {Saulson}, and {Smith}}}]{Josh:scatter2012}
\bibinfo{author}{\bibnamefont{{Maga{\~n}a-Sandoval}}, \bibfnamefont{F.}},
  \bibinfo{author}{\bibfnamefont{R.~X.} \bibnamefont{{Adhikari}}},
  \bibinfo{author}{\bibfnamefont{V.}~\bibnamefont{{Frolov}}},
  \bibinfo{author}{\bibfnamefont{J.}~\bibnamefont{{Harms}}},
  \bibinfo{author}{\bibfnamefont{J.}~\bibnamefont{{Lee}}},
  \bibinfo{author}{\bibfnamefont{S.}~\bibnamefont{{Sankar}}},
  \bibinfo{author}{\bibfnamefont{P.~R.} \bibnamefont{{Saulson}}}, and
  \bibinfo{author}{\bibfnamefont{J.~R.} \bibnamefont{{Smith}}},
  \bibinfo{year}{2012}, \bibinfo{journal}{Journal of the Optical Society of
  America A} \textbf{\bibinfo{volume}{29}}, \bibinfo{pages}{1722}.

\bibitem[{\citenamefont{Mandic and Buonanno}(2006)}]{Vuk:PreBigBang}
\bibinfo{author}{\bibnamefont{Mandic}, \bibfnamefont{V.}}, and
  \bibinfo{author}{\bibfnamefont{A.}~\bibnamefont{Buonanno}},
  \bibinfo{year}{2006}, \bibinfo{journal}{Phys. Rev. D}
  \textbf{\bibinfo{volume}{73}}, \bibinfo{pages}{063008},
  \urlprefix\url{http://link.aps.org/doi/10.1103/PhysRevD.73.063008}.

\bibitem[{\citenamefont{M{\'a}rka} \emph{et~al.}(2002)\citenamefont{M{\'a}rka,
  Takamori, Ando, Bertolini, Cella, DeSalvo, Fukushima, Iida, Jacquier,
  Kawamura, Nishi, Numata} \emph{et~al.}}]{Szabi:TAMASAS}
\bibinfo{author}{\bibnamefont{M{\'a}rka}, \bibfnamefont{S.}},
  \bibinfo{author}{\bibfnamefont{A.}~\bibnamefont{Takamori}},
  \bibinfo{author}{\bibfnamefont{M.}~\bibnamefont{Ando}},
  \bibinfo{author}{\bibfnamefont{A.}~\bibnamefont{Bertolini}},
  \bibinfo{author}{\bibfnamefont{G.}~\bibnamefont{Cella}},
  \bibinfo{author}{\bibfnamefont{R.}~\bibnamefont{DeSalvo}},
  \bibinfo{author}{\bibfnamefont{M.}~\bibnamefont{Fukushima}},
  \bibinfo{author}{\bibfnamefont{Y.}~\bibnamefont{Iida}},
  \bibinfo{author}{\bibfnamefont{F.}~\bibnamefont{Jacquier}},
  \bibinfo{author}{\bibfnamefont{S.}~\bibnamefont{Kawamura}},
  \bibinfo{author}{\bibfnamefont{Y.}~\bibnamefont{Nishi}},
  \bibinfo{author}{\bibfnamefont{K.}~\bibnamefont{Numata}}, \emph{et~al.},
  \bibinfo{year}{2002}, \bibinfo{journal}{Classical and Quantum Gravity}
  \textbf{\bibinfo{volume}{19}}(\bibinfo{number}{7}), \bibinfo{pages}{1605}.

\bibitem[{\citenamefont{McClelland}
  \emph{et~al.}(2011)\citenamefont{McClelland, Mavalvala, Chen, and
  Schnabel}}]{McClelland:2011fe}
\bibinfo{author}{\bibnamefont{McClelland}, \bibfnamefont{D.~E.}},
  \bibinfo{author}{\bibfnamefont{N.}~\bibnamefont{Mavalvala}},
  \bibinfo{author}{\bibfnamefont{Y.}~\bibnamefont{Chen}}, and
  \bibinfo{author}{\bibfnamefont{R.}~\bibnamefont{Schnabel}},
  \bibinfo{year}{2011}, \bibinfo{journal}{Laser {\&} Photonics Reviews}
  \textbf{\bibinfo{volume}{5}}(\bibinfo{number}{5}), \bibinfo{pages}{677}.

\bibitem[{\citenamefont{McClelland}
  \emph{et~al.}(1993)\citenamefont{McClelland, Savage, Tridgell, and
  Mavaddat}}]{DMC:healing93}
\bibinfo{author}{\bibnamefont{McClelland}, \bibfnamefont{D.~E.}},
  \bibinfo{author}{\bibfnamefont{C.~M.} \bibnamefont{Savage}},
  \bibinfo{author}{\bibfnamefont{A.~J.} \bibnamefont{Tridgell}}, and
  \bibinfo{author}{\bibfnamefont{R.}~\bibnamefont{Mavaddat}},
  \bibinfo{year}{1993}, \bibinfo{journal}{Phys. Rev. D}
  \textbf{\bibinfo{volume}{48}}, \bibinfo{pages}{5475},
  \urlprefix\url{http://link.aps.org/doi/10.1103/PhysRevD.48.5475}.

\bibitem[{\citenamefont{Mckenzie}(2008)}]{KirksThesis}
\bibinfo{author}{\bibnamefont{Mckenzie}, \bibfnamefont{K.}},
  \bibinfo{year}{2008}, \emph{\bibinfo{title}{{Squeezing in the Audio
  Gravitational Wave Detection Band}}}, Ph.D. thesis,
  \bibinfo{school}{Australian National University}.

\bibitem[{\citenamefont{McKenzie} \emph{et~al.}(2004)\citenamefont{McKenzie,
  Grosse, Bowen, Whitcomb, Gray, McClelland, and Lam}}]{Kirk:2004}
\bibinfo{author}{\bibnamefont{McKenzie}, \bibfnamefont{K.}},
  \bibinfo{author}{\bibfnamefont{N.}~\bibnamefont{Grosse}},
  \bibinfo{author}{\bibfnamefont{W.~P.} \bibnamefont{Bowen}},
  \bibinfo{author}{\bibfnamefont{S.~E.} \bibnamefont{Whitcomb}},
  \bibinfo{author}{\bibfnamefont{M.~B.} \bibnamefont{Gray}},
  \bibinfo{author}{\bibfnamefont{D.~E.} \bibnamefont{McClelland}}, and
  \bibinfo{author}{\bibfnamefont{P.~K.} \bibnamefont{Lam}},
  \bibinfo{year}{2004}, \bibinfo{journal}{Phys. Rev. Lett.}
  \textbf{\bibinfo{volume}{93}}, \bibinfo{pages}{161105},
  \urlprefix\url{http://link.aps.org/doi/10.1103/PhysRevLett.93.161105}.

\bibitem[{\citenamefont{Meers}(1988)}]{Mee1988}
\bibinfo{author}{\bibnamefont{Meers}, \bibfnamefont{B.~J.}},
  \bibinfo{year}{1988}, \bibinfo{journal}{Phys. Rev. D}
  \textbf{\bibinfo{volume}{38}}, \bibinfo{pages}{2317},
  \urlprefix\url{http://link.aps.org/doi/10.1103/PhysRevD.38.2317}.

\bibitem[{\citenamefont{{Melchior}}(1983)}]{Melchior:Tides}
\bibinfo{author}{\bibnamefont{{Melchior}}, \bibfnamefont{P.}},
  \bibinfo{year}{1983}, \emph{\bibinfo{title}{{The tides of the planet earth
  /2nd edition/}}}.

\bibitem[{\citenamefont{Miao} \emph{et~al.}(2013)\citenamefont{Miao, Chen, and
  Adhikari}}]{Haixing:CC2013}
\bibinfo{author}{\bibnamefont{Miao}, \bibfnamefont{H.}},
  \bibinfo{author}{\bibfnamefont{Y.}~\bibnamefont{Chen}}, and
  \bibinfo{author}{\bibfnamefont{R.}~\bibnamefont{Adhikari}},
  \bibinfo{year}{2013}, \bibinfo{title}{Comparison of quantum noise for optical
  configurations of laser-interferometric gravitational-wave detectors},
  \bibinfo{note}{in prep.}

\bibitem[{\citenamefont{Miller} \emph{et~al.}(2011)\citenamefont{Miller, Evans,
  Barsotti, Fritschel, MacInnis, Mittleman, Shapiro, Soto, and
  Torrie}}]{Miller:ESD}
\bibinfo{author}{\bibnamefont{Miller}, \bibfnamefont{J.}},
  \bibinfo{author}{\bibfnamefont{M.}~\bibnamefont{Evans}},
  \bibinfo{author}{\bibfnamefont{L.}~\bibnamefont{Barsotti}},
  \bibinfo{author}{\bibfnamefont{P.}~\bibnamefont{Fritschel}},
  \bibinfo{author}{\bibfnamefont{M.}~\bibnamefont{MacInnis}},
  \bibinfo{author}{\bibfnamefont{R.}~\bibnamefont{Mittleman}},
  \bibinfo{author}{\bibfnamefont{B.}~\bibnamefont{Shapiro}},
  \bibinfo{author}{\bibfnamefont{J.}~\bibnamefont{Soto}}, and
  \bibinfo{author}{\bibfnamefont{C.}~\bibnamefont{Torrie}},
  \bibinfo{year}{2011}, \bibinfo{journal}{Physics Letters A}
  \textbf{\bibinfo{volume}{375}}(\bibinfo{number}{3}), \bibinfo{pages}{788 },
  ISSN \bibinfo{issn}{0375-9601},
  \urlprefix\url{http://www.sciencedirect.com/science/article/pii/S0375960110015719}.

\bibitem[{\citenamefont{Miller} \emph{et~al.}(2008)\citenamefont{Miller,
  Willems, Yamamoto, Agresti, and DeSalvo}}]{JM:Mesa2008}
\bibinfo{author}{\bibnamefont{Miller}, \bibfnamefont{J.}},
  \bibinfo{author}{\bibfnamefont{P.}~\bibnamefont{Willems}},
  \bibinfo{author}{\bibfnamefont{H.}~\bibnamefont{Yamamoto}},
  \bibinfo{author}{\bibfnamefont{J.}~\bibnamefont{Agresti}}, and
  \bibinfo{author}{\bibfnamefont{R.}~\bibnamefont{DeSalvo}},
  \bibinfo{year}{2008}, \bibinfo{journal}{Classical and Quantum Gravity}
  \textbf{\bibinfo{volume}{25}}(\bibinfo{number}{23}), \bibinfo{pages}{235016}.

\bibitem[{\citenamefont{Misner} \emph{et~al.}(1973)\citenamefont{Misner,
  Thorne, and Wheeler}}]{MTW1973}
\bibinfo{author}{\bibnamefont{Misner}, \bibfnamefont{C.~W.}},
  \bibinfo{author}{\bibfnamefont{K.~S.} \bibnamefont{Thorne}}, and
  \bibinfo{author}{\bibfnamefont{J.~A.} \bibnamefont{Wheeler}},
  \bibinfo{year}{1973}, \emph{\bibinfo{title}{Gravitation}}
  (\bibinfo{publisher}{Freeman}).

\bibitem[{\citenamefont{{Mitrofanov}}
  \emph{et~al.}(2004)\citenamefont{{Mitrofanov}, {Prokhorov}, {Tokmakov}, and
  {Willems}}}]{Charge:2004}
\bibinfo{author}{\bibnamefont{{Mitrofanov}}, \bibfnamefont{V.}},
  \bibinfo{author}{\bibfnamefont{L.}~\bibnamefont{{Prokhorov}}},
  \bibinfo{author}{\bibfnamefont{K.}~\bibnamefont{{Tokmakov}}}, and
  \bibinfo{author}{\bibfnamefont{P.}~\bibnamefont{{Willems}}},
  \bibinfo{year}{2004}, \bibinfo{journal}{Classical and Quantum Gravity}
  \textbf{\bibinfo{volume}{21}}, \bibinfo{pages}{1083}.

\bibitem[{\citenamefont{Miyakawa} \emph{et~al.}(2006)\citenamefont{Miyakawa,
  Ward, Adhikari, Evans, Abbott, Bork, Busby, Heefner, Ivanov, Smith, Taylor,
  Vass} \emph{et~al.}}]{Osamu:2006}
\bibinfo{author}{\bibnamefont{Miyakawa}, \bibfnamefont{O.}},
  \bibinfo{author}{\bibfnamefont{R.}~\bibnamefont{Ward}},
  \bibinfo{author}{\bibfnamefont{R.}~\bibnamefont{Adhikari}},
  \bibinfo{author}{\bibfnamefont{M.}~\bibnamefont{Evans}},
  \bibinfo{author}{\bibfnamefont{B.}~\bibnamefont{Abbott}},
  \bibinfo{author}{\bibfnamefont{R.}~\bibnamefont{Bork}},
  \bibinfo{author}{\bibfnamefont{D.}~\bibnamefont{Busby}},
  \bibinfo{author}{\bibfnamefont{J.}~\bibnamefont{Heefner}},
  \bibinfo{author}{\bibfnamefont{A.}~\bibnamefont{Ivanov}},
  \bibinfo{author}{\bibfnamefont{M.}~\bibnamefont{Smith}},
  \bibinfo{author}{\bibfnamefont{R.}~\bibnamefont{Taylor}},
  \bibinfo{author}{\bibfnamefont{S.}~\bibnamefont{Vass}}, \emph{et~al.},
  \bibinfo{year}{2006}, \bibinfo{journal}{Phys. Rev. D}
  \textbf{\bibinfo{volume}{74}}, \bibinfo{pages}{022001},
  \urlprefix\url{http://link.aps.org/doi/10.1103/PhysRevD.74.022001}.

\bibitem[{\citenamefont{Mizuno}(1995)}]{Miz1995}
\bibinfo{author}{\bibnamefont{Mizuno}, \bibfnamefont{J.}},
  \bibinfo{year}{1995}, \emph{\bibinfo{title}{Comparison of optical
  configurations for laser-interferometric gravitational-wave detectors}},
  Ph.D. thesis, \bibinfo{school}{{Universit\"at Hannover and
  Max-Planck-Institut f\"ur Quantenoptik, Garching}}.

\bibitem[{\citenamefont{{Mizuno}} \emph{et~al.}(1993)\citenamefont{{Mizuno},
  {Strain}, {Nelson}, {Chen}, {Schilling}, {R{\"u}diger}, {Winkler}, and
  {Danzmann}}}]{Mizuno:RSE1993}
\bibinfo{author}{\bibnamefont{{Mizuno}}, \bibfnamefont{J.}},
  \bibinfo{author}{\bibfnamefont{K.~A.} \bibnamefont{{Strain}}},
  \bibinfo{author}{\bibfnamefont{P.~G.} \bibnamefont{{Nelson}}},
  \bibinfo{author}{\bibfnamefont{J.~M.} \bibnamefont{{Chen}}},
  \bibinfo{author}{\bibfnamefont{R.}~\bibnamefont{{Schilling}}},
  \bibinfo{author}{\bibfnamefont{A.}~\bibnamefont{{R{\"u}diger}}},
  \bibinfo{author}{\bibfnamefont{W.}~\bibnamefont{{Winkler}}}, and
  \bibinfo{author}{\bibfnamefont{K.}~\bibnamefont{{Danzmann}}},
  \bibinfo{year}{1993}, \bibinfo{journal}{Physics Letters A}
  \textbf{\bibinfo{volume}{175}}, \bibinfo{pages}{273}.

\bibitem[{\citenamefont{Morrison}
  \emph{et~al.}(1994{\natexlab{a}})\citenamefont{Morrison, Meers, Robertson,
  and Ward}}]{Morrison:94theory}
\bibinfo{author}{\bibnamefont{Morrison}, \bibfnamefont{E.}},
  \bibinfo{author}{\bibfnamefont{B.~J.} \bibnamefont{Meers}},
  \bibinfo{author}{\bibfnamefont{D.~I.} \bibnamefont{Robertson}}, and
  \bibinfo{author}{\bibfnamefont{H.}~\bibnamefont{Ward}},
  \bibinfo{year}{1994}{\natexlab{a}}, \bibinfo{journal}{Applied Optics}
  \textbf{\bibinfo{volume}{33}}(\bibinfo{number}{22}), \bibinfo{pages}{5041},
  \urlprefix\url{http://ao.osa.org/abstract.cfm?URI=ao-33-22-5041}.

\bibitem[{\citenamefont{Morrison}
  \emph{et~al.}(1994{\natexlab{b}})\citenamefont{Morrison, Meers, Robertson,
  and Ward}}]{Morrison:94exp}
\bibinfo{author}{\bibnamefont{Morrison}, \bibfnamefont{E.}},
  \bibinfo{author}{\bibfnamefont{B.~J.} \bibnamefont{Meers}},
  \bibinfo{author}{\bibfnamefont{D.~I.} \bibnamefont{Robertson}}, and
  \bibinfo{author}{\bibfnamefont{H.}~\bibnamefont{Ward}},
  \bibinfo{year}{1994}{\natexlab{b}}, \bibinfo{journal}{Applied optics}
  \textbf{\bibinfo{volume}{33}}(\bibinfo{number}{22}), \bibinfo{pages}{5037}.

\bibitem[{\citenamefont{Mueller} \emph{et~al.}(2000)\citenamefont{Mueller,
  ze~Shu, Adhikari, Tanner, Reitze, Sigg, Mavalvala, and
  Camp}}]{Mueller:2000ul}
\bibinfo{author}{\bibnamefont{Mueller}, \bibfnamefont{G.}},
  \bibinfo{author}{\bibfnamefont{Q.}~\bibnamefont{ze~Shu}},
  \bibinfo{author}{\bibfnamefont{R.}~\bibnamefont{Adhikari}},
  \bibinfo{author}{\bibfnamefont{D.~B.} \bibnamefont{Tanner}},
  \bibinfo{author}{\bibfnamefont{D.}~\bibnamefont{Reitze}},
  \bibinfo{author}{\bibfnamefont{D.}~\bibnamefont{Sigg}},
  \bibinfo{author}{\bibfnamefont{N.}~\bibnamefont{Mavalvala}}, and
  \bibinfo{author}{\bibfnamefont{J.}~\bibnamefont{Camp}}, \bibinfo{year}{2000},
  \bibinfo{journal}{Opt. Lett.}
  \textbf{\bibinfo{volume}{25}}(\bibinfo{number}{4}), \bibinfo{pages}{266},
  \urlprefix\url{http://ol.osa.org/abstract.cfm?URI=ol-25-4-266}.

\bibitem[{\citenamefont{{Nawrodt}} \emph{et~al.}(2010)\citenamefont{{Nawrodt},
  {Schwarz}, {Kroker}, {Martin}, {Br{\"u}ckner}, {Cunningham}, {Gro{\ss}e},
  {Grib}, {Heinert}, {Hough}, {K{\"a}sebier}, {Kley}}
  \emph{et~al.}}]{Ronny:Surface2010}
\bibinfo{author}{\bibnamefont{{Nawrodt}}, \bibfnamefont{R.}},
  \bibinfo{author}{\bibfnamefont{C.}~\bibnamefont{{Schwarz}}},
  \bibinfo{author}{\bibfnamefont{S.}~\bibnamefont{{Kroker}}},
  \bibinfo{author}{\bibfnamefont{I.~W.} \bibnamefont{{Martin}}},
  \bibinfo{author}{\bibfnamefont{F.}~\bibnamefont{{Br{\"u}ckner}}},
  \bibinfo{author}{\bibfnamefont{L.}~\bibnamefont{{Cunningham}}},
  \bibinfo{author}{\bibfnamefont{V.}~\bibnamefont{{Gro{\ss}e}}},
  \bibinfo{author}{\bibfnamefont{A.}~\bibnamefont{{Grib}}},
  \bibinfo{author}{\bibfnamefont{D.}~\bibnamefont{{Heinert}}},
  \bibinfo{author}{\bibfnamefont{J.}~\bibnamefont{{Hough}}},
  \bibinfo{author}{\bibfnamefont{T.}~\bibnamefont{{K{\"a}sebier}}},
  \bibinfo{author}{\bibfnamefont{E.~B.} \bibnamefont{{Kley}}}, \emph{et~al.},
  \bibinfo{year}{2010}, \bibinfo{journal}{ArXiv e-prints} \eprint{1003.2893}.

\bibitem[{\citenamefont{Newell} \emph{et~al.}(1997)\citenamefont{Newell,
  Richman, Nelson, Stebbins, Bender, Faller, and Mason}}]{Richman:1997}
\bibinfo{author}{\bibnamefont{Newell}, \bibfnamefont{D.~B.}},
  \bibinfo{author}{\bibfnamefont{S.~J.} \bibnamefont{Richman}},
  \bibinfo{author}{\bibfnamefont{P.~G.} \bibnamefont{Nelson}},
  \bibinfo{author}{\bibfnamefont{R.~T.} \bibnamefont{Stebbins}},
  \bibinfo{author}{\bibfnamefont{P.~L.} \bibnamefont{Bender}},
  \bibinfo{author}{\bibfnamefont{J.~E.} \bibnamefont{Faller}}, and
  \bibinfo{author}{\bibfnamefont{J.}~\bibnamefont{Mason}},
  \bibinfo{year}{1997}, \bibinfo{journal}{Review of Scientific Instruments}
  \textbf{\bibinfo{volume}{68}}(\bibinfo{number}{8}), \bibinfo{pages}{3211}.

\bibitem[{\citenamefont{NGO}(2013)}]{NGO:YellowBook}
\bibinfo{author}{\bibnamefont{NGO}}, \bibinfo{year}{2013},
  \emph{\bibinfo{title}{Yellow Book}}, \bibinfo{type}{Technical Report},
  \bibinfo{institution}{NGO},
  \urlprefix\url{http://elisa-ngo.org/publications/publications-yellow-book}.

\bibitem[{\citenamefont{Ott}(2009)}]{Ott:SN2009}
\bibinfo{author}{\bibnamefont{Ott}, \bibfnamefont{C.~D.}},
  \bibinfo{year}{2009}, \bibinfo{journal}{Classical and Quantum Gravity}
  \textbf{\bibinfo{volume}{26}}(\bibinfo{number}{6}), \bibinfo{pages}{063001},
  \urlprefix\url{http://stacks.iop.org/0264-9381/26/i=6/a=063001}.

\bibitem[{\citenamefont{{Ottaway}} \emph{et~al.}(2006)\citenamefont{{Ottaway},
  {Betzwieser}, {Ballmer}, {Waldman}, and {Kells}}}]{Ottaway:2006vb}
\bibinfo{author}{\bibnamefont{{Ottaway}}, \bibfnamefont{D.}},
  \bibinfo{author}{\bibfnamefont{J.}~\bibnamefont{{Betzwieser}}},
  \bibinfo{author}{\bibfnamefont{S.}~\bibnamefont{{Ballmer}}},
  \bibinfo{author}{\bibfnamefont{S.}~\bibnamefont{{Waldman}}}, and
  \bibinfo{author}{\bibfnamefont{W.}~\bibnamefont{{Kells}}},
  \bibinfo{year}{2006}, \bibinfo{journal}{Optics Letters}
  \textbf{\bibinfo{volume}{31}}, \bibinfo{pages}{450}.

\bibitem[{\citenamefont{Ottaway} \emph{et~al.}(2012)\citenamefont{Ottaway,
  Fritschel, and Waldman}}]{Sam:Scatter2012}
\bibinfo{author}{\bibnamefont{Ottaway}, \bibfnamefont{D.~J.}},
  \bibinfo{author}{\bibfnamefont{P.}~\bibnamefont{Fritschel}}, and
  \bibinfo{author}{\bibfnamefont{S.~J.} \bibnamefont{Waldman}},
  \bibinfo{year}{2012}, \bibinfo{journal}{Opt. Express}
  \textbf{\bibinfo{volume}{20}}(\bibinfo{number}{8}), \bibinfo{pages}{8329},
  \urlprefix\url{http://www.opticsexpress.org/abstract.cfm?URI=oe-20-8-8329}.

\bibitem[{\citenamefont{Owen}(2005)}]{Owen:2005}
\bibinfo{author}{\bibnamefont{Owen}, \bibfnamefont{B.~J.}},
  \bibinfo{year}{2005}, \bibinfo{journal}{Phys. Rev. Lett.}
  \textbf{\bibinfo{volume}{95}}, \bibinfo{pages}{211101},
  \urlprefix\url{http://link.aps.org/doi/10.1103/PhysRevLett.95.211101}.

\bibitem[{\citenamefont{{Owen}}(2006)}]{Owen:2006}
\bibinfo{author}{\bibnamefont{{Owen}}, \bibfnamefont{B.~J.}},
  \bibinfo{year}{2006}, \bibinfo{journal}{Classical and Quantum Gravity}
  \textbf{\bibinfo{volume}{23}}, \bibinfo{pages}{1}.

\bibitem[{\citenamefont{{Pan}}(2006)}]{YiPan:SRC}
\bibinfo{author}{\bibnamefont{{Pan}}, \bibfnamefont{Y.}}, \bibinfo{year}{2006},
  \bibinfo{journal}{ArXiv General Relativity and Quantum Cosmology e-prints}
  \eprint{arXiv:gr-qc/0608128}.

\bibitem[{\citenamefont{Peebles}(1993)}]{Peebles:Cosmo}
\bibinfo{author}{\bibnamefont{Peebles}, \bibfnamefont{P.~J.~E.}},
  \bibinfo{year}{1993}, \emph{\bibinfo{title}{{Principles of Physical
  Cosmology}}}, Princeton series in physics (\bibinfo{publisher}{Princeton
  University Press}), ISBN \bibinfo{isbn}{9780691019338},
  \urlprefix\url{http://books.google.com/books?id=AmlEt6TJ6jAC}.

\bibitem[{\citenamefont{Penn} \emph{et~al.}(2006)\citenamefont{Penn, Ageev,
  Busby, Harry, Gretarsson, Numata, and Willems}}]{Penn20063}
\bibinfo{author}{\bibnamefont{Penn}, \bibfnamefont{S.~D.}},
  \bibinfo{author}{\bibfnamefont{A.}~\bibnamefont{Ageev}},
  \bibinfo{author}{\bibfnamefont{D.}~\bibnamefont{Busby}},
  \bibinfo{author}{\bibfnamefont{G.~M.} \bibnamefont{Harry}},
  \bibinfo{author}{\bibfnamefont{A.~M.} \bibnamefont{Gretarsson}},
  \bibinfo{author}{\bibfnamefont{K.}~\bibnamefont{Numata}}, and
  \bibinfo{author}{\bibfnamefont{P.}~\bibnamefont{Willems}},
  \bibinfo{year}{2006}, \bibinfo{journal}{Physics Letters A}
  \textbf{\bibinfo{volume}{352}}(\bibinfo{number}{1{\^a}€``2}),
  \bibinfo{pages}{3 }, ISSN \bibinfo{issn}{0375-9601},
  \urlprefix\url{http://www.sciencedirect.com/science/article/pii/S0375960105017846}.

\bibitem[{\citenamefont{Penn} \emph{et~al.}(2003)\citenamefont{Penn, Sneddon,
  Armandula, Betzwieser, Cagnoli, Camp, Crooks, Fejer, Gretarsson, Harry,
  Hough, Kittelberger} \emph{et~al.}}]{Penn:CQG2003}
\bibinfo{author}{\bibnamefont{Penn}, \bibfnamefont{S.~D.}},
  \bibinfo{author}{\bibfnamefont{P.~H.} \bibnamefont{Sneddon}},
  \bibinfo{author}{\bibfnamefont{H.}~\bibnamefont{Armandula}},
  \bibinfo{author}{\bibfnamefont{J.~C.} \bibnamefont{Betzwieser}},
  \bibinfo{author}{\bibfnamefont{G.}~\bibnamefont{Cagnoli}},
  \bibinfo{author}{\bibfnamefont{J.}~\bibnamefont{Camp}},
  \bibinfo{author}{\bibfnamefont{D.~R.~M.} \bibnamefont{Crooks}},
  \bibinfo{author}{\bibfnamefont{M.~M.} \bibnamefont{Fejer}},
  \bibinfo{author}{\bibfnamefont{A.~M.} \bibnamefont{Gretarsson}},
  \bibinfo{author}{\bibfnamefont{G.~M.} \bibnamefont{Harry}},
  \bibinfo{author}{\bibfnamefont{J.}~\bibnamefont{Hough}},
  \bibinfo{author}{\bibfnamefont{S.~E.} \bibnamefont{Kittelberger}},
  \emph{et~al.}, \bibinfo{year}{2003}, \bibinfo{journal}{Classical and Quantum
  Gravity} \textbf{\bibinfo{volume}{20}}(\bibinfo{number}{13}),
  \bibinfo{pages}{2917}.

\bibitem[{\citenamefont{Peterson}(1993)}]{Pet1993}
\bibinfo{author}{\bibnamefont{Peterson}, \bibfnamefont{J.}},
  \bibinfo{year}{1993}, \bibinfo{journal}{Open-file report {U. S. Geological
  Survey}} \textbf{\bibinfo{volume}{93-322}}.

\bibitem[{\citenamefont{Phillips}(1987)}]{Phillips:1987}
\bibinfo{author}{\bibnamefont{Phillips}, \bibfnamefont{W.~A.}},
  \bibinfo{year}{1987}, \bibinfo{journal}{Reports on Progress in Physics}
  \textbf{\bibinfo{volume}{50}}(\bibinfo{number}{12}), \bibinfo{pages}{1657},
  \urlprefix\url{http://stacks.iop.org/0034-4885/50/i=12/a=003}.

\bibitem[{\citenamefont{{Phinney}}(1991)}]{Sterl:Minimal}
\bibinfo{author}{\bibnamefont{{Phinney}}, \bibfnamefont{E.~S.}},
  \bibinfo{year}{1991}, \bibinfo{journal}{Astrophysical Journal Letters}
  \textbf{\bibinfo{volume}{380}}, \bibinfo{pages}{L17}.

\bibitem[{\citenamefont{Phinney}(2003)}]{Phi2003}
\bibinfo{author}{\bibnamefont{Phinney}, \bibfnamefont{E.~S.}},
  \bibinfo{year}{2003}, \bibinfo{journal}{{NASA Mission Concept Study}} .

\bibitem[{\citenamefont{Pitkin} \emph{et~al.}(2011)\citenamefont{Pitkin, Reid,
  Rowan, and Hough}}]{lrr-2011-5}
\bibinfo{author}{\bibnamefont{Pitkin}, \bibfnamefont{M.}},
  \bibinfo{author}{\bibfnamefont{S.}~\bibnamefont{Reid}},
  \bibinfo{author}{\bibfnamefont{S.}~\bibnamefont{Rowan}}, and
  \bibinfo{author}{\bibfnamefont{J.}~\bibnamefont{Hough}},
  \bibinfo{year}{2011}, \bibinfo{journal}{Living Reviews in Relativity}
  \textbf{\bibinfo{volume}{14}}(\bibinfo{number}{5}).

\bibitem[{\citenamefont{{Planck Collaboration}}
  \emph{et~al.}(2013)\citenamefont{{Planck Collaboration}, {Ade}, {Aghanim},
  {Armitage-Caplan}, {Arnaud}, {Ashdown}, {Atrio-Barandela}, {Aumont},
  {Baccigalupi}, {Banday}, and et~al.}}]{Planck:2013}
\bibinfo{author}{\bibnamefont{{Planck Collaboration}}},
  \bibinfo{author}{\bibfnamefont{P.~A.~R.} \bibnamefont{{Ade}}},
  \bibinfo{author}{\bibfnamefont{N.}~\bibnamefont{{Aghanim}}},
  \bibinfo{author}{\bibfnamefont{C.}~\bibnamefont{{Armitage-Caplan}}},
  \bibinfo{author}{\bibfnamefont{M.}~\bibnamefont{{Arnaud}}},
  \bibinfo{author}{\bibfnamefont{M.}~\bibnamefont{{Ashdown}}},
  \bibinfo{author}{\bibfnamefont{F.}~\bibnamefont{{Atrio-Barandela}}},
  \bibinfo{author}{\bibfnamefont{J.}~\bibnamefont{{Aumont}}},
  \bibinfo{author}{\bibfnamefont{C.}~\bibnamefont{{Baccigalupi}}},
  \bibinfo{author}{\bibfnamefont{A.~J.} \bibnamefont{{Banday}}}, and
  \bibinfo{author}{\bibnamefont{et~al.}}, \bibinfo{year}{2013},
  \bibinfo{journal}{ArXiv e-prints} \eprint{1303.5062}.

\bibitem[{\citenamefont{Plissi} \emph{et~al.}(1998)\citenamefont{Plissi,
  Strain, Torrie, Robertson, Killbourn, Rowan, Twyford, Ward, Skeldon, and
  Hough}}]{plissi:3055}
\bibinfo{author}{\bibnamefont{Plissi}, \bibfnamefont{M.~V.}},
  \bibinfo{author}{\bibfnamefont{K.~A.} \bibnamefont{Strain}},
  \bibinfo{author}{\bibfnamefont{C.~I.} \bibnamefont{Torrie}},
  \bibinfo{author}{\bibfnamefont{N.~A.} \bibnamefont{Robertson}},
  \bibinfo{author}{\bibfnamefont{S.}~\bibnamefont{Killbourn}},
  \bibinfo{author}{\bibfnamefont{S.}~\bibnamefont{Rowan}},
  \bibinfo{author}{\bibfnamefont{S.~M.} \bibnamefont{Twyford}},
  \bibinfo{author}{\bibfnamefont{H.}~\bibnamefont{Ward}},
  \bibinfo{author}{\bibfnamefont{K.~D.} \bibnamefont{Skeldon}}, and
  \bibinfo{author}{\bibfnamefont{J.}~\bibnamefont{Hough}},
  \bibinfo{year}{1998}, \bibinfo{journal}{Review of Scientific Instruments}
  \textbf{\bibinfo{volume}{69}}(\bibinfo{number}{8}), \bibinfo{pages}{3055},
  \urlprefix\url{http://link.aip.org/link/?RSI/69/3055/1}.

\bibitem[{\citenamefont{Poenaru and Greiner}(1997)}]{Gruyter:1997}
\bibinfo{author}{\bibnamefont{Poenaru}, \bibfnamefont{D.}}, and
  \bibinfo{author}{\bibfnamefont{W.}~\bibnamefont{Greiner}},
  \bibinfo{year}{1997}, \emph{\bibinfo{title}{Experimental techniques in
  nuclear physics}} (\bibinfo{publisher}{Walter de Gruyter}), ISBN
  \bibinfo{isbn}{9783110144673},
  \urlprefix\url{http://books.google.com/books?id=7yq7ZV2G5D8C}.

\bibitem[{\citenamefont{Pohl} \emph{et~al.}(2002)\citenamefont{Pohl, Liu, and
  Thompson}}]{Pohl:RMP}
\bibinfo{author}{\bibnamefont{Pohl}, \bibfnamefont{R.~O.}},
  \bibinfo{author}{\bibfnamefont{X.}~\bibnamefont{Liu}}, and
  \bibinfo{author}{\bibfnamefont{E.}~\bibnamefont{Thompson}},
  \bibinfo{year}{2002}, \bibinfo{journal}{Rev. Mod. Phys.}
  \textbf{\bibinfo{volume}{74}}, \bibinfo{pages}{991}.

\bibitem[{\citenamefont{Pollack} \emph{et~al.}(2010)\citenamefont{Pollack,
  Turner, Schlamminger, Hagedorn, and Gundlach}}]{UW:Charging2010}
\bibinfo{author}{\bibnamefont{Pollack}, \bibfnamefont{S.~E.}},
  \bibinfo{author}{\bibfnamefont{M.~D.} \bibnamefont{Turner}},
  \bibinfo{author}{\bibfnamefont{S.}~\bibnamefont{Schlamminger}},
  \bibinfo{author}{\bibfnamefont{C.~A.} \bibnamefont{Hagedorn}}, and
  \bibinfo{author}{\bibfnamefont{J.~H.} \bibnamefont{Gundlach}},
  \bibinfo{year}{2010}, \bibinfo{journal}{Phys. Rev. D}
  \textbf{\bibinfo{volume}{81}}, \bibinfo{pages}{021101},
  \urlprefix\url{http://link.aps.org/doi/10.1103/PhysRevD.81.021101}.

\bibitem[{\citenamefont{{Ponslet} and {Miller}}(1998)}]{ponslet:432}
\bibinfo{author}{\bibnamefont{{Ponslet}}, \bibfnamefont{E.~R.}}, and
  \bibinfo{author}{\bibfnamefont{W.~O.} \bibnamefont{{Miller}}},
  \bibinfo{year}{1998}, in \emph{\bibinfo{booktitle}{Smart Structures and
  Materials 1998: Passive Damping and Isolation}}, edited by
  \bibinfo{editor}{\bibfnamefont{L.~P.} \bibnamefont{{Davis}}}, volume
  \bibinfo{volume}{3327} of \emph{\bibinfo{series}{Society of Photo-Optical
  Instrumentation Engineers (SPIE) Conference Series}}, pp.
  \bibinfo{pages}{432--443}.

\bibitem[{\citenamefont{Prince and the {LISA Science Team}}(2009)}]{LISA:2009}
\bibinfo{author}{\bibnamefont{Prince}, \bibfnamefont{T.}}, and
  \bibinfo{author}{\bibnamefont{the {LISA Science Team}}},
  \bibinfo{year}{2009}, \bibinfo{title}{{LISA}: Probing the universe with
  gravitational waves}.

\bibitem[{\citenamefont{Prince} \emph{et~al.}(2002)\citenamefont{Prince, Tinto,
  Larson, and Armstrong}}]{PrEA2002}
\bibinfo{author}{\bibnamefont{Prince}, \bibfnamefont{T.~A.}},
  \bibinfo{author}{\bibfnamefont{M.}~\bibnamefont{Tinto}},
  \bibinfo{author}{\bibfnamefont{S.~L.} \bibnamefont{Larson}}, and
  \bibinfo{author}{\bibfnamefont{J.~W.} \bibnamefont{Armstrong}},
  \bibinfo{year}{2002}, \bibinfo{journal}{Phys.~Rev.D}
  \textbf{\bibinfo{volume}{66}}, \bibinfo{pages}{122002}.

\bibitem[{\citenamefont{{PRISM Collaboration}}
  \emph{et~al.}(2013)\citenamefont{{PRISM Collaboration}, {Andre},
  {Baccigalupi}, {Barbosa}, {Bartlett}, {Bartolo}, {Battistelli}, {Battye},
  {Bendo}, {Bernard}, {Bersanelli}, {Bethermin}} \emph{et~al.}}]{PRISM:concept}
\bibinfo{author}{\bibnamefont{{PRISM Collaboration}}},
  \bibinfo{author}{\bibfnamefont{P.}~\bibnamefont{{Andre}}},
  \bibinfo{author}{\bibfnamefont{C.}~\bibnamefont{{Baccigalupi}}},
  \bibinfo{author}{\bibfnamefont{D.}~\bibnamefont{{Barbosa}}},
  \bibinfo{author}{\bibfnamefont{J.}~\bibnamefont{{Bartlett}}},
  \bibinfo{author}{\bibfnamefont{N.}~\bibnamefont{{Bartolo}}},
  \bibinfo{author}{\bibfnamefont{E.}~\bibnamefont{{Battistelli}}},
  \bibinfo{author}{\bibfnamefont{R.}~\bibnamefont{{Battye}}},
  \bibinfo{author}{\bibfnamefont{G.}~\bibnamefont{{Bendo}}},
  \bibinfo{author}{\bibfnamefont{J.-P.} \bibnamefont{{Bernard}}},
  \bibinfo{author}{\bibfnamefont{M.}~\bibnamefont{{Bersanelli}}},
  \bibinfo{author}{\bibfnamefont{M.}~\bibnamefont{{Bethermin}}}, \emph{et~al.},
  \bibinfo{year}{2013}, \bibinfo{journal}{ArXiv e-prints} \eprint{1306.2259}.

\bibitem[{\citenamefont{Punturo} \emph{et~al.}(2010)\citenamefont{Punturo,
  Abernathy, Acernese, Allen, Andersson, Arun, Barone, Barr, Barsuglia, Beker,
  Beveridge, Birindelli} \emph{et~al.}}]{3G:Science}
\bibinfo{author}{\bibnamefont{Punturo}, \bibfnamefont{M.}},
  \bibinfo{author}{\bibfnamefont{M.}~\bibnamefont{Abernathy}},
  \bibinfo{author}{\bibfnamefont{F.}~\bibnamefont{Acernese}},
  \bibinfo{author}{\bibfnamefont{B.}~\bibnamefont{Allen}},
  \bibinfo{author}{\bibfnamefont{N.}~\bibnamefont{Andersson}},
  \bibinfo{author}{\bibfnamefont{K.}~\bibnamefont{Arun}},
  \bibinfo{author}{\bibfnamefont{F.}~\bibnamefont{Barone}},
  \bibinfo{author}{\bibfnamefont{B.}~\bibnamefont{Barr}},
  \bibinfo{author}{\bibfnamefont{M.}~\bibnamefont{Barsuglia}},
  \bibinfo{author}{\bibfnamefont{M.}~\bibnamefont{Beker}},
  \bibinfo{author}{\bibfnamefont{N.}~\bibnamefont{Beveridge}},
  \bibinfo{author}{\bibfnamefont{S.}~\bibnamefont{Birindelli}}, \emph{et~al.},
  \bibinfo{year}{2010}, \bibinfo{journal}{Classical and Quantum Gravity}
  \textbf{\bibinfo{volume}{27}}(\bibinfo{number}{8}), \bibinfo{pages}{084007}.

\bibitem[{\citenamefont{Purdue and Chen}(2002)}]{PuCh2002}
\bibinfo{author}{\bibnamefont{Purdue}, \bibfnamefont{P.}}, and
  \bibinfo{author}{\bibfnamefont{Y.}~\bibnamefont{Chen}}, \bibinfo{year}{2002},
  \bibinfo{journal}{Phys. Rev. D} \textbf{\bibinfo{volume}{66}},
  \bibinfo{pages}{122004},
  \urlprefix\url{http://link.aps.org/doi/10.1103/PhysRevD.66.122004}.

\bibitem[{\citenamefont{Regehr}(1995)}]{Regehr:PhD}
\bibinfo{author}{\bibnamefont{Regehr}, \bibfnamefont{M.~W.}},
  \bibinfo{year}{1995}, \bibinfo{journal}{Thesis (Ph.D.) - California Institute
  of Technology} , \bibinfo{pages}{3}.

\bibitem[{\citenamefont{Rehbein} \emph{et~al.}(2008)\citenamefont{Rehbein,
  M\"uller-Ebhardt, Somiya, Danilishin, Schnabel, Danzmann, and
  Chen}}]{Yanbei:DoubleSpring}
\bibinfo{author}{\bibnamefont{Rehbein}, \bibfnamefont{H.}},
  \bibinfo{author}{\bibfnamefont{H.}~\bibnamefont{M\"uller-Ebhardt}},
  \bibinfo{author}{\bibfnamefont{K.}~\bibnamefont{Somiya}},
  \bibinfo{author}{\bibfnamefont{S.~L.} \bibnamefont{Danilishin}},
  \bibinfo{author}{\bibfnamefont{R.}~\bibnamefont{Schnabel}},
  \bibinfo{author}{\bibfnamefont{K.}~\bibnamefont{Danzmann}}, and
  \bibinfo{author}{\bibfnamefont{Y.}~\bibnamefont{Chen}}, \bibinfo{year}{2008},
  \bibinfo{journal}{Phys. Rev. D} \textbf{\bibinfo{volume}{78}},
  \bibinfo{pages}{062003},
  \urlprefix\url{http://link.aps.org/doi/10.1103/PhysRevD.78.062003}.

\bibitem[{\citenamefont{Rehbein} \emph{et~al.}(2007)\citenamefont{Rehbein,
  M\"uller-Ebhardt, Somiya, Li, Schnabel, Danzmann, and Chen}}]{Yanbei:Local}
\bibinfo{author}{\bibnamefont{Rehbein}, \bibfnamefont{H.}},
  \bibinfo{author}{\bibfnamefont{H.}~\bibnamefont{M\"uller-Ebhardt}},
  \bibinfo{author}{\bibfnamefont{K.}~\bibnamefont{Somiya}},
  \bibinfo{author}{\bibfnamefont{C.}~\bibnamefont{Li}},
  \bibinfo{author}{\bibfnamefont{R.}~\bibnamefont{Schnabel}},
  \bibinfo{author}{\bibfnamefont{K.}~\bibnamefont{Danzmann}}, and
  \bibinfo{author}{\bibfnamefont{Y.}~\bibnamefont{Chen}}, \bibinfo{year}{2007},
  \bibinfo{journal}{Phys. Rev. D} \textbf{\bibinfo{volume}{76}},
  \bibinfo{pages}{062002},
  \urlprefix\url{http://link.aps.org/doi/10.1103/PhysRevD.76.062002}.

\bibitem[{\citenamefont{{Rempe}} \emph{et~al.}(1992)\citenamefont{{Rempe},
  {Thompson}, {Kimble}, and {Lalezari}}}]{Ramin:Loss}
\bibinfo{author}{\bibnamefont{{Rempe}}, \bibfnamefont{G.}},
  \bibinfo{author}{\bibfnamefont{R.~J.} \bibnamefont{{Thompson}}},
  \bibinfo{author}{\bibfnamefont{H.~J.} \bibnamefont{{Kimble}}}, and
  \bibinfo{author}{\bibfnamefont{R.}~\bibnamefont{{Lalezari}}},
  \bibinfo{year}{1992}, \bibinfo{journal}{Optics Letters}
  \textbf{\bibinfo{volume}{17}}, \bibinfo{pages}{363}.

\bibitem[{\citenamefont{Ringler and Hutt}(2010)}]{RiHu2010}
\bibinfo{author}{\bibnamefont{Ringler}, \bibfnamefont{A.}}, and
  \bibinfo{author}{\bibfnamefont{C.}~\bibnamefont{Hutt}}, \bibinfo{year}{2010},
  \bibinfo{journal}{Seismological Research Letters}
  \textbf{\bibinfo{volume}{81}}(\bibinfo{number}{6}), \bibinfo{pages}{972}.

\bibitem[{\citenamefont{{Robertson}}
  \emph{et~al.}(2002)\citenamefont{{Robertson}, {Cagnoli}, {Crooks}, {Elliffe},
  {Faller}, {Fritschel}, {Go{\ss}ler}, {Grant}, {Heptonstall}, {Hough},
  {L{\"u}ck}, {Mittleman}} \emph{et~al.}}]{SUS:2002}
\bibinfo{author}{\bibnamefont{{Robertson}}, \bibfnamefont{N.~A.}},
  \bibinfo{author}{\bibfnamefont{G.}~\bibnamefont{{Cagnoli}}},
  \bibinfo{author}{\bibfnamefont{D.~R.~M.} \bibnamefont{{Crooks}}},
  \bibinfo{author}{\bibfnamefont{E.}~\bibnamefont{{Elliffe}}},
  \bibinfo{author}{\bibfnamefont{J.~E.} \bibnamefont{{Faller}}},
  \bibinfo{author}{\bibfnamefont{P.}~\bibnamefont{{Fritschel}}},
  \bibinfo{author}{\bibfnamefont{S.}~\bibnamefont{{Go{\ss}ler}}},
  \bibinfo{author}{\bibfnamefont{A.}~\bibnamefont{{Grant}}},
  \bibinfo{author}{\bibfnamefont{A.}~\bibnamefont{{Heptonstall}}},
  \bibinfo{author}{\bibfnamefont{J.}~\bibnamefont{{Hough}}},
  \bibinfo{author}{\bibfnamefont{H.}~\bibnamefont{{L{\"u}ck}}},
  \bibinfo{author}{\bibfnamefont{R.}~\bibnamefont{{Mittleman}}}, \emph{et~al.},
  \bibinfo{year}{2002}, \bibinfo{journal}{Classical and Quantum Gravity}
  \textbf{\bibinfo{volume}{19}}, \bibinfo{pages}{4043}.

\bibitem[{\citenamefont{Rocchi} \emph{et~al.}(2012)\citenamefont{Rocchi,
  Coccia, Fafone, Malvezzi, Minenkov, and Sperandio}}]{aVirgo:TCS}
\bibinfo{author}{\bibnamefont{Rocchi}, \bibfnamefont{A.}},
  \bibinfo{author}{\bibfnamefont{E.}~\bibnamefont{Coccia}},
  \bibinfo{author}{\bibfnamefont{V.}~\bibnamefont{Fafone}},
  \bibinfo{author}{\bibfnamefont{V.}~\bibnamefont{Malvezzi}},
  \bibinfo{author}{\bibfnamefont{Y.}~\bibnamefont{Minenkov}}, and
  \bibinfo{author}{\bibfnamefont{L.}~\bibnamefont{Sperandio}},
  \bibinfo{year}{2012}, \bibinfo{journal}{Journal of Physics: Conference
  Series} \textbf{\bibinfo{volume}{363}}(\bibinfo{number}{1}),
  \bibinfo{pages}{2016}.

\bibitem[{\citenamefont{{Romero} and {Dehnen}}(1981)}]{Romero:1981}
\bibinfo{author}{\bibnamefont{{Romero}}, \bibfnamefont{F.}}, and
  \bibinfo{author}{\bibfnamefont{H.}~\bibnamefont{{Dehnen}}},
  \bibinfo{year}{1981}, \bibinfo{journal}{Zeitschrift Naturforschung Teil A}
  \textbf{\bibinfo{volume}{36}}, \bibinfo{pages}{948}.

\bibitem[{\citenamefont{{Rowan}} \emph{et~al.}(2005)\citenamefont{{Rowan},
  {Hough}, and {Crooks}}}]{Sheila:2005}
\bibinfo{author}{\bibnamefont{{Rowan}}, \bibfnamefont{S.}},
  \bibinfo{author}{\bibfnamefont{J.}~\bibnamefont{{Hough}}}, and
  \bibinfo{author}{\bibfnamefont{D.~R.~M.} \bibnamefont{{Crooks}}},
  \bibinfo{year}{2005}, \bibinfo{journal}{Physics Letters A}
  \textbf{\bibinfo{volume}{347}}, \bibinfo{pages}{25}.

\bibitem[{\citenamefont{{Rubakov}} \emph{et~al.}(1982)\citenamefont{{Rubakov},
  {Sazhin}, and {Veryaskin}}}]{Rubakov:1982}
\bibinfo{author}{\bibnamefont{{Rubakov}}, \bibfnamefont{V.~A.}},
  \bibinfo{author}{\bibfnamefont{M.~V.} \bibnamefont{{Sazhin}}}, and
  \bibinfo{author}{\bibfnamefont{A.~V.} \bibnamefont{{Veryaskin}}},
  \bibinfo{year}{1982}, \bibinfo{journal}{Physics Letters B}
  \textbf{\bibinfo{volume}{115}}, \bibinfo{pages}{189}.

\bibitem[{\citenamefont{Saccorotti}
  \emph{et~al.}(2011)\citenamefont{Saccorotti, Piccinini, Cauchie, and
  Fiori}}]{Saccorotti:2011hm}
\bibinfo{author}{\bibnamefont{Saccorotti}, \bibfnamefont{G.}},
  \bibinfo{author}{\bibfnamefont{D.}~\bibnamefont{Piccinini}},
  \bibinfo{author}{\bibfnamefont{L.}~\bibnamefont{Cauchie}}, and
  \bibinfo{author}{\bibfnamefont{I.}~\bibnamefont{Fiori}},
  \bibinfo{year}{2011}, \bibinfo{journal}{Bulletin of the Seismological Society
  of America} \textbf{\bibinfo{volume}{101}}(\bibinfo{number}{2}),
  \bibinfo{pages}{568}.

\bibitem[{\citenamefont{Sathyaprakash}
  \emph{et~al.}(2012)\citenamefont{Sathyaprakash, Abernathy, Acernese, Ajith,
  Allen, Amaro-Seoane, Andersson, Aoudia, Arun, Astone, Krishnan, Barack}
  \emph{et~al.}}]{Sathyaprakash:2012jt}
\bibinfo{author}{\bibnamefont{Sathyaprakash}, \bibfnamefont{B.}},
  \bibinfo{author}{\bibfnamefont{M.}~\bibnamefont{Abernathy}},
  \bibinfo{author}{\bibfnamefont{F.}~\bibnamefont{Acernese}},
  \bibinfo{author}{\bibfnamefont{P.}~\bibnamefont{Ajith}},
  \bibinfo{author}{\bibfnamefont{B.}~\bibnamefont{Allen}},
  \bibinfo{author}{\bibfnamefont{P.}~\bibnamefont{Amaro-Seoane}},
  \bibinfo{author}{\bibfnamefont{N.}~\bibnamefont{Andersson}},
  \bibinfo{author}{\bibfnamefont{S.}~\bibnamefont{Aoudia}},
  \bibinfo{author}{\bibfnamefont{K.}~\bibnamefont{Arun}},
  \bibinfo{author}{\bibfnamefont{P.}~\bibnamefont{Astone}},
  \bibinfo{author}{\bibfnamefont{B.}~\bibnamefont{Krishnan}},
  \bibinfo{author}{\bibfnamefont{L.}~\bibnamefont{Barack}}, \emph{et~al.},
  \bibinfo{year}{2012}, \bibinfo{journal}{Classical and Quantum Gravity}
  \textbf{\bibinfo{volume}{29}}(\bibinfo{number}{12}), \bibinfo{pages}{124013}.

\bibitem[{\citenamefont{Sathyaprakash and Schutz}(2009)}]{Sathya:LRR}
\bibinfo{author}{\bibnamefont{Sathyaprakash}, \bibfnamefont{B.}}, and
  \bibinfo{author}{\bibfnamefont{B.~F.} \bibnamefont{Schutz}},
  \bibinfo{year}{2009}, \bibinfo{journal}{Living Reviews in Relativity}
  \textbf{\bibinfo{volume}{12}}(\bibinfo{number}{2}),
  \urlprefix\url{http://www.livingreviews.org/lrr-2009-2}.

\bibitem[{\citenamefont{Sato} \emph{et~al.}(2009)\citenamefont{Sato, Kawamura,
  Ando, Nakamura, Tsubono, Araya, Funaki, Ioka, Kanda, Moriwaki, Musha,
  Nakazawa} \emph{et~al.}}]{Sato:DECIGO2009}
\bibinfo{author}{\bibnamefont{Sato}, \bibfnamefont{S.}},
  \bibinfo{author}{\bibfnamefont{S.}~\bibnamefont{Kawamura}},
  \bibinfo{author}{\bibfnamefont{M.}~\bibnamefont{Ando}},
  \bibinfo{author}{\bibfnamefont{T.}~\bibnamefont{Nakamura}},
  \bibinfo{author}{\bibfnamefont{K.}~\bibnamefont{Tsubono}},
  \bibinfo{author}{\bibfnamefont{A.}~\bibnamefont{Araya}},
  \bibinfo{author}{\bibfnamefont{I.}~\bibnamefont{Funaki}},
  \bibinfo{author}{\bibfnamefont{K.}~\bibnamefont{Ioka}},
  \bibinfo{author}{\bibfnamefont{N.}~\bibnamefont{Kanda}},
  \bibinfo{author}{\bibfnamefont{S.}~\bibnamefont{Moriwaki}},
  \bibinfo{author}{\bibfnamefont{M.}~\bibnamefont{Musha}},
  \bibinfo{author}{\bibfnamefont{K.}~\bibnamefont{Nakazawa}}, \emph{et~al.},
  \bibinfo{year}{2009}, \bibinfo{journal}{Journal of Physics: Conference
  Series} \textbf{\bibinfo{volume}{154}}(\bibinfo{number}{1}),
  \bibinfo{pages}{012040},
  \urlprefix\url{http://stacks.iop.org/1742-6596/154/i=1/a=012040}.

\bibitem[{\citenamefont{{Sato}} \emph{et~al.}(1999)\citenamefont{{Sato},
  {Miyoki}, {Ohashi}, {Fujimoto}, {Yamazaki}, {Fukushima}, {Ueda}, {Ueda},
  {Watanabe}, {Nakamura}, {Etoh}, {Kitajima}} \emph{et~al.}}]{TAMA:Loss}
\bibinfo{author}{\bibnamefont{{Sato}}, \bibfnamefont{S.}},
  \bibinfo{author}{\bibfnamefont{S.}~\bibnamefont{{Miyoki}}},
  \bibinfo{author}{\bibfnamefont{M.}~\bibnamefont{{Ohashi}}},
  \bibinfo{author}{\bibfnamefont{M.-K.} \bibnamefont{{Fujimoto}}},
  \bibinfo{author}{\bibfnamefont{T.}~\bibnamefont{{Yamazaki}}},
  \bibinfo{author}{\bibfnamefont{M.}~\bibnamefont{{Fukushima}}},
  \bibinfo{author}{\bibfnamefont{A.}~\bibnamefont{{Ueda}}},
  \bibinfo{author}{\bibfnamefont{K.-I.} \bibnamefont{{Ueda}}},
  \bibinfo{author}{\bibfnamefont{K.}~\bibnamefont{{Watanabe}}},
  \bibinfo{author}{\bibfnamefont{K.}~\bibnamefont{{Nakamura}}},
  \bibinfo{author}{\bibfnamefont{K.}~\bibnamefont{{Etoh}}},
  \bibinfo{author}{\bibfnamefont{N.}~\bibnamefont{{Kitajima}}}, \emph{et~al.},
  \bibinfo{year}{1999}, \bibinfo{journal}{\ao} \textbf{\bibinfo{volume}{38}},
  \bibinfo{pages}{2880}.

\bibitem[{\citenamefont{Saulson}(1994)}]{Saulson:book}
\bibinfo{author}{\bibnamefont{Saulson}, \bibfnamefont{P.}},
  \bibinfo{year}{1994}, \emph{\bibinfo{title}{Fundamentals of Interferometric
  Gravitational Wave Detectors}} (\bibinfo{publisher}{World Scientific
  Publishing Company, Incorporated}), ISBN \bibinfo{isbn}{9789810218201},
  \urlprefix\url{http://books.google.com/books?id=4JyGQgAACAAJ}.

\bibitem[{\citenamefont{Saulson}(1984)}]{Saulson:NN}
\bibinfo{author}{\bibnamefont{Saulson}, \bibfnamefont{P.~R.}},
  \bibinfo{year}{1984}, \bibinfo{journal}{Phys. Rev. D}
  \textbf{\bibinfo{volume}{30}}, \bibinfo{pages}{732}.

\bibitem[{\citenamefont{Saulson}(1990)}]{Saulson:ThermalNoise}
\bibinfo{author}{\bibnamefont{Saulson}, \bibfnamefont{P.~R.}},
  \bibinfo{year}{1990}, \bibinfo{journal}{Phys. Rev. D}
  \textbf{\bibinfo{volume}{42}}, \bibinfo{pages}{2437}.

\bibitem[{\citenamefont{Sayed}(2003)}]{Say2003}
\bibinfo{author}{\bibnamefont{Sayed}, \bibfnamefont{A.~H.}},
  \bibinfo{year}{2003}, \emph{\bibinfo{title}{{Fundamentals of Adaptive
  Filtering}}} (\bibinfo{publisher}{John Wiley \& Sons}).

\bibitem[{\citenamefont{{Sazhin}}(1978)}]{Sazhin}
\bibinfo{author}{\bibnamefont{{Sazhin}}, \bibfnamefont{M.~V.}},
  \bibinfo{year}{1978}, \bibinfo{journal}{Soviet Astronomy}
  \textbf{\bibinfo{volume}{22}}, \bibinfo{pages}{36}.

\bibitem[{\citenamefont{Scheel} \emph{et~al.}(2009)\citenamefont{Scheel, Boyle,
  Chu, Kidder, Matthews, and Pfeiffer}}]{Scheel:2009}
\bibinfo{author}{\bibnamefont{Scheel}, \bibfnamefont{M.~A.}},
  \bibinfo{author}{\bibfnamefont{M.}~\bibnamefont{Boyle}},
  \bibinfo{author}{\bibfnamefont{T.}~\bibnamefont{Chu}},
  \bibinfo{author}{\bibfnamefont{L.~E.} \bibnamefont{Kidder}},
  \bibinfo{author}{\bibfnamefont{K.~D.} \bibnamefont{Matthews}}, and
  \bibinfo{author}{\bibfnamefont{H.~P.} \bibnamefont{Pfeiffer}},
  \bibinfo{year}{2009}, \bibinfo{journal}{Phys. Rev. D}
  \textbf{\bibinfo{volume}{79}}, \bibinfo{pages}{024003},
  \urlprefix\url{http://link.aps.org/doi/10.1103/PhysRevD.79.024003}.

\bibitem[{\citenamefont{Schiller} \emph{et~al.}(1992)\citenamefont{Schiller,
  Yu, Fejer, and Byer}}]{Schiller:92}
\bibinfo{author}{\bibnamefont{Schiller}, \bibfnamefont{S.}},
  \bibinfo{author}{\bibfnamefont{I.~I.} \bibnamefont{Yu}},
  \bibinfo{author}{\bibfnamefont{M.~M.} \bibnamefont{Fejer}}, and
  \bibinfo{author}{\bibfnamefont{R.~L.} \bibnamefont{Byer}},
  \bibinfo{year}{1992}, \bibinfo{journal}{Optics Letters}
  \textbf{\bibinfo{volume}{17}}(\bibinfo{number}{5}), \bibinfo{pages}{378}.

\bibitem[{\citenamefont{Schilling} \emph{et~al.}(1981)\citenamefont{Schilling,
  Schnupp, Winkler, Billing, Maischberger, and Rudiger}}]{Schilling:1981}
\bibinfo{author}{\bibnamefont{Schilling}, \bibfnamefont{R.}},
  \bibinfo{author}{\bibfnamefont{L.}~\bibnamefont{Schnupp}},
  \bibinfo{author}{\bibfnamefont{W.}~\bibnamefont{Winkler}},
  \bibinfo{author}{\bibfnamefont{H.}~\bibnamefont{Billing}},
  \bibinfo{author}{\bibfnamefont{K.}~\bibnamefont{Maischberger}}, and
  \bibinfo{author}{\bibfnamefont{A.}~\bibnamefont{Rudiger}},
  \bibinfo{year}{1981}, \bibinfo{journal}{Journal of Physics E: Scientific
  Instruments} \textbf{\bibinfo{volume}{14}}(\bibinfo{number}{1}),
  \bibinfo{pages}{65},
  \urlprefix\url{http://stacks.iop.org/0022-3735/14/i=1/a=018}.

\bibitem[{\citenamefont{{Schnabel}}
  \emph{et~al.}(2010)\citenamefont{{Schnabel}, {Mavalvala}, {McClelland}, and
  {Lam}}}]{Schnabel:2010review}
\bibinfo{author}{\bibnamefont{{Schnabel}}, \bibfnamefont{R.}},
  \bibinfo{author}{\bibfnamefont{N.}~\bibnamefont{{Mavalvala}}},
  \bibinfo{author}{\bibfnamefont{D.~E.} \bibnamefont{{McClelland}}}, and
  \bibinfo{author}{\bibfnamefont{P.~K.} \bibnamefont{{Lam}}},
  \bibinfo{year}{2010}, \bibinfo{journal}{Nature Communications}
  \textbf{\bibinfo{volume}{1}}, \bibinfo{eid}{121}.

\bibitem[{\citenamefont{Schnupp} \emph{et~al.}(1985)\citenamefont{Schnupp,
  Winkler, Maischberger, Rudiger, and Schilling}}]{Schnupp:1985}
\bibinfo{author}{\bibnamefont{Schnupp}, \bibfnamefont{L.}},
  \bibinfo{author}{\bibfnamefont{W.}~\bibnamefont{Winkler}},
  \bibinfo{author}{\bibfnamefont{K.}~\bibnamefont{Maischberger}},
  \bibinfo{author}{\bibfnamefont{A.}~\bibnamefont{Rudiger}}, and
  \bibinfo{author}{\bibfnamefont{R.}~\bibnamefont{Schilling}},
  \bibinfo{year}{1985}, \bibinfo{journal}{Journal of Physics E: Scientific
  Instruments} \textbf{\bibinfo{volume}{18}}(\bibinfo{number}{6}),
  \bibinfo{pages}{482}.

\bibitem[{\citenamefont{Schofield}(2010)}]{Robert:Upconv}
\bibinfo{author}{\bibnamefont{Schofield}, \bibfnamefont{R.}},
  \bibinfo{year}{2010}, \bibinfo{howpublished}{personal communication}.

\bibitem[{\citenamefont{Shoemaker} \emph{et~al.}(1988)\citenamefont{Shoemaker,
  Schilling, Schnupp, Winkler, Maischberger, and R\"udiger}}]{DHS:Garching}
\bibinfo{author}{\bibnamefont{Shoemaker}, \bibfnamefont{D.}},
  \bibinfo{author}{\bibfnamefont{R.}~\bibnamefont{Schilling}},
  \bibinfo{author}{\bibfnamefont{L.}~\bibnamefont{Schnupp}},
  \bibinfo{author}{\bibfnamefont{W.}~\bibnamefont{Winkler}},
  \bibinfo{author}{\bibfnamefont{K.}~\bibnamefont{Maischberger}}, and
  \bibinfo{author}{\bibfnamefont{A.}~\bibnamefont{R\"udiger}},
  \bibinfo{year}{1988}, \bibinfo{journal}{Phys. Rev. D}
  \textbf{\bibinfo{volume}{38}}, \bibinfo{pages}{423}.

\bibitem[{\citenamefont{{Sidles} and {Sigg}}(2006)}]{Sidles:2006un}
\bibinfo{author}{\bibnamefont{{Sidles}}, \bibfnamefont{J.~A.}}, and
  \bibinfo{author}{\bibfnamefont{D.}~\bibnamefont{{Sigg}}},
  \bibinfo{year}{2006}, \bibinfo{journal}{Physics Letters A}
  \textbf{\bibinfo{volume}{354}}, \bibinfo{pages}{167}.

\bibitem[{\citenamefont{Siegman}(1986)}]{Siegman:Lasers}
\bibinfo{author}{\bibnamefont{Siegman}, \bibfnamefont{A.}},
  \bibinfo{year}{1986}, \emph{\bibinfo{title}{Lasers}}
  (\bibinfo{publisher}{University Science Books}), ISBN
  \bibinfo{isbn}{9780935702118}.

\bibitem[{\citenamefont{Smith-Lefebvre}
  \emph{et~al.}(2011)\citenamefont{Smith-Lefebvre, Ballmer, Evans, Waldman,
  Kawabe, Frolov, and Mavalvala}}]{Nic:2011}
\bibinfo{author}{\bibnamefont{Smith-Lefebvre}, \bibfnamefont{N.}},
  \bibinfo{author}{\bibfnamefont{S.}~\bibnamefont{Ballmer}},
  \bibinfo{author}{\bibfnamefont{M.}~\bibnamefont{Evans}},
  \bibinfo{author}{\bibfnamefont{S.}~\bibnamefont{Waldman}},
  \bibinfo{author}{\bibfnamefont{K.}~\bibnamefont{Kawabe}},
  \bibinfo{author}{\bibfnamefont{V.}~\bibnamefont{Frolov}}, and
  \bibinfo{author}{\bibfnamefont{N.}~\bibnamefont{Mavalvala}},
  \bibinfo{year}{2011}, \bibinfo{journal}{Opt. Lett.}
  \textbf{\bibinfo{volume}{36}}(\bibinfo{number}{22}), \bibinfo{pages}{4365},
  \urlprefix\url{http://ol.osa.org/abstract.cfm?URI=ol-36-22-4365}.

\bibitem[{\citenamefont{{Somiya}}(2011)}]{Somiya:2011tb}
\bibinfo{author}{\bibnamefont{{Somiya}}, \bibfnamefont{K.}},
  \bibinfo{year}{2011}, \bibinfo{journal}{ArXiv e-prints} \eprint{1111.7185}.

\bibitem[{\citenamefont{Somiya}(2012)}]{somiya2012detector}
\bibinfo{author}{\bibnamefont{Somiya}, \bibfnamefont{K.}},
  \bibinfo{year}{2012}, \bibinfo{journal}{Classical and Quantum Gravity}
  \textbf{\bibinfo{volume}{29}}(\bibinfo{number}{12}), \bibinfo{pages}{124007}.

\bibitem[{\citenamefont{{Starobinskii}}(1979)}]{Staro:1979}
\bibinfo{author}{\bibnamefont{{Starobinskii}}, \bibfnamefont{A.~A.}},
  \bibinfo{year}{1979}, \bibinfo{journal}{JETP Letters}
  \textbf{\bibinfo{volume}{30}}, \bibinfo{pages}{719}.

\bibitem[{\citenamefont{{Stefszky}}
  \emph{et~al.}(2012)\citenamefont{{Stefszky}, {Mow-Lowry}, {Y Chua},
  {Shaddock}, {Buchler}, {Vahlbruch}, {Khalaidovski}, {Schnabel}, {Lam}, and
  {McClelland}}}]{Stefszky:Balanced2012}
\bibinfo{author}{\bibnamefont{{Stefszky}}, \bibfnamefont{M.~S.}},
  \bibinfo{author}{\bibfnamefont{C.~M.} \bibnamefont{{Mow-Lowry}}},
  \bibinfo{author}{\bibfnamefont{S.~S.} \bibnamefont{{Y Chua}}},
  \bibinfo{author}{\bibfnamefont{D.~A.} \bibnamefont{{Shaddock}}},
  \bibinfo{author}{\bibfnamefont{B.~C.} \bibnamefont{{Buchler}}},
  \bibinfo{author}{\bibfnamefont{H.}~\bibnamefont{{Vahlbruch}}},
  \bibinfo{author}{\bibfnamefont{A.}~\bibnamefont{{Khalaidovski}}},
  \bibinfo{author}{\bibfnamefont{R.}~\bibnamefont{{Schnabel}}},
  \bibinfo{author}{\bibfnamefont{P.~K.} \bibnamefont{{Lam}}}, and
  \bibinfo{author}{\bibfnamefont{D.~E.} \bibnamefont{{McClelland}}},
  \bibinfo{year}{2012}, \bibinfo{journal}{Classical and Quantum Gravity}
  \textbf{\bibinfo{volume}{29}}(\bibinfo{number}{14}), \bibinfo{pages}{145015}.

\bibitem[{\citenamefont{Strain}(2012)}]{Ken:GEOseismic}
\bibinfo{author}{\bibnamefont{Strain}, \bibfnamefont{K.~A.}},
  \bibinfo{year}{2012}, \bibinfo{howpublished}{personal communication}.

\bibitem[{\citenamefont{Strain} \emph{et~al.}(1994)\citenamefont{Strain,
  Danzmann, Mizuno, Nelson, R{\"u}diger, Schilling, and Winkler}}]{Strain:TCS}
\bibinfo{author}{\bibnamefont{Strain}, \bibfnamefont{K.~A.}},
  \bibinfo{author}{\bibfnamefont{K.}~\bibnamefont{Danzmann}},
  \bibinfo{author}{\bibfnamefont{J.}~\bibnamefont{Mizuno}},
  \bibinfo{author}{\bibfnamefont{P.~G.} \bibnamefont{Nelson}},
  \bibinfo{author}{\bibfnamefont{A.}~\bibnamefont{R{\"u}diger}},
  \bibinfo{author}{\bibfnamefont{R.}~\bibnamefont{Schilling}}, and
  \bibinfo{author}{\bibfnamefont{W.}~\bibnamefont{Winkler}},
  \bibinfo{year}{1994}, \bibinfo{journal}{Physics Letters A}
  \textbf{\bibinfo{volume}{194}}(\bibinfo{number}{1}), \bibinfo{pages}{124}.

\bibitem[{\citenamefont{Strain and Meers}(1991)}]{StMe1991}
\bibinfo{author}{\bibnamefont{Strain}, \bibfnamefont{K.~A.}}, and
  \bibinfo{author}{\bibfnamefont{B.~J.} \bibnamefont{Meers}},
  \bibinfo{year}{1991}, \bibinfo{journal}{Phys. Rev. Lett.}
  \textbf{\bibinfo{volume}{66}}, \bibinfo{pages}{1391},
  \urlprefix\url{http://link.aps.org/doi/10.1103/PhysRevLett.66.1391}.

\bibitem[{\citenamefont{{Strain}} \emph{et~al.}(2003)\citenamefont{{Strain},
  {M{\"u}ller}, {Delker}, {Reitze}, {Tanner}, {Mason}, {Willems}, {Shaddock},
  {Gray}, {Mow-Lowry}, and {McClelland}}}]{SensingStrain:2003}
\bibinfo{author}{\bibnamefont{{Strain}}, \bibfnamefont{K.~A.}},
  \bibinfo{author}{\bibfnamefont{G.}~\bibnamefont{{M{\"u}ller}}},
  \bibinfo{author}{\bibfnamefont{T.}~\bibnamefont{{Delker}}},
  \bibinfo{author}{\bibfnamefont{D.~H.} \bibnamefont{{Reitze}}},
  \bibinfo{author}{\bibfnamefont{D.~B.} \bibnamefont{{Tanner}}},
  \bibinfo{author}{\bibfnamefont{J.~E.} \bibnamefont{{Mason}}},
  \bibinfo{author}{\bibfnamefont{P.~A.} \bibnamefont{{Willems}}},
  \bibinfo{author}{\bibfnamefont{D.~A.} \bibnamefont{{Shaddock}}},
  \bibinfo{author}{\bibfnamefont{M.~B.} \bibnamefont{{Gray}}},
  \bibinfo{author}{\bibfnamefont{C.}~\bibnamefont{{Mow-Lowry}}}, and
  \bibinfo{author}{\bibfnamefont{D.~E.} \bibnamefont{{McClelland}}},
  \bibinfo{year}{2003}, \bibinfo{journal}{Applied Optics}
  \textbf{\bibinfo{volume}{42}}, \bibinfo{pages}{1244}.

\bibitem[{\citenamefont{{Strigin} and {Vyatchanin}}(2007)}]{StVy2007}
\bibinfo{author}{\bibnamefont{{Strigin}}, \bibfnamefont{S.~E.}}, and
  \bibinfo{author}{\bibfnamefont{S.~P.} \bibnamefont{{Vyatchanin}}},
  \bibinfo{year}{2007}, \bibinfo{journal}{Physics Letters A}
  \textbf{\bibinfo{volume}{365}}, \bibinfo{pages}{10}.

\bibitem[{\citenamefont{{Sun}} \emph{et~al.}(2006)\citenamefont{{Sun},
  {Allard}, {Buchman}, {Williams}, and {Byer}}}]{Sun:2006}
\bibinfo{author}{\bibnamefont{{Sun}}, \bibfnamefont{K.-X.}},
  \bibinfo{author}{\bibfnamefont{B.}~\bibnamefont{{Allard}}},
  \bibinfo{author}{\bibfnamefont{S.}~\bibnamefont{{Buchman}}},
  \bibinfo{author}{\bibfnamefont{S.}~\bibnamefont{{Williams}}}, and
  \bibinfo{author}{\bibfnamefont{R.~L.} \bibnamefont{{Byer}}},
  \bibinfo{year}{2006}, \bibinfo{journal}{Classical and Quantum Gravity}
  \textbf{\bibinfo{volume}{23}}, \bibinfo{pages}{141}.

\bibitem[{\citenamefont{Sun} \emph{et~al.}(1996)\citenamefont{Sun, Fejer,
  Gustafson, and Byer}}]{Stanford:Sagnac1996}
\bibinfo{author}{\bibnamefont{Sun}, \bibfnamefont{K.-X.}},
  \bibinfo{author}{\bibfnamefont{M.~M.} \bibnamefont{Fejer}},
  \bibinfo{author}{\bibfnamefont{E.}~\bibnamefont{Gustafson}}, and
  \bibinfo{author}{\bibfnamefont{R.~L.} \bibnamefont{Byer}},
  \bibinfo{year}{1996}, \bibinfo{journal}{Phys. Rev. Lett.}
  \textbf{\bibinfo{volume}{76}}, \bibinfo{pages}{3053}.

\bibitem[{\citenamefont{{Sykora} and {de Groot}}(2011)}]{Zygo:2011}
\bibinfo{author}{\bibnamefont{{Sykora}}, \bibfnamefont{D.~M.}}, and
  \bibinfo{author}{\bibfnamefont{P.}~\bibnamefont{{de Groot}}},
  \bibinfo{year}{2011}, in \emph{\bibinfo{booktitle}{Society of Photo-Optical
  Instrumentation Engineers (SPIE) Conference Series}}, volume
  \bibinfo{volume}{8126} of \emph{\bibinfo{series}{Society of Photo-Optical
  Instrumentation Engineers (SPIE) Conference Series}}.

\bibitem[{\citenamefont{Szil\'agyi}
  \emph{et~al.}(2009)\citenamefont{Szil\'agyi, Lindblom, and
  Scheel}}]{Bela:2009Sim}
\bibinfo{author}{\bibnamefont{Szil\'agyi}, \bibfnamefont{B.}},
  \bibinfo{author}{\bibfnamefont{L.}~\bibnamefont{Lindblom}}, and
  \bibinfo{author}{\bibfnamefont{M.~A.} \bibnamefont{Scheel}},
  \bibinfo{year}{2009}, \bibinfo{journal}{Phys. Rev. D}
  \textbf{\bibinfo{volume}{80}}, \bibinfo{pages}{124010},
  \urlprefix\url{http://link.aps.org/doi/10.1103/PhysRevD.80.124010}.

\bibitem[{\citenamefont{Takahashi}(2012)}]{Takahashi:email}
\bibinfo{author}{\bibnamefont{Takahashi}, \bibfnamefont{R.}},
  \bibinfo{year}{2012}, \bibinfo{howpublished}{personal communication}.

\bibitem[{\citenamefont{Takahashi} \emph{et~al.}(2008)\citenamefont{Takahashi,
  Arai, Tatsumi, Fukushima, Yamazaki, Fujimoto, Agatsuma, Arase, Nakagawa,
  Takamori, Tsubono, DeSalvo} \emph{et~al.}}]{TAMA:2008}
\bibinfo{author}{\bibnamefont{Takahashi}, \bibfnamefont{R.}},
  \bibinfo{author}{\bibfnamefont{K.}~\bibnamefont{Arai}},
  \bibinfo{author}{\bibfnamefont{D.}~\bibnamefont{Tatsumi}},
  \bibinfo{author}{\bibfnamefont{M.}~\bibnamefont{Fukushima}},
  \bibinfo{author}{\bibfnamefont{T.}~\bibnamefont{Yamazaki}},
  \bibinfo{author}{\bibfnamefont{M.-K.} \bibnamefont{Fujimoto}},
  \bibinfo{author}{\bibfnamefont{K.}~\bibnamefont{Agatsuma}},
  \bibinfo{author}{\bibfnamefont{Y.}~\bibnamefont{Arase}},
  \bibinfo{author}{\bibfnamefont{N.}~\bibnamefont{Nakagawa}},
  \bibinfo{author}{\bibfnamefont{A.}~\bibnamefont{Takamori}},
  \bibinfo{author}{\bibfnamefont{K.}~\bibnamefont{Tsubono}},
  \bibinfo{author}{\bibfnamefont{R.}~\bibnamefont{DeSalvo}}, \emph{et~al.},
  \bibinfo{year}{2008}, \bibinfo{journal}{Classical and Quantum Gravity}
  \textbf{\bibinfo{volume}{25}}(\bibinfo{number}{11}), \bibinfo{pages}{114036}.

\bibitem[{\citenamefont{Takahashi} \emph{et~al.}(2002)\citenamefont{Takahashi,
  Saito, Fukushima, Ando, Arai, Tatsumi, Heinzel, Kawamura, Yamazaki, and
  Moriwaki}}]{TAMA:Gas}
\bibinfo{author}{\bibnamefont{Takahashi}, \bibfnamefont{R.}},
  \bibinfo{author}{\bibfnamefont{Y.}~\bibnamefont{Saito}},
  \bibinfo{author}{\bibfnamefont{M.}~\bibnamefont{Fukushima}},
  \bibinfo{author}{\bibfnamefont{M.}~\bibnamefont{Ando}},
  \bibinfo{author}{\bibfnamefont{K.}~\bibnamefont{Arai}},
  \bibinfo{author}{\bibfnamefont{D.}~\bibnamefont{Tatsumi}},
  \bibinfo{author}{\bibfnamefont{G.}~\bibnamefont{Heinzel}},
  \bibinfo{author}{\bibfnamefont{S.}~\bibnamefont{Kawamura}},
  \bibinfo{author}{\bibfnamefont{T.}~\bibnamefont{Yamazaki}}, and
  \bibinfo{author}{\bibfnamefont{S.}~\bibnamefont{Moriwaki}},
  \bibinfo{year}{2002}, \bibinfo{journal}{Journal of Vacuum Science and
  Technology A: Vacuum, Surfaces, and Films} .

\bibitem[{\citenamefont{Tarallo} \emph{et~al.}(2007)\citenamefont{Tarallo,
  Miller, Agresti, D'Ambrosio, DeSalvo, Forest, Lagrange, Mackowsky, Michel,
  Montorio, Morgado, Pinard} \emph{et~al.}}]{JM:Mesa2007}
\bibinfo{author}{\bibnamefont{Tarallo}, \bibfnamefont{M.~G.}},
  \bibinfo{author}{\bibfnamefont{J.}~\bibnamefont{Miller}},
  \bibinfo{author}{\bibfnamefont{J.}~\bibnamefont{Agresti}},
  \bibinfo{author}{\bibfnamefont{E.}~\bibnamefont{D'Ambrosio}},
  \bibinfo{author}{\bibfnamefont{R.}~\bibnamefont{DeSalvo}},
  \bibinfo{author}{\bibfnamefont{D.}~\bibnamefont{Forest}},
  \bibinfo{author}{\bibfnamefont{B.}~\bibnamefont{Lagrange}},
  \bibinfo{author}{\bibfnamefont{J.~M.} \bibnamefont{Mackowsky}},
  \bibinfo{author}{\bibfnamefont{C.}~\bibnamefont{Michel}},
  \bibinfo{author}{\bibfnamefont{J.~L.} \bibnamefont{Montorio}},
  \bibinfo{author}{\bibfnamefont{N.}~\bibnamefont{Morgado}},
  \bibinfo{author}{\bibfnamefont{L.}~\bibnamefont{Pinard}}, \emph{et~al.},
  \bibinfo{year}{2007}, \bibinfo{journal}{Appl. Opt.}
  \textbf{\bibinfo{volume}{46}}(\bibinfo{number}{26}), \bibinfo{pages}{6648}.

\bibitem[{\citenamefont{Tatsumi}(2008)}]{Tat2008}
\bibinfo{author}{\bibnamefont{Tatsumi}, \bibfnamefont{D.}},
  \bibinfo{year}{2008}, \bibinfo{journal}{J.~Phys.~Conf.}
  \textbf{\bibinfo{volume}{120}}, \bibinfo{pages}{032011}.

\bibitem[{\citenamefont{{Tatsumi}} \emph{et~al.}(2006)\citenamefont{{Tatsumi},
  {Arai}, and {TAMA Collaboration}}}]{Koji:glitch2006}
\bibinfo{author}{\bibnamefont{{Tatsumi}}, \bibfnamefont{D.}},
  \bibinfo{author}{\bibfnamefont{K.}~\bibnamefont{{Arai}}}, and
  \bibinfo{author}{\bibnamefont{{TAMA Collaboration}}}, \bibinfo{year}{2006},
  \bibinfo{journal}{Journal of Physics Conference Series}
  \textbf{\bibinfo{volume}{32}}, \bibinfo{pages}{94}.

\bibitem[{\citenamefont{Taylor} \emph{et~al.}(1979)\citenamefont{Taylor,
  Fowler, and McCulloch}}]{TFM1979}
\bibinfo{author}{\bibnamefont{Taylor}, \bibfnamefont{J.~H.}},
  \bibinfo{author}{\bibfnamefont{L.~A.} \bibnamefont{Fowler}}, and
  \bibinfo{author}{\bibfnamefont{P.~M.} \bibnamefont{McCulloch}},
  \bibinfo{year}{1979}, \bibinfo{journal}{Nature}
  \textbf{\bibinfo{volume}{277}}, \bibinfo{pages}{437}.

\bibitem[{\citenamefont{{The Virgo Collaboration}}(2008)}]{Virgo:FI2008}
\bibinfo{author}{\bibnamefont{{The Virgo Collaboration}}},
  \bibinfo{year}{2008}, \bibinfo{journal}{Appl. Opt.}
  \textbf{\bibinfo{volume}{47}}(\bibinfo{number}{31}), \bibinfo{pages}{5853},
  \urlprefix\url{http://ao.osa.org/abstract.cfm?URI=ao-47-31-5853}.

\bibitem[{\citenamefont{{The Virgo Collaboration}}(2009)}]{aVir2009}
\bibinfo{author}{\bibnamefont{{The Virgo Collaboration}}},
  \bibinfo{year}{2009}, \emph{\bibinfo{title}{{Advanced Virgo Baseline
  Design}}}, \bibinfo{type}{Technical Report} \bibinfo{number}{{VIR 027A 09}},
  \bibinfo{institution}{Virgo},
  \urlprefix\url{https://wwwcascina.virgo.infn.it/advirgo/}.

\bibitem[{\citenamefont{{The Virgo Collaboration}}(2010)}]{Virgo:FI2010}
\bibinfo{author}{\bibnamefont{{The Virgo Collaboration}}},
  \bibinfo{year}{2010}, \bibinfo{journal}{Appl. Opt.}
  \textbf{\bibinfo{volume}{49}}(\bibinfo{number}{25}), \bibinfo{pages}{4780},
  \urlprefix\url{http://dx.doi.org/10.1364/ao.49.004780}.

\bibitem[{\citenamefont{{The Virgo Collaboration}}(2012)}]{aVirgo:TDR}
\bibinfo{author}{\bibnamefont{{The Virgo Collaboration}}},
  \bibinfo{year}{2012}, \emph{\bibinfo{title}{Advanced Virgo Technical Design
  Report}}, \bibinfo{type}{Technical Report},
  \urlprefix\url{https://tds.ego-gw.it/ql/?c=8940}.

\bibitem[{\citenamefont{Thorne}(1989)}]{Kip:scatter1989}
\bibinfo{author}{\bibnamefont{Thorne}, \bibfnamefont{K.~S.}},
  \bibinfo{year}{1989}, \emph{\bibinfo{title}{Light Scattering and Proposed
  Baffle Configuration for the {LIGO}}}, \bibinfo{type}{LIGO Technical Report},
  \bibinfo{institution}{Caltech},
  \bibinfo{address}{\url{http://www.ligo.caltech.edu/docs/T/T890017-00.pdf}}.

\bibitem[{\citenamefont{Thorne and Winstein}(1999)}]{Kip:Dancing}
\bibinfo{author}{\bibnamefont{Thorne}, \bibfnamefont{K.~S.}}, and
  \bibinfo{author}{\bibfnamefont{C.~J.} \bibnamefont{Winstein}},
  \bibinfo{year}{1999}, \bibinfo{journal}{Phys. Rev. D}
  \textbf{\bibinfo{volume}{60}}, \bibinfo{pages}{082001}.

\bibitem[{\citenamefont{Thorpe} \emph{et~al.}(2010)\citenamefont{Thorpe,
  Leibrandt, Fortier, and Rosenband}}]{NIST:Wiener}
\bibinfo{author}{\bibnamefont{Thorpe}, \bibfnamefont{M.~J.}},
  \bibinfo{author}{\bibfnamefont{D.~R.} \bibnamefont{Leibrandt}},
  \bibinfo{author}{\bibfnamefont{T.~M.} \bibnamefont{Fortier}}, and
  \bibinfo{author}{\bibfnamefont{T.}~\bibnamefont{Rosenband}},
  \bibinfo{year}{2010}, \bibinfo{journal}{Opt. Express}
  \textbf{\bibinfo{volume}{18}}(\bibinfo{number}{18}), \bibinfo{pages}{18744}.

\bibitem[{\citenamefont{Tinto and Dhurandhar}(2005)}]{lrr-2005-4}
\bibinfo{author}{\bibnamefont{Tinto}, \bibfnamefont{M.}}, and
  \bibinfo{author}{\bibfnamefont{S.~V.} \bibnamefont{Dhurandhar}},
  \bibinfo{year}{2005}, \bibinfo{journal}{Living Reviews in Relativity}
  \textbf{\bibinfo{volume}{8}}(\bibinfo{number}{4}),
  \urlprefix\url{http://www.livingreviews.org/lrr-2005-4}.

\bibitem[{\citenamefont{{Tomaru}} \emph{et~al.}(2002)\citenamefont{{Tomaru},
  {Suzuki}, {Uchiyama}, {Yamamoto}, {Shintomi}, {Taylor}, {Yamamoto}, {Miyoki},
  {Ohashi}, and {Kuroda}}}]{LCGT:heatextract}
\bibinfo{author}{\bibnamefont{{Tomaru}}, \bibfnamefont{T.}},
  \bibinfo{author}{\bibfnamefont{T.}~\bibnamefont{{Suzuki}}},
  \bibinfo{author}{\bibfnamefont{T.}~\bibnamefont{{Uchiyama}}},
  \bibinfo{author}{\bibfnamefont{A.}~\bibnamefont{{Yamamoto}}},
  \bibinfo{author}{\bibfnamefont{T.}~\bibnamefont{{Shintomi}}},
  \bibinfo{author}{\bibfnamefont{C.~T.} \bibnamefont{{Taylor}}},
  \bibinfo{author}{\bibfnamefont{K.}~\bibnamefont{{Yamamoto}}},
  \bibinfo{author}{\bibfnamefont{S.}~\bibnamefont{{Miyoki}}},
  \bibinfo{author}{\bibfnamefont{M.}~\bibnamefont{{Ohashi}}}, and
  \bibinfo{author}{\bibfnamefont{K.}~\bibnamefont{{Kuroda}}},
  \bibinfo{year}{2002}, \bibinfo{journal}{Physics Letters A}
  \textbf{\bibinfo{volume}{301}}, \bibinfo{pages}{215}.

\bibitem[{\citenamefont{Turner}(1997)}]{Turner:1997cs}
\bibinfo{author}{\bibnamefont{Turner}, \bibfnamefont{M.~S.}},
  \bibinfo{year}{1997}, \bibinfo{journal}{Phys. Rev. D}
  \textbf{\bibinfo{volume}{55}}, \bibinfo{pages}{R435},
  \urlprefix\url{http://link.aps.org/doi/10.1103/PhysRevD.55.R435}.

\bibitem[{\citenamefont{{Tyson} and {Giffard}}(1978)}]{Giffard:1978}
\bibinfo{author}{\bibnamefont{{Tyson}}, \bibfnamefont{J.~A.}}, and
  \bibinfo{author}{\bibfnamefont{R.~P.} \bibnamefont{{Giffard}}},
  \bibinfo{year}{1978}, \bibinfo{journal}{Annual Review of Astronomy and
  Astrophysics} \textbf{\bibinfo{volume}{16}}, \bibinfo{pages}{521}.

\bibitem[{\citenamefont{Uchiyama} \emph{et~al.}(1998)\citenamefont{Uchiyama,
  Tatsumi, Tomaru, Tobar, Kuroda, Suzuki, Sato, Yamamoto, Haruyama, and
  Shintomi}}]{uchiyama1998cryogenic}
\bibinfo{author}{\bibnamefont{Uchiyama}, \bibfnamefont{T.}},
  \bibinfo{author}{\bibfnamefont{D.}~\bibnamefont{Tatsumi}},
  \bibinfo{author}{\bibfnamefont{T.}~\bibnamefont{Tomaru}},
  \bibinfo{author}{\bibfnamefont{M.}~\bibnamefont{Tobar}},
  \bibinfo{author}{\bibfnamefont{K.}~\bibnamefont{Kuroda}},
  \bibinfo{author}{\bibfnamefont{T.}~\bibnamefont{Suzuki}},
  \bibinfo{author}{\bibfnamefont{N.}~\bibnamefont{Sato}},
  \bibinfo{author}{\bibfnamefont{A.}~\bibnamefont{Yamamoto}},
  \bibinfo{author}{\bibfnamefont{T.}~\bibnamefont{Haruyama}}, and
  \bibinfo{author}{\bibfnamefont{T.}~\bibnamefont{Shintomi}},
  \bibinfo{year}{1998}, \bibinfo{journal}{Physics Letters A}
  \textbf{\bibinfo{volume}{242}}(\bibinfo{number}{4}), \bibinfo{pages}{211}.

\bibitem[{\citenamefont{Uchiyama} \emph{et~al.}(1999)\citenamefont{Uchiyama,
  Tomaru, Tobar, Tatsumi, Miyoki, Ohashi, Kuroda, Suzuki, Sato, Haruyama}
  \emph{et~al.}}]{uchiyama1999mechanical}
\bibinfo{author}{\bibnamefont{Uchiyama}, \bibfnamefont{T.}},
  \bibinfo{author}{\bibfnamefont{T.}~\bibnamefont{Tomaru}},
  \bibinfo{author}{\bibfnamefont{M.}~\bibnamefont{Tobar}},
  \bibinfo{author}{\bibfnamefont{D.}~\bibnamefont{Tatsumi}},
  \bibinfo{author}{\bibfnamefont{S.}~\bibnamefont{Miyoki}},
  \bibinfo{author}{\bibfnamefont{M.}~\bibnamefont{Ohashi}},
  \bibinfo{author}{\bibfnamefont{K.}~\bibnamefont{Kuroda}},
  \bibinfo{author}{\bibfnamefont{T.}~\bibnamefont{Suzuki}},
  \bibinfo{author}{\bibfnamefont{N.}~\bibnamefont{Sato}},
  \bibinfo{author}{\bibfnamefont{T.}~\bibnamefont{Haruyama}}, \emph{et~al.},
  \bibinfo{year}{1999}, \bibinfo{journal}{Physics Letters A}
  \textbf{\bibinfo{volume}{261}}(\bibinfo{number}{1}), \bibinfo{pages}{5}.

\bibitem[{\citenamefont{Uehara} \emph{et~al.}(1995)\citenamefont{Uehara, Ueda,
  Ueda, Sekiguchi, Mitake, Nakamura, Kitajima, and
  Kataoka}}]{uehara1995ultralow}
\bibinfo{author}{\bibnamefont{Uehara}, \bibfnamefont{N.}},
  \bibinfo{author}{\bibfnamefont{A.}~\bibnamefont{Ueda}},
  \bibinfo{author}{\bibfnamefont{K.}~\bibnamefont{Ueda}},
  \bibinfo{author}{\bibfnamefont{H.}~\bibnamefont{Sekiguchi}},
  \bibinfo{author}{\bibfnamefont{T.}~\bibnamefont{Mitake}},
  \bibinfo{author}{\bibfnamefont{K.}~\bibnamefont{Nakamura}},
  \bibinfo{author}{\bibfnamefont{N.}~\bibnamefont{Kitajima}}, and
  \bibinfo{author}{\bibfnamefont{I.}~\bibnamefont{Kataoka}},
  \bibinfo{year}{1995}, \bibinfo{journal}{Optics letters}
  \textbf{\bibinfo{volume}{20}}(\bibinfo{number}{6}), \bibinfo{pages}{530}.

\bibitem[{\citenamefont{{Ugolini}} \emph{et~al.}(2008)\citenamefont{{Ugolini},
  {Girard}, {Harry}, and {Mitrofanov}}}]{Ugolini:2008}
\bibinfo{author}{\bibnamefont{{Ugolini}}, \bibfnamefont{D.}},
  \bibinfo{author}{\bibfnamefont{M.}~\bibnamefont{{Girard}}},
  \bibinfo{author}{\bibfnamefont{G.~M.} \bibnamefont{{Harry}}}, and
  \bibinfo{author}{\bibfnamefont{V.~P.} \bibnamefont{{Mitrofanov}}},
  \bibinfo{year}{2008}, \bibinfo{journal}{Physics Letters A}
  \textbf{\bibinfo{volume}{372}}, \bibinfo{pages}{5741}.

\bibitem[{\citenamefont{{Ushomirsky}}
  \emph{et~al.}(2000)\citenamefont{{Ushomirsky}, {Cutler}, and
  {Bildsten}}}]{Bildsten:2000}
\bibinfo{author}{\bibnamefont{{Ushomirsky}}, \bibfnamefont{G.}},
  \bibinfo{author}{\bibfnamefont{C.}~\bibnamefont{{Cutler}}}, and
  \bibinfo{author}{\bibfnamefont{L.}~\bibnamefont{{Bildsten}}},
  \bibinfo{year}{2000}, \bibinfo{journal}{Monthly Notices of the Royal
  Astronomical Society} \textbf{\bibinfo{volume}{319}}, \bibinfo{pages}{902}.

\bibitem[{\citenamefont{Vahlbruch} \emph{et~al.}(2007)\citenamefont{Vahlbruch,
  Chelkowski, Danzmann, and Schnabel}}]{Vahlbruch:2007da}
\bibinfo{author}{\bibnamefont{Vahlbruch}, \bibfnamefont{H.}},
  \bibinfo{author}{\bibfnamefont{S.}~\bibnamefont{Chelkowski}},
  \bibinfo{author}{\bibfnamefont{K.}~\bibnamefont{Danzmann}}, and
  \bibinfo{author}{\bibfnamefont{R.}~\bibnamefont{Schnabel}},
  \bibinfo{year}{2007}, \bibinfo{journal}{New Journal of Physics}
  \textbf{\bibinfo{volume}{9}}(\bibinfo{number}{10}), \bibinfo{pages}{371}.

\bibitem[{\citenamefont{Vahlbruch} \emph{et~al.}(2006)\citenamefont{Vahlbruch,
  Chelkowski, Hage, Franzen, Danzmann, and Schnabel}}]{Henning:PRL2006}
\bibinfo{author}{\bibnamefont{Vahlbruch}, \bibfnamefont{H.}},
  \bibinfo{author}{\bibfnamefont{S.}~\bibnamefont{Chelkowski}},
  \bibinfo{author}{\bibfnamefont{B.}~\bibnamefont{Hage}},
  \bibinfo{author}{\bibfnamefont{A.}~\bibnamefont{Franzen}},
  \bibinfo{author}{\bibfnamefont{K.}~\bibnamefont{Danzmann}}, and
  \bibinfo{author}{\bibfnamefont{R.}~\bibnamefont{Schnabel}},
  \bibinfo{year}{2006}, \bibinfo{journal}{Phys. Rev. Lett.}
  \textbf{\bibinfo{volume}{97}}, \bibinfo{pages}{011101},
  \urlprefix\url{http://link.aps.org/doi/10.1103/PhysRevLett.97.011101}.

\bibitem[{\citenamefont{Van~Haasteren}
  \emph{et~al.}(2009)\citenamefont{Van~Haasteren, Levin, McDonald, and
  Lu}}]{Yuri:PulsarTiming2009}
\bibinfo{author}{\bibnamefont{Van~Haasteren}, \bibfnamefont{R.}},
  \bibinfo{author}{\bibfnamefont{Y.}~\bibnamefont{Levin}},
  \bibinfo{author}{\bibfnamefont{P.}~\bibnamefont{McDonald}}, and
  \bibinfo{author}{\bibfnamefont{T.}~\bibnamefont{Lu}}, \bibinfo{year}{2009},
  \bibinfo{journal}{Monthly Notices of the Royal Astronomical Society}
  \textbf{\bibinfo{volume}{395}}(\bibinfo{number}{2}), \bibinfo{pages}{1005},
  ISSN \bibinfo{issn}{1365-2966},
  \urlprefix\url{http://dx.doi.org/10.1111/j.1365-2966.2009.14590.x}.

\bibitem[{\citenamefont{Varvella} \emph{et~al.}(2004)\citenamefont{Varvella,
  Calloni, Fiore, Milano, and Arnaud}}]{Monica:Magnetic}
\bibinfo{author}{\bibnamefont{Varvella}, \bibfnamefont{M.}},
  \bibinfo{author}{\bibfnamefont{E.}~\bibnamefont{Calloni}},
  \bibinfo{author}{\bibfnamefont{L.~D.} \bibnamefont{Fiore}},
  \bibinfo{author}{\bibfnamefont{L.}~\bibnamefont{Milano}}, and
  \bibinfo{author}{\bibfnamefont{N.}~\bibnamefont{Arnaud}},
  \bibinfo{year}{2004}, \bibinfo{journal}{Astroparticle Physics}
  \textbf{\bibinfo{volume}{21}}(\bibinfo{number}{3}), \bibinfo{pages}{325 },
  ISSN \bibinfo{issn}{0927-6505},
  \urlprefix\url{http://www.sciencedirect.com/science/article/pii/S0927650504000180}.

\bibitem[{\citenamefont{de~Vine} \emph{et~al.}(2010)\citenamefont{de~Vine,
  Ware, McKenzie, Spero, Klipstein, and Shaddock}}]{Glenn:TDI2010}
\bibinfo{author}{\bibnamefont{de~Vine}, \bibfnamefont{G.}},
  \bibinfo{author}{\bibfnamefont{B.}~\bibnamefont{Ware}},
  \bibinfo{author}{\bibfnamefont{K.}~\bibnamefont{McKenzie}},
  \bibinfo{author}{\bibfnamefont{R.~E.} \bibnamefont{Spero}},
  \bibinfo{author}{\bibfnamefont{W.~M.} \bibnamefont{Klipstein}}, and
  \bibinfo{author}{\bibfnamefont{D.~A.} \bibnamefont{Shaddock}},
  \bibinfo{year}{2010}, \bibinfo{journal}{Phys. Rev. Lett.}
  \textbf{\bibinfo{volume}{104}}, \bibinfo{pages}{211103},
  \urlprefix\url{http://link.aps.org/doi/10.1103/PhysRevLett.104.211103}.

\bibitem[{\citenamefont{Vinet}(2010)}]{Vinet:LG}
\bibinfo{author}{\bibnamefont{Vinet}, \bibfnamefont{J.-Y.}},
  \bibinfo{year}{2010}, \bibinfo{journal}{Phys. Rev. D}
  \textbf{\bibinfo{volume}{82}}, \bibinfo{pages}{042003}.

\bibitem[{\citenamefont{Vinet} \emph{et~al.}(1996)\citenamefont{Vinet, Brisson,
  and Braccini}}]{Stefano:scatter}
\bibinfo{author}{\bibnamefont{Vinet}, \bibfnamefont{J.-Y.}},
  \bibinfo{author}{\bibfnamefont{V.}~\bibnamefont{Brisson}}, and
  \bibinfo{author}{\bibfnamefont{S.}~\bibnamefont{Braccini}},
  \bibinfo{year}{1996}, \bibinfo{journal}{Phys. Rev. D}
  \textbf{\bibinfo{volume}{54}}, \bibinfo{pages}{1276}.

\bibitem[{\citenamefont{Volland}(1995)}]{AE:handbook}
\bibinfo{author}{\bibnamefont{Volland}, \bibfnamefont{H.}},
  \bibinfo{year}{1995}, \emph{\bibinfo{title}{Handbook of Atmospheric
  Electrodynamics}}, volume~\bibinfo{volume}{1} of
  \emph{\bibinfo{series}{Handbook of Atmospheric Electrodynamics}}
  (\bibinfo{publisher}{CRC Press}, \bibinfo{address}{Boca Raton, FL}), ISBN
  \bibinfo{isbn}{9780849386473}.

\bibitem[{\citenamefont{Vorvick}(2012)}]{Cheryl:Picture}
\bibinfo{author}{\bibnamefont{Vorvick}, \bibfnamefont{C.}},
  \bibinfo{year}{2012}, \bibinfo{howpublished}{personal communication}.

\bibitem[{\citenamefont{Vyatchanin and Matsko}(1996)}]{VyMa1996a}
\bibinfo{author}{\bibnamefont{Vyatchanin}, \bibfnamefont{S.~P.}}, and
  \bibinfo{author}{\bibfnamefont{A.~B.} \bibnamefont{Matsko}},
  \bibinfo{year}{1996}, \bibinfo{journal}{JETP} \textbf{\bibinfo{volume}{82}},
  \bibinfo{pages}{1007}.

\bibitem[{\citenamefont{Wahlquist}(1987)}]{Wah1987}
\bibinfo{author}{\bibnamefont{Wahlquist}, \bibfnamefont{H.}},
  \bibinfo{year}{1987}, \bibinfo{journal}{Gen.~Rel.~Gravitation}
  \textbf{\bibinfo{volume}{19}}, \bibinfo{pages}{1101}.

\bibitem[{\citenamefont{Walsh} \emph{et~al.}(1999)\citenamefont{Walsh,
  Leistner, and Oreb}}]{Walsh:99}
\bibinfo{author}{\bibnamefont{Walsh}, \bibfnamefont{C.~J.}},
  \bibinfo{author}{\bibfnamefont{A.~J.} \bibnamefont{Leistner}}, and
  \bibinfo{author}{\bibfnamefont{B.~F.} \bibnamefont{Oreb}},
  \bibinfo{year}{1999}, \bibinfo{journal}{Appl. Opt.}
  \textbf{\bibinfo{volume}{38}}(\bibinfo{number}{22}), \bibinfo{pages}{4790},
  \urlprefix\url{http://ao.osa.org/abstract.cfm?URI=ao-38-22-4790}.

\bibitem[{\citenamefont{Webb}(1992)}]{uSeism:1992}
\bibinfo{author}{\bibnamefont{Webb}, \bibfnamefont{S.}}, \bibinfo{year}{1992},
  \bibinfo{journal}{The Journal of the Acoustical Society of America}
  \textbf{\bibinfo{volume}{92}}, \bibinfo{pages}{2141}.

\bibitem[{\citenamefont{Weber}(1960)}]{Weber:GW}
\bibinfo{author}{\bibnamefont{Weber}, \bibfnamefont{J.}}, \bibinfo{year}{1960},
  \bibinfo{journal}{Phys. Rev.} \textbf{\bibinfo{volume}{117}},
  \bibinfo{pages}{306}.

\bibitem[{\citenamefont{Weber}(1970)}]{Weber:Bars}
\bibinfo{author}{\bibnamefont{Weber}, \bibfnamefont{J.}}, \bibinfo{year}{1970},
  \bibinfo{journal}{Phys. Rev. Lett.} \textbf{\bibinfo{volume}{24}},
  \bibinfo{pages}{276}.

\bibitem[{\citenamefont{Weinberg}(2004)}]{Weinberg:GW2003}
\bibinfo{author}{\bibnamefont{Weinberg}, \bibfnamefont{S.}},
  \bibinfo{year}{2004}, \bibinfo{journal}{Phys. Rev. D}
  \textbf{\bibinfo{volume}{69}}, \bibinfo{pages}{023503},
  \urlprefix\url{http://link.aps.org/doi/10.1103/PhysRevD.69.023503}.

\bibitem[{\citenamefont{Weiss}(1972)}]{Rai:QPR}
\bibinfo{author}{\bibnamefont{Weiss}, \bibfnamefont{R.}}, \bibinfo{year}{1972},
  \emph{\bibinfo{title}{Electromagnetically Coupled Broadband Gravitational
  Antenna}}, \bibinfo{type}{Technical Report},
  \bibinfo{institution}{Massachusetts Institute of Technology},
  \urlprefix\url{{https://dcc.ligo.org/cgi-bin/DocDB/ShowDocument?docid=38618}}.

\bibitem[{\citenamefont{Weiss}(1999)}]{Rai:RMP}
\bibinfo{author}{\bibnamefont{Weiss}, \bibfnamefont{R.}}, \bibinfo{year}{1999},
  \bibinfo{journal}{Rev. Mod. Phys.} \textbf{\bibinfo{volume}{71}},
  \bibinfo{pages}{S187}.

\bibitem[{\citenamefont{Wenzel}(2012)}]{Wenzel:web}
\bibinfo{author}{\bibnamefont{Wenzel}}, \bibinfo{year}{2012},
  \urlprefix\url{http://www.wenzel.com/oscillators.htm}.

\bibitem[{\citenamefont{Willemenot and Touboul}(2000)}]{onera:2000}
\bibinfo{author}{\bibnamefont{Willemenot}, \bibfnamefont{E.}}, and
  \bibinfo{author}{\bibfnamefont{P.}~\bibnamefont{Touboul}},
  \bibinfo{year}{2000}, \bibinfo{journal}{Review of Scientific Instruments}
  \textbf{\bibinfo{volume}{71}}(\bibinfo{number}{1}), \bibinfo{pages}{310},
  \urlprefix\url{http://link.aip.org/link/?RSI/71/310/1}.

\bibitem[{\citenamefont{{Willke}} \emph{et~al.}(2008)\citenamefont{{Willke},
  {Danzmann}, {Frede}, {King}, {Kracht}, {Kwee}, {Puncken}, {Savage}, {Schulz},
  {Seifert}, {Veltkamp}, {Wagner}} \emph{et~al.}}]{Benno:2008}
\bibinfo{author}{\bibnamefont{{Willke}}, \bibfnamefont{B.}},
  \bibinfo{author}{\bibfnamefont{K.}~\bibnamefont{{Danzmann}}},
  \bibinfo{author}{\bibfnamefont{M.}~\bibnamefont{{Frede}}},
  \bibinfo{author}{\bibfnamefont{P.}~\bibnamefont{{King}}},
  \bibinfo{author}{\bibfnamefont{D.}~\bibnamefont{{Kracht}}},
  \bibinfo{author}{\bibfnamefont{P.}~\bibnamefont{{Kwee}}},
  \bibinfo{author}{\bibfnamefont{O.}~\bibnamefont{{Puncken}}},
  \bibinfo{author}{\bibfnamefont{R.~L.} \bibnamefont{{Savage}},
  \bibfnamefont{Jr.}},
  \bibinfo{author}{\bibfnamefont{B.}~\bibnamefont{{Schulz}}},
  \bibinfo{author}{\bibfnamefont{F.}~\bibnamefont{{Seifert}}},
  \bibinfo{author}{\bibfnamefont{C.}~\bibnamefont{{Veltkamp}}},
  \bibinfo{author}{\bibfnamefont{S.}~\bibnamefont{{Wagner}}}, \emph{et~al.},
  \bibinfo{year}{2008}, \bibinfo{journal}{Classical and Quantum Gravity}
  \textbf{\bibinfo{volume}{25}}(\bibinfo{number}{11}), \bibinfo{pages}{114040}.

\bibitem[{\citenamefont{{Winkler}} \emph{et~al.}(1991)\citenamefont{{Winkler},
  {Danzmann}, {R{\"u}diger}, and {Schilling}}}]{Winkler:TCS}
\bibinfo{author}{\bibnamefont{{Winkler}}, \bibfnamefont{W.}},
  \bibinfo{author}{\bibfnamefont{K.}~\bibnamefont{{Danzmann}}},
  \bibinfo{author}{\bibfnamefont{A.}~\bibnamefont{{R{\"u}diger}}}, and
  \bibinfo{author}{\bibfnamefont{R.}~\bibnamefont{{Schilling}}},
  \bibinfo{year}{1991}, \bibinfo{journal}{\pra} \textbf{\bibinfo{volume}{44}},
  \bibinfo{pages}{7022}.

\bibitem[{\citenamefont{Winkler} \emph{et~al.}(1994)\citenamefont{Winkler,
  Schilling, Danzmann, Mizuno, R\"{u}diger, and Strain}}]{Winkler:94}
\bibinfo{author}{\bibnamefont{Winkler}, \bibfnamefont{W.}},
  \bibinfo{author}{\bibfnamefont{R.}~\bibnamefont{Schilling}},
  \bibinfo{author}{\bibfnamefont{K.}~\bibnamefont{Danzmann}},
  \bibinfo{author}{\bibfnamefont{J.}~\bibnamefont{Mizuno}},
  \bibinfo{author}{\bibfnamefont{A.}~\bibnamefont{R\"{u}diger}}, and
  \bibinfo{author}{\bibfnamefont{K.~A.} \bibnamefont{Strain}},
  \bibinfo{year}{1994}, \bibinfo{journal}{Appl. Opt.}
  \textbf{\bibinfo{volume}{33}}(\bibinfo{number}{31}), \bibinfo{pages}{7547},
  \urlprefix\url{http://ao.osa.org/abstract.cfm?URI=ao-33-31-7547}.

\bibitem[{\citenamefont{Wise} \emph{et~al.}(2005)\citenamefont{Wise, Quetschke,
  Deshpande, Mueller, Reitze, Tanner, Whiting, Chen, T\"unnermann, Kley, and
  Clausnitzer}}]{Wise:2005}
\bibinfo{author}{\bibnamefont{Wise}, \bibfnamefont{S.}},
  \bibinfo{author}{\bibfnamefont{V.}~\bibnamefont{Quetschke}},
  \bibinfo{author}{\bibfnamefont{A.~J.} \bibnamefont{Deshpande}},
  \bibinfo{author}{\bibfnamefont{G.}~\bibnamefont{Mueller}},
  \bibinfo{author}{\bibfnamefont{D.~H.} \bibnamefont{Reitze}},
  \bibinfo{author}{\bibfnamefont{D.~B.} \bibnamefont{Tanner}},
  \bibinfo{author}{\bibfnamefont{B.~F.} \bibnamefont{Whiting}},
  \bibinfo{author}{\bibfnamefont{Y.}~\bibnamefont{Chen}},
  \bibinfo{author}{\bibfnamefont{A.}~\bibnamefont{T\"unnermann}},
  \bibinfo{author}{\bibfnamefont{E.}~\bibnamefont{Kley}}, and
  \bibinfo{author}{\bibfnamefont{T.}~\bibnamefont{Clausnitzer}},
  \bibinfo{year}{2005}, \bibinfo{journal}{Phys. Rev. Lett.}
  \textbf{\bibinfo{volume}{95}}, \bibinfo{pages}{013901},
  \urlprefix\url{http://link.aps.org/doi/10.1103/PhysRevLett.95.013901}.

\bibitem[{\citenamefont{{Woodard}} \emph{et~al.}(2011)\citenamefont{{Woodard},
  {Romania}, and {Tsamis}}}]{Woodard:HFGW}
\bibinfo{author}{\bibnamefont{{Woodard}}, \bibfnamefont{R.~P.}},
  \bibinfo{author}{\bibfnamefont{M.~G.} \bibnamefont{{Romania}}}, and
  \bibinfo{author}{\bibfnamefont{N.~C.} \bibnamefont{{Tsamis}}},
  \bibinfo{year}{2011}, \bibinfo{journal}{Classical and Quantum Gravity}
  \textbf{\bibinfo{volume}{28}}(\bibinfo{number}{7}), \bibinfo{pages}{075013}.

\bibitem[{\citenamefont{{Woods}} \emph{et~al.}(1994)\citenamefont{{Woods},
  {Wrigley}, {Rottman}, and {Haring}}}]{Grating:Scatter94}
\bibinfo{author}{\bibnamefont{{Woods}}, \bibfnamefont{T.~N.}},
  \bibinfo{author}{\bibfnamefont{R.~T.} \bibnamefont{{Wrigley}},
  \bibfnamefont{III}}, \bibinfo{author}{\bibfnamefont{G.~J.}
  \bibnamefont{{Rottman}}}, and \bibinfo{author}{\bibfnamefont{R.~E.}
  \bibnamefont{{Haring}}}, \bibinfo{year}{1994}, \bibinfo{journal}{\ao}
  \textbf{\bibinfo{volume}{33}}, \bibinfo{pages}{4273}.

\bibitem[{\citenamefont{Wu} \emph{et~al.}(1986)\citenamefont{Wu, Kimble, Hall,
  and Wu}}]{Kimble:1986}
\bibinfo{author}{\bibnamefont{Wu}, \bibfnamefont{L.-A.}},
  \bibinfo{author}{\bibfnamefont{H.~J.} \bibnamefont{Kimble}},
  \bibinfo{author}{\bibfnamefont{J.~L.} \bibnamefont{Hall}}, and
  \bibinfo{author}{\bibfnamefont{H.}~\bibnamefont{Wu}}, \bibinfo{year}{1986},
  \bibinfo{journal}{Phys. Rev. Lett.} \textbf{\bibinfo{volume}{57}},
  \bibinfo{pages}{2520},
  \urlprefix\url{http://link.aps.org/doi/10.1103/PhysRevLett.57.2520}.

\bibitem[{\citenamefont{{Yagi} and {Tanaka}}(2010)}]{Yagi:DECIGOalt}
\bibinfo{author}{\bibnamefont{{Yagi}}, \bibfnamefont{K.}}, and
  \bibinfo{author}{\bibfnamefont{T.}~\bibnamefont{{Tanaka}}},
  \bibinfo{year}{2010}, \bibinfo{journal}{Progress of Theoretical Physics}
  \textbf{\bibinfo{volume}{123}}, \bibinfo{pages}{1069}.

\bibitem[{\citenamefont{Yamamoto}(2007)}]{Hiro}
\bibinfo{author}{\bibnamefont{Yamamoto}, \bibfnamefont{H.}},
  \bibinfo{year}{2007}, \bibinfo{howpublished}{personal communication}.

\bibitem[{\citenamefont{Yamamoto} \emph{et~al.}(2008)\citenamefont{Yamamoto,
  Hayakawa, Okada, Uchiyama, Miyoki, Ohashi, Kuroda, Kanda, Tatsumi, and
  Tsunesada}}]{Kazuhiro:Cosmic}
\bibinfo{author}{\bibnamefont{Yamamoto}, \bibfnamefont{K.}},
  \bibinfo{author}{\bibfnamefont{H.}~\bibnamefont{Hayakawa}},
  \bibinfo{author}{\bibfnamefont{A.}~\bibnamefont{Okada}},
  \bibinfo{author}{\bibfnamefont{T.}~\bibnamefont{Uchiyama}},
  \bibinfo{author}{\bibfnamefont{S.}~\bibnamefont{Miyoki}},
  \bibinfo{author}{\bibfnamefont{M.}~\bibnamefont{Ohashi}},
  \bibinfo{author}{\bibfnamefont{K.}~\bibnamefont{Kuroda}},
  \bibinfo{author}{\bibfnamefont{N.}~\bibnamefont{Kanda}},
  \bibinfo{author}{\bibfnamefont{D.}~\bibnamefont{Tatsumi}}, and
  \bibinfo{author}{\bibfnamefont{Y.}~\bibnamefont{Tsunesada}},
  \bibinfo{year}{2008}, \bibinfo{journal}{Phys. Rev. D}
  \textbf{\bibinfo{volume}{78}}, \bibinfo{pages}{022004}.

\bibitem[{\citenamefont{Zucker and Whitcomb}(1996)}]{Zucker:Gas}
\bibinfo{author}{\bibnamefont{Zucker}, \bibfnamefont{M.}}, and
  \bibinfo{author}{\bibfnamefont{S.~E.} \bibnamefont{Whitcomb}},
  \bibinfo{year}{1996}, in \emph{\bibinfo{booktitle}{Proceedings of the Seventh
  Marcel Grossman Meeting on recent developments in theoretical and
  experimental general relativity, gravitation, and relativistic field
  theories}}, volume~\bibinfo{volume}{1}, p. \bibinfo{pages}{1434}.

\bibitem[{\citenamefont{ZygoEPO}(2011)}]{Gari:Zygo}
\bibinfo{author}{\bibnamefont{ZygoEPO}}, \bibinfo{year}{2011},
  \bibinfo{title}{{ITM04} phasemap}, \bibinfo{howpublished}{Metrology Report}.

\end{thebibliography}

\end{document}